%% file: PopovPhD.tex
\documentclass[a4paper,12pt]{book}
\input{preambula}
\begin{document}

\bibliographystyle{iopart-custom}
%\bibliographystyle{iopart-num}
%\bibliographystyle{revtex}
%\bibliographystyle{rpm}
%\bibliographystyle{prsty}
%\bibliographystyle{unsrt}
%\bibliographystyle{abbrv}
%bibliographystyle{ieeetr}

%\linenumbers 

\pagenumbering{roman}

\onehalfspacing

% \input{naslovnica_hr}
% \newpage

\input{naslovnica_eng}
\newpage

\input{izjava}
\newpage

\input{amdg}
\newpage
 
\input{zahvala}
\newpage

% \input{dok_kartica_hr}
% \newpage

\input{dok_kartica_eng}
\newpage

%\singlespacing

\begin{spacing}{0.8}
\tableofcontents
\end{spacing}

\input{introduction}

\input{exp_survey}

\input{theory}

\input{CLFV}

\input{moments}

\input{conclusions}

\input{appendices}

%\onehalfspacing

\input{sazetak}

%\singlespacing

\bibliography{HEP_bibtex}

\input{cv}

\end{document}

%% file: preambula.tex
\pdfoutput=1
\usepackage[font=small,labelfont=bf]{caption}
\usepackage[cp1250]{inputenc}
\usepackage[dvips]{graphicx}
\usepackage[croatian,english]{babel}
\usepackage[]{amssymb}
\usepackage{amsmath}
\usepackage{cancel}
\usepackage[]{mathrsfs}
\usepackage[]{tensor}
\usepackage[]{textcomp}
\usepackage[]{accents}
\usepackage{tocbibind}
\usepackage[symbol]{footmisc}
\usepackage[sf,bf,center]{titlesec}
\titleformat{\section}[hang]{\Large\bfseries\filcenter}{\S\,\oldstylenums{\thesection}}{10pt}{}[]
\titleformat{\subsection}[hang]{\large\bfseries\filcenter}{\S\,\oldstylenums{\thesubsection}}{10pt}{}[]
\usepackage[fit]{truncate}
\usepackage{fancyhdr}
\usepackage[toc,page]{appendix}

\DeclareMathAlphabet{\mathpzc}{OT1}{pzc}{m}{it}

\newcommand{\be}{\begin{equation}}
\newcommand{\ee}{\end{equation}}
\newcommand{\bc}{\begin{center}}
\newcommand{\ec}{\end{center}}

\def\Slash#1{\setbox0=\hbox{$#1$}  % set a box for #1
   \dimen0=\wd0     % and get its size
   \setbox1=\hbox{/} \dimen1=\wd1  % get size of /
   \ifdim\dimen0>\dimen1   % #1 is bigger
      \rlap{\hbox to \dimen0{\hfil/\hfil}} % so center / in box
      #1     % and print #1
   \else     % / is bigger
      \rlap{\hbox to \dimen1{\hfil$#1$\hfil}} % so center #1
      /      % and print /
   \fi}

\renewcommand {\c} {\'{c}}

\newcommand {\cc} {\v{c}}

\newcommand {\s} {\v{s}}

\newcommand {\z} {\v{z}}

\renewcommand {\dj} {\hbox{d\hskip-1.1ex{\raise0.640ex\hbox{--}}\skip 0.70ex}}
\newcommand {\Dj} {\hbox{D\hskip-1.9ex{\raise0.175ex\hbox{--}}\hskip 0.85ex}}

\makeatletter
\def\overbracket#1{\mathop{\vbox{\ialign{##\crcr\noalign{\kern3\p@}
\downbracketfill\crcr\noalign{\kern3\p@\nointerlineskip}
$\hfil\displaystyle{#1}\hfil$\crcr}}}\limits}
\def\underbracket#1{\mathop{\vtop{\ialign{##\crcr
$\hfil\displaystyle{#1}\hfil$\crcr\noalign{\kern3\p@\nointerlineskip}
\upbracketfill\crcr\noalign{\kern3\p@}}}}\limits}
\def\upbracketfill{$\m@th\makesm@sh{\llap{\vrule\@height3\p@\@width.7\p@}}%
\leaders\vrule\@height.7\p@\hfill
\makesm@sh{\rlap{\vrule\@height3\p@\@width.7\p@}}$}
\def\downbracketfill{$\m@th
\makesm@sh{\llap{\vrule\@height.7\p@\@depth2.3\p@\@width.7\p@}}%
\leaders\vrule\@height.7\p@\hfill
\makesm@sh{\rlap{\vrule\@height.7\p@\@depth2.3\p@\@width.7\p@}}$}
\makeatother

\usepackage{lipsum}

\usepackage{cite}

\usepackage{url}

\usepackage{caption}

\usepackage{color}

\makeatletter
\renewcommand*{\cleardoublepage}{\clearpage\if@twoside \ifodd\c@page\else
\hbox{}%
\thispagestyle{empty}%
\newpage%
\if@twocolumn\hbox{}\newpage\fi\fi\fi}
\makeatother

\usepackage{appendix}

\usepackage{epstopdf}

\newcommand{\Tr}{\mathop{\mathrm{Tr}}} % trag, trace - Tr( )

\RequirePackage{lineno}

\usepackage{setspace}

%% file: naslovnica_eng.tex
\thispagestyle{empty}

\bc

{\large \textsc{University of Zagreb}}\\
\vspace{0.2cm}
{\large \textsc{Faculty of Science}}\\
\vspace{0.2cm}
{\large \textsc{Physics Department}}\\
\vspace{2.5cm}
{\Large Luka Popov}\\

\vspace{2.8cm}

%%%%%%%%%%%%%%%%%%%%%%%%%%%%%%%%%%%%%%%%%%%%%%%%%%
{\huge \textbf{Lepton flavor violation}}\\
\vspace{0.3cm}
{\huge \textbf{in supersymmetric}}\\
\vspace{0.3cm}
{\huge \textbf{low-scale seesaw models}}\\
%%%%%%%%%%%%%%%%%%%%%%%%%%%%%%%%%%%%%%%%%%%%%%%%%%

\vspace{2.7cm}

{\large
Doctoral Thesis submitted to the Physics Department\\
Faculty of Science, University of Zagreb\\
for the academic degree of\\
Doctor of Natural Sciences (Physics)\\

\vspace{2.5cm}

Zagreb, 2013.}

\vspace{2.0cm}

% \scriptsize{revised version \ $5 \ | \ 1 \ | \ 2013$}\\

\ec

\newpage
\thispagestyle{empty}
\mbox{}

%% file: izjava.tex
\thispagestyle{empty}

\hspace{1.0cm}\\

\vspace{1.0cm}

\begin{minipage}[t]{11.0cm}
\noindent
This thesis was made under the mentorship of Prof Amon Ilakovac, 
within University doctoral study at Physics Department of 
Faculty of Science of University of Zagreb.\\

\noindent
Ova disertacija izra\dj ena je pod vodstvom prof.~dr.~sc.~Amona Ilakovca, 
u sklopu Sveu\v cili\v snog doktrorskog studija pri Fizi\v ckom 
odsjeku Prirodoslovno-matemati\v ckoga fakulteta Sveu\v cili\v sta u Zagrebu.\\
\end{minipage}

\newpage
\thispagestyle{empty}
\mbox{}

%% file: amdg.tex
\thispagestyle{empty}

\hspace{1.0cm}\\

\vspace{3.0cm}

%\begin{minipage}[t]{11.0cm}
\noindent
\bc
{\large A. M. D. G.}
\ec
%\end{minipage}

\newpage
\thispagestyle{empty}
\mbox{}

%% file: zahvala.tex
\thispagestyle{empty}

{\huge \textbf{Acknowledgments}}\\

\vspace{1.0cm}

First and above all, I wish to thank my beloved wife Mirela,
for illuminating our home with the warmth of her heart.

I would also like to thank my parents Danka and Bojan who never ceased 
to give me moral support throughout my work and studies.

My sincere gratitude goes to my colleagues Branimir, Mirko, Ivica and
Sanjin for all the discussions regarding physics and beyond.
I was honoured with your friendship.

I use this opportunity to say thanks to all my professors for
generously sharing their knowledge with us.

Special gratitude goes to Mrs Marina Kavur from students' office,
for always being on our side during the never-ending bureaucratic 
battles.

\hyphenation{Pilaftsis}
I also which to express my gratitude to our collaborator 
Prof Apostolos Pilaftsis from University of Manchester for his
participation on this project.

Last but not least, I thank my mentor Prof Amon Ilakovac, without
whose patient guidance and support this thesis would never see 
the light of day.

\newpage
\thispagestyle{empty}
\mbox{}

\newpage 

%% file: dok_kartica_eng.tex
\thispagestyle{empty}

\bc
{\Large \textsc{Basic documentation card}}\\
\ec

\vspace{1.0cm}

\noindent
{\large University of Zagreb \hspace{4.0cm} Doctoral Thesis\\
Faculty of Science\\
Physics Department}\\

\vspace{1.0cm}

\bc
{\Large \textbf{Lepton flavor violation in supersymmetric low-scale seesaw models}}\\

\vspace{1.0cm}

{\large \textsc{Luka Popov}\\

\vspace{0.2cm}

Faculty of Science, Zagreb}\\
\ec

\vspace{0.5cm}

%\lipsum[1-2]

The minimal supersymmetric standard model with a low scale see-saw mechanism
is presented. Within this framework, the lepton flavour violation in the charged
lepton sector is thoroughly studied. Special attention is paid to the 
individual loop contributions due to the heavy neutrinos $N_{1,2,3}$,
sneutrinos $\widetilde{N}_{1,2,3}$ and soft SUSY-breaking terms. For the first time,
the complete set of box diagrams is included, in addition to the photon and
$Z$-boson mediated interactions. The complete set of chiral amplitudes 
and their associate form-factors related to the neutrinoless 
three-body charged lepton flavor violating
decays of the muon and tau, such as $\mu \to eee$,
$\tau \to \mu\mu\mu$, $\tau \to e\mu\mu$ and $\tau \to ee\mu$, as well as
the coherent $\mu \to e$ conversion in nuclei, were derived. The obtained
analytical results are general and can be applied to most of the New Physics
models with charged lepton flavor violation. This systematic analysis 
has revealed the existence of two new box form factors, which have not 
been considered before in the existing literature in this area of physics.

In the same model, the systematic study of one-loop contributions to the 
muon anomalous magnetic dipole moment $a_\mu$ and the electron electric
dipole moment $d_e$ is performed. Special attention is paid to the 
effect of the sneutrino soft SUSY-breaking parameters, $B_\nu$ and $A_\nu$,
and their universal CP phases ($\theta$ and $\phi$) on $a_\mu$ and $d_e$.

\vspace{0.5cm}

\bc
(135 pages, 169 references, original in English)\\
\ec

\vspace{0.5cm}


\begin{tabular}{ll}
 & \\
Supervisor: &  \hskip-4ex Prof~Amon Ilakovac, University of Zagreb\\
 & \\
Committee:
	&  \hskip-4ex 1. Prof Kre\s imir Kumeri\cc ki, University of Zagreb \\
	&  \hskip-4ex 2. Dr Vuko Brigljevi\'{c}, Senior scientist, IRB Zagreb \\
	&  \hskip-4ex 3. Prof Amon Ilakovac, University of Zagreb \\
	&  \hskip-4ex 4. Prof Apostolos Pilaftsis, University of Machester \\
	&  \hskip-4ex 5. Prof Mirko Planini\'{c}, University of Zagreb \\
 & \\
Replacements:
	&  \hskip-3ex 1. Prof Dubravko Klabu\cc ar, University of Zagreb \\
	&  \hskip-3ex 2. Dr Kre\s o Kadija, Senior scientist, IRB Zagreb \\
& \\
Thesis accepted: & 2013
\end{tabular}

\newpage

%% file: introduction.tex
\chapter*{Introduction}
\addcontentsline{toc}{chapter}
{\protect\numberline{}Introduction}
\markboth{Introduction}{}

\pagenumbering{arabic}

\vspace{0.5cm}

In the first part of this thesis the study of charged lepton flavor violation
(CLFV) is performed in low-scale seesaw model of minimal supersymmetric standard
model ($\nu_R$MSSM) within the framework of minimal supergravity (mSUGRA).
There are two dominant sources of CLFV: one
originating from the usual soft supersymmetry-breaking sector, and other 
entirely supersymmetric coming from the supersymmetric neutrino Yukawa
sector. Both sources are taken into account within this framework, and
number of possible lepton-flavor-violating transitions are calculated. 
Supersymmetric low-scale seesaw models offer distinct correlated predictions 
for lepton flavor violating signatures, which might be discovered in current
and projected experiments.

In the second part, the same model is used to study the anomalous magnetic
and electric dipole moments of charged leptons. The numerical estimates of the
muon anomalous magnetic moment and the electron electric dipole moment will 
be given as a function of key parameters. The electron electric dipole moment
is found to be naturally small in this model, and can be probed in the present
and future experiments.

The thesis is organized as follows:
\begin{itemize}

\item The first chapter gives a brief experimental survey, the current
  and projected experiments regarding the detection of charged lepton
  flavor violation and anomalous dipole moments of charged leptons. 

\item The second chapter presents the theoretical framework
  which underlines the study of lepton flavor violation and anomalous dipole
  moments given in the thesis.

\item The third chapter exposes the analytic and numerical results for
  various lepton flavor violating transitions, as well as some important
  physical implications which follow.

\item The fourth chapter gives the analysis of the muon anomalous 
  magnetic moment and the electron electric dipole moment in supersymmetric
  low-scale seesaw models with right-handed neutrino superfields.

\item Concluding remarks are given in the fifth chapter.
%In the final, fifth chapter the concluding remarks are given.

\item The appendices contain technical details regarding the relevant
  interaction vertices, loop functions and formfactors.

\end{itemize}

The main results of the thesis are the following:
\begin{itemize}
\item The soft SUSY-breaking effects in the $Z$-boson-mediated graphs dominate the
  CLFV observables for appreciable regions of the $\nu_R$MSSM parameter space in mSUGRA.
  But for $m_N \lesssim 1 \textrm{ TeV}$ the box diagrams involving heavy neutrinos in
  the loop can be comparable to, or even greater than the corresponding $Z$-boson-mediated
  diagrams in $\mu\to eee$ and $\mu\to e$ conversion in nuclei. Therefore, the usual
  paradigm with the photon dipole-moment operators dominating the CLFV observables in
  high-scale seesaw models have to be radically modified.

\item Heavy singlet neutrino and sneutrino contributions to anomalous magnetic dipole
  moment of the muon are small, typically one to two orders of magnitude below the 
  muon anomaly $\Delta a_\mu$. The largest effect on $\Delta a_\mu$ instead comes from
  left-handed sneutrinos and sleptons, exactly as is the case in the MSSM without 
  right-handed neutrinos. Heavy singlet neutrinos do not contribute to the electric
  dipole moment (EDM) of the electron either. The main contribution to EDM comes from
  SUSY-breaking terms, but only if one of the CP phases ($\theta$ and/or $\phi$)
  introduced to SUSY-breaking sector is nonvanishing.

\end{itemize}

%% file: exp_survey.tex
\chapter{Experimental survey}
\label{expsurvey}

Neutrino oscillation experiments have provided undisputed evidence of
lepton flavor violation (LFV) in the neutrino sector, pointing towards
physics beyond the Standard Model (SM). Nevertheless, no evidence of
LFV has been found in the charged lepton sector of SM, implying conservation
of the individual lepton number associated with the electron $e$, the muon
$\mu$ and the tau lepton $\tau$. All past and current experiments were only
able to report upper limits on observables of charged lepton flavor
violation (CLFV). The experimental detection of CLFV would certenaly pave
the way to the New Physics.

Measurements of the anomalous magnetic dipole moment of the muon
(i.e.\ its deviation form the SM prediction, $\Delta a_\mu$) can give an
important constraint on model-building, since any New Physics contribution
must remain within $\Delta a_\mu$ limit. Study of the electric dipole
moment of the electron $d_e$ is even more compelling, since the observation
of non-zero (i.e.\ $\gtrsim 10^{-33}\; e\,\textrm{cm}$) value for $d_e$ 
would signify the existence of CP-violating physics beyond the Standard Model.
%\\

\newpage

\section{Neutrino oscillations}

When a neutrino is produced in some weak interaction process, and it
propagates through some finite distance, there is a non-zero probability
that it will change its flavor. This well established and observed
fact is known as \emph{neutrino oscillation}
\cite{Pontecorvo1957,Pontecorvo1958,Maki1962}, due to the oscillatory
dependence of the flavor change probability with respect to the neutrino
energy and the distance of the propagation.

There are numerous neutrino experiments which report the lepton flavor
violation in the neutrino sector, by observing the disappearances or the
appearances of a particular neutrino flavor.

In solar neutrino experiments, first by Homestake \cite{Cleveland1998} and
later confirmed by others \cite{Fukuda1996,Abdurashitov2009,
Anselmann1992,Hampel1999,Altmann2005,Fukuda2002,Ahmad2001,Ahmad2002},
the disappearance of the solar electron
neutrino $\nu_e$ is observed. Atmospheric muon neutrinos $\nu_\mu$
and antineutrinos $\bar{\nu}_\mu$ disappeared in Super-Kamiokande experiment
\cite{Fukuda1998,Ashie2004}.
The disappearance of reactor electron antineutrinos $\bar{\nu}_e$ is
observed in Kam-LAND reactor \cite{Eguchi2003,Araki2005} and in
DOUBLE-CHOOZ experiment \cite{Abe2012}.
Muon neutrinos $\nu_\mu$ disappeared in
the long-baseline accelerator neutrino experiments MINOS
\cite{Michael2006,Adamson2008}
and K2K \cite{Ahn2006}. Short-baseline reactor experiments Daya Bay
\cite{An2012,An2013}
and RENO \cite{Ahn2012} report the disappearance of the reactor electron antineutrinos
$\bar{\nu}_e$.

The appearance of electron neutrino $\nu_e$ in a beam of muon neutrinos $\nu_\mu$
in long-baseline accelerator is reported by T2K \cite{Abe2011}
and MINOS \cite{Adamson2011} experiments.

All these experiments have provided undisputed evidence for neutrino oscillations
caused by finite (non-zero) neutrino masses and, consequently, neutrino mixing
parameters. Since neutrinos are massive, the transition from the neutrino
flavor eigenstate fields ($\nu_e$, $\nu_\mu$, $\nu_\tau$) which makes
the lepton charged current in weak interactions to the neutrino mass
eigenstate fields ($\nu_1$, $\nu_2$, $\nu_3$) is non-trivial:
\begin{equation}
\nu_l (x) = \sum_{i=1}^3 U_{li} \nu_i (x) \;, \quad l=e,\mu,\tau \,.
\end{equation}
Unitary matrix $U$ is known as Pontecorvo-Maki-Nakagawa-Sakata matrix
\cite{Pontecorvo1957,Pontecorvo1958,Maki1962} and is usually parametrized as
\begin{equation}\label{pmns}
U_{PMNS} = \begin{pmatrix}
  c_{12} c_{13} & s_{12} c_{13} & s_{13} e^{-i \delta} \\
  -s_{12} c_{23}-c_{12} s_{23} s_{13}e^{i \delta}
    & c_{12} c_{23}-s_{12} s_{23} s_{13} e^{i \delta} & s_{23} c_{13} \\
  s_{12} s_{23}-c_{12} c_{23} s_{13}e^{i \delta}
    & -c_{12} s_{23}-s_{12} c_{23} s_{13} e^{i \delta} & c_{23} c_{13} \\
\end{pmatrix} \cdot P \,,
\end{equation}
where $P=\textrm{diag}(1,e^{i\alpha},e^{i\beta})$, $c_{ij}\equiv\cos\theta_{ij}$ and
$s_{ij}\equiv\sin\theta_{ij}$. $\theta_{12}$ denotes solar mixing angle, $\theta_{23}$
atmospheric mixing angle and $\theta_{13}$ reactor mixing angle. Phases $\delta$,
$\alpha$ and $\beta$ stand for Dirac CP violating phase and two Majorana CP
violating phases, respectively.

Nonzero values of $\theta_{13}$ reported in recent reactor neutrino oscillation
experiments \cite{Abe2012,An2012,Ahn2012} strongly indicate a nontrivial
neutrino-flavor structure and possibly CP violation.

\section{Searching for CLFV}

The existence of lepton flavor violation (LFV) in the neutrino sector implies
the possibility of LFV in the charged sector as well. However, in spite of
intense experimental searches
\cite{Adam2011,Bellgardt1988,Dohmen1993,Bertl2006, Miyazaki2011,Miyazaki2013,Aubert2009,
Hayasaka2010,Lees2010,Miyazaki2010,Aubert2010,Beringer2012}
no evidence of LFV in the charged lepton sector of the Standard Model (SM) has
yet been found.

All past and current experiments searching for the charged lepton flavor violation
(CLFV) were only able to report upper limits on the observables associated with
CLFV. Recently, the MEG collaboration \cite{Adam2011} has announced an improved
upper limit on the branching ratio of the CLFV decay $\mu\to e\gamma$, with
$B(\mu\to  e\gamma) < 2.4\times 10^{-12}$ at the 90\% confidence
level (CL). As  also shown in  Table \ref{TableI}, future experiments
searching for the CLFV processes, $\mu\to e\gamma$, $\mu\to eee$, coherent
$\mu \to e$ conversion in nuclei, $\tau\to e\gamma/\mu\gamma$,
$\tau\to 3\ \mbox{leptons}$ and $\tau\to \mbox{lepton $+$ light  meson}$,
are  expected to  reach branching-ratio sensitivities to the level of
$10^{-13}$ \cite{Golden2012,Adam2012} ($10^{-14}$ \cite{Hewett2012}),
$10^{-16}$ \cite{Berger2013} ($10^{-17}$ \cite{Hewett2012}),
$10^{-17}$~\cite{Kurup2011,Abrams2012,Kutschke2011,Dukes2011}
($10^{-18}$~\cite{Hewett2012,Kuno2005,Barlow2011}),
$10^{-9}$~\cite{Hayasaka2009,Bona2007}, $10^{-10}$~\cite{Hayasaka2009}
and $10^{-10}$~\cite{Hayasaka2009},  respectively. The values in parentheses
indicate the sensitivities that are expected to  be achieved by  the new
generation CLFV  experiments in  the next decade. Most interestingly,  the
projected  sensitivity  for $\mu\to eee$ and $\mu  \to e$ conversion in nuclei
is  expected to increase by five and  six orders of  magnitude, respectively.
The history and current status of the experimental search for CLFV is
very nicely exposed in Ref~\cite{Bernstein2013}, which is highly 
recommended for further reading.

\renewcommand{\arraystretch}{1.2}

\begin{table}[ht]
\captionsetup{font=footnotesize}
\centering
\footnotesize
 \begin{tabular}{rlll}
\hline\hline
 No.~ & Observable & Upper Limit & Future Sensitivity \\
\hline
 1.~  & $B (\mu \to e \gamma)$ & $2.4 \times 10^{-12}$ \cite{Adam2011}
      &$1$--$2\times 10^{-13}$ \cite{Golden2012,Adam2012}, $10^{-14}$ \cite{Hewett2012}\\
 2.~  & $B (\mu \to eee)$ & $10^{-12}$ \cite{Bellgardt1988}
      &$10^{-16}$ \cite{Berger2013}, $10^{-17}$  \cite{Hewett2012} \\
 3.~  & $R_{\mu e}^{\rm Ti}$ & $4.3\times 10^{-12}$ \cite{Dohmen1993}
      &$3$--$7\times 10^{-17}$ \cite{Kurup2011,Abrams2012,Kutschke2011,Dukes2011},
         $10^{-18}$ \cite{Hewett2012,Kuno2005,Barlow2011}\\
 4.~  & $R_{\mu e}^{\rm Au}$ & $7 \times 10^{-13}$ \cite{Bertl2006}
      &$3$--$7\times 10^{-17}$ \cite{Kurup2011,Abrams2012,Kutschke2011,Dukes2011},
         $10^{-18}$ \cite{Hewett2012,Kuno2005,Barlow2011}\\
 5.~  & $B(\tau \to e \gamma)$ & $3.3 \times 10^{-8}$
    \cite{Miyazaki2011,Miyazaki2013,Aubert2009,
    Hayasaka2010,Lees2010,Miyazaki2010,Aubert2010,Beringer2012}
      & $1$--$2\times 10^{-9}$ \cite{Bona2007,Hayasaka2009}\\
 6.~  & $B(\tau \to \mu \gamma)$ & $4.4 \times 10^{-8}$
    \cite{Miyazaki2011,Miyazaki2013,Aubert2009,
    Hayasaka2010,Lees2010,Miyazaki2010,Aubert2010,Beringer2012}
      & $2\times 10^{-9}$ \cite{Bona2007,Hayasaka2009} \\
 7.~  & $B(\tau \to eee)$ & $2.7 \times 10^{-8}$
    \cite{Miyazaki2011,Miyazaki2013,Aubert2009,
    Hayasaka2010,Lees2010,Miyazaki2010,Aubert2010,Beringer2012}
      & $2 \times 10^{-10}$ \cite{Bona2007,Hayasaka2009} \\
 8.~  & $B(\tau \to e\mu\mu)$ & $2.7 \times 10^{-8}$
    \cite{Miyazaki2011,Miyazaki2013,Aubert2009,
    Hayasaka2010,Lees2010,Miyazaki2010,Aubert2010,Beringer2012}
      & $10^{-10}$ \cite{Hayasaka2009} \\
 9.~  & $B(\tau \to \mu\mu\mu)$ & $2.1 \times 10^{-8}$
    \cite{Miyazaki2011,Miyazaki2013,Aubert2009,
    Hayasaka2010,Lees2010,Miyazaki2010,Aubert2010,Beringer2012}
      & $2 \times 10^{-10}$ \cite{Bona2007,Hayasaka2009} \\
10.~  & $B(\tau \to \mu ee)$ & $1.8 \times 10^{-8}$
    \cite{Miyazaki2011,Miyazaki2013,Aubert2009,
    Hayasaka2010,Lees2010,Miyazaki2010,Aubert2010,Beringer2012}
      & $10^{-10}$ \cite{Hayasaka2009}\\
\hline\hline
\end{tabular}
\footnotesize \caption{Current upper limits and future sensitivities of CLFV observables
under study.}\label{TableI}
\end{table}

Given that  CLFV is forbidden in the SM, its observation would constitute
a clear signature for New Physics, which makes this field of investigation
ever more exciting.

\section{Measuring lepton dipole moments}

The anomalous magnetic dipole moment (MDM) of the muon, $a_\mu$ is a high
precision observable extremely sensitive to physics beyond the Standard Model.
Its current experimental value, according to PDG \cite{Beringer2012}, is
\begin{equation}
a_\mu^\textrm{exp} = (116592089 \pm 63) \times 10^{-11} \,.
\end{equation}

The Standard Model prediction of this observable reads
\begin{equation}
a_\mu^\textrm{SM} = (116591802 \pm 49) \times 10^{-11} \,.
\end{equation}

The difference between measured and predicted value,
\begin{equation}
\Delta a_\mu \equiv a_\mu^\textrm{exp} - a_\mu^\textrm{SM}
  = (287 \pm 80) \times 10^{-11}
\end{equation}
is at the $3.6 \,\sigma$ confidence level (CL) and has therefore been
called \emph{the muon anomaly}. This value limits the allowed contributions
of New Physics to MDM and consequently can be used as a strong
constraint on model-building, or even eliminate some of the proposed
New Physics models.

Likewise, the electric dipole moment (EDM) of the electron, $d_e$, constitutes
a very sensitive probe for CP violation induced by new CP phases present in the
physics beyond the Standard Model. The present upper limit on $d_e$ is reported
to be \cite{Beringer2012,Hudson2011,Jung2013}
\begin{equation}
\label{deUB}
d_e < 10.5 \times 10^{-28} \;e\,\textrm{cm} \,.
\end{equation}

Future projected experiments utilizing paramagnetic systems, such as Cesium,
Rubidium and Francium, may extend the current sensitivity to the
$10^{-29} - 10^{-31}\; e\,\textrm{cm}$ level
\cite{Jung2013,Amini2008,Kittle2004,Weiss2003,Sakemi2011,Wundt2012,
Kara2012,Raidal2008}. In the Standard Model,
the predictions for $d_e$ range from $10^{-38}\; e\,\textrm{cm}$ to
$10^{-33}\; e\,\textrm{cm}$ depending on whether the Dirac CP phase in
light neutrino mixing is zero or not (for detalis see Ref
\cite{Pospelov2005}). Therefore,
any observation of non-zero value of $d_e$, i.e.\ value larger than
$10^{-33}\; e\,\textrm{cm}$, would signify the existence of CP-violating
physics beyond the Standard Model.

For that reason, these observables are of great interest for the
investigation of possible scenarios for the New Physics. The announced
higher-precision measurement of $a_\mu$ by a factor of 4 in the future
Fermilab experiment E989
\cite{Roberts2011,Roberts2006,Venanzoni2012,Venanzoni2012a,E989} as well
as the expected future sensitivities of the electron EDM down to the
level of $\sim 10^{-31}\; e\,\textrm{cm}$ \cite{Jung2013}, renders the
study of the dipole moments even more actual and interesting.

For further reading, the reader is encouraged to the excellent reviews
provided by Refs~\cite{Fukuyama2012,Jung2013a,Raidal2008}.

%% file: theory.tex
\chapter{Theoretical framework} \label{theory}
\allowdisplaybreaks

In this chapter we will expose some basic features of the theoretical framework
which underlines the study of lepton flavor violation and anomalous dipole
moments given in the thesis.

In the first section, we will give the basic structure of the
Minimal Supersymmetric Standard Model (MSSM), as well as some main
features regarding the Soft Supersymmetry Breaking in the MSSM.
The notation used when discussing the Supersymmetry (SUSY) will correspond to the
one used in Drees et al.\ \cite{Drees2004}, adapted to Petcov et al.
\cite{Petcov2004}. For further reading regarding SUSY in general
and MSSM in particular, the reader is encouraged to consult 
Refs~\cite{Drees2004,Nilles1984,Haber1985,Derendinger,Wess1992}.

Second section is dedicated to the seesaw mechanisms, with the main focus
on the low-scale version of the seesaw mechanism type I.

Finally, the the MSSM extended by low-scale right handed
neutrinos (or $\nu_R$MSSM) is introduced.

\newpage

\section{Basic features of the MSSM}

The basic idea behind all supersymmetric models is that there is a symmetry
(conveniently called \emph{supersymmetry}) which transforms a fermion into
the boson and vice versa. The \emph{Minimal Supersymmetric Standard Model}
supersymmetrizes the SM with minimal extension of the SM particle spectrum:
every SM particle is accompanied by one \emph{superparticle} or
a \emph{superpartner}. The superpartners of matter fermions are
spin zero particles, called \emph{sfermions}. They can be further
classified into the scalar leptons or \emph{sleptons}
and scalar quarks or \emph{squarks}. Matter fermions and their superpartners
are described by \emph{chiral superfields}. The superpartners of
SM gauge bosons are spin one-half particles called \emph{gauginos}. They
can be further classified into the stronlgy interacting \emph{gluinos}
and electroweak \emph{zino} and \emph{winos} (superpartners of $Z$ and
$W$ bosons, respectively). Together with SM gauge bosons, they are described
by \emph{vector superfields}. Superpartners of Higgs bosons are spin
one-half particles called \emph{higgsinos} and, along with the latter,
are described by chiral superfields. The electroweak symmetry breaking
mixes the electroweak gauginos with higgsinos resulting in physical
particles referred to as \emph{charginos} and \emph{neutralinos}. Table
\ref{mssm} displays full filed contents of the MSSM, with the corresponding
quantum numbers.

\begin{table}[h!]
\centering
\tiny
\begin{tabular}{cccccccc}
\hline
\multicolumn{8}{c}{\bf Field contents of the MSSM} \\
\hline
Superfield & \multicolumn{2}{c}{Bosons} & \multicolumn{2}{c}{Fermions} &
	$SU_c(3)$ & $SU_L(2)$ & $U_Y(1)$ \\
\hline\hline
gauge \\
$\mathbf{G^a}$ & gluon & $g^a$ & gluino & $\tilde{g}^a$ & 8 & 0 & 0 \\
$\mathbf{V^k}$ & electroweak & $W^k\,\,(W^\pm,Z)$ & wino, zino & $\tilde{\lambda}^k\,\,
	(\tilde{w}^\pm,\tilde{z})$ & 1 & 3 & 0 \\
$\mathbf{V'}$ & hypercharge & $B\,\,(\gamma)$ & bino & $\tilde{\lambda}_0\,\,
	(\tilde{\gamma})$ & 1 & 1 & 0 \\
\hline
matter \\
$\mathbf{L_i}$ & {} & \multicolumn{1}{l}{
	$\tilde{L}_i = (\tilde{\nu},\tilde{e})_L$} & {} &
	\multicolumn{1}{l}{$L_i = (\nu,e)_L$} & 1 & 2 & -1 \\
$\mathbf{E_i}$ & \raisebox{1.5ex}[0pt]{sleptons} & \multicolumn{1}{l}{
	$\tilde{E}_i = \tilde{e}_R$} & \raisebox{1.5ex}[0pt]{leptons}
	& \multicolumn{1}{l}{$E_i = e_R$} & 1 & 1 & 2 \\
$\mathbf{Q_i}$ & {} & \multicolumn{1}{l}{
	$\tilde{Q}_i = (\tilde{u},\tilde{d})_L$} & {} &
	\multicolumn{1}{l}{$Q_i = (u,d)_L$}  & 3 & 2 & 1/3 \\
$\mathbf{U_i}$ & squarks & \multicolumn{1}{l}{
	$\tilde{U}_i = \tilde{u}_R$}
	& quarks &
	\multicolumn{1}{l}{$U_i = u_R^c$} & $3^*$ & 1 & -4/3 \\
$\mathbf{D_i}$ & {} & \multicolumn{1}{l}{
	$\tilde{D}_i = \tilde{d}_R$} & {} &
	\multicolumn{1}{l}{$D_i = d_R^c$} & $3^*$ & 1 & 2/3 \\
\hline
Higgs \\
$\mathbf{H_1}$ & {} & \multicolumn{1}{l}{$H_1$} & {} & $\tilde{H}_1$ &
	1 & 2 & -1 \\
$\mathbf{H_2}$ & \raisebox{1.5ex}[0pt]{Higgs bosons}
	& \multicolumn{1}{l}{$H_2$} & \raisebox{1.5ex}[0pt]{Higgsinos} &
	$\bar{H}_2$ & 1 & 2 & 1 \\
\hline
\end{tabular}
\captionsetup{font=footnotesize}
\caption{Superfields of the MSSM} \label{mssm}
\end{table}

As can be seen from Table \ref{mssm}, there are two Higgs superfields in
the MSSM. These can be written as
\begin{equation}
 H_1 = \begin{pmatrix} H_1^1 \\ H_1^2 \end{pmatrix} \;, \quad
 H_2 = \begin{pmatrix} H_2^1 \\ H_2^2 \end{pmatrix} \,.
\end{equation}
$H_1$ field is sometimes referred to as the down type Higgs ($Y=-1$),
superfield containing $h_1$ and $\tilde{h}_{1L}$, while $H_2$ is referred
to as the up type Higgs superfield containing $h_2$ and $\tilde{h}_{2L}$.
The component fields denoted by lower case letters are given by
\begin{eqnarray}
h_1 \equiv \begin{pmatrix} h_1^1 \\ h_1^2 \end{pmatrix}
  = \begin{pmatrix} h_1^0 \\ h_1^- \end{pmatrix} & \; ; \quad &
h_2 \equiv \begin{pmatrix} h_2^1 \\ h_2^2 \end{pmatrix}
  = \begin{pmatrix} h_2^+ \\ h_2^0 \end{pmatrix} \,, \\[1.5ex]
\tilde{h}_{1L} \equiv \begin{pmatrix} \tilde{h}_1^1 \\ \tilde{h}_1^2 \end{pmatrix}
  = \begin{pmatrix} \tilde{h}_1^0 \\ \tilde{h}_1^- \end{pmatrix}_L & \; ; \quad &
\tilde{h}_{2L} \equiv \begin{pmatrix} \tilde{h}_2^1 \\ \tilde{h}_2^2 \end{pmatrix}
  = \begin{pmatrix} \tilde{h}_2^+ \\ \tilde{h}_2^0 \end{pmatrix}_L \,.
\end{eqnarray}

After the spontaneous breakdown of electroweak symmetry, the Higgs vacuum expectation
values (VEVs) are given by real, positive quantities $v_1$ and $v_2$,
\begin{equation}
\left< h_1 \right> = \frac{1}{\sqrt{2}}
  \begin{pmatrix} v_1 \\ 0 \end{pmatrix} \,; \quad
\left< h_2 \right> = \frac{1}{\sqrt{2}}
  \begin{pmatrix} 0 \\ v_2 \end{pmatrix} \,,
\end{equation}
which arise from the minimization of the Higgs potential.
The ratio of these values,
\begin{equation} \label{tb}
 \frac{v_2}{v_1} \equiv \tan\beta
\end{equation}
is considered to be a free parameter of the theory, at least regarding
the fermion masses.

Let us proceed to the interaction and mass terms in the Lagrangian density
$\mathcal{L}_{\textrm{MSSM}}$ which partly comes from the exact supersymmetrization
of the SM. Full MSSM Lagrangian can be written as the sum of two parts,
\begin{equation}\label{susy+ssb}
 \mathcal{L}_{\textrm{MSSM}} = \mathcal{L}_{\textrm{SUSY}}
 + \mathcal{L}_{\textrm{SSB}} \,.
\end{equation}

While $\mathcal{L}_{\textrm{SUSY}}$ is fully supersymmetric, the
$\mathcal{L}_{\textrm{SSB}}$ contains terms which explicitly break
the supersymmetry (acronym SSB stands for SuperSymmetry Breakdown).

% soft interaction terms with
% mass dimensions less than four as well as mass terms, which describe
% the heavier masses of sparticles as different from those of their
% particle partners. These terms arise from supersymmetry breaking terms.

Let's first take a look to the contents of $\mathcal{L}_{\textrm{SUSY}}$.
The supersymmetric part of the MSSM Lagrangian can be further decomposed as
\begin{equation}
\mathcal{L}_{\textrm{SUSY}} = \mathcal{L}_{g} +
  \mathcal{L}_{M} + \mathcal{L}_{H} \,,
\end{equation}
where $\mathcal{L}_{g}$, $\mathcal{L}_{M}$ and $\mathcal{L}_{H}$ are pure
gauge, matter and Higgs-Yukawa parts, respectively. Detailed expressions
for these terms can be found in the literature \cite[pp 171-172]{Drees2004}.
The part which is most interesting for the purposes of this thesis is
the \emph{superpotential}, which constitutes important part of
$\mathcal{L}_{H}$, and reads
% ----- NOTACIJA !!!
\begin{equation}\label{WMSSM}
\mathcal{W}_{\textrm{MSSM}} = \mu \, H_1 \cdot H_2 
  + \bar{E}_i \, {\mathbf h}_{ij}^e\, H_1 \cdot L_j 
  + \bar{D}_i \, {\mathbf h}_{ij}^d\, H_1 \cdot Q_j  
  + \bar{U}_i \, {\mathbf h}_{ij}^u\, H_2 \cdot Q_j \,,
\end{equation}
where ${\mathbf h}$ matrices are given by
\begin{eqnarray}
{\mathbf h}_{ij}^{e\,\dagger} & = & \frac{g_2}{\sqrt{2} M_W \cos\beta}
  \left( \mathbf{m}_e \right)_{ij} \,,\\
{\mathbf h}_{ij}^{d\,\dagger} & = & \frac{g_2}{\sqrt{2} M_W \cos\beta}
  \left( \mathbf{m}_d \right)_{ij} \,,\\
{\mathbf h}_{ij}^{u\,\dagger} & = & \frac{g_2}{\sqrt{2} M_W \cos\beta}
  \left( \mathbf{m}_u \right)_{ij} \,.
\end{eqnarray}
Here, $\mathbf{m}_e$, $\mathbf{m}_d$ and $\mathbf{m}_u$ represent
$3 \times 3$ lepton, down-quark and up-quark mass matrices, respectively.
The dot products are defined in two-component notation \cite{Derendinger,Bajc}
as $A \cdot B \equiv \epsilon_{\alpha\beta} A^\alpha B^\beta$ 
($\epsilon_{12} \equiv +1$). Second, third
and fourth terms in right-hand side of Eq~\eqref{WMSSM} are just supersymmetric
generalization of the Yukawa couplings in the Standard Model Lagrangian
(for this and other aspects of the SM see Ref \cite{Herrero1998}). The first term
is however new, and can be thought of as a supersymmetric generalization of
a higgsino mass term. It can be shown that the consistent incorporation of
spontaneus electroweak symmetry breakdown requires $\mu$ to be of the order
of the weak scale.

One more thing needs to be adressed at this point, and that is the implicit
assumption of the conservation of $R$-parity defined by a quantum number
$R_p$ given by
\begin{equation}
R_p = (-1)^{3(B-L) + 2S} \,,
\end{equation}
where $B$, $L$ and $S$ stand for barion number, lepton number and spin of the
particle, respectively. The conservation of $R_p$ in the MSSM may be posited
as a natural assumption in a minimal supersymmetric extensions of the SM,
due to the barion and lepton number conservations in the SM Lagrangian.

Let's now turn back to \eqref{susy+ssb} and analyse the contents of the
$\mathcal{L}_{\textrm{SSB}}$. There are several constraints which need
to be put upon the supersymmetry breaking terms. First, they need to be
``small'' compared to the fully supersymmetric part
$\mathcal{L}_{\textrm{SUSY}}$.
Second, and most important, they must obey certain mass dimensional
constrains in order to preserve the desired convergent behavior of
the supersymmetric theory at high energies as well as the
nonrenormalization of its superpotential couplings. According to the
Symanzik's rule \cite[pp 107-8]{Coleman1988} this turns out to
be possible in all orders in perturbation theory only if the explicit
supersymmetry breaking terms are \emph{soft}
\cite{Witten1981,Dimopoulos1981,Sakai1981,Kaul1982}, i.e.\ that
every field operator occuring in $\mathcal{L}_{\textrm{SSB}}$ has
mass dimension strictly less then four. The Eq~\eqref{susy+ssb} is
therefore usually written as
\begin{equation}\label{susy+soft}
 \mathcal{L}_{\textrm{MSSM}} = \mathcal{L}_{\textrm{SUSY}}
 + \mathcal{L}_{\textrm{SOFT}} \,.
\end{equation}

Taking all this into account, one can write down the expression
for $\mathcal{L}_{\textrm{SOFT}}$, by collecting all allowed soft
SUSY-breaking terms \cite[p 185]{Drees2004},
\begin{eqnarray}
- {\cal L}_{SOFT} & = &
 \tilde{q}_{iL}^* ( \mathcal{M}_{\tilde{q}}^2 )_{ij} \tilde{q}_{jL}
    + \tilde{u}_{iR}^* ( \mathcal{M}_{\tilde{u}}^2 )_{ij} \tilde{u}_{jR}
    + \tilde{d}_{iR}^* ( \mathcal{M}_{\tilde{d}}^2 )_{ij} \tilde{d}_{jR} \nonumber \\
 && {} + \tilde{l}_{iL}^* ( \mathcal{M}_{\tilde{l}}^2 )_{ij} \tilde{l}_{jL}
    + \tilde{e}_{iR}^* ( \mathcal{M}_{\tilde{e}}^2 )_{ij} \tilde{e}_{jR} \nonumber \\
 && {} + \Big[ h_1 \cdot \tilde{l}_{iL} (A^e)^T_{ij} \tilde{e}_{jR}^* +
    h_1 \cdot \tilde{q}_{iL} (A^d)^T_{ij} \tilde{d}_{jR}^* \nonumber \\
 && {} + \tilde{q}_{iL} \cdot h_2 (A^u)^T_{ij} \tilde{u}_{jR}^* + \textrm{h.c.} \Big]
  \nonumber \\
 && {} + m_1^2 |h_1|^2 + m_2^2 |h_2|^2 + (B \mu h_1 \cdot h_2 + \textrm{h.c.}) \nonumber \\
 && {} + \frac{1}{2} ( M_1 \bar{\tilde{\lambda}}_0 P_L \tilde{\lambda}_0
    + M_1^* \bar{\tilde{\lambda}}_0 P_R \tilde{\lambda}_0 ) \nonumber \\
 && {} + \frac{1}{2} ( M_2 \bar{\vec{\tilde{\lambda}}} P_L \vec{\tilde{\lambda}}
    + M_2^* \bar{\vec{\tilde{\lambda}}} P_R \vec{\tilde{\lambda}} ) \nonumber \\
 && {} + \frac{1}{2} ( M_3 \bar{\tilde{g}}^a P_L \tilde{g}^a
    + M_3^* \bar{\tilde{g}}^a P_R \tilde{g}^a )
\label{L_SOFT}
\end{eqnarray}

Practical calculations within the MSSM usually include several simplifying
assumptions in order to drastically reduce the number of additional parameters
in the model. Different assumptions result in different versions of the
\emph{Constrained Minimal Supersymmetric Standard Model} or CMSSM.

\begin{sloppypar}
In this thesis, we will adopt the framework of \emph{Minimal Super Gravity}
(mSUGRA) model. Since MSSM fields alone cannot break supersymmetry spontaneously
at the weak scale \cite[pp 183-5]{Drees2004}, spontaneous supersymmetry breakdown
needs to be effected in a sector of fields which are singlets with respect
to the SM gauge group. This sector is known as the \emph{hidden} or
\emph{secluded sector}. SUSY breaking is then transmitted to the gauge nonsinglet
\emph{observable} or \emph{visible sector} by a messenger sector associated
by a typical mass scale $M_M$. Unlike the details of the spontaneous SUSY-breaking
in the hidden sector, the mechanism of its transmission from hidden sector to the
MSSM fields does have an immediate impact on the observable sparticle spectrum and
then also on the SUSY phenomenology. The most economical mechanism of this kind
uses gravitational strength interactions based on local supersymmetry also known as
\emph{supergravity} \cite{Nilles1984,Arnowitt1984}.
\end{sloppypar}

The great benefit in using the mSUGRA model is the fact that it reduces the extra
one hundred and five parameters (compared to the nineteen parameters of the SM)
to the set $\{p\}$ of just five parameters,
\begin{equation} \label{mSUGRAparam}
\{p\} = \{\textrm{sign}(\mu), m_0, M_{1/2}, A_0, \tan\beta \} \,,
\end{equation}
where $\textrm{sign}(\mu)$ stands for the sign of the $\mu$ parameter in superpotential
\eqref{WMSSM}, $m_0$ constitute masses of the scalars ($m_{ij} = m_0 \delta_{ij}$),
$M_{1/2}$ is common mass of all three MSSM gauginos,
$A_0$ is common trilinear coupling constant (higgs-sfermion-sfermion)
and $\tan\beta$ is ratio of VEVs defined by Eq~\eqref{tb}.
These parameters are also referred to as the
\emph{supersymmetry breaking parameters}. Their values are usually imposed
on the scale of Grand Unification (GUT), and then via Renormalization Group Equations
(RGE) \cite{Petcov2004} transmitted down to the weak scale.

There are quite a few reasons to work in the framework of the MSSM with $R$-parity
conserved. The MSSM provides a quantum-mechanically stable solution to the guage
hierarchy problem and predicts rather accurate unification of the SM gauge
couplings close to the grand unified theory (GUT) scale. The lightest supersymmetric
particle (LSP) is stable and, if neutral, such as the neutralino, could represent a
good candidate for the dark matter in the Universe. Besides that, the MSSM typically
predicts a SM-like Higgs boson lighter than 135 GeV, in agreement with the recent
observations for a $\sim 125 \textrm{ GeV}$ Higgs boson, made by ATLAS
\cite{Aad2012} and CMS \cite{Chatrchyan2012,Chatrchyan2013} Collaborations.

%%%%%%%%%%%%%%%%%%%%%%%%%%%%%%%%%%%%%%%%%%%%%%%%%%%%%%%%%%%%%%%%%%%%%%%%%%%%%%%%%%%%%%%
%%%%%%%%%%%%%%%%%%%%%%%%%%%%%%%%%%%%%%%%%%%%%%%%%%%%%%%%%%%%%%%%%%%%%%%%%%%%%%%%%%%%%%%
%%%%%%%%%%%%%%%%%%%%%%%%%%%%%%%%%%%%%%%%%%%%%%%%%%%%%%%%%%%%%%%%%%%%%%%%%%%%%%%%%%%%%%%

\section{Seesaw mechanism}

Neutrino oscillation experiments (see Chapter \ref{expsurvey}) have indisputably
shown that neutrinos are not massless, as was once believed to be. This imposes
the necessity to extend the Standard Model (as well as the MSSM) in a way that will
consistently allow the existence of massive neutrinos.
One of the most interesting extensions in that sense is provided by so-called
seesaw mechanism. There are three realizations of the seesaw mechanism:
the seesaw type one
\cite{Minkowski1977,Gell-Mann1979,Yanagida1979,Glashow1979,Mohapatra1980,Schechter1980},
the seesaw type two
\cite{Konetschny1977,Schechter1980,Lazarides1981,Mohapatra1981,Cheng1980,Schechter1982}
and the seesaw type three \cite{Foot1989}. These three scenarios differ by the nature
of their seesaw messengers needed to explain the small neutrino masses. For the purpose
of this thesis, we will explain and adopt a low-scale variant of the seesaw type-I
realization, whose messengers are three singlet neutrinos $N_{1,2,3}$. But first
let us examine the usual, high-scale variant, seesaw type-I mechanism in order to
detect its weaknesses and to demonstrate how low-scale variant can overcome them.
%In this short review we will mostly use notation used in Ref \cite{Grimus2006}.

The leptonic Yukawa sector of the SM with massless neutrinos is described by
\begin{equation}
 \mathcal{L}_Y^{(SM)} =
 - \begin{pmatrix} \overline{\nu'_i} & \overline{l'_i} \end{pmatrix}_L
  {\mathbf h}_{ij}^{(l)\,\dagger}
  \begin{pmatrix} \phi^+ \\ \phi^0 \end{pmatrix} l'_{jR} \; + \textrm{h.c.}
\end{equation}

Here, the primes indicate that the fields are not written in the mass
basis (so-called \emph{physical states}), but rather in the interaction basis.
${\mathbf h}^{(l)}$ and ${\mathbf h}^{(\nu)}$ are $3 \times 3$ lepton and 
neutrino Yukawa matrices, respectively.

The consistent and straightforward extension of this sector by a right-handed
neutrinos includes both the extra Yukawa neutrino term and the
mass term which is singlet under the SM gauge group ,
\begin{eqnarray}
 \mathcal{L}_Y^{(SM+\nu_R)} & = &
 {} - \begin{pmatrix} \overline{\nu'_i} & \overline{l'_i} \end{pmatrix}_L
  {\mathbf h}_{ij}^{(l)\,\dagger}
  \begin{pmatrix} \phi^+ \\ \phi^0 \end{pmatrix} l'_{jR} \nonumber \\
 && {} - \begin{pmatrix} \overline{\nu'_i} & \overline{l'_i} \end{pmatrix}_L
  {\mathbf h}_{ij}^{(\nu)\,\dagger}
  \begin{pmatrix} \phi^{0\dagger} \\ - \phi^{+\dagger} \end{pmatrix} \nu'_{jR} \nonumber \\
 && {} - \frac{1}{2} \, M \, \overline{(\nu'_R)^C} \, \nu'_R \; + \textrm{h.c.}
\end{eqnarray}

After the spontaneous breakdowns of the electroweak symmetry,
\begin{equation}
\Phi(x) \to \frac{1}{\sqrt{2}}
  \begin{pmatrix} 0 \\ v \end{pmatrix} \,,
\end{equation}
one ends with the well-known expression for lepton masses,
\begin{equation}
 (m_l)_{ij} = \frac{v}{\sqrt{2}} \, {\mathbf h}_{ij}^{(l)\,\dagger} \,,
  \quad
 (m_D)_{ij} = \frac{v}{\sqrt{2}} \, {\mathbf h}_{ij}^{(\nu)\,\dagger} \,,
  \quad M \,.
\end{equation}
Here $m_l$ represents masses of the charged leptons, $m_D$ stands for
the Dirac mass matrix, and $M$ is the Majorana mass matrix.
The former two make the mass term for neutrinos,
\begin{equation} \label{Lnumass}
 \mathcal{L}_\nu^{(mass)} = - \frac{1}{2}
  \begin{pmatrix} \overline{\nu'_L} & \overline{(\nu'_R)^C} \end{pmatrix}
  \underbrace{\begin{pmatrix} 0 & m_D \\ m_D^T & M \end{pmatrix}}_{\mathcal{M}_{D+M}}
  \begin{pmatrix} (\nu'_L)^C \\ \nu'_R  \end{pmatrix} \,.
\end{equation}
In order to get from the interaction to mass basis, i.e.\ to write the Lagrangian
in terms of physical states, one needs to diagonalize the $\mathcal{M}_{D+M}$ matrix.
This is performed with unitary $6 \times 6$ matrix $W$,
\begin{equation} \label{D+M}
 W^T \mathcal{M}_{D+M} W =
  \begin{pmatrix} \mathcal{M}_\nu & 0 \\ 0 & \mathcal{M}_N \end{pmatrix} \,.
\end{equation}
This matrix equation is solved by Taylor expansion, order by order \cite{Grimus2000}.
Keeping only the leading term, the solutions of Eq~\eqref{D+M} read \cite{Grimus2006}
\begin{equation} \label{ssmasses}
 \mathcal{M}_\nu \simeq - m_D^T M^{-1} m_D \,, \quad
 \mathcal{M}_N \simeq M  \,,
\end{equation}
\begin{equation} \label{ssW}
 W \simeq \begin{pmatrix} \mathbf{1}_{3\times 3} & (M^{-1} m_D)^\dagger \\
    - M^{-1} m_d & \mathbf{1}_{3\times 3} \end{pmatrix}  \sim
  \begin{pmatrix} 1 & \sqrt{m_\nu / m_N} \\ \sqrt{m_\nu / m_N} & 1 \end{pmatrix}  \,.
\end{equation}
Matrix $W$ transforms fields written in the interaction basis to the
one written in the mass basis,
\begin{equation}
 \begin{pmatrix} (\nu'_L)^C \\ \nu'_R \end{pmatrix} = 
   W \begin{pmatrix} \nu_L^C \\ \nu_R \end{pmatrix}
\end{equation}
Finally one can re-write the Lagrangian \eqref{Lnumass} in the mass
basis,
\begin{equation}
 \mathcal{L}_\nu^{(mass)} = - \frac{1}{2}
  \begin{pmatrix} \bar\nu_L & \bar\nu_R^C \end{pmatrix}
  \begin{pmatrix} \mathcal{M}_\nu & 0 \\ 0 & \mathcal{M}_N \end{pmatrix}
  \begin{pmatrix} \nu_L^C \\ \nu_R \end{pmatrix} \,.
\end{equation}

If we allow the Yukawa matrices to be of arbitrary form, we have to
face two unpleasant consequences:
\begin{enumerate}
 \item From Eq~\eqref{ssmasses} we see that mass of light neutrinos is
  roughly given by $m_\nu \sim m_D^2/M$. Since the light neutrino masses
  are of the order $m_\nu \sim 0.1 \textrm{ eV}$, and if we assume
  that Yukawa couplings are of order $\sim 0.1$, it follows that
  the heavy singlet neutrinos must assume masses of order
  $\sim 10^{12-14} \textrm{ GeV}$. That is inconvenient by itself, since
  its direct detection is way beyond the reach of experiments in high
  energy physics.
 \item From Eq~\eqref{ssW} we see that the mixing between light and heavy
  neutrinos is of the order $\xi_{\nu N} \sim \sqrt{m_\nu / m_N} \sim 10^{-12}$,
  for light neutrino masses $m_\nu \sim 0.1 \textrm{ eV}$. That means that the
  heavy neutrinos decouple form low-energy processes of CLFV in the SM with
  right-handed neutrinos, giving rise to extremely suppressed and unobservable
  rates.
\end{enumerate}

One way to overcome these difficulties is to impose the presence of the
approximate lepton flavor symmetries
\cite{Wyler1983,Mohapatra1986,Mohapatra1986a,Nandi1986,Branco1989,
Pilaftsis1992,Dev2012} in the theory. These symmetries result in a
specific structure of Yukawa matrices which, if exact, can provide
massless light neutrinos regardless of the masses of heavy neutrinos,
so that
\begin{equation}
 \mathcal{M}_\nu =- m_D^T M^{-1} m_D + \ldots \equiv 0 \,.
\end{equation}
Small neutrino masses can then be reproduced by breaking the imposed
symmetry by just the right amount. This scenario allows the heavy neutrino
mass scale to be as low as $100 \textrm{ GeV}$. Unlike in the usual seesaw
scenario, the light-to-heavy neutrino mixings $\xi_{\nu N}$ are not
correlated to the light neutrino masses $m_\nu$. Instead, $\xi_{\nu N}$
are free parameters, constrained by experimental limits on the deviations
of the $W^\pm$ and $Z$-boson couplings to leptons with respect to their SM
values \cite{Nardi1994,Antusch2009,Bergmann1999,Aguila2008}.

Approximate lepton flavor symmetries do not restrict the size of the LFV,
and so potentially large phenomena of CLFV may be predicted. This feature
is quite generic both in the SM \cite{Ilakovac1995} and in the MSSM
\cite{Ilakovac2009,Ilakovac2011} extended with low-scale right-handed
neutrinos. This new source of LFV, in addition to the one resulting
from the frequently considered soft SUSY breaking sector
\cite{Borzumati1986,Hisano1996,Hisano1995,Hisano1999,Carvalho2001,
Hisano2009,Ellis2002},
will be in particular interest in the study provided in this thesis.

%%%%%%%%%%%%%%%%%%%%%%%%%%%%%%%%%%%%%%%%%%%%%%%%%%%%%%%%%%%%%%%%%%%%%%%%%%%%%%%%%%%%%%%
%%%%%%%%%%%%%%%%%%%%%%%%%%%%%%%%%%%%%%%%%%%%%%%%%%%%%%%%%%%%%%%%%%%%%%%%%%%%%%%%%%%%%%%
%%%%%%%%%%%%%%%%%%%%%%%%%%%%%%%%%%%%%%%%%%%%%%%%%%%%%%%%%%%%%%%%%%%%%%%%%%%%%%%%%%%%%%%

\section{MSSM extended with right-handed neutrinos}

The SM and the MSSM extended by low-scale right-handed neutrinos in the presence
of the approximate lepton-number symmetries will be denoted by $\nu_R$SM and
$\nu_R$MSSM, respectively. Although some of the results displayed in this thesis
may be applicable to the more general soft SUSY breaking scenarios, this study
will be performed within the mSUGRA framework.

The $\nu_R$MSSM has some interesting features compared with the MSSM. In particular,
the heavy singlet sneutrinos may emerge as a new viable candidates of cold dark
matter \cite{Arina2008,Deppisch2008,Josse-Michaux2011,An2012a,Dumont2012}. In
addition, the mechanism of low-scale resonant leptogenesis
\cite{Pilaftsis2005,Pilaftsis2005a,Deppisch2011,Pilaftsis1997,Pilaftsis2004}
could provide a possible explanation for the observed baryon asymmetry in the
Universe, as the parameter space for successful electroweak baryogenesis
gets squeezed by the current LHC data \cite{Cohen2012,Carena2013}.

Given the multitude of quantum states mediating LFV in  the $\nu_R$MSSM,
the predicted values for observables of CLFV in this model turn out to be
generically larger than the corresponding ones in the $\nu_R$SM, except
possibly for $B(l\to l'\gamma)$ \cite{Ilakovac2009,Ilakovac2011},
where $l, l' = e, \mu, \tau$. The origin of suppression for the
latter  branching  ratios  may partially be attributed to the SUSY no-go
theorem due to Ferrara and Remiddi \cite{Ferrara1974}, which states that
the magnetic dipole moment operator necessarily violates SUSY and it must therefore
vanish in the supersymmetric limit of the theory.

In this section, we will describe the leptonic sector of the $\nu_R$MSSM and
introduce the neutrino Yukawa structure of two baseline scenarios based on
approximate lepton-number symmetries and universal Majorana masses at the
GUT scale. These scenarios will be used to present generic predictions
of the CLFV within the framework of mSUGRA, and to analyze the
anomalous magnetic and electric dipole moments within the same framework.

The leptonic superpotential part of the $\nu_R$MSSM reads:
\begin{equation}
  \label{Wpot}
W_{\rm lepton} =  \widehat{E}^C {\bf h}_e \widehat{H}_d
\widehat{L} + \widehat{N}^C {\bf h}_\nu \widehat{L} \widehat{H}_u
+ \frac{1}{2}\,\widehat{N}^C {\bf m}_M \widehat{N}^C \,,
\end{equation}
where $\widehat{H}_{u,d}$, $\widehat{L}$, $\widehat{E}$ and $\widehat{N}^C$
denote the two Higgs-doublet superfields, the three left-  and  right-handed
charged-lepton superfields and the three right-handed neutrino superfields,
respectively. The  Yukawa couplings ${\bf  h}_{e,\nu}$ and the Majorana mass
parameters ${\bf m}_M$ form $3\times 3$ complex matrices. Here, the  Majorana mass
matrix ${\bf  m}_M$ is taken to be SO(3)-symmetric at the $m_N$ scale,
i.e. $\mathbf{m}_M = m_N\, {\bf 1}_3$.

In the low-scale seesaw models models with the presence of approximate lepton
symmetries, the neutrino induced LFV transitions from a charged lepton
$l=\mu\,, \tau$ to another charged lepton $l'\neq l$ are functions of the
ratios \cite{Ilakovac1995,Bernabeu1987,Korner1993,Bernabeu1993,Deppisch2005}
\begin{equation}
  \label{Omega}
{\bf  \Omega}_{l'l} = \frac{v^2_u}{2  m^2_N} ({\bf  h}^\dagger_\nu
{\bf h}_\nu)_{l'l} = \sum_{i=1}^3 B_{l'N_i} B_{l N_i} \,,
\end{equation}
and are not constrained by the usual seesaw factor $m_\nu/m_N$, where
$v_u/\sqrt{2} \equiv \langle H_u\rangle$ is the vacuum expectation value
(VEV) of the Higgs doublet $H_u$, with
$\tan\beta \equiv \langle H_u\rangle/\langle H_d\rangle$. The
mixing matrix $B_{lN_i}$ that occurs in the interaction of the $W^\pm$ bosons
with the charged leptons $l = e, \mu, \tau$ and the three heavy neutrinos
$N_{1,2,3}$ is defined  in Appendix \ref{vertices}. It is important to note that the
LFV parameters $\Omega_{l'l}$ do not directly depend on the RGE evolution of the soft
SUSY-breaking parameters, except through the VEV $v_u$ defined at the minimum of 
the Higgs potential.

In the electroweak interaction basis $\{ \nu_{e,\mu,\tau\,L},  \nu_{1,2,3\,R}^C\}$,
the neutrino mass matrix in the $\nu_R$MSSM takes on the standard seesaw type-I form:
\begin{eqnarray}
  \label{Mnu}
{\bf M}_\nu &=& \left(\begin{array}{cc} 0 & {\bf m}_D \\ {\bf m}_D^T &
  {\bf m}_M^* \end{array}\right) \ ,
\end{eqnarray}
where ${\bf m}_D = \sqrt{2}M_W\sin\beta\, g^{-1}_w {\bf h}_\nu^\dagger$
and ${\bf m}_M$ are the Dirac- and Majorana-neutrino mass matrices, respectively.
Complex conjugation of ${\bf m}_M$ matrix is a consequence of the Majorana mass 
term in the superpotential $W_{\rm lepton}$ \eqref{Wpot}. 
In this thesis, we consider two baseline scenarios of neutrino Yukawa couplings.
The first one realizes a U(1) leptonic symmetry
\cite{Pilaftsis2005,Pilaftsis2005a,Deppisch2011} and is given by
\begin{eqnarray}
  \label{YU1}
{\bf h}_\nu &=&
 \left(\begin{array}{lll}
 0 & 0 & 0 \\
 a\, e^{-\frac{i\pi}{4}} & b\, e^{-\frac{i\pi}{4}} & c\, e^{-\frac{i\pi}{4}} \\
 a\, e^{ \frac{i\pi}{4}} & b\, e^{ \frac{i\pi}{4}} & c\, e^{ \frac{i\pi}{4}}
 \end{array}\right)\; .
\end{eqnarray}

In the second scenario, the structure of the neutrino Yukawa matrix
${\bf h}_\nu$ is motivated by  the discrete symmetry group $A_4$ and
has the following form \cite{Kersten2007}:
\begin{eqnarray}
  \label{YA4}
{\bf h}_\nu &=&
 \left(\begin{array}{lll}
 a & b & c \\
 a e^{-\frac{2\pi i}{3}} & b\, e^{-\frac{2\pi i}{3}} & c\, e^{-\frac{2\pi i}{3}} \\
 a e^{ \frac{2\pi i}{3}} & b\, e^{ \frac{2\pi i}{3}} & c\, e^{ \frac{2\pi i}{3}}
\end{array}\right)\;.
\end{eqnarray}

In Eqs~\eqref{YU1} and \eqref{YA4}, the Yukawa parameters $a$, $b$ and $c$
are  assumed to be real. As was explained in the previous section, the small
neutrino masses can be obtained by adding small symmetry-breaking terms into
these matrices thus making the above mentioned symmetries approximate rather
than exact. The predictions for CLFV observables, however, remain independent
of the flavor structure of these small terms, needed to fit the low-energy
neutrino data. For this reason, the particular symmetry breaking patterns of
the above two baseline Yukawa scenarios will not be discussed in this thesis.

Another source of LFV in the models under consideration comes
from sneutrino interactions. Specifically, the sneutrino mass Lagrangian in
flavor and mass bases is given by
\begin{eqnarray}
{\cal L}^{(\tilde{\nu})}
 &=&
 (\tilde{\nu}_L^\dagger,\tilde{\nu}_R^{C\,\dagger},\tilde{\nu}_L^T,\tilde{\nu}_R^{C\,T})
 \,{\bf M}^2_{\tilde{\nu}}\,
 \left(
 \begin{array}{c}
  \tilde{\nu}_L\\ \tilde{\nu}_R^{C}\\ \tilde{\nu}_L^*\\ \tilde{\nu}_R^{C*}
 \end{array}
 \right) \\[1.5ex]
 & = &
  \tilde{N}^\dagger {\cal U}^{\tilde{\nu}\dagger}
  {\bf M}^2_{\tilde{\nu}}\, {\cal U}^{\tilde{\nu}} \tilde{N}
 \ =\
 \tilde{N}^\dagger \hat{{\bf M}}^2_{\tilde{\nu}} \tilde{N} \,,
\end{eqnarray}
where  ${\bf M}^2_{\tilde{\nu}}$ is a $12\times  12$ Hermitian mass
matrix in the flavor basis and $\hat{{\bf M}}^2_{\tilde{\nu}}$ is
the corresponding diagonal mass  matrix in the mass basis. More
explicitly, in the flavor basis $\{\tilde{\nu}_{e,\mu,\tau\, L},
\tilde{\nu}_{1,2,3\,R}^C,\tilde{\nu}_{e,\mu,\tau\,L}^*,
\tilde{\nu}_{1,2,3\,R}^{C*}\}$, the sneutrino mass matrix
${\bf M}^2_{\tilde{\nu}}$ may be cast into the following form:
\begin{eqnarray}
   \label{M2snu1}
{\bf  M}^2_{\tilde{\nu}}
 &=&
 \left(
 \begin{array}{cccc}
 {\bf  H}_1 & {\bf  N} & {\bf  0} & {\bf  M} \\
 {\bf  N}^\dagger & {\bf  H}_2^T & {\bf  M}^T & {\bf  0} \\
 {\bf  0} & {\bf  M}^* & {\bf  H}_1^T & {\bf  N}^* \\
 {\bf  M}^\dagger & {\bf  0} & {\bf  N}^T & {\bf  H}_2
 \end{array}
 \right)\; ,
\end{eqnarray}
where the block entries are the $3\times 3$ matrices, namely
\begin{eqnarray}
  \label{M2snu2}
{\bf  H}_1 &=& {\bf  m}^2_{\tilde{L}} + {\bf  m}_D {\bf  m}_D^\dagger
+ \frac{1}{2} M_Z^2 \cos 2\beta
\nonumber\\
{\bf  H}_2 &=& {\bf  m}^2_{\tilde{\nu}} + {\bf  m}_D^\dagger {\bf
  m}_D + {\bf  m}_M {\bf  m}_M^\dagger
\nonumber\\
{\bf  M} &=& {\bf  m}_D ({\bf  A}_\nu - \mu \cot\beta)
\nonumber\\
{\bf  N} &=& {\bf  m}_D {\bf  m}_M \, .
\end{eqnarray}
Here, ${\bf m}^2_{\tilde{L}}$,  ${\bf m}^2_{\tilde{\nu}}$ and
${\bf A}_\nu$ are $3\times 3$ soft SUSY-breaking matrices associated with
the left-handed slepton doublets, the right-handed sneutrinos and their
trilinear couplings, respectively.

In \emph{the supersymmetric limit}, all the soft SUSY-breaking matrices are equal
to zero, $\tan\beta=1$ and $\mu=0$. As a consequence, the sneutrino mass matrix
${\bf M}^2_{\tilde{\nu}}$ can be expressed in terms of the neutrino mass matrix
${\bf M}_\nu$ in~(\ref{Mnu}) as follows:
\begin{eqnarray}
{\bf M}^2_{\tilde{\nu}}
 &\stackrel{\rm SUSY}{\longrightarrow} &
 \left(
 \begin{array}{cc}
 {\bf M}_\nu {\bf M}_\nu^\dagger & {\bf 0}_{6\times 6} \\
 {\bf 0}_{6\times 6} & {\bf M}_\nu^\dagger {\bf M}_\nu
 \end{array}
 \right) \,,
\label{MnutMnu}
\end{eqnarray}
resulting with the expected equality between neutrino and  sneutrino
mixings. Sneutrino LFV mixings do depend on the RGE evolution of the
$\nu_R$MSSM parameters, but unlike the LFV mixings induced by soft SUSY-breaking
terms, the sneutrino LFV mixings do not vanish at the GUT scale.

The sneutrino LFV  mixings are obtained as combinations of unitary matrices
which diagonalize  the sneutrino,  slepton and  chargino mass matrices. It is
interesting to notice that in the diagonalization of the sneutrino mass matrix
${\bf M}^2_{\tilde{\nu}}$ in \eqref{M2snu1}, the sneutrino fields
$\tilde{\nu}_{e,\mu,\tau\,L}$, $\tilde{\nu}_{1,2,3\,R}^C$ and their complex
conjugates $\tilde{\nu}_{e,\mu,\tau\,L}^*$, $\tilde{\nu}_{1,2,3\,R}^{C*}$ are
treated independently. As a result, the expressions for $\tilde{\nu}_{e,\mu,\tau\, L}$
and $\tilde{\nu}_{1,2,3\,R}^C$, in terms of the real-valued mass eigen\-states
$\widetilde{N}_{1,2,\dots,12}$, are not manifestly complex conjugates to
$\tilde{\nu}_{e,\mu,\tau L}^*$ and $\tilde{\nu}_{1,2,3R}^{C*}$, thus leading to a
two-fold interpretation of the flavor basis fields,
\begin{eqnarray}
\tilde{\nu}^*_{i}  &=& (\tilde{\nu}_i)^*
 \  =\ {\cal U}^{\tilde{\nu}*}_{i A} \widetilde{N}_A\;,\nonumber\\
\tilde{\nu}^*_{i} &=&
 {\cal U}^{\tilde{\nu}}_{i + 6\, A} \widetilde{N}_A\ ,
\end{eqnarray}
where  $\tilde{\nu}_{1,2,3} \equiv \tilde{\nu}_{e,\mu,\tau\,L}$  and
$\tilde{\nu}_{4,5,6} \equiv \tilde{\nu}^C_{1,2,3\,R}$, with $i=1,2,\dots,6$
and $A=1,2,\dots,12$. For this reason, in Appendix \ref{vertices}
we include all equivalent forms in which Lagrangians, such as
${\cal L}_{\overline{e}\tilde{\chi}^-\tilde{N}}$ and
${\cal L}_{\tilde{N}\tilde{N}Z}$, can be written down.

Finally, a third source of LFV in the $\nu_R$MSSM comes from soft SUSY-breaking
LFV terms~\cite{Borzumati1986,Hisano1995}. 
These LFV terms are induced by RGE running and, in the mSUGRA framework,
vanish at the GUT scale. Their size strongly depends on the interval of the RGE
evolution from the GUT scale to the universal heavy neutrino mass scale $m_N$.

All the three different mechanisms of LFV, mediated by heavy neutrinos, heavy
sneutrinos and soft SUSY-breaking terms, depend explicitly on the neutrino Yukawa
matrix ${\bf h}_\nu$ and vanish in the limit ${\bf h}_\nu \to 0$.

We will end this chapter with a technical remark. The diagonalization of
$12\times 12$ sneutrino mass matrix ${\bf M}^2_{\tilde{\nu}}$ and the resulting
interaction vertices will be evaluated  numerically, without  approximations. To
perform the diagonalization of ${\bf M}^2_{\tilde{\nu}}$ numerically, the method developed in
Ref~\cite{Pilaftsis2008} for the neutrino mass matrix will be used. This method becomes very
efficient if one of the diagonal submatrices has eigenvalues larger than the 
entries in all other submatrices. 
It will therefore be assumed that the heavy neutrino mass scale $m_N$ is of the
order of, or larger than the scale of the other mass parameters in the $\nu_R$MSSM.

%% file: CLFV.tex
\chapter[CLFV observables]{Charged lepton flavor violation} \label{CLFV}

In this chapter, the results and key details regarding the
calculations for a number of CLFV observables in the $\nu_R$MSSM
will be presented.

In the first section, the analytical results for the amplitudes of
CLFV decays $l\to l'\gamma$ and $Z\to l\,l'^C$, as well as
their branching ratios will be given. Second section gives analytical
expressions for the neutrinoless three-body decays $l\to l' l_1  l_2^C$
pertinent to muon and tau decays. Third section will deal with coherent
$\mu\to e$ conversion in nuclei, giving analytical results for transition
amplitudes. All analytical results are expressed in terms of one-loop
functions and composite form factors defined in the appendices at the
end of this thesis.

Finally, last section will present the numerical results for above
mentioned processes, accompanied by the brief description of the numerical
methods used and corresponding discussion regarding the very results.

These results are presented in Ref \cite{Popov2013}.

\newpage

\section{The Decays $l\to l'\gamma$ and $Z\to l l'^C$} \label{sec:Zgll}

\begin{figure}[h!]
 \centering
  \includegraphics[clip,
  width=0.7\textwidth,height=0.5\textheight,
     angle=0]{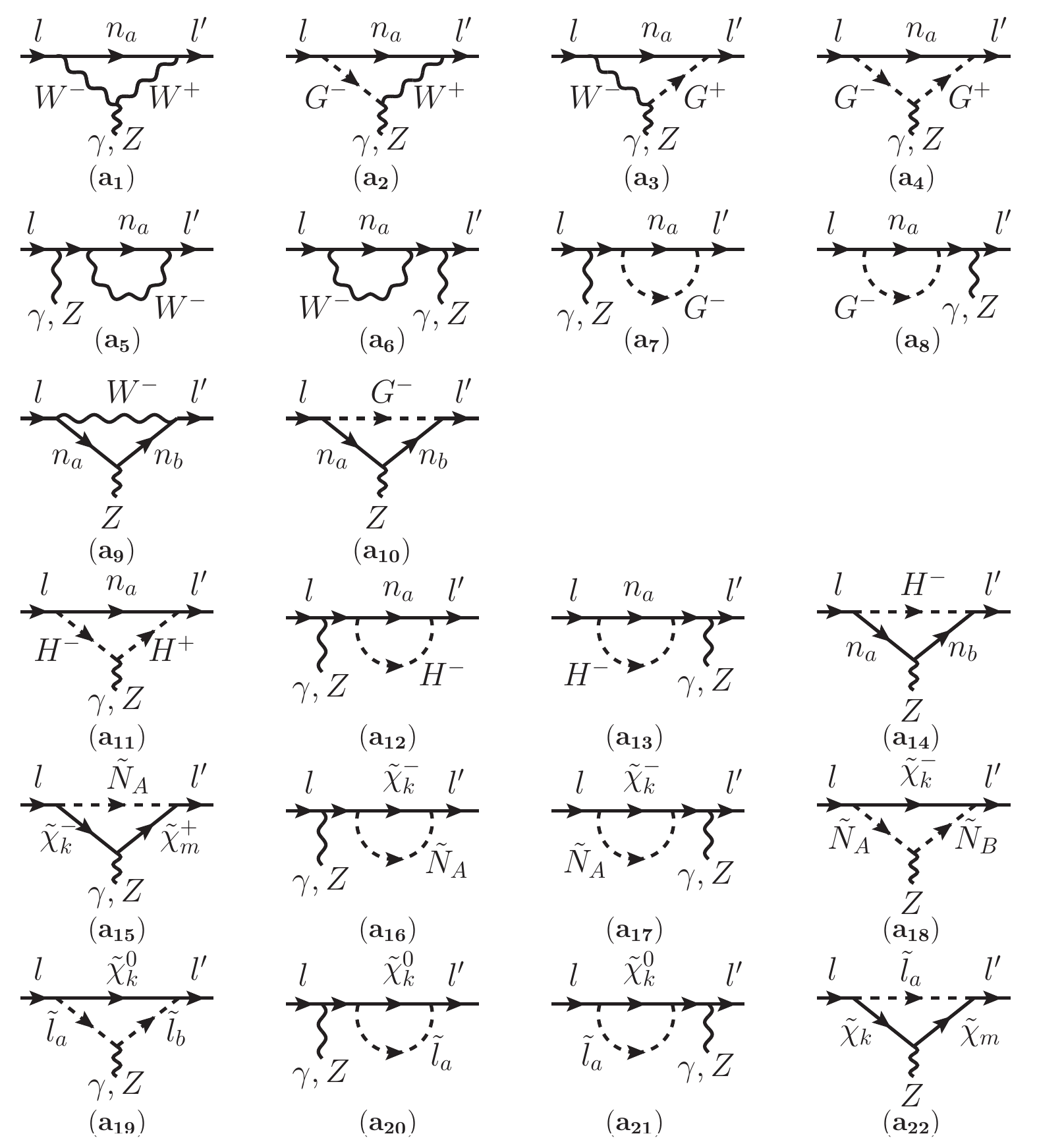}
\caption{Feynman  graphs contributing to $l\to l'\gamma$ and $Z\to l^Cl'$
($l\to Z l'$) amplitudes. Here $n_a$ ($a=1\ldots 6$) and $\widetilde{N}_A$
($A=1\ldots 12$) stand for neutrinos and snutrinos in mass basis, respectively.} 
\label{f1}
\end{figure}

At the one-loop level, the effective $\gamma l'l$ and $Zl'l$  couplings
are generated by the Feynman graphs shown in Fig~\ref{f1}. The general
form of the transition amplitudes associated with these effective couplings
is given by
\begin{eqnarray}
  \label{Tll'g}
{\cal T}^{\gamma l'l}_\mu
 \!&=&\!
 \frac{e\, \alpha_w}{8\pi M^2_W}\: \bar{l}'
 \Big[ (F^L_\gamma)_{l'l}\, (q^2\gamma_\mu-\Slash{q}q_\mu) P_L
     + (F^R_\gamma)_{l'l}\, (q^2\gamma_\mu-\Slash{q}q_\mu) P_R
\nonumber\\&&
     +\  (G^L_\gamma)_{l'l}\, i \sigma_{\mu\nu}q^\nu P_L
     +  (G^R_\gamma)_{l'l}\, i \sigma_{\mu\nu}q^\nu P_R \Big]\: l,
\\
  \label{Tll'Z}
{\cal T}^{Z l'l}_\mu
 \!&=&\!
 \frac{g_w\, \alpha_w}{8\pi \cos\theta_w}\: \bar{l}'
 \Big[ (F_Z^L)_{l'l}\, \gamma_\mu P_L
     + (F_Z^R)_{l'l}\, \gamma_\mu P_R \Big]\: l,
\end{eqnarray}
where  $P_{L(R)}   =  \frac{1}{2}\,[1-\!(+)\,\gamma_5]$,  $\alpha_w  =
g^2_w/(4\pi)$, $e$  is the  electromagnetic coupling constant,  $M_W =
g_w \sqrt{v^2_u  +v^2_d}/2$ is the  $W$-boson mass, $\theta_w$  is the
weak mixing angle and $q =  p_{l'} - p_l$ is the photon momentum.  The
form      factors      $(F^L_\gamma)_{l'l}$,      $(F^R_\gamma)_{l'l}$
$(G^L_\gamma)_{l'l}$,    $(G^R_\gamma)_{l'l}$,   $(F_Z^L)_{l'l}$   and
$(F_Z^R)_{l'l}$    receive   contributions   from    heavy   neutrinos
$N_{1,2,3}$, heavy  sneutrinos $\widetilde{N}_{1,2,3}$ and  RGE induced
soft SUSY-breaking terms.  The  analytical expressions for these three
individual contributions  are given in  Appendix~\ref{sec:olff}.
Note that, according to the normalization used, the  composite  form factors
$(G^L_\gamma)_{l'l}$ and $(G^R_\gamma)_{l'l}$  have dimensions of mass,
whilst all other form factors are dimensionless.

It is important to remark that  the  transition  amplitudes
(\ref{Tll'g})  and (\ref{Tll'Z})  are  also constituent  parts of  the
leptonic  amplitudes  $l\to  l'l_1l_2^C$ and  semileptonic  amplitudes
$l\to  l'q_1 \bar{q}_2$,  which will  be discussed  in more  detail in
Sections \ref{sec:lll}  and \ref{sec:mueff}.   To  calculate the  CLFV
decay  $l\to l'\gamma$,  we only  need to  consider the  dipole moment
operators  associated with  the  form factors $(G^L_\gamma)_{l'l}$  and
$(G^R_\gamma)_{l'l}$  in  (\ref{Tll'g}). Taking  this  last fact  into
account, the branching ratios for $l\to  l'\gamma$ and $Z \to l\, l'^C +
l^C l'$ are given by
\begin{eqnarray}
  \label{Bllpg}
B(l\to l'\gamma)
 \!&=&\!
 \frac{\alpha_w^3 s_w^2}{256\pi^2} \frac{m_l^3}{M_W^4\Gamma_l}
 \Big( |(G^L_\gamma)_{l'l}|^2 + |(G^R_\gamma)_{l'l}|^2 \Big) \,,\\
  \label{Blplc}
B(Z\to l'l^C+l'^Cl)
 \!&=&\!
 \frac{\alpha_w^3 M_W}{768 \pi^2 c_w^3\Gamma_Z} \Big(|(F_Z^L)_{l'l}|^2 +
 |(F_Z^R)_{l'l}|^2\Big) \,.
\end{eqnarray}
The above expressions are valid up to the leading order  in
external charged lepton masses and external momenta, which constitutes
an excellent  approximation for our  purposes.  Thus, in~(\ref{Blplc})
we have assumed that the $Z$-boson mass $M_Z$ is much smaller than the
SUSY and heavy neutrino mass  scales, $M_{\rm SUSY}$ and $m_N$, and we
have kept the leading term in an expansion of small momenta and masses
for  the external  particles. In  the  decoupling regime  of all  soft
SUSY-breaking and charged Higgs-boson masses, the low-energy sector of
the $\nu_R$MSSM becomes the $\nu_R$SM.  In this $\nu_R$SM limit of the
theory, the analytical expressions  for $B(l\to l'\gamma)$ and $B(Z\to
l'l^C+l'^Cl)$ take on the forms given in Refs~\cite{Cheng1980a} and
\cite{Bernabeu1987,Korner1993,Bernabeu1993}, respectively.

\section{Three-Body Leptonic Decays $l\to l'l_1l_2^C$} \label{sec:lll}
    
We now study the three-body CLFV decays $ l\to   l' l_1 l_2^C$,
where $ l$ can be the muon or tau lepton, and $ l',\,  l_1,\,  l_2$
denote other charged leptons to which $ l$ is allowed to decay kinematically.

\begin{figure}[h!]
 \centering
  \includegraphics[clip,
  width=0.7\textwidth,height=0.45\textheight,
     angle=0]{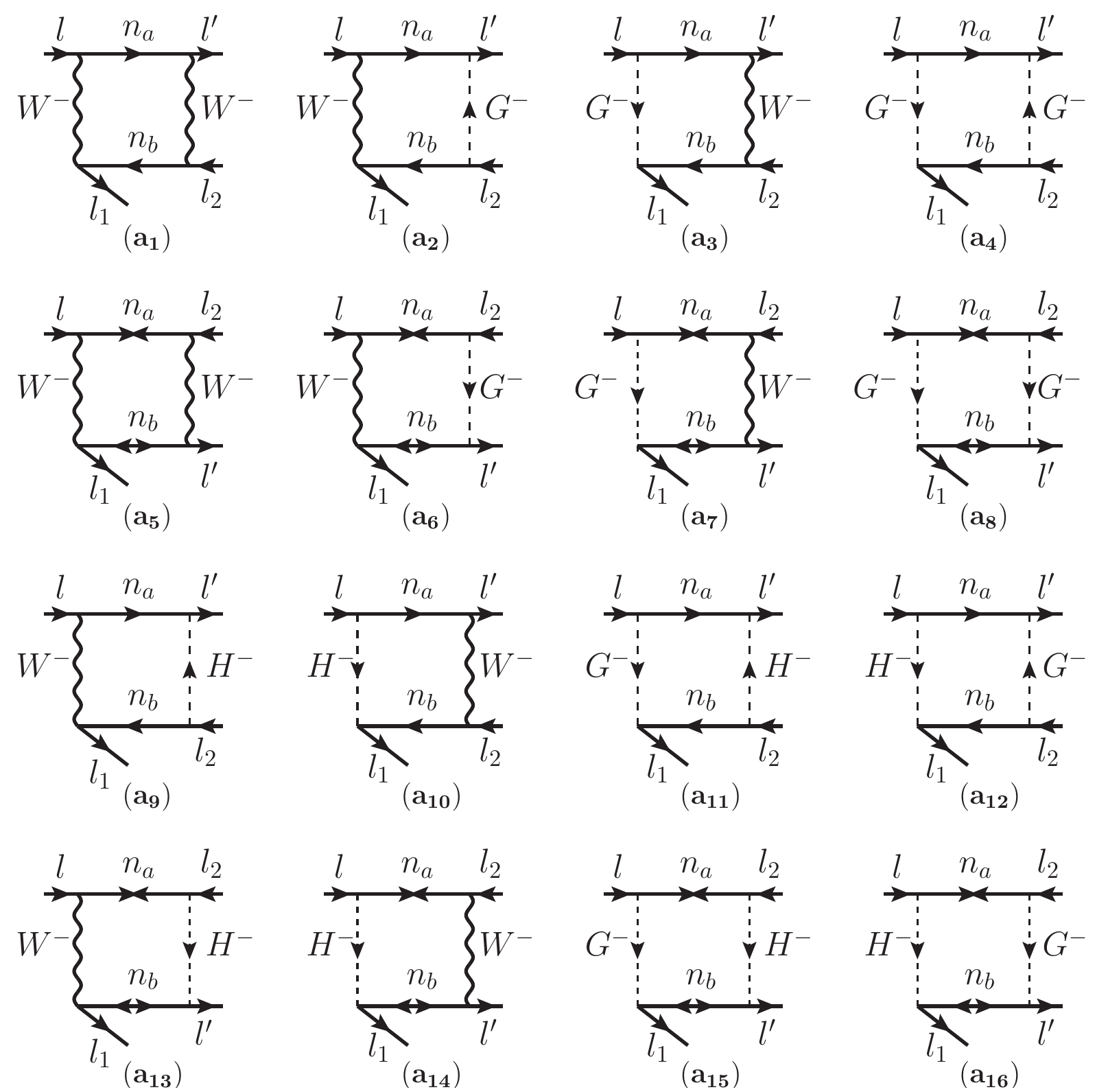}
  \includegraphics[clip,
  width=0.7\textwidth,height=0.27\textheight,
     angle=0]{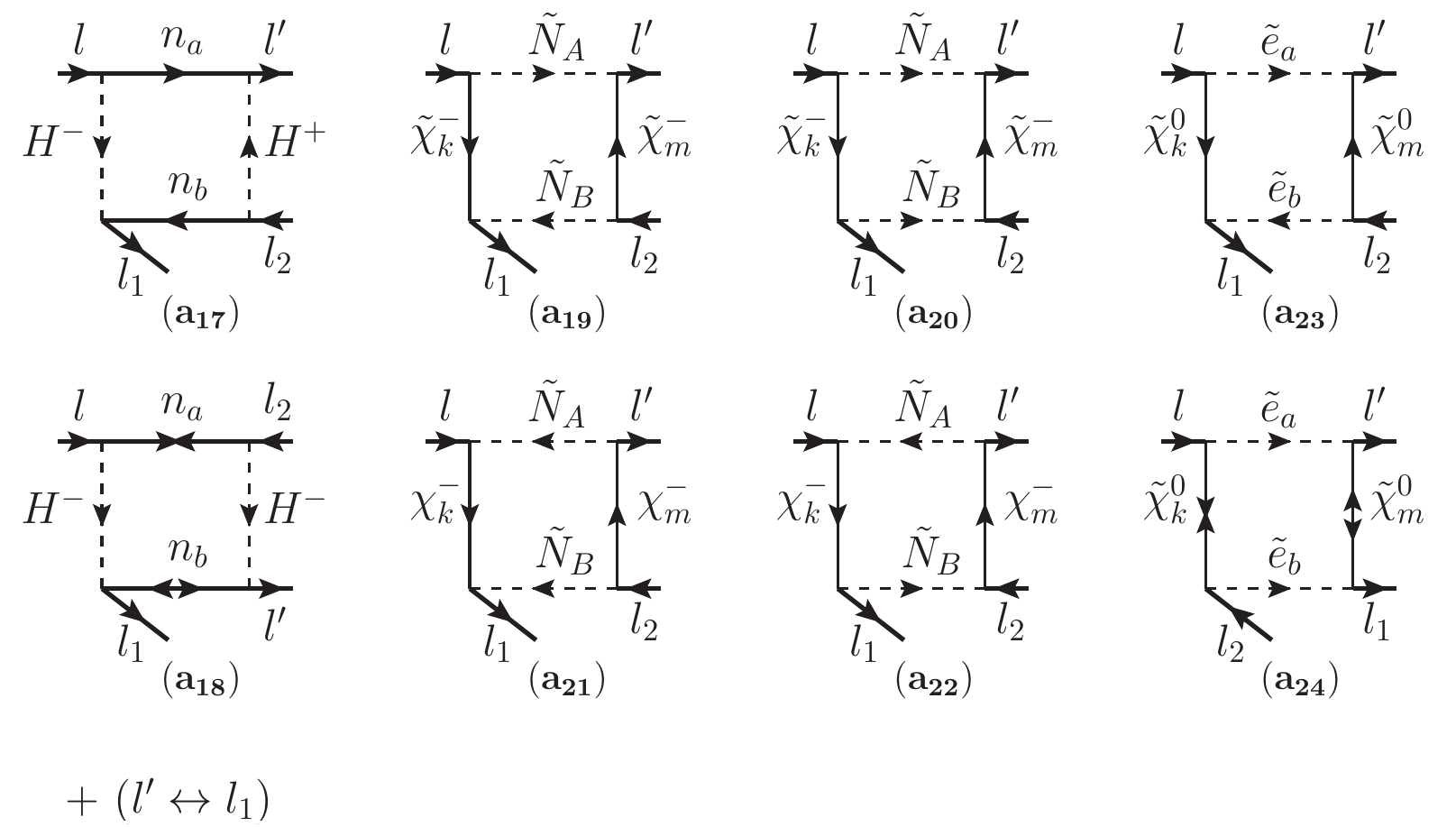}
\caption{Feynman  graphs contributing to the box $l\to l'l_1l_2^C$
amplitudes.} \label{f2}
\end{figure}

The transition amplitude for $ l\to  l' l_1 l_2^C$ receives contributions
from $\gamma$- and $Z$-boson-mediated graphs shown in Fig~\ref{f1} and from
box graphs displayed in Fig~\ref{f2}. The amplitudes for these three
contributions are:
\begin{eqnarray}
  \label{Tgl3l}
{\cal T}_{\gamma}^{ll'l_1l_2}
 \!&=&\!
   \label{Tllpg}
 \frac{\alpha_w^2 s_w^2}{2 M_W^2}
 \Big\{ \delta_{l_1l_2}\, \bar{l}'\,
  \Big[ (F_\gamma^L)_{l'l}\, \gamma_\mu P_L + (F_\gamma^R)_{l'l}\,
    \gamma_\mu P_R  +
 \frac{(\Slash{p}-\Slash{p}')}{(p-p')^2}
  \nonumber\\
 & \cdot &
         \Big( (G_\gamma^L)_{l'l}\, \gamma_\mu P_L +
         (G_\gamma^R)_{l'l}\, \gamma_\mu P_R \Big) \Big]\, l
      \, \bar{l}_1\gamma^\mu l_2^C
  - [ l'\leftrightarrow l_1 ] \Big\} \,,\\
  \label{TZl3l}
%%%
{\cal T}_Z^{ll'l_1l_2}
   \label{TllpZ}
 \!&=&\!
 \frac{\alpha_w^2}{2 M_W^2}
 \Big[ \delta_{l_1l_2}\, \bar{l}'\Big( (F_Z^L)_{l'l}\,\gamma_\mu P_L
+ (F_Z^R)_{l'l}\, \gamma_\mu P_R\Big) l \nonumber\\
 & \cdot & \bar{l}_1 \Big(g_L^l\, \gamma^\mu P_L + g_R^l\, \gamma^\mu P_R\Big) l_2^C
  - ( l'\leftrightarrow l_1 ) \Big] \,,\\
  \label{TBl3l}
{\cal T}_{\rm box}^{ll'l_1l_2}
  \label{Tbl_1}
 \!&=&\!
 - \frac{\alpha_w^2}{4 M_W^2}
 \Big(
    B_{\ell V}^{LL}\, \bar{l}'\gamma_\mu P_L l\ \bar{l}_1\gamma^\mu P_L l_2^C
  + B_{\ell V}^{RR}\, \bar{l}'\gamma_\mu P_R l\ \bar{l}_1\gamma^\mu P_R l_2^C
\nonumber\\
  &+& B_{\ell V}^{LR}\, \bar{l}'\gamma_\mu P_L l\ \bar{l}_1\gamma^\mu P_R l_2^C
  + B_{\ell V}^{RL}\, \bar{l}'\gamma_\mu P_R l\ \bar{l}_1\gamma^\mu P_L l_2^C
  \nonumber\\
  &+& B_{\ell S}^{LL}\, \bar{l}' P_L l\ \bar{l}_1 P_L l_2^C
  + B_{\ell S}^{RR}\, \bar{l}' P_R l\ \bar{l}_1 P_R l_2^C
  \nonumber\\
  &+& B_{\ell S}^{LR}\, \bar{l}' P_L l\ \bar{l}_1 P_R l_2^C
  + B_{\ell S}^{RL}\, \bar{l}' P_R l\ \bar{l}_1 P_L l_2^C
\nonumber\\
  &+& B_{\ell T}^{LL}\, \bar{l}' \sigma_{\mu\nu} P_L l\ \bar{l}_1
  \sigma^{\mu\nu} P_L l_2^C
  + B_{\ell T}^{RR}\, \bar{l}' \sigma_{\mu\nu} P_R l\ \bar{l}_1
  \sigma^{\mu\nu} P_R l_2^C  \Big) \\
\!&\equiv&\!
  \label{Tbl_2}
- \frac{\alpha_w^2}{4 M_W^2} \sum_{X,Y=L,R}\ \sum_{A=V,S,T}
    B_{\ell A}^{XY}\, \bar{l}'\Gamma^X_{A} l\ \bar{l}_1\Gamma^Y_{A}
    l_2^C \,,
\end{eqnarray}
where  $g_L^l =  -1/2+s_w^2$ and  $g_R^l=s_w^2$  are $Z$-boson--lepton
couplings  and  $s_w=\sin\theta_w$.   The  composite  box  form factors
$B_{\ell  A}^{XY}$
%,   $A=V,S,T$,  $X,Y=L,R$
are   given  in  Appendix~\ref{sec:olff}.  
The labels $V$, $S$ and $T$ denote the form factors of
the vector, scalar and tensor  combinations of the currents, while $L$
and  $R$  distinguish between  left  and  right  chiralities of  those
currents. The box form factors contain both direct and Fierz-transformed
contributions (see Appendix~\ref{app:ff}).  
Equation \eqref{Tbl_2} represents a
shorthand   expression   that   takes   account  of   all   individual
contributions  to  the   amplitude  ${\cal  T}_{\rm  box}^{ll'l_1l_2}$
induced  by   box  graphs.   Explicitly,   the  matrices  $\Gamma^X_A$
appearing in \eqref{Tbl_2} read:
\begin{equation}
\left(\Gamma^L_V,\Gamma^R_V,\Gamma^L_S,\Gamma^R_S,\Gamma^L_T,
\Gamma^R_T \right)
 =
\left(\gamma_\mu P_L, \gamma_\mu P_R, P_L, P_R,
\sigma_{\mu\nu}P_L, \sigma_{\mu\nu}P_R \right) \,.
\end{equation}

As   a  consequence   of  the   identity   $\sigma^{\mu\nu}\gamma_5  =
-\frac{i}{2}\varepsilon^{\mu\nu\rho\tau}  \sigma_{\rho\tau}$,
the tensor form factors  $B_{\ell T}^{LR}$ and $B_{\ell T}^{RL}$
vanish  in the sum  \eqref{Tbl_2}, i.e.\ $B_{\ell T}^{LR}=B_{\ell T}^{RL}=0$.
A very similar  chiral structure is found in the  semileptonic box
amplitudes defined in the next  section as well. It should also be
said that the previous studies of these processes
\cite{Hisano1996,Arganda2006} do not include in their calculations the
chiral structures  $P_L\times  P_R$  and $P_R\times  P_L$ and their
corresponding form factors $B_{\ell S}^{LR}$ and $B_{\ell S}^{RL}$.

In a three-generation model, the transition amplitude for the decays
$l\to l'l_1l_2^C$ may fall in one of the following three classes or categories
\cite{Ilakovac1995}: (i)~$l'\neq l_1=l_2$, (ii)~$l'= l_1=l_2$, and
(iii)~$l'= l_1\neq l_2$. In the first two classes, total lepton number is
conserved, whilst in the third class the total lepton number 
is violated by two units on the current level.
Since the predictions for the observables in class (iii) turn
out to be unobservably small in the $\nu_R$MSSM, these processes will be
ignored. Moreover, the universal indices $l'l$ which appear in the
photon  and  $Z$-boson  form factors, i.e.~$F_\gamma^L$,  $F_\gamma^R$,
$F_Z^L$ and  $F_Z^R$, will be dropped out for the sake of readability.
Given the above simplification and the notation of the box form factors
\eqref{Tbl_2}, the branching ratios for the class (i) and (ii) of CLFV
three-body decays are given by
\begin{eqnarray}
  \label{Bl3l_1}
B(l\to l'l_1l_1^C)
 &=&
 \frac{m_l^5\alpha_w^4}{24576\pi^3 M_W^4\Gamma_l}
 \Bigg\{\bigg[
  \Big|2s_w^2 (F_\gamma^L + F_Z^L) - F_Z^L - B_{\ell V}^{LL}\Big|^2 \nonumber\\
 &+& \Big|2s_w^2 (F_\gamma^R + F_Z^R) - B_{\ell V}^{RR}\Big|^2
  + \Big|2s_w^2 (F_\gamma^L + F_Z^L) - B_{\ell V}^{LR} \Big|^2
  \nonumber\\
 &+&\Big|2s_w^2 (F_\gamma^R + F_Z^R) - F_Z^R - B_{\ell V}^{RL} \Big|^2
  \bigg]
\nonumber\\
 &+& \frac{1}{4} \Big( |B_{\ell S}^{LL} |^2 + |B_{\ell S}^{RR} |^2 +
  |B_{\ell S}^{LR} |^2 + |B_{\ell S}^{RL} |^2 \Big)
\nonumber\\
 &+& 12 \Big( |B_{\ell T}^{LL}|^2  +  |B_{\ell T}^{RR}|^2 \Big)
\nonumber\\
 &+& \frac{32s_w^4}{m_l} \Big[
 \mbox{Re} \Big( (F_\gamma^R + F_Z^R ) G_\gamma^{L*}\Big)
 + \mbox{Re} \Big( (F_\gamma^L + F_Z^L ) G_\gamma^{R*}\Big)
 \Big]
\nonumber\\
 &-& \frac{8s_w^2}{m_l} \Big[
 \mbox{Re}\Big( (F_Z^R + B_{\ell V}^{RR}  + B_{\ell V}^{RL} ) G_\gamma^{L*}\Big)
\nonumber\\
 &+& \mbox{Re}\Big( (F_Z^L + B_{\ell V}^{LL} + B_{\ell V}^{LR} )
  G_\gamma^{R*}\Big)\Big]
\nonumber\\
 &-& \frac{32 s_w^4}{m_l^2} \Big(|G_\gamma^L|^2 + |G_\gamma^R|^2\Big)
 \bigg(\ln\frac{m^2_l}{m^2_{l'}} - 3\bigg)
\Bigg\}\,,
\end{eqnarray}
\begin{eqnarray}
  \label{Bl3l_2}
B(l\to l'l'l'^C)
 &=&
 \frac{m_l^5\alpha_w^4}{24576\pi^3 M_W^4\Gamma_l}\:
 \Bigg\{ 2 \bigg[
  \Big|2s_w^2 (F_\gamma^L + F_Z^L) - F_Z^L - \frac{1}{2}B_{\ell V}^{LL} \Big|^2
  \nonumber\\
  &+& \Big|2s_w^2 (F_\gamma^R + F_Z^R) - \frac{1}{2}B_{\ell V}^{RR} \Big|^2 \bigg]
  + \Big|2s_w^2 (F_\gamma^L + F_Z^L) - B_{\ell V}^{LR} \Big|^2
  \nonumber\\
  &+& \Big|2s_w^2 (F_\gamma^R + F_Z^R) - (F_Z^R + B_{\ell V}^{RL} )\Big|^2
  + \frac{1}{8} \Big( |B_{\ell S}^{LL}|^2 + |B_{\ell S}^{RR}|^2 \Big)
\nonumber\\
 &+& 6 \Big( |B_{\ell T}^{LL}|^2 +  |B_{\ell T}^{RR}|^2 \Big)
\nonumber\\
 &+& \frac{48s_w^4}{m_l} \Big[
 \mbox{Re} \Big( ( F_\gamma^R + F_Z^R ) G_\gamma^{L*}\Big)
 + \mbox{Re} \Big( ( F_\gamma^L + F_Z^L ) G_\gamma^{R*}\Big)
 \Big]
\nonumber\\
 &-& \frac{8s_w^2}{m_l} \Big[
 \mbox{Re} \Big( \big(F_Z^R + B_{\ell V}^{RR}
     + B_{\ell V}^{RL} \big) G_\gamma^{L*}\Big)
\nonumber\\
 &+& \mbox{Re} \Big( \big(2 F_Z^L + B_{\ell V}^{LL}
     + B_{\ell V}^{LR} \big) G_\gamma^{R*}\Big)\Big]
\nonumber\\
 &+& \frac{32 s_w^4}{m_l^2} \Big(|G_\gamma^L|^2 + |G_\gamma^R|^2\Big)
 \bigg(\ln\frac{m^2_l}{m^2_{l'}} - \frac{11}{4}\bigg)
 \Bigg\}\; ,
\end{eqnarray}
where $m_l$ and $m_{l'}$, $m_{l_1}$, $m_{l_2}$ are the masses of the
initial- and  final-state charged leptons and $\Gamma_l$  is the decay
width of  the charged lepton $l$.  It should be emphasized that the
transition amplitudes \eqref{Tgl3l}, \eqref{TZl3l} and \eqref{TBl3l} 
as well as the
branching  ratios \eqref{Bl3l_1} and \eqref{Bl3l_2}  have the most
general chiral  and form factor structure to the leading  order in the external
masses and  momenta, which makes them applicable to most models of the New
Physics containing CLFV. Even more general result can be found in the
Appendix \ref{app:ff}.

These results have been checked in the $\nu_R$SM limit of the theory in
which the branching ratios \eqref{Bl3l_1} and \eqref{Bl3l_2} go over
to the results presented in Ref \cite{Ilakovac1995}.

\section{Coherent $\mu\to e$ Conversion in a Nucleus} \label{sec:mueff}

The coherent  $\mu\to e$  conversion in a  nucleus corresponds  to the
process  $J_\mu\to e^-J^+$, where  $J_\mu$ is  an atom  of nucleus $J$
with  one  orbital  electron  replaced  by a  muon  and $J^+$  is  the
corresponding ion without the muon. The transition amplitude for such a
CLFV process,
\begin{eqnarray}
\label{TmueJ}
{\cal T}^{\mu e;J}
 &=&
   \langle J^+ e^- | {\cal T}^{d\mu \to de} | J_\mu\rangle
 + \langle J^+ e^- | {\cal T}^{u\mu \to ue} | J_\mu\rangle\; ,
\end{eqnarray}
depends on two effective box operators,
\begin{eqnarray}
  \label{Tdmude}
{\cal T}_{\rm box}^{d\mu \to de}
 &=&
  - \frac{\alpha_w^2}{4 M_W^2} \sum_{X,Y=L,R}\ \sum_{A=V,S,T}
   B_{dA}^{XY}\: \overline{e}\, \Gamma^X_A \mu\ \bar{d}\, \Gamma^X_A d
\nonumber\\
 &=&
 - \frac{\alpha_w^2}{2 M_W^2}\,
 (d^\dagger d)\;\, \bar{e}\, ( V_d^R\, P_R + V_d^L\, P_L )\, \mu\; ,\\
  \label{Tumuue}
{\cal T}_{\rm box}^{u\mu \to ue}
 &=&
 - \frac{\alpha_w^2}{4 M_W^2} \sum_{X,Y=L,R}\ \sum_{A=V,S,T}
   B_{u A}^{XY}\: \bar{e}\, \Gamma^X_A \mu\ \bar{u}\, \Gamma^X_A u
\nonumber\\
 &=&
 - \frac{\alpha_w^2}{2 M_W^2}\,
 (u^\dagger u)\;\, \bar{e}\, (V_u^R\, P_R + V_u^L\, P_L)\, \mu\; .
\end{eqnarray}
Here $\mu$ and $e$ are the  muon and electron wave functions and $d$
and $u$  are field operators acting  on the $J_\mu$  and $J^+$ states,
respectively. The form factors $B_{dA}^{XY}$ and  $B_{uA}^{XY}$ are given
in the Appendix \ref{sec:olff}. The composite form factors $V_d^{L}$,  $V_u^{L}$,
$V_d^{R}$, $V_u^{R}$ may be written as
\begin{eqnarray}
\label{V_ff}
V_d^L
 &=&
 - \frac{1}{3} s_w^2 \Big(F^L_\gamma - \frac{1}{m_\mu} G^R_\gamma\Big)
 + \Big(\frac{1}{4} - \frac{1}{3} s_w^2\Big) F_Z^L \nonumber\\
 &&{}+\, \frac{1}{4}\Big(B_{d V}^{LL}+B_{d V}^{LR} + B_{d S}^{RR} + B_{d S}^{RL}\Big)\,,
 \\[1.5ex]
V_d^R
 &=&
 -  \frac{1}{3} s_w^2 \Big(F^R_\gamma - \frac{1}{m_\mu} G^L_\gamma\Big)
 + \Big(\frac{1}{4} - \frac{1}{3} s_w^2\Big) F_Z^R \nonumber\\
 &&{}+\, \frac{1}{4}\Big(B_{d V}^{RR}+B_{d V}^{RL} + B_{d S}^{LL} + B_{d S}^{LR}\Big)\,,
 \\[1.5ex]
V_u^L
 &=&
 \frac{2}{3} s_w^2 \Big(F^L_\gamma - \frac{1}{m_\mu} G^R_\gamma\Big)
 + \Big(- \frac{1}{4} + \frac{2}{3} s_w^2\Big) F_Z^L \nonumber\\
 &&{}+ \frac{1}{4}\Big(B_{u V}^{LL}+B_{u V}^{LR} + B_{u S}^{RR} + B_{u S}^{RL}\Big)\,,
 \\[1.5ex]
V_u^R
 &=&
  \frac{2}{3} s_w^2 \Big(F^R_\gamma - \frac{1}{m_\mu} G^L_\gamma\Big)
 + \Big(- \frac{1}{4} + \frac{2}{3} s_w^2\Big) F_Z^R \nonumber\\
 &&{}+ \frac{1}{4}\Big(B_{u V}^{RR}+B_{u V}^{RL} + B_{u S}^{LL} + B_{u S}^{LR}\Big)\,,
\end{eqnarray}
where  $F_\gamma^L$, $F_\gamma^R$,  $F_Z^L$, $F_Z^R$  is  the shorthand
notation     for     $(F_\gamma^L)_{e\mu}$,     $(F_\gamma^R)_{e\mu}$,
$(F_Z^L)_{e\mu}$,  $(F_Z^R)_{e\mu}$.

The next step aims to  determine the  nucleon matrix  elements  of the
operators $u^\dagger u$ and $d^\dagger d$. These are given by
\begin{eqnarray}
\langle J^+e^- |u^\dagger u| J_\mu \rangle &=& (2Z+N) F(-m_\mu^2)\,,
\nonumber\\
\langle J^+e^- |d^\dagger d| J_\mu \rangle &=& (Z+2N) F(-m_\mu^2)\,,
\end{eqnarray}
where  the form factor $F(q^2)$  incorporates the  recoil of  the $J^+$
ion \cite{Chiang1993}, and the factors $2Z+N$ and $Z+2N$ count the number
of $u$  and $d$  quarks in the  nucleus $J$, respectively.  Hence, the
matrix element for $J_\mu\to J^+\mu^-$ can be written down as
\begin{eqnarray}
T^{J_\mu\to J^+ e^-}
 &=&
 -\frac{\alpha_w^2}{2 M_W^2}\, F(-m_\mu^2)\: \bar{e}
 \,(Q_W^L\, P_R + Q_W^R\, P_L )\,
 \mu\,,
\label{TJmue}
\end{eqnarray}
with
\begin{eqnarray}
\label{QWLR}
Q_W^L &=& (2Z+N) V_u^L +(Z+2N)V_d^L\, ,
\nonumber\\
Q_W^R &=& (2Z+N) V_u^R +(Z+2N)V_d^R\, .
\end{eqnarray}

Given the transition amplitude \eqref{TJmue}, the decay rate $J_\mu\to
J^+e^-$ is found to be
\begin{eqnarray}
\label{RmueJ}
R^J_{\mu    e}     &=&    \frac{\alpha^3\alpha_w^4    m_\mu^5}{16\pi^2
  M_W^4\Gamma_{{\rm  capture}}} \frac{Z^4_{\rm  eff}}{Z} |F(-m_\mu^2)|^2
\: \Big(\,|Q_W^L|^2+|Q_W^R|^2\,\Big)\, ,
\end{eqnarray}
where $\Gamma_{{\rm capture}}$ is the  capture rate of the muon by the
nucleus, and  $Z_{\rm eff}$ is  the effective charge which  takes into
account coherent effects which can occur  in the nucleus $J$ due to its
finite  size. In  this analysis, the values  of  $Z_{\rm eff}$ quoted
in Ref \cite{Kitano2002} are used.  Like before, the  branching
ratio \eqref{RmueJ} possesses the most general form factor structure to
the leading order in  external masses and momenta and is relevant to most
models of New  Physics with CLFV.

The analytical results presented in this section are found to be consistent 
with the results given in Refs \cite{Ilakovac2000,Ilakovac2009,Alonso2013} 
in the $\nu_R$SM limit of the theory.

\section{Numerical Results}\label{numerics}

In this section, the numerical analysis of CLFV observables in the $\nu_R$MSSM
will be presented. In order to reduce the number of independent parameters, we
adopt the constrained framework of mSUGRA, discussed in Chapter \ref{theory}.
In detail, the model parameters are: (i) the usual SM parameters, such as gauge
coupling constants, the quark and charged-lepton Yukawa matrices inputted at the
scale $M_Z$, (ii) the heavy neutrino mass $m_N$ and the neutrino Yukawa matrix
${\bf h}_\nu$ evaluated at $m_N$, (iii) the universal mSUGRA parameters $m_0$,
$M_{1/2}$ and $A_0$ inputted at the GUT scale, and (iv) the ratio $\tan\beta$
of the Higgs VEVs and the sign of the superpotential Higgs-mixing parameter $\mu$.

The allowed ranges of the soft SUSY-breaking parameters $m_0$, $M_{1/2}$, $A_0$
and $\tan\beta$ are strongly constrained by a number of accelerator and cosmological
data \cite{Aad2012,Chatrchyan2012,Aad2012a,Chatrchyan2012a,Ellis2012,Chatrchyan2013}.
For definiteness, we consider the following set of input parameters:
\begin{equation}
\begin{array}{ll}
  \label{mSUGRA}
 \tan\beta = 10\,, \qquad &  m_0 = 1000~{\rm GeV}\,, \\
 A_0 = -3000~{\rm GeV}\,, \qquad & M_{1/2} = 1000~{\rm GeV}\,.
\end{array}
\end{equation}
Here the $\mu$ parameter is taken to be positive, whilst its absolute value
$|\mu |$ is derived form the minimization of the Higgs potential at the scale
$M_Z$. Using Refs
\cite{Carena2000,Lee2013,Lee2004,Heinemeyer2011},
one can verify that the parameter set \eqref{mSUGRA} predicts a SM-like Higgs
boson with $m_H\approx 125 \textrm{ GeV}$, in agreement with the recent discovery
at the LHC \cite{Aad2012,Chatrchyan2012a,Chatrchyan2013}, and is compatible with the current lower
limits on gluino and squark masses \cite{Aad2012a,Chatrchyan2012a,Chatrchyan2013}. The set \eqref{mSUGRA} is also
in agreement with having the lightest neutralino as the Dark Matter in the
Universe \cite{Ellis2012}.

We employ the one-loop RGE equations given in Refs \cite{Chankowski2002,Petcov2004}
to evolve the gauge coupling constants and the quark and charged lepton Yukawa
matrices from $M_Z$ to the GUT scale, while the heavy neutrino mass matrix ${\bf  m}_M$
and the neutrino Yukawa matrix ${\bf h}_\nu$ are evolved from the heavy neutrino mass
threshold $m_N$ to the GUT scale. Furthermore, we assume that the heavy neutrino-sneutrino
sector is approximately supersymmetric above $m_N$. For purposes of RGE 
evolution, this is a good
approximation for $m_N$ larger than the typical soft SUSY-breaking scale \cite{Ilakovac2009}.
At the GUT scale, the mSUGRA universality conditions are used to express the soft
SUSY-breaking masses, in terms of $m_0$, $M_{1/2}$ and $A_0$. Hence, all scalar masses
receive a soft SUSY-breaking mass  $m_0$, all gauginos are mass-degenerate to $M_{1/2}$,
and all scalar trilinear couplings are of the form ${\bf h}_x A_0$, with $x=u,d,l,\nu$,
where ${\bf h}_x$ are the Yukawa matrices at the GUT scale. The sneutrino mass matrix
acquires additional contributions from the heavy neutrino mass matrix. The sparticle mass
matrices and trilinear couplings are evolved from the GUT scale to $M_Z$, except for the
sneutrino masses which are evolved to the heavy neutrino threshold $m_N$. Having thus
obtained all sparticle and sneutrino mass matrices, one can numerically evaluate
all particle masses and interaction  vertices in the $\nu_R$MSSM, without approximations.

To simplify our numerical analysis, two representative scenarios of Yukawa textures
discussed in Chapter \ref{theory} are considered. Specifically, the first scenario
realizes the  U(1)-symmetric Yukawa texture in \eqref{YU1}, for  which we take either
$a=b$ and $c=0$, or $a=c$  and $b=0$, or $b=c$ and $a=0$, thus giving rise to CLFV
processes $\mu\to e X$, $\tau\to e X$ and $\tau\to\mu X$, respectively. Here $X$ stands
for the state(s) with zero net lepton number, e.g.\ $X=\gamma,\, e^+e^-,\,  \mu^+\mu^-,\,
q\overline{q}$. The second scenario is motivated by the $A_4$ group and uses the Yukawa
texture \eqref{YA4}, where the parameters $a$, $b$ and $c$ are taken to be all equal,
i.e.\ $a=b=c$.

The heavy neutrino mass scale $m_N$ strongly depends on the size of the symmetry-breaking
terms in the Yukawa matrix ${\bf  h}_\nu$. For instance, for the model described by Eq
\eqref{YU1}, the typical values of the $U(1)$-lepton-symmetry-breaking parameters
$\epsilon_l \equiv \epsilon_{e,\mu,\tau}$ consistent with low-scale resonant leptogenesis
is $\epsilon \stackrel{<}{{}_\sim} 10^{-5}$ \cite{Pilaftsis2005,Pilaftsis2005a,Deppisch2011},
leading to light-neutrino masses
\begin{eqnarray}
  \label{mNmax}
m_\nu\ \sim\ \frac{\epsilon^2_l v^2}{m_N} &\sim& 10^{-2}\, \textrm{eV}
 \Big(\frac{\epsilon_l}{10^{-6}}\Big)^2\
                            \Big(\frac{1\,\textrm{TeV}}{m_N}\Big)\ .
\end{eqnarray}

Taking into account the constraint $m_\nu  \stackrel{>}{{}_\sim} 10^{-1} \textrm{ eV}$
generically derived from neutrino oscillation data, we may estimate that the heavy
neutrino mass scale $m_N$ is typically restricted to be  less than $10 \textrm{ TeV}$,
for $\epsilon_l  = 10^{-5}$. If the assumption of successful low-scale leptogenesis is
relaxed, the symmetry-breaking parameters $\epsilon_l$ has only to be couple of orders
in magnitude smaller than the Yukawa parameters $a$, $b$ and $c$, with
$a,\,b,\,c \stackrel{<}{{}_\sim} 10$.  Thus, for $\epsilon_l <10^{-3}-10^{-2}$, the heavy
neutrino mass scale $m_N$ may be as large as $10^7-10^9 \textrm{ TeV}$, leading to the
decoupling of heavy neutrinos from low-energy observables. As the main interest of this
thesis is in the interplay between heavy neutrino, sneutrino and soft SUSY-breaking
contributions to CLFV observables, the focus will be only on the parameter space in which
$m_N < 10 \textrm{ TeV}$.

In the present analysis, we consider that the symmetry-preserving Yukawa parameters
$a$, $b$ and $c$ are limited by the perturbativity condition:
$\mbox{Tr}\,  {\bf h}^\dagger_\nu {\bf h}_\nu < 4\pi$, which is required to hold true for
the entire interval of the RGE evolution:
$\ln(M_Z/\textrm{TeV})< t < \ln(M_{\rm
  GUT}/\textrm{ TeV})$. For  the model described by Eq~\eqref{YU1}, this condition
translates into the constraint: $a<0.34$, and for the model described by Eq~\eqref{YA4},
to: $a<0.23$. For that reason, the numerical values for points in parameter space for which
the aforementioned perturbativity condition gets violated will not be displayed.

%%%%%%%%%%%%%%%%%%%%%%%%%%%%%%%%%%%%%%%%%%%%%%%%%%%%%%%%%%%%%%%%%%%%%%%
\begin{figure}[!ht]
 \centering
 \includegraphics[clip,width=0.40\textwidth]{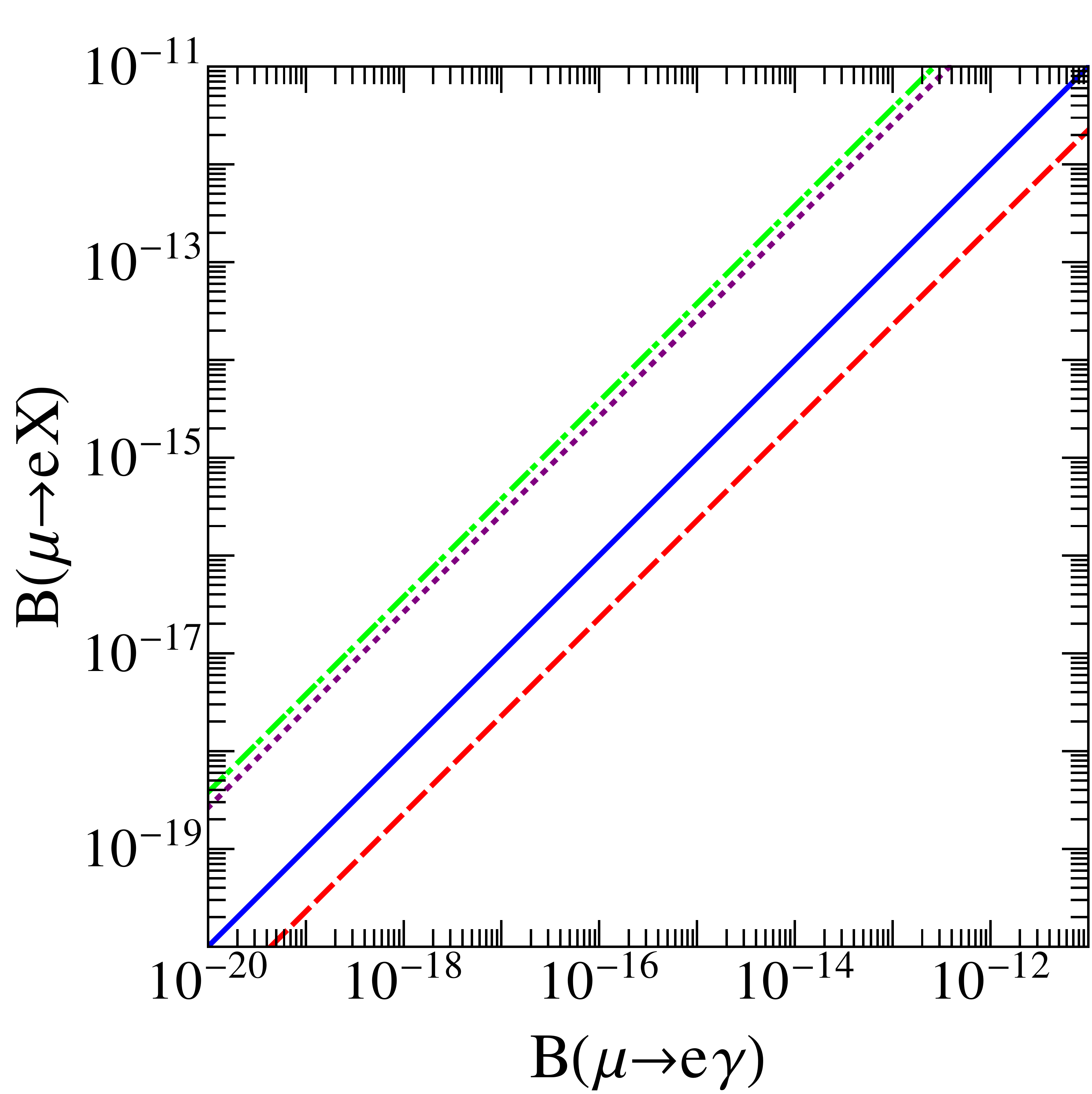}\hspace{1cm}
 \includegraphics[clip,width=0.39\textwidth]{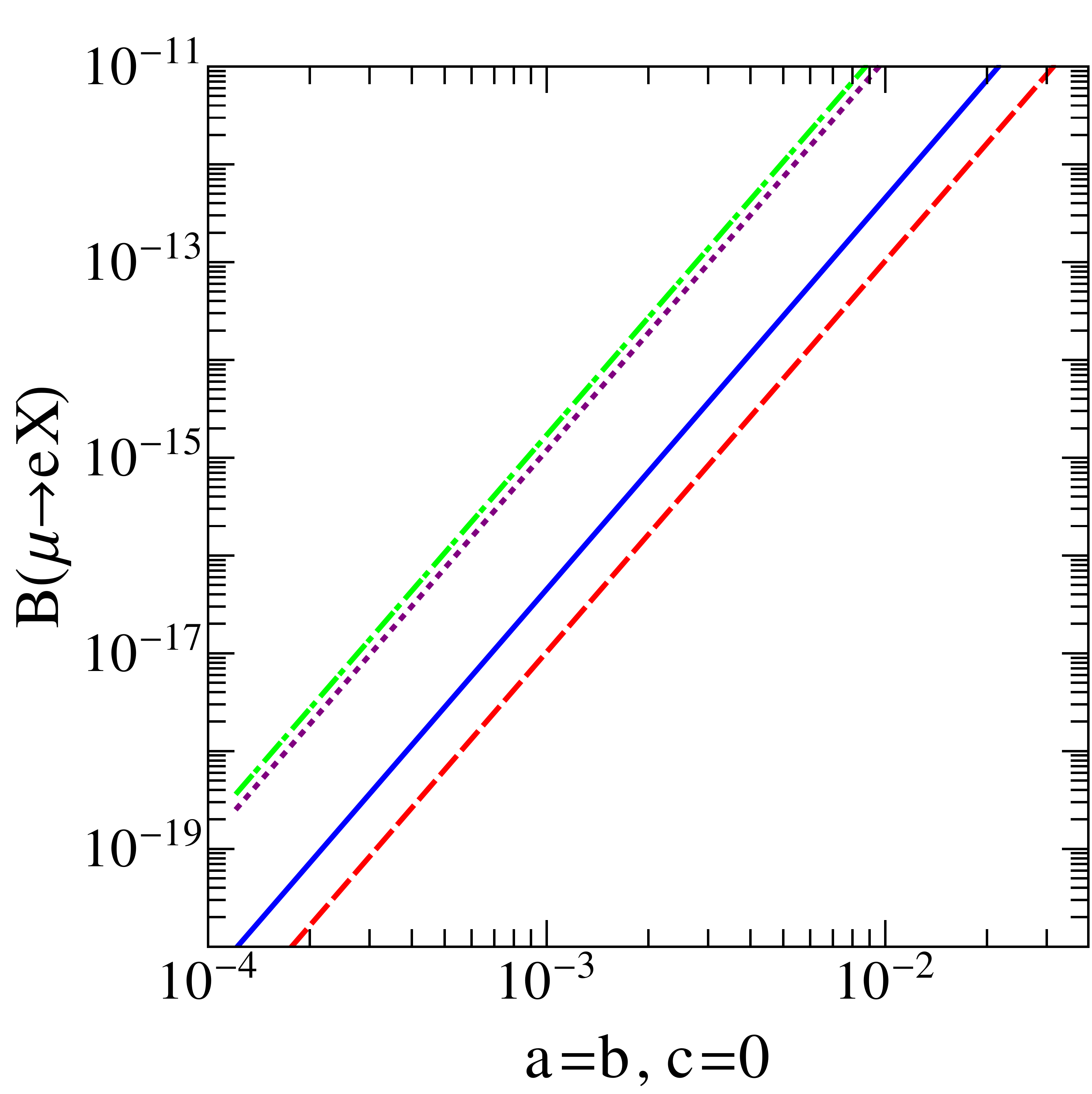}
 \\[.02\textwidth]
 \includegraphics[clip,width=0.40\textwidth]{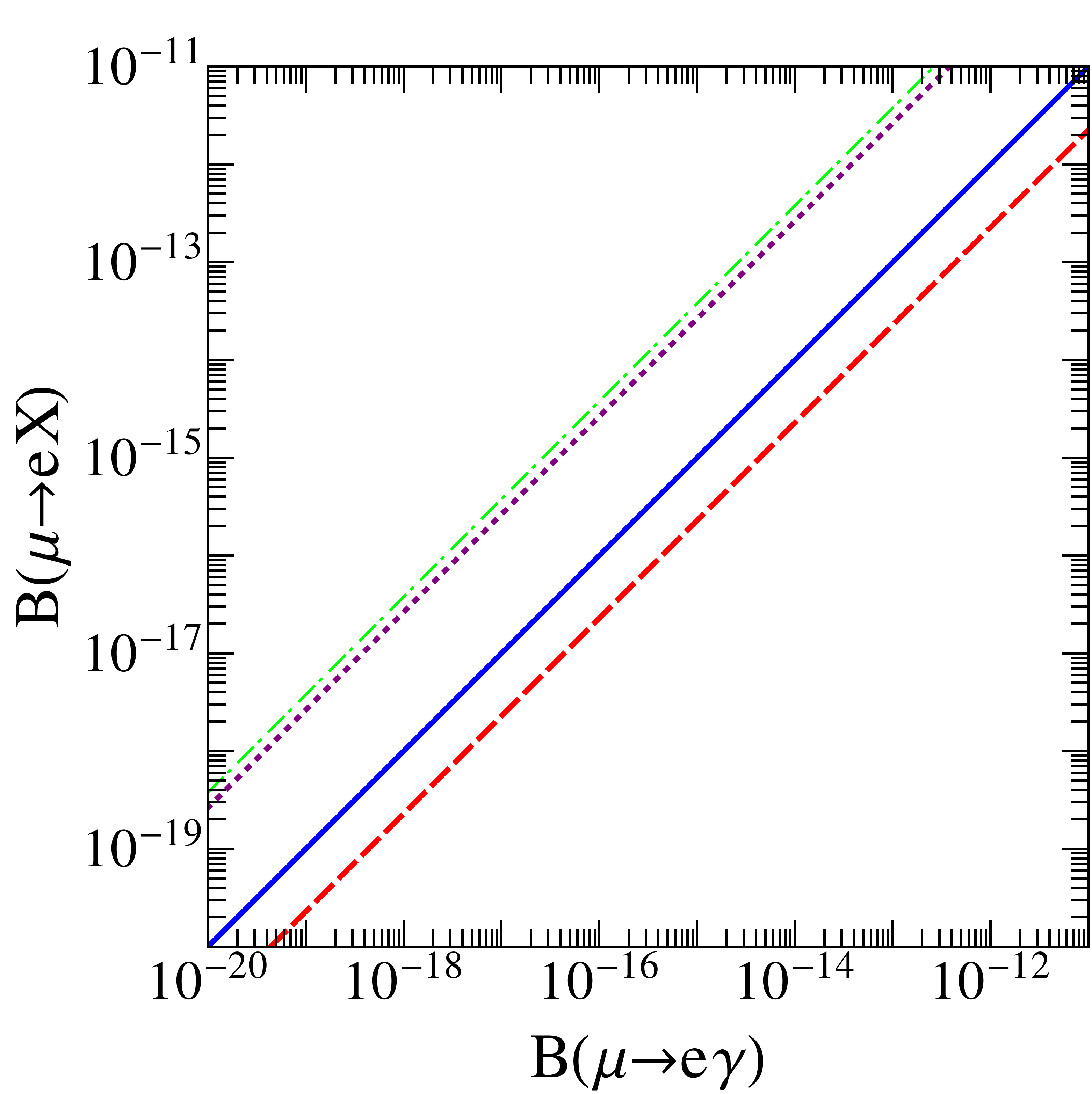}\hspace{1cm}
 \includegraphics[clip,width=0.39\textwidth]{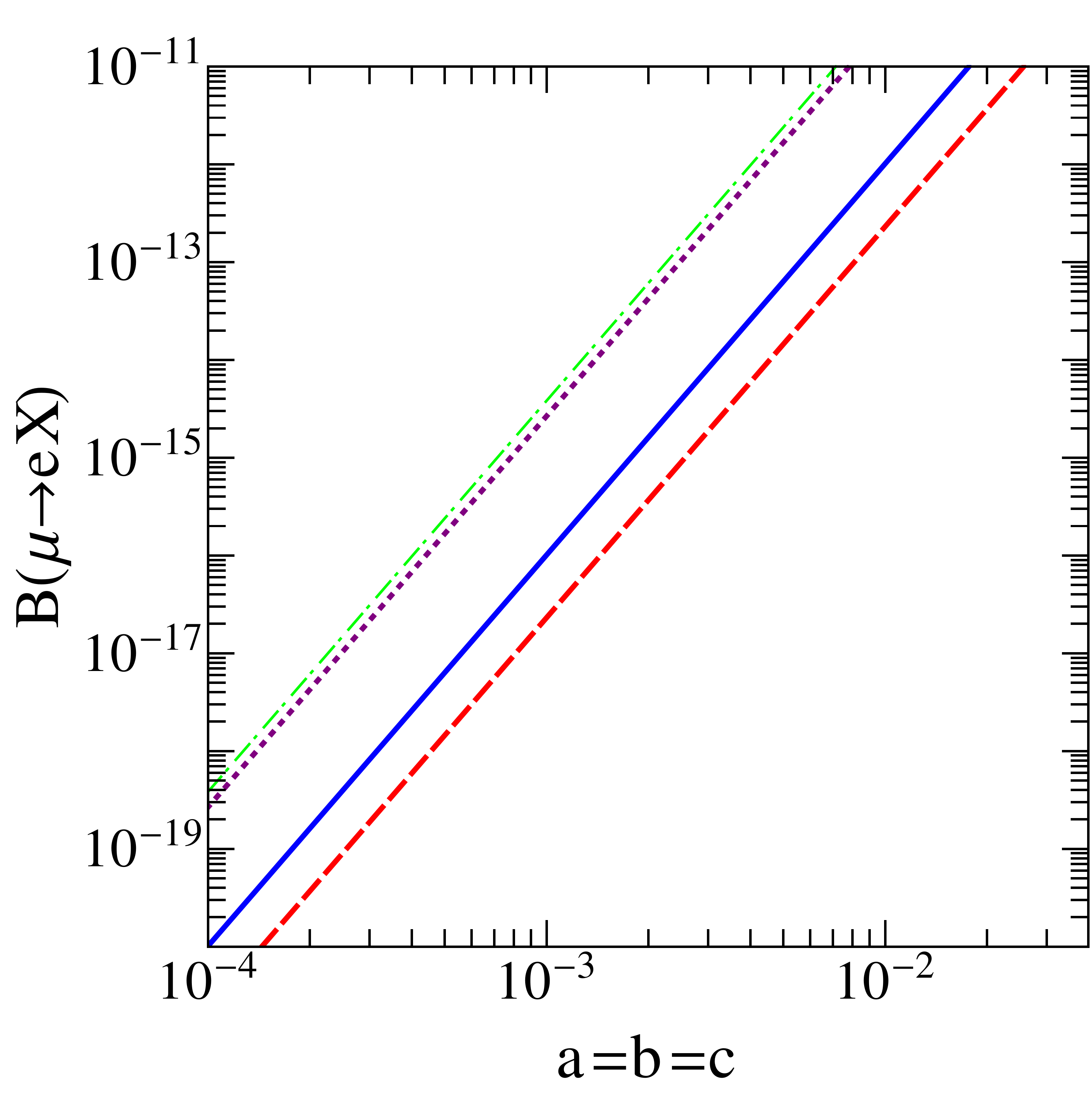}
\caption{Numerical  estimates of  $B(\mu\to e\gamma)$  [blue (solid)],
  $B(\mu\to  eee)$   [red  (dashed)],  $R^{\rm   Ti}_{\mu  e}$~[violet
    (dotted)]  and  $R^{\rm  Au}_{\mu  e}$~[green  (dash-dotted)],  as
  functions  of  $B(\mu\to e\gamma)$  (left  pannels)  and the  Yukawa
  parameter  $a$  (right pannels),  for  $m_N=400$~GeV and  $\tan\beta
  =10$.   The  upper two  pannels  correspond  to  the Yukawa  texture
  (\ref{YU1}), with $a=b$ and $c=0$,  and the lower two pannels to the
  Yukawa texture (\ref{YA4}), with $a=b=c$.}
\label{Fig3}
\end{figure}

\begin{figure}[!ht]
 \centering
 \includegraphics[clip,width=0.40\textwidth]{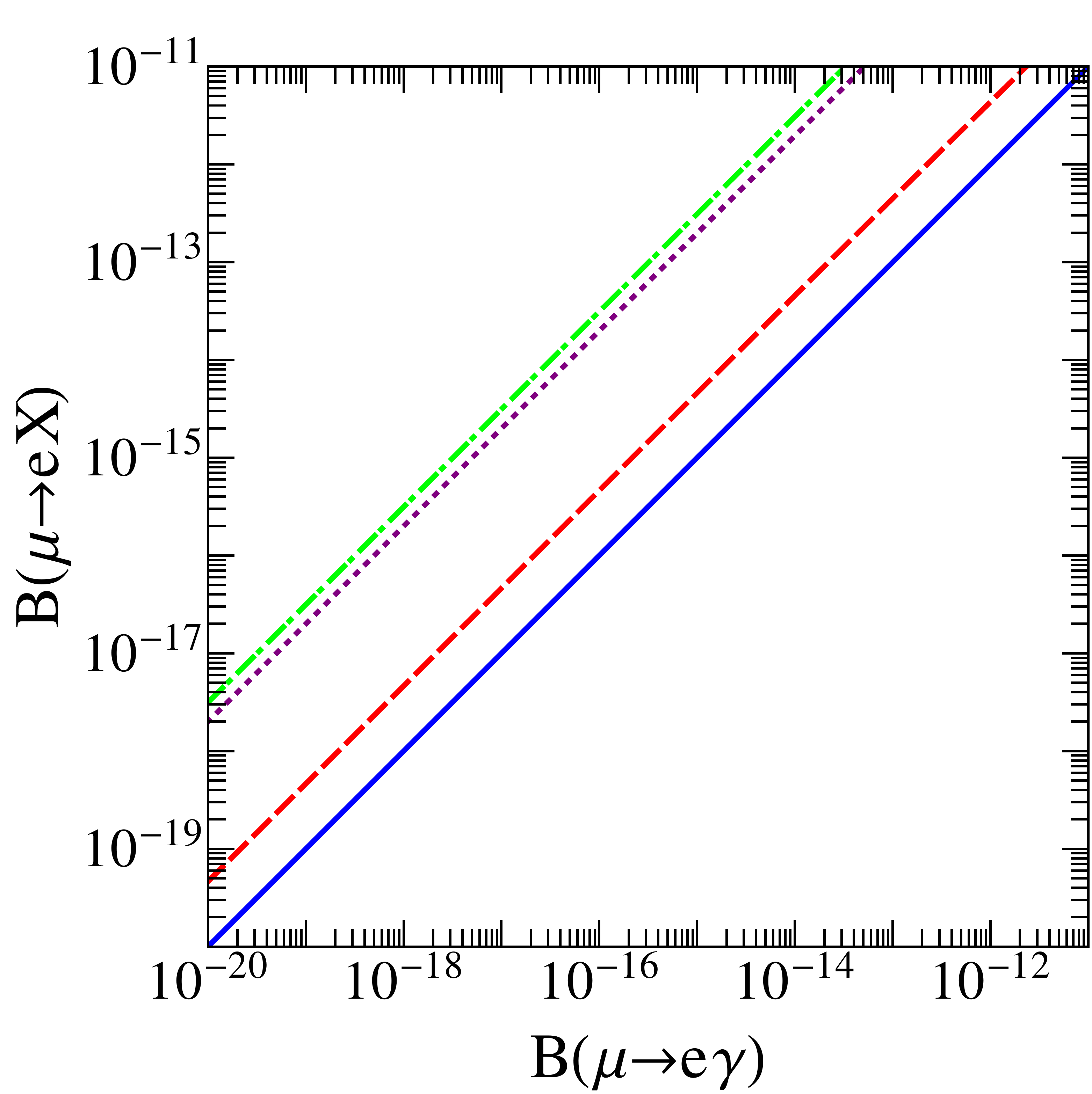}\hspace{1cm}
 \includegraphics[clip,width=0.39\textwidth]{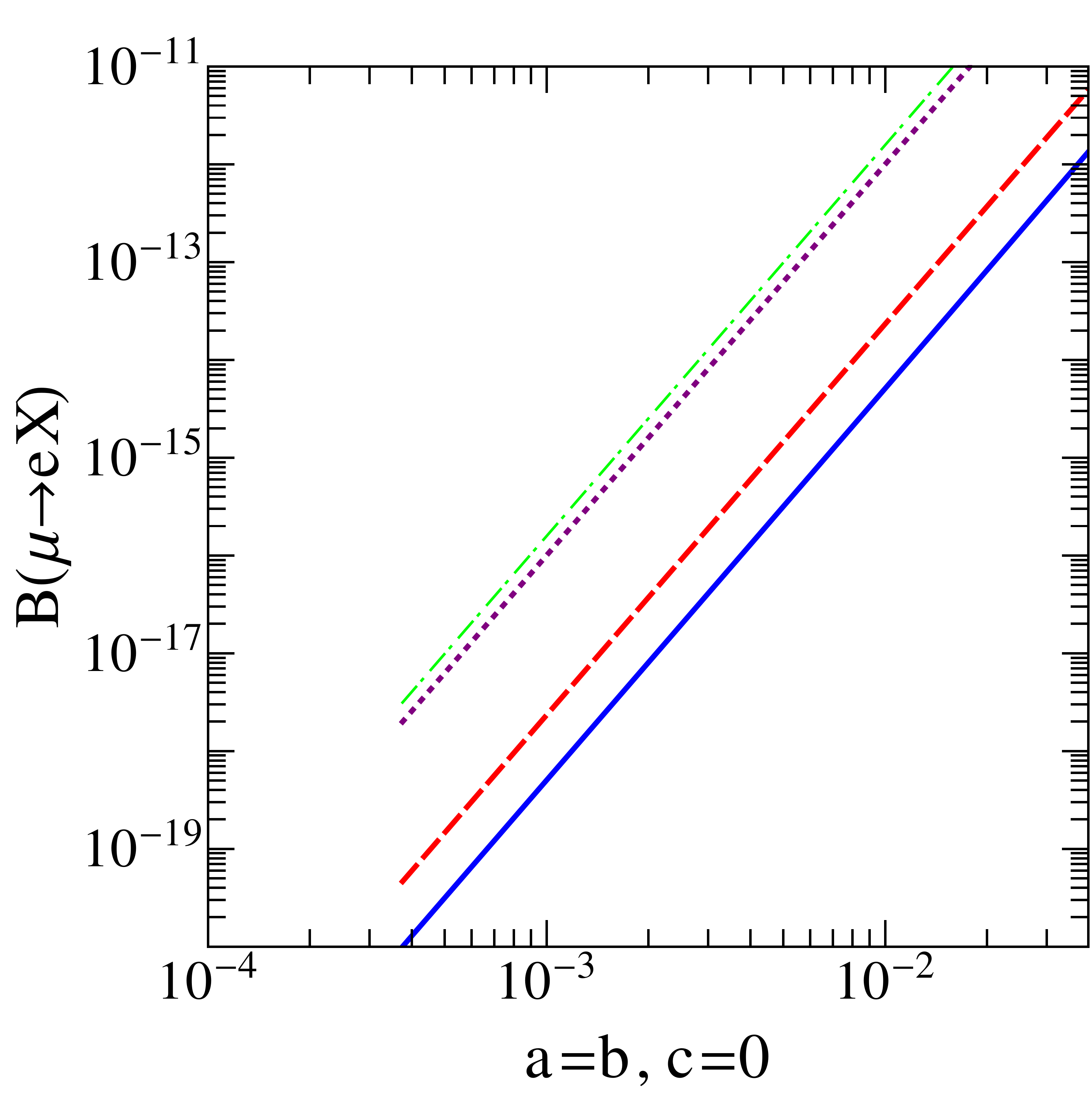}
 \\[.02\textwidth]
 \includegraphics[clip,width=0.40\textwidth]{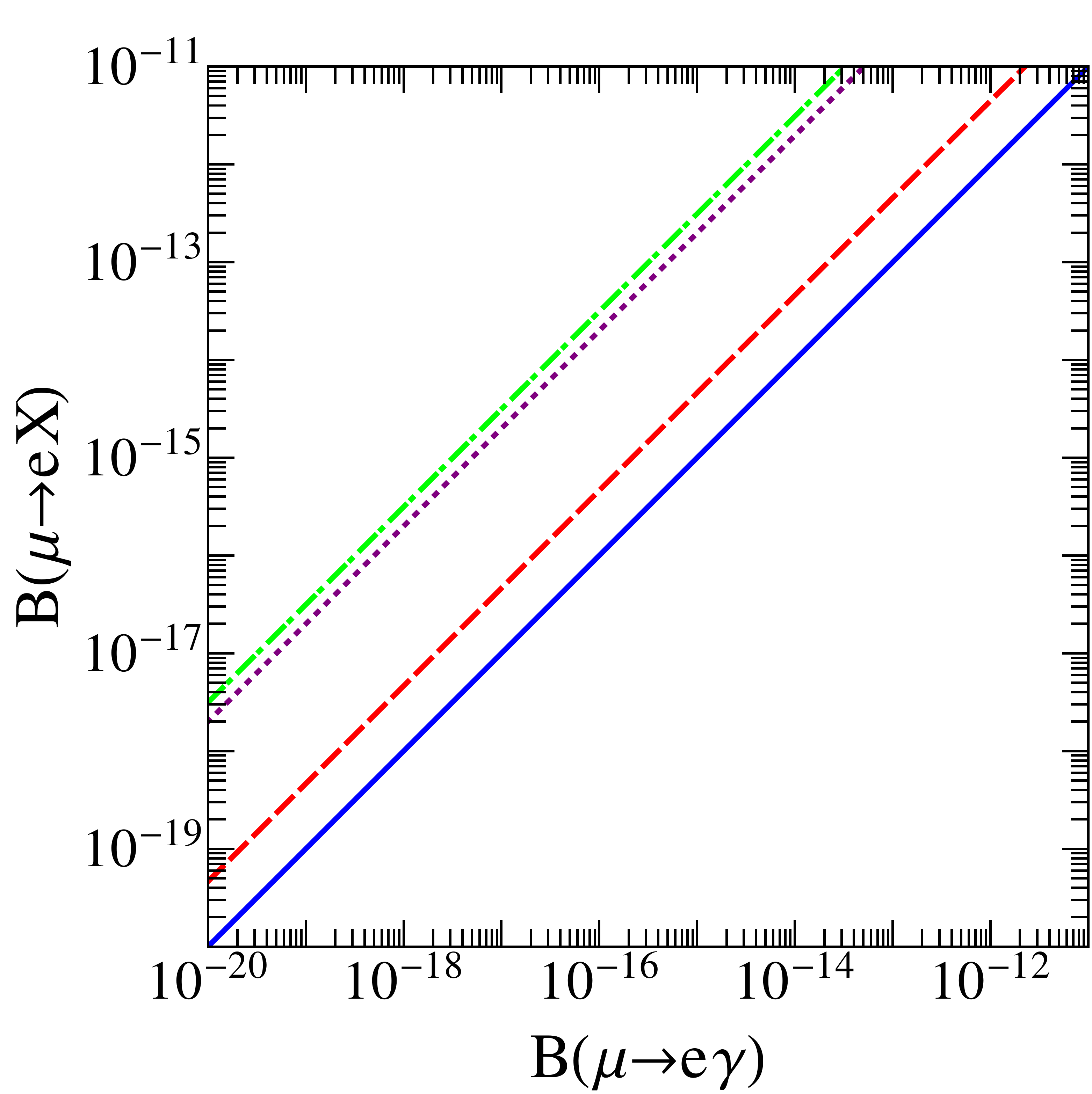}\hspace{1cm}
 \includegraphics[clip,width=0.39\textwidth]{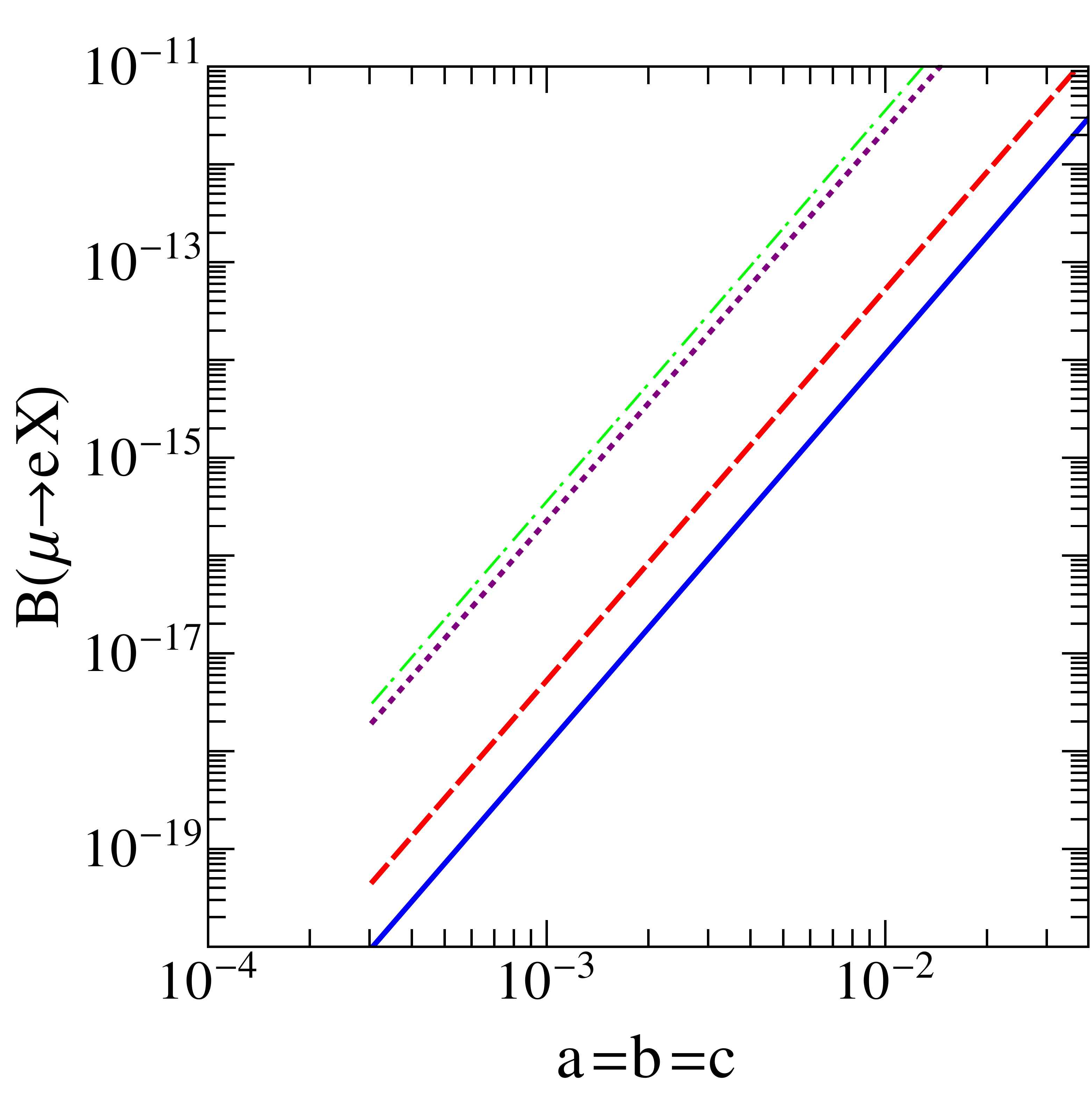}
\caption{The same as in Fig~\ref{Fig3}, but for $m_N=1$~TeV.}
\label{Fig4}
\end{figure}
%%%%%%%%%%%%%%%%%%%%%%%%%%%%%%%%%%%%%%%%%%%%%%%%%%%%%%%%%%%%%%%%%%%%%%%

In Fig~\ref{Fig3} are displayed numerical predictions for the $\mu$-LFV
observables $B(\mu \to e X)$: $B(\mu\to e\gamma)$ [blue (solid) line],
$B(\mu\to  eee)$ [red (dashed)  line], $R^{\rm Ti}_{\mu  e}$~[violet (dotted)
line] and $R^{\rm Au}_{\mu e}$ [green (dash-dotted) line], as functions of
$B(\mu\to e\gamma)$  (left pannels) and the Yukawa parameter $a$ (right pannels),
for $m_N=400$~GeV and $\tan\beta =10$. The upper two pannels assume the Yukawa
texture in \eqref{YU1}, with $a=b$ and $c=0$, whilst the lower two pannels correspond
to the Yukawa texture  in \eqref{YA4}, with $a=b=c$. In Fig~\ref{Fig4}, we give
numerical estimates for the same set of $\mu$-LFV observables, but for $m_N = 1$~TeV.
In Figs~\ref{Fig3} and \ref{Fig4}, the Yukawa parameter $a$ has been chosen, such that
$10^{-20}<B(\mu\to e\gamma)<10^{-10}$. Such a range of values includes both the present
\cite{Adam2011,Bellgardt1988,Dohmen1993,Bertl2006,Hayasaka2010,Aubert2010}
and future \cite{Abrams2012,Kutschke2011,Dukes2011,Kuno2005,Barlow2011,Bona2007,
Mu2e,XProject,Ritt2006}
experimental limits. As can be seen from  Figs~\ref{Fig3} and \ref{Fig4},the CLFV
observables under study depend quadratically on the Yukawa parameter $a$, namely they
are proportional to $a^2$. Instead, the quartic Yukawa terms proportional to $a^4$
\cite{Ilakovac2009} remain always small, which is a consequence of the imposed
perturbativity constraint: $\textrm{Tr}({\bf  h}_\nu^\dagger{\bf  h}_\nu)<4\pi$,
up to the GUT scale.

%%%%%%%%%%%%%%%%%%%%%%%%%%%%%%%%%%%%%%%%%%%%%%%%%%%%%%%%%%%%%%%%%%%%%%
\begin{figure}[!ht]
 \centering
 \includegraphics[clip,width=0.4\textwidth]{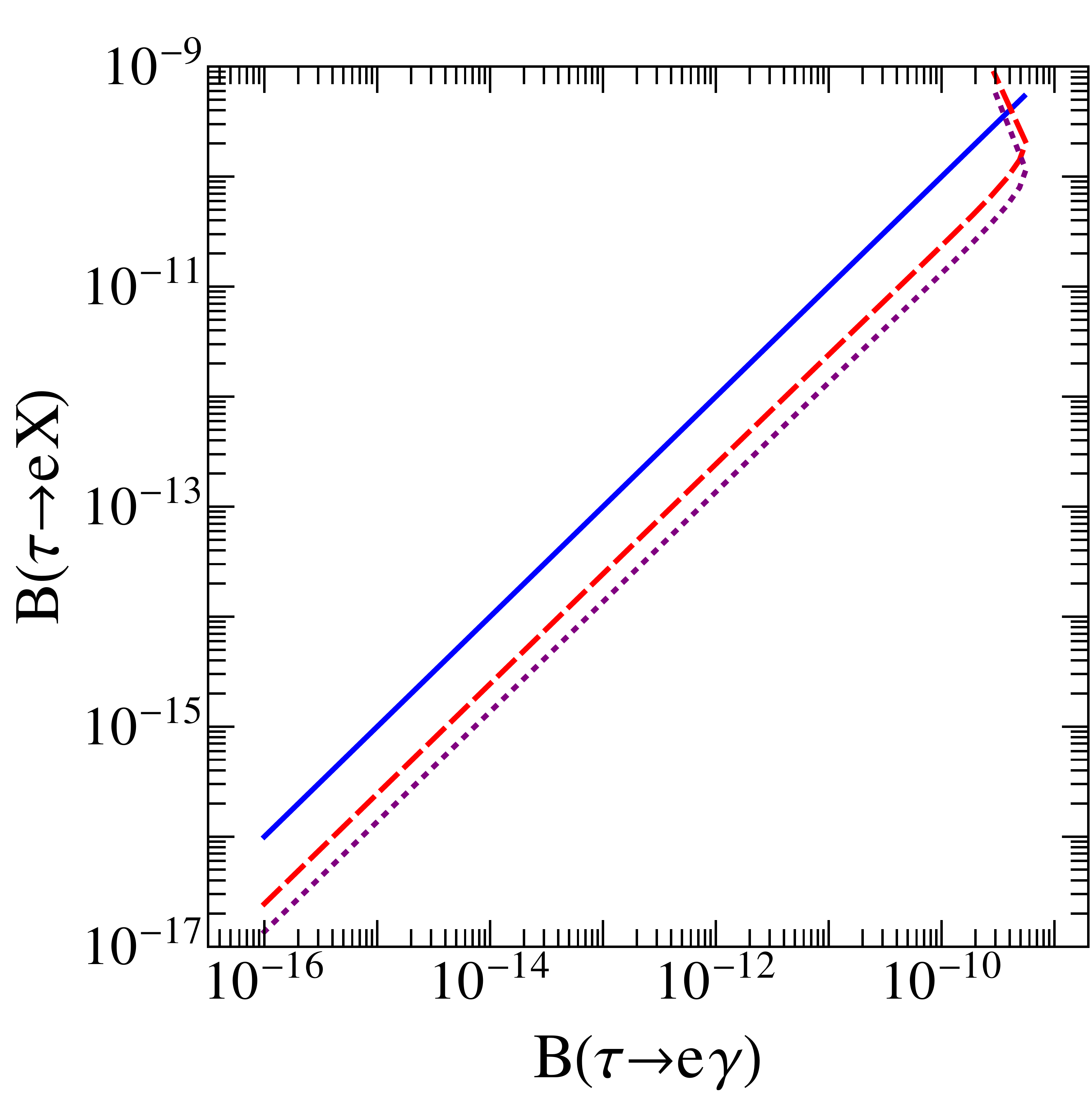}\hspace{1cm}
 \includegraphics[clip,width=0.4\textwidth]{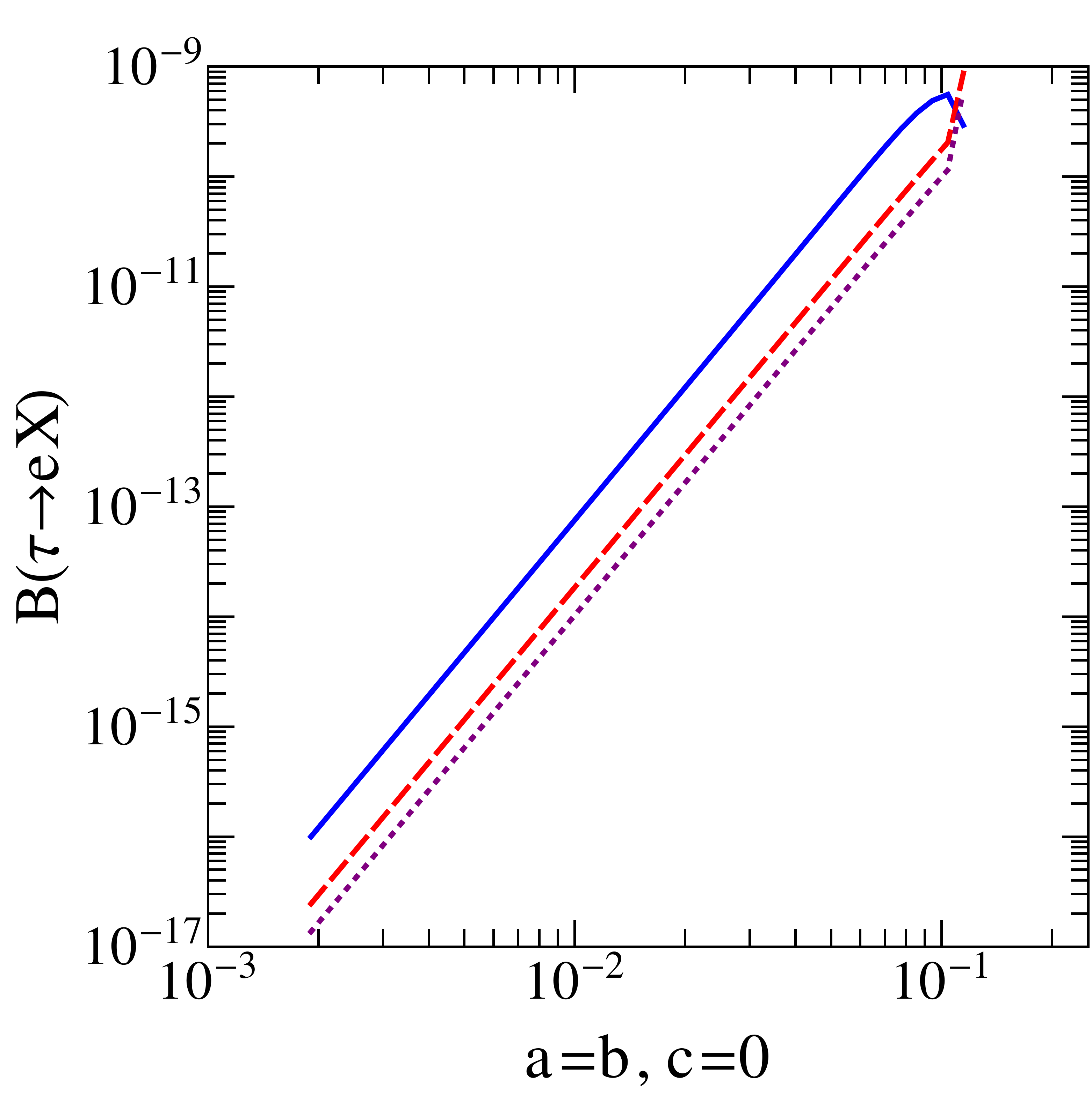}
 \\[.02\textwidth]
 \includegraphics[clip,width=0.4\textwidth]{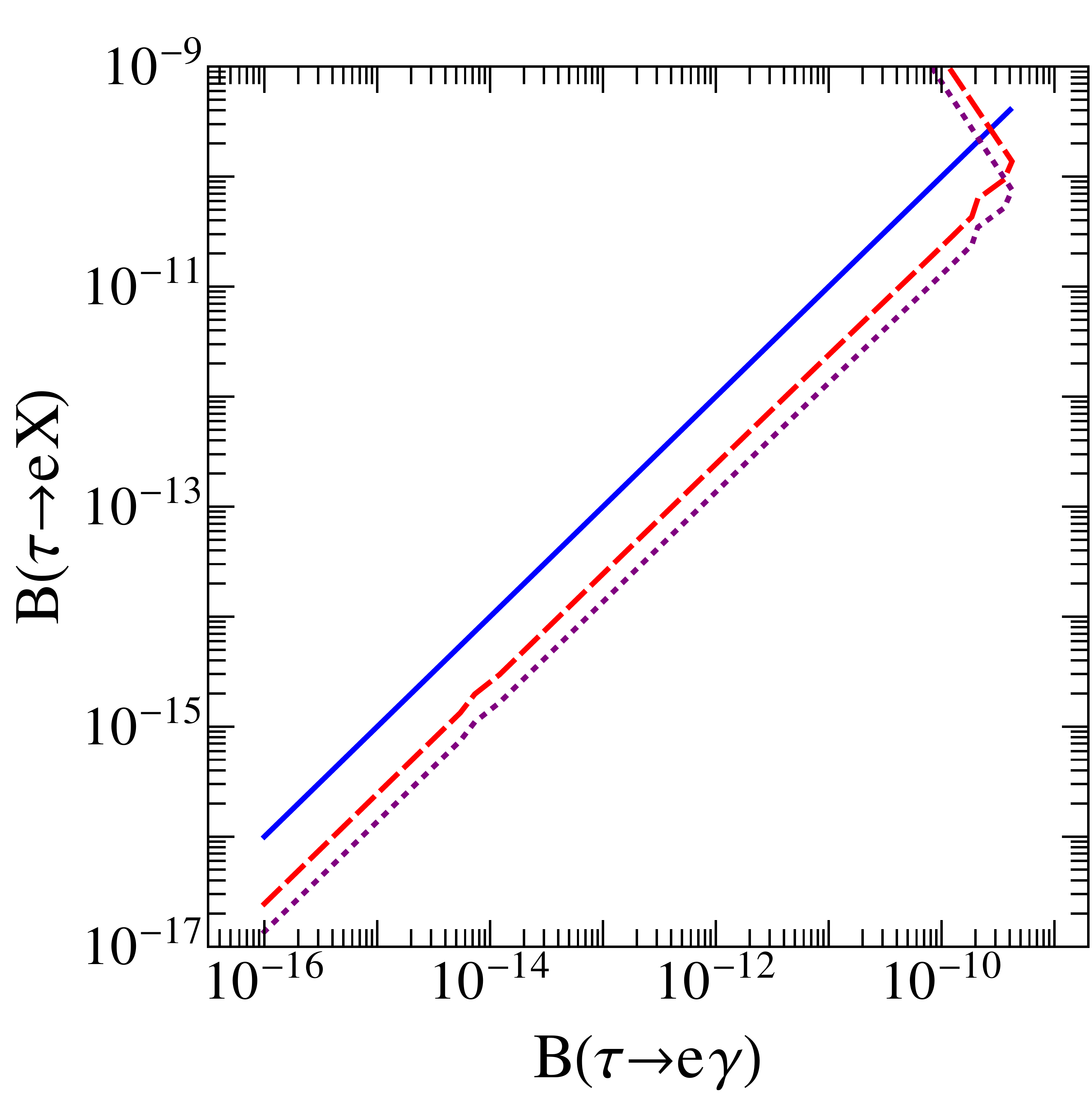}\hspace{1cm}
 \includegraphics[clip,width=0.4\textwidth]{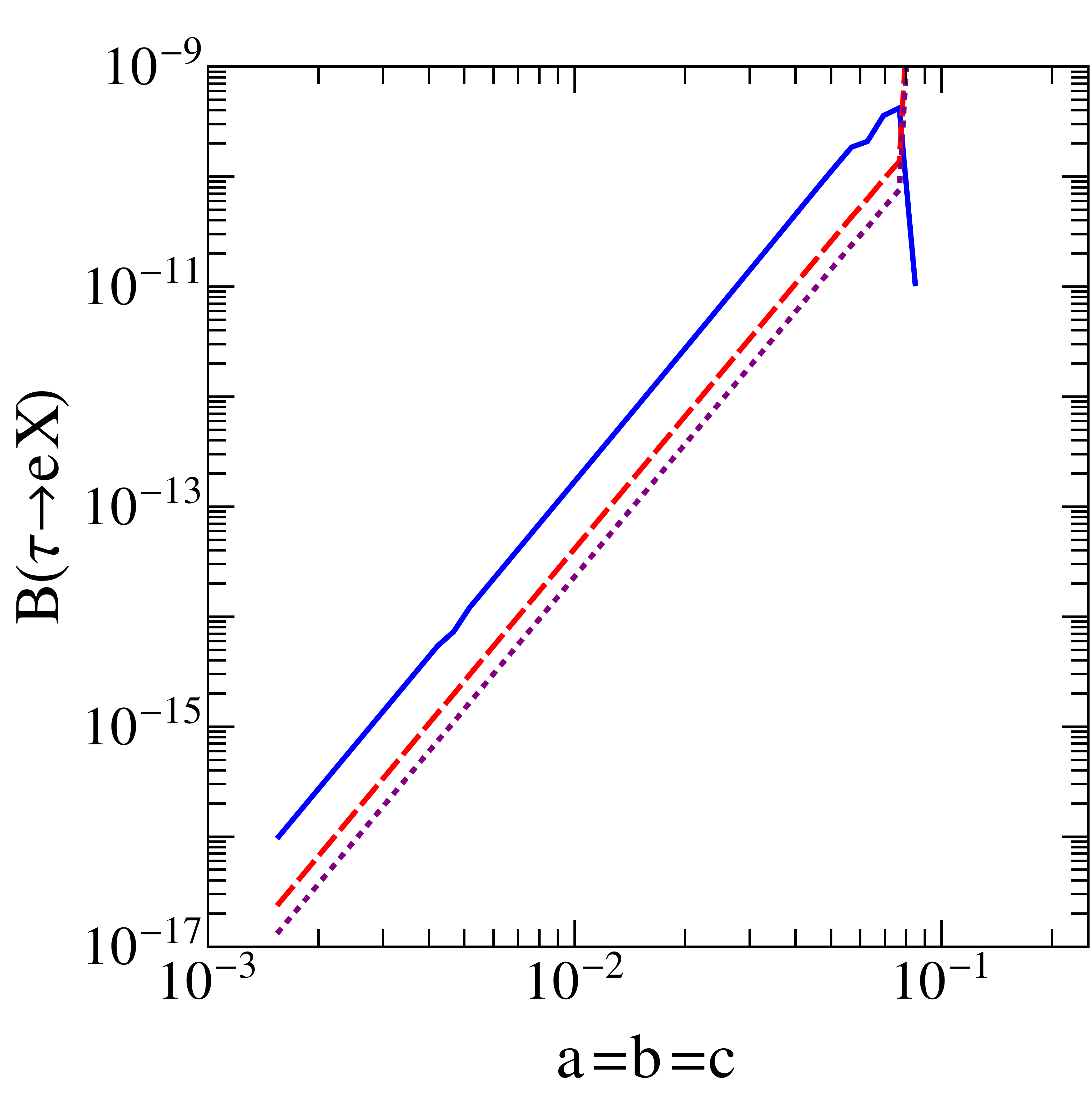}
\caption{Numerical estimates of $B(\tau\to e\gamma)$ [blue (solid)],
  $B(\tau\to eee)$ [red (dashed)] and $B(\tau\to e\mu\mu)$ [violet
    (dotted)], as functions of $B(\tau\to e\gamma)$ (left pannels) and
  the Yukawa parameter $a$ (right pannels), for $m_N=400$~GeV and
  $\tan\beta=10$.  The upper pannels present predictions for the Yukawa
  texture (\ref{YU1}), with $a=c$ and $b=0$, and the lower pannels for
  the Yukawa texture (\ref{YA4}), with $a=b=c$.}
\label{Fig5}
\end{figure}

\begin{figure}[!ht]
 \centering
 \includegraphics[clip,width=0.4\textwidth]{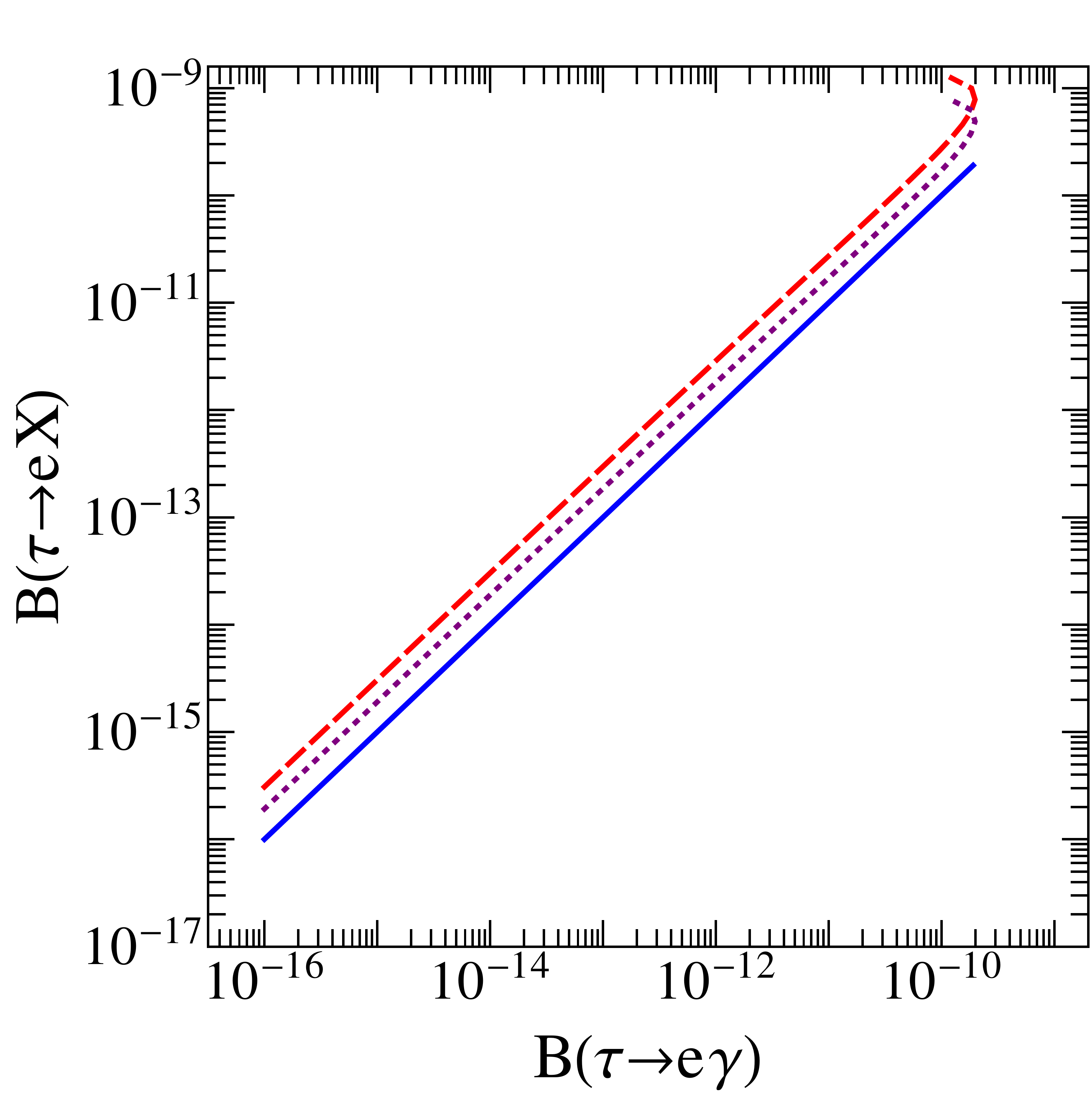}\hspace{1cm}
 \includegraphics[clip,width=0.4\textwidth]{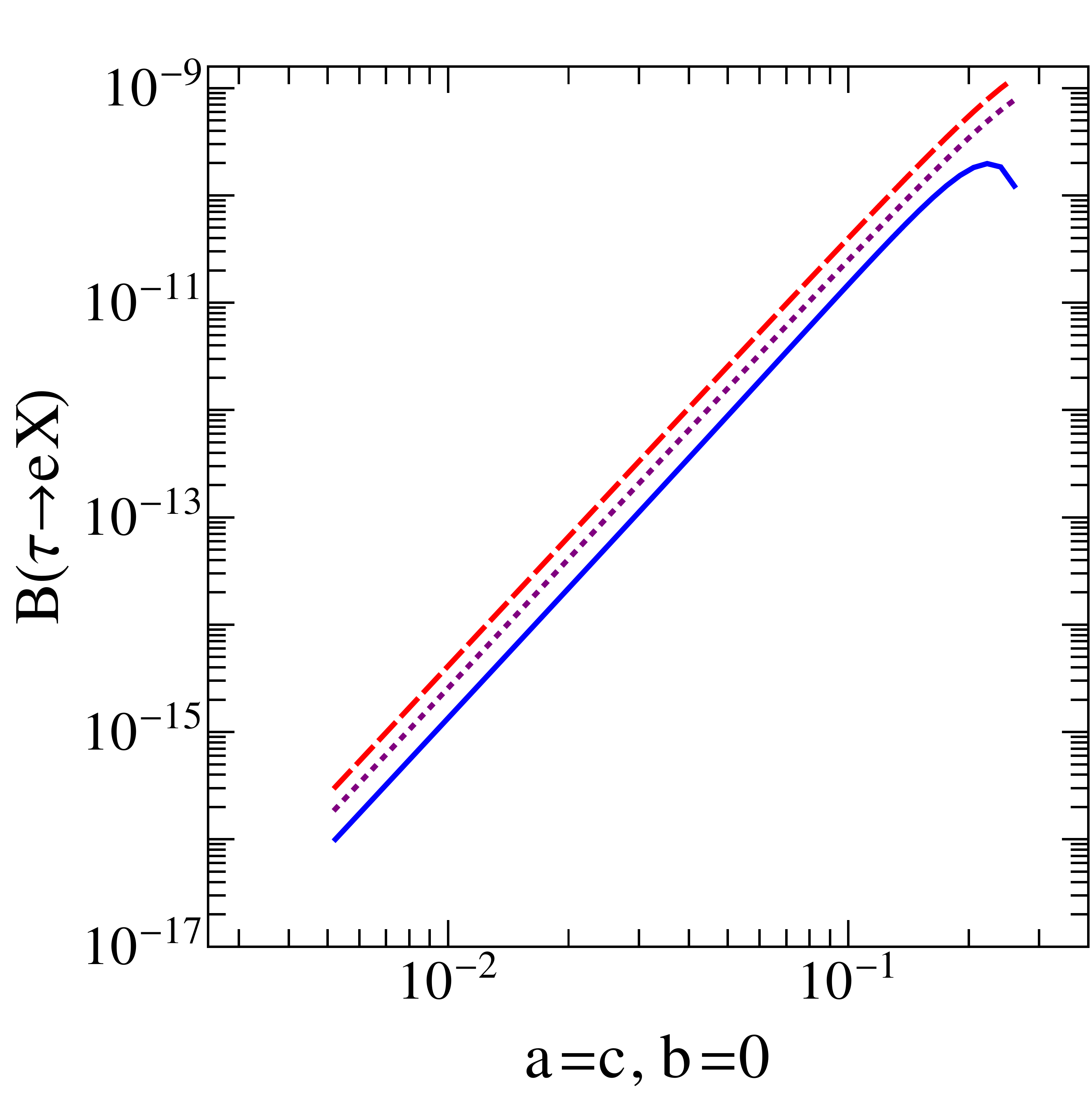}
 \\[.02\textwidth]
 \includegraphics[clip,width=0.4\textwidth]{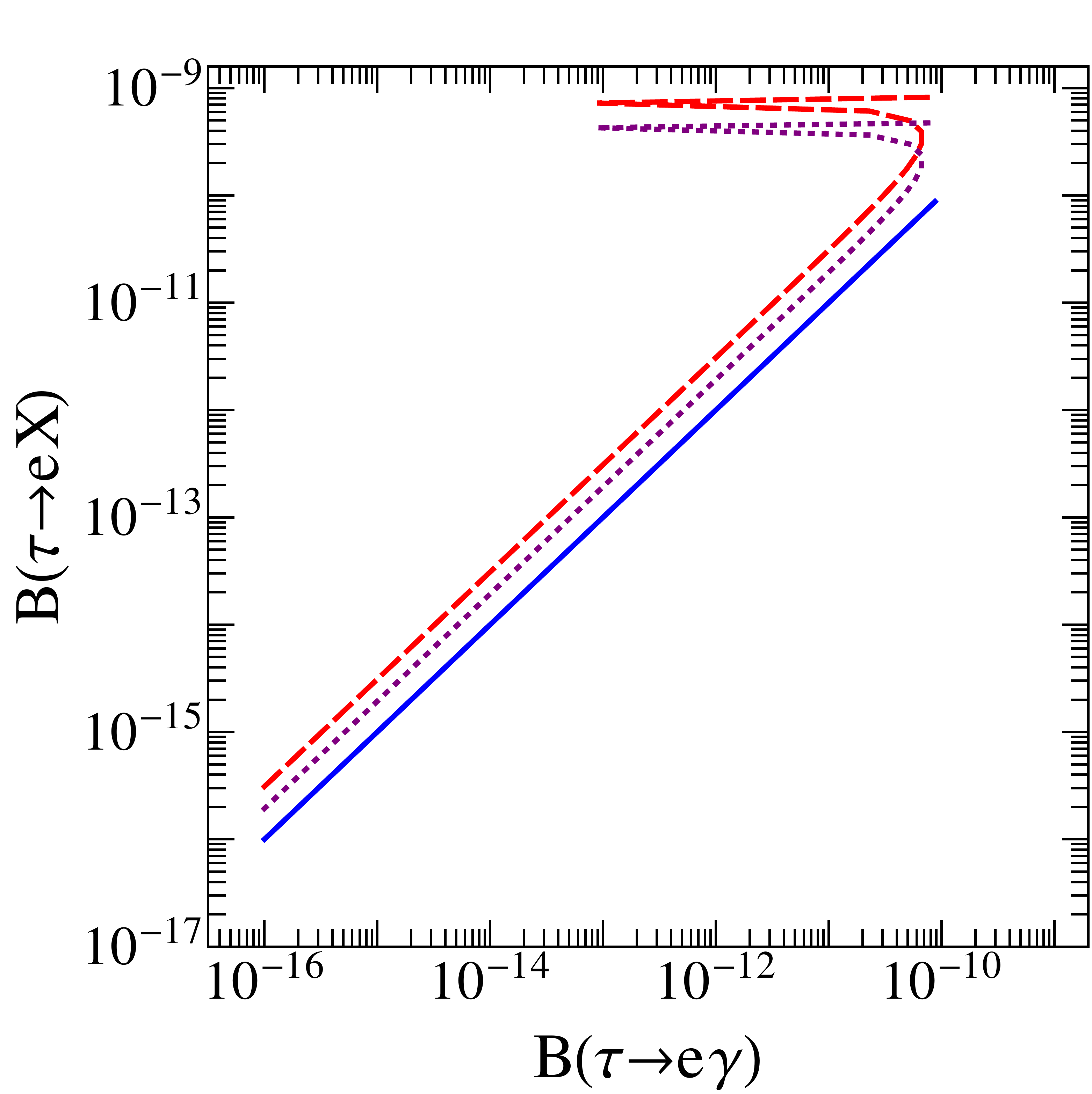}\hspace{1cm}
 \includegraphics[clip,width=0.4\textwidth]{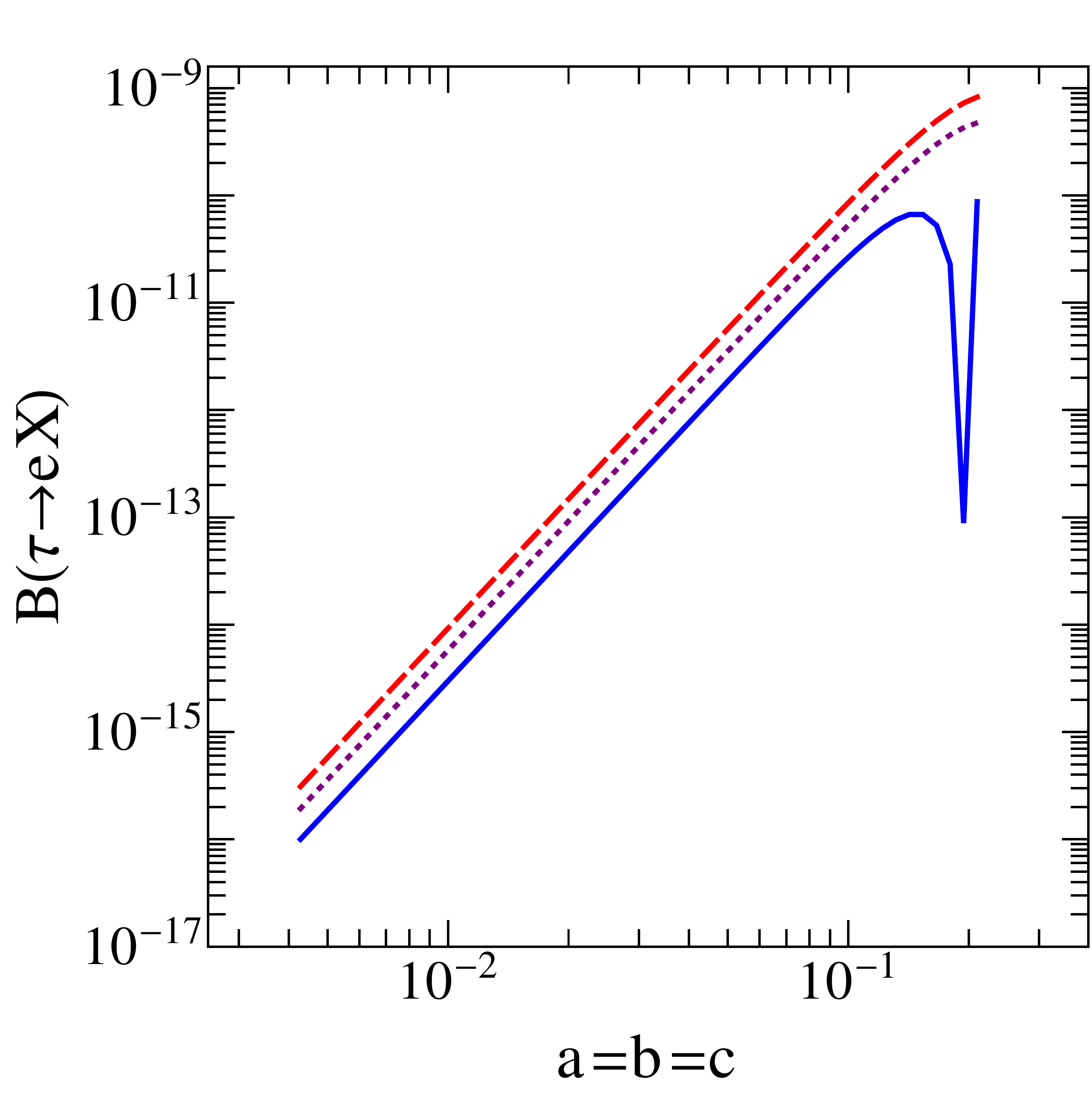}
\caption{The same as in Fig~\ref{Fig5}, but for $m_N=1$~TeV.}
\label{Fig6}
\end{figure}
%%%%%%%%%%%%%%%%%%%%%%%%%%%%%%%%%%%%%%%%%%%%%%%%%%%%%%%%%%%%%%%%%%%%%%

By analogy,  Figs~\ref{Fig5} and \ref{Fig6} present numerical estimates of the
$\tau$-LFV  observables $B(\tau \to e X)$: $B(\tau\to e\gamma)$
[blue (solid) lines],  $B(\tau\to eee)$ [red (dashed) lines] and $B(\tau\to e\mu\mu)$
[violet  (dotted) lines], as functions of $B(\tau\to  e\gamma)$ (left pannels) and
the Yukawa parameter $a$ (right pannels), for $m_N=400 \textrm{ GeV}$  and $m_N=1 \textrm{ TeV}$,
respectively. The predictions for the fully complementary observables $B(\tau \to \mu  X)$:
$B(\tau\to  \mu\gamma)$, $B(\tau\to \mu\mu\mu)$  and $B(\tau\to  \mu  ee)$ are not
displayed. The upper pannels give our predictions for the Yukawa texture \eqref{YU1},
with $a=c$ and $b=0$, and the lower pannels for the Yukawa texture \eqref{YA4}, with $a=b=c$.
In  both  Figs~\ref{Fig5} and \ref{Fig6}, the  Yukawa parameter $a$ has been chosen, such that  
$10^{-16}<B(\tau\to e\gamma)<10^{-7}$. 
As can be seen from Figs~\ref{Fig5} and \ref{Fig6},
all observables $B(\tau \to e X)$ of $\tau$-LFV (with $X = \gamma,\, ee,\,  \mu\mu$) exhibit
similar quadratic dependence on the small Yukawa parameter $a$. However, close to the largest
perturbatively  allowed values of $a$, i.e.\ $a\stackrel{<}{{}_\sim}0.34$ for the  model
\eqref{YU1} and $a\stackrel{<}{{}_\sim} 0.23$ for the model \eqref{YA4}, some of the observables
of $\tau$-LFV exhibit either numerical instability, or the existence of a cancellation region in
parameter space, as will be seen below.

%%%%%%%%%%%%%%%%%%%%%%%%%%%%%%%%%%%%%%%%%%%%%%%%%%%%%%%%%%%%%%%%%%%%%%
\begin{figure}[!ht]
 \centering
 \includegraphics[clip,width=0.4\textwidth]{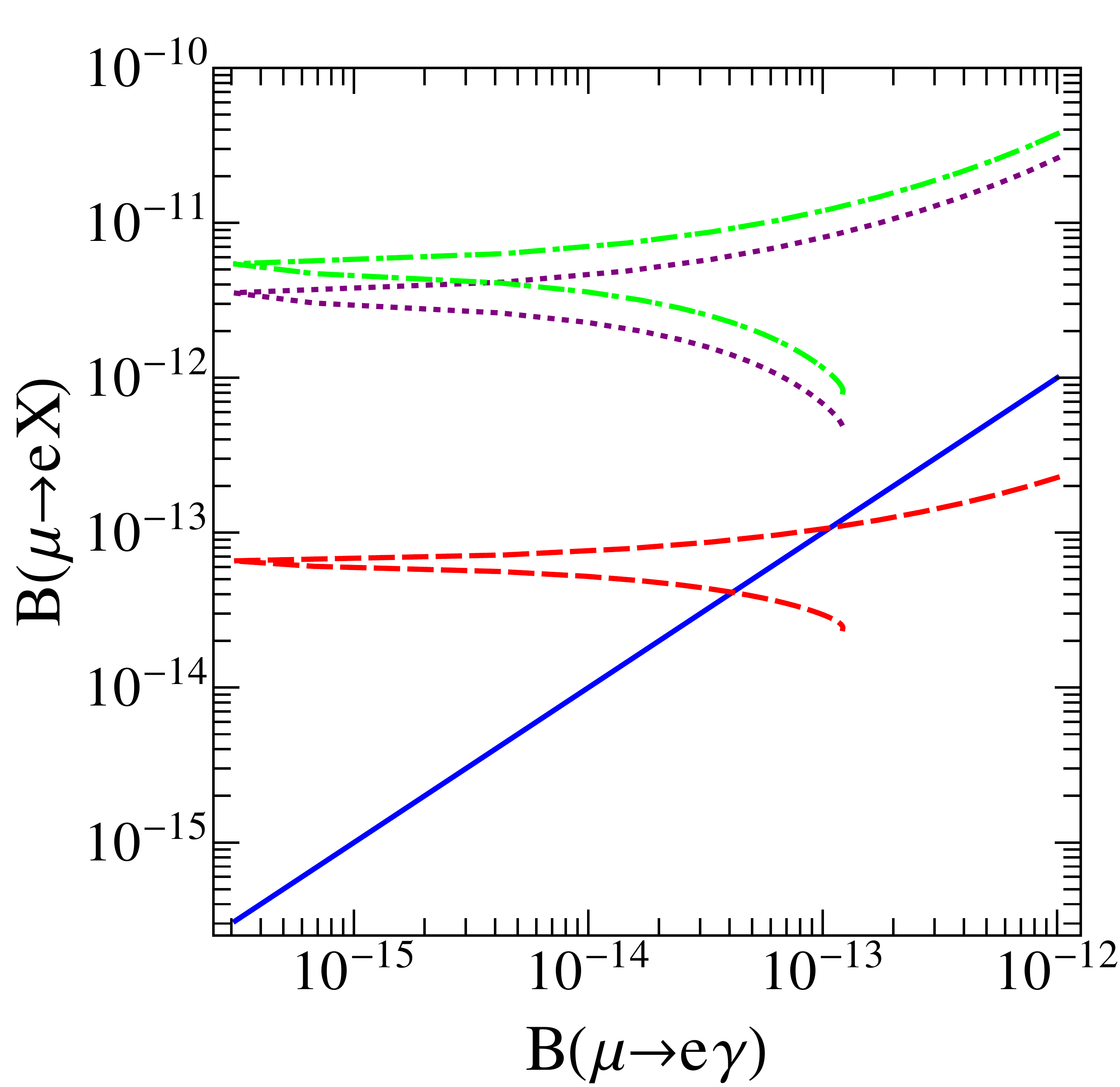}\hspace{1cm}
 \includegraphics[clip,width=0.4\textwidth]{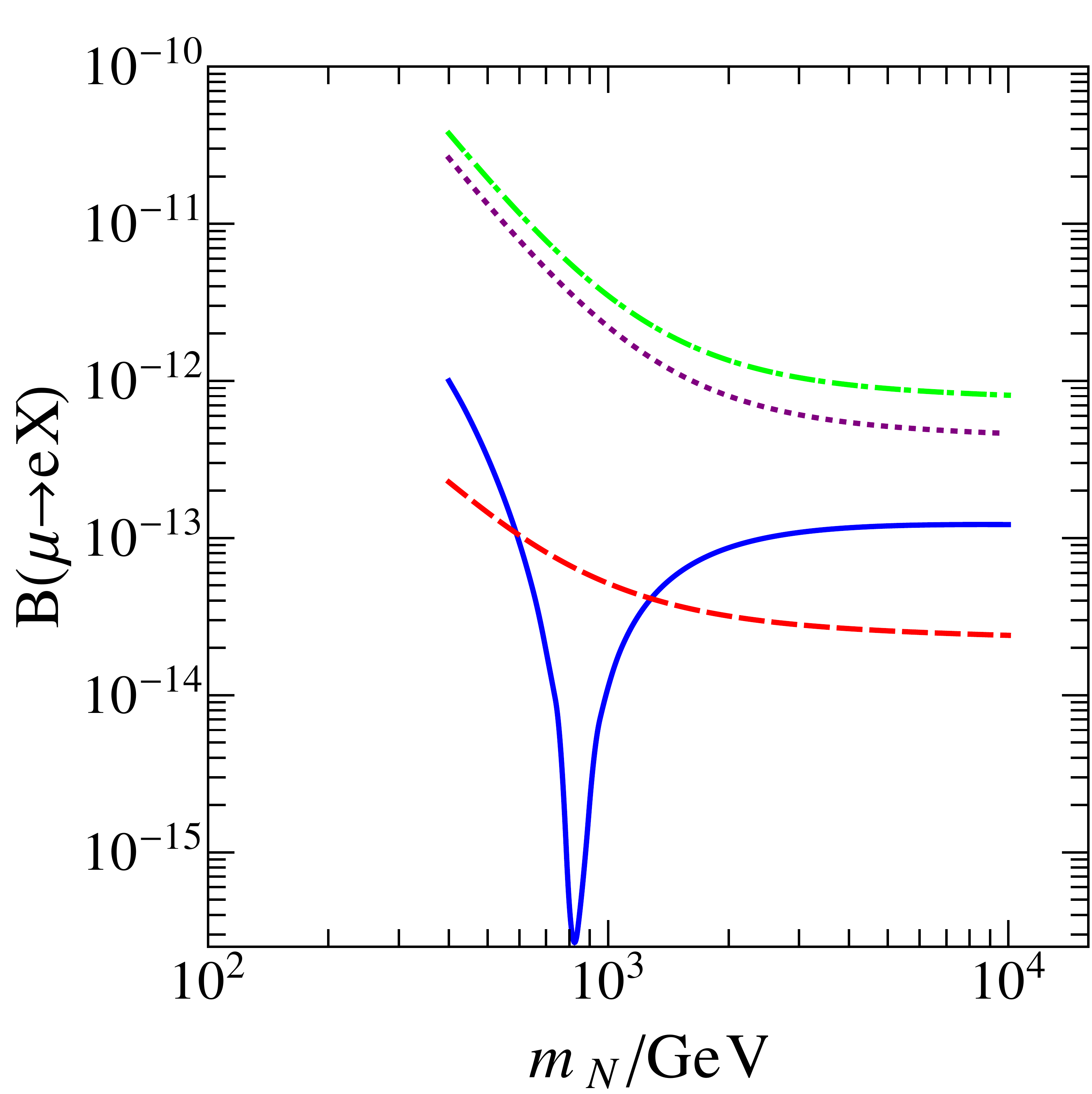}
 \\[.02\textwidth]
 \includegraphics[clip,width=0.4\textwidth]{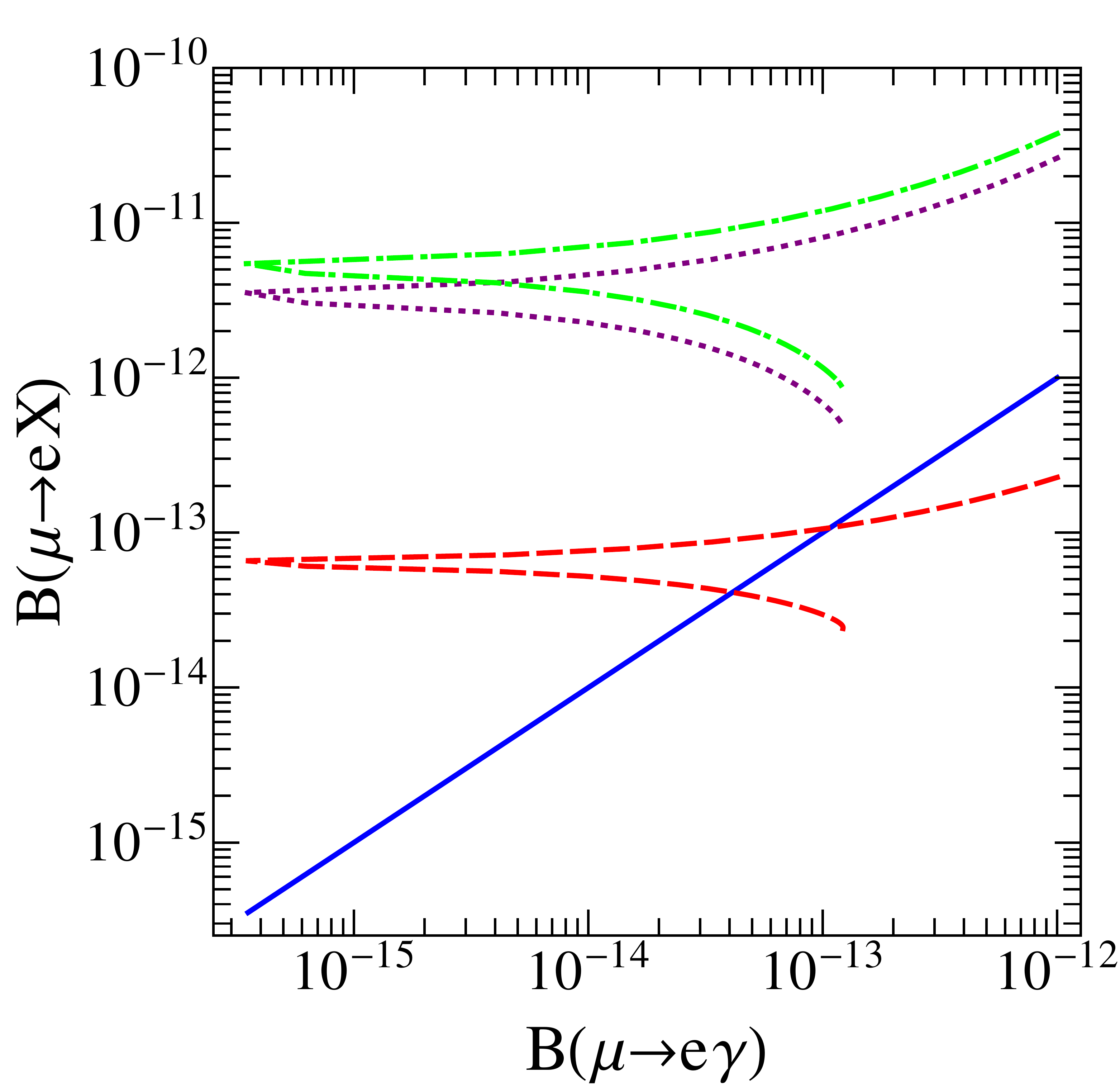}\hspace{1cm}
 \includegraphics[clip,width=0.4\textwidth]{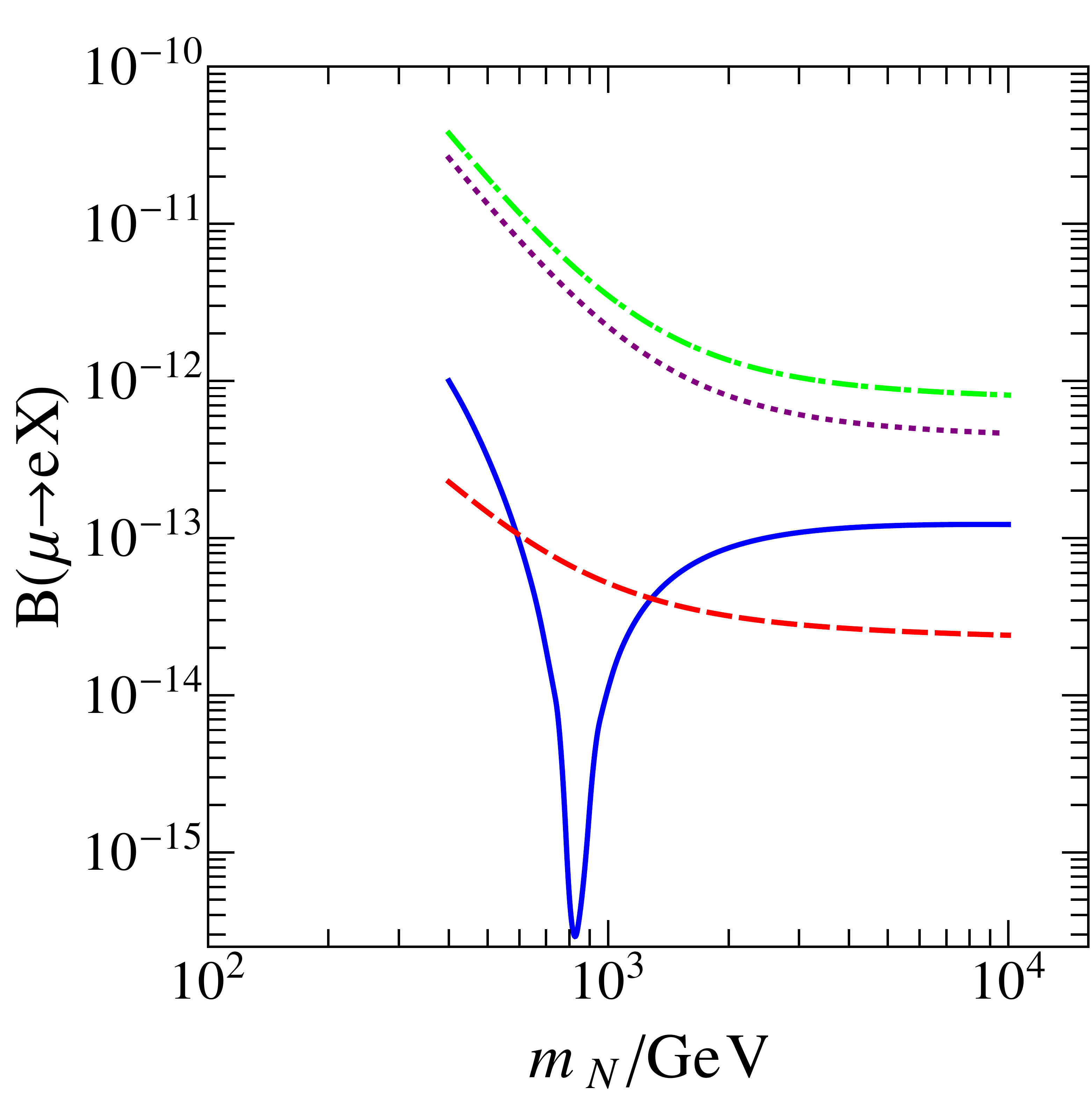}
\caption{Numerical  estimates of  $B(\mu\to e\gamma)$  [blue (solid)],
  $B(\mu\to  eee)$   [red  (dashed)],  $R^{\rm   Ti}_{\mu  e}$~[violet
    (dotted)]  and  $R^{\rm  Au}_{\mu  e}$~[green  (dash-dotted)],  as
  functions  of  $B(\mu\to  e\gamma)$  (left pannels)  and  the  heavy
  neutrino  mass scale  $m_N$ (right  pannels).  In  all  pannels, the
  Yukawa parameter  $a$ was  kept fixed by  the condition  $B(\mu\to e
  \gamma)=10^{-12}$  for $m_N  = 400$~GeV,  and $\tan\beta  =  10$ was
  used.   The upper pannels  display numerical  values for  the Yukawa
  texture~(\ref{YU1}), with $a=b$ and $c=0$, and the lower pannels for
  the Yukawa texture (\ref{YA4}), with $a=b=c$.}
\label{Fig7}
\end{figure}
%%%%%%%%%%%%%%%%%%%%%%%%%%%%%%%%%%%%%%%%%%%%%%%%%%%%%%%%%%%%%%%%%%%%%%

Figure \ref{Fig7} presents numerical estimates of $B(\mu\to e\gamma)$
[blue (solid) line], $B(\mu\to  eee)$ [red  (dashed) line],
$R^{\rm Ti}_{\mu e}$ [violet (dotted) line] and $R^{\rm Au}_{\mu e}$
[green (dash-dotted) line], as functions of $B(\mu\to  e\gamma)$ (left pannels)
and the heavy neutrino mass scale $m_N$ (right pannels). In all pannels, the Yukawa
parameter $a$ is fixed by the condition $B(\mu\to e \gamma)=10^{-12}$ for
$m_N=400 \textrm{ GeV}$, using the benchmark value $\tan\beta=10$. The upper pannels
display numerical values for the Yukawa texture \eqref{YU1}, with $a=b$ and $c=0$,
and the lower pannels for the Yukawa texture \eqref{YA4},  with $a=b=c$. The heavy
neutrino mass is varied within the LHC explorable range: $400\textrm{ GeV}<m_N<10\textrm{ TeV}$.
All observables $B(\mu \to e X)$ of $\mu$-LFV (with $X = \gamma,\, ee,\,  {\rm Ti},\, {\rm Au}$)
exhibit a non-trivial dependence on $m_N$. The branching ratio $B(\mu\to e\gamma)$ shows a dip
at $m_N\approx 800\textrm{ GeV}$ in both models \eqref{YU1} and \eqref{YA4}, signifying the
existence of a cancellation region in parameter space, due to the loops involving heavy
neutrino, sneutrino and soft SUSY-breaking terms. For $m_N \stackrel{>}{{}_\sim} 3\textrm{ TeV}$,
all observables tend to a constant value, as a result of the dominance of the soft
SUSY-breaking contributions.

%%%%%%%%%%%%%%%%%%%%%%%%%%%%%%%%%%%%%%%%%%%%%%%%%%%%%%%%%%%%%%%%%%%%%%
\begin{figure}[!ht]
 \centering
 \includegraphics[clip,width=0.4\textwidth]{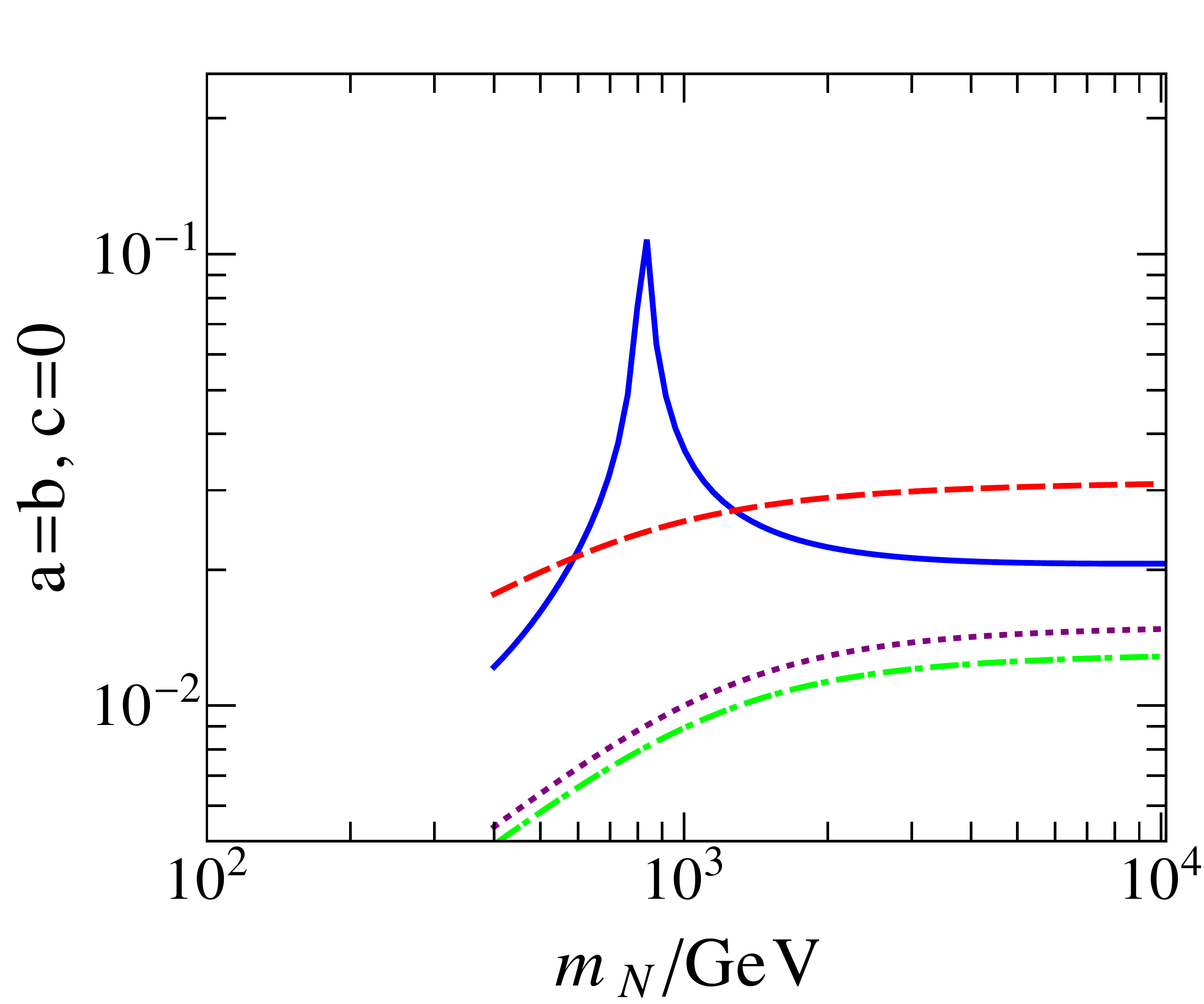}\hspace{1cm}
 \includegraphics[clip,width=0.4\textwidth]{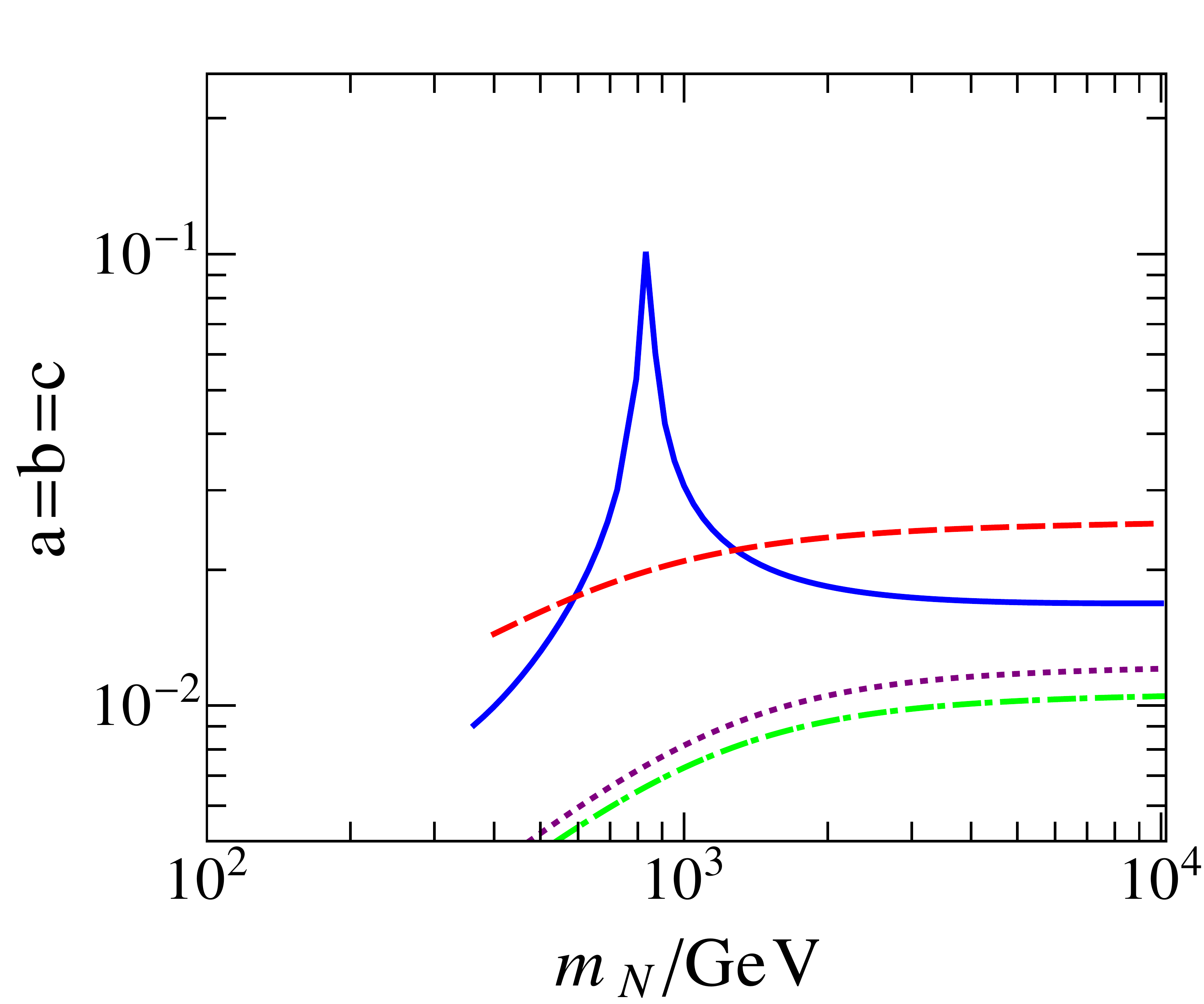}
\caption{Contours of the Yukawa  parameters $(a,b,c)$ versus $m_N$, for
  $B(\mu\to e\gamma)$ [blue  (solid)], $B(\mu\to eee)$ [red (dashed)],
  $R^{\rm   Ti}_{\mu  e}$~[violet   (dotted)]  and   $R^{\rm  Au}_{\mu
    e}$~[green (dash-dotted)],  where $a$ and $m_N$  are determined by
  the  condition  $B(\mu\to e\gamma)  =  10^{-12}$.  All contours  are
  evaluated with $\tan\beta=10$ and  for different Yukawa textures, as
  indicated by the vertical axes labels.}
\label{Fig8}
\end{figure}
%%%%%%%%%%%%%%%%%%%%%%%%%%%%%%%%%%%%%%%%%%%%%%%%%%%%%%%%%%%%%%%%%%%%%%

In Fig~\ref{Fig8} we show contours of the Yukawa parameters $(a,b,c)$ versus the
heavy neutrino mass scale $m_N$, for  $B(\mu\to e\gamma)$ [blue  (solid) line],
$B(\mu\to  eee)$ [red  (dashed) line],  $R^{\rm Ti}_{\mu e}$ [violet (dotted)  line]
and $R^{\rm Au}_{\mu e}$ [green (dash-dotted) line]. The Yukawa parameter $a$ and
$m_N$ are determined by the condition $B(\mu\to  e\gamma) = 10^{-12}$. The labels in
the vertical axes indicate the two Yukawa textures in \eqref{YU1} and \eqref{YA4}, which
we have adopted in our analysis. The contours for $B(\mu\to  e\gamma)$ display a maximum
for $m_N \approx  800 \textrm{ GeV}$, as a consequence of cancellations between heavy
neutrino, sneutrino and soft SUSY-breaking contributions (cf Fig~\ref{Fig7}).

%%%%%%%%%%%%%%%%%%%%%%%%%%%%%%%%%%%%%%%%%%%%%%%%%%%%%%%%%%%%%%%%%%%%%%

\begin{figure}[!ht]
 \centering
 \includegraphics[clip,width=0.4\textwidth]{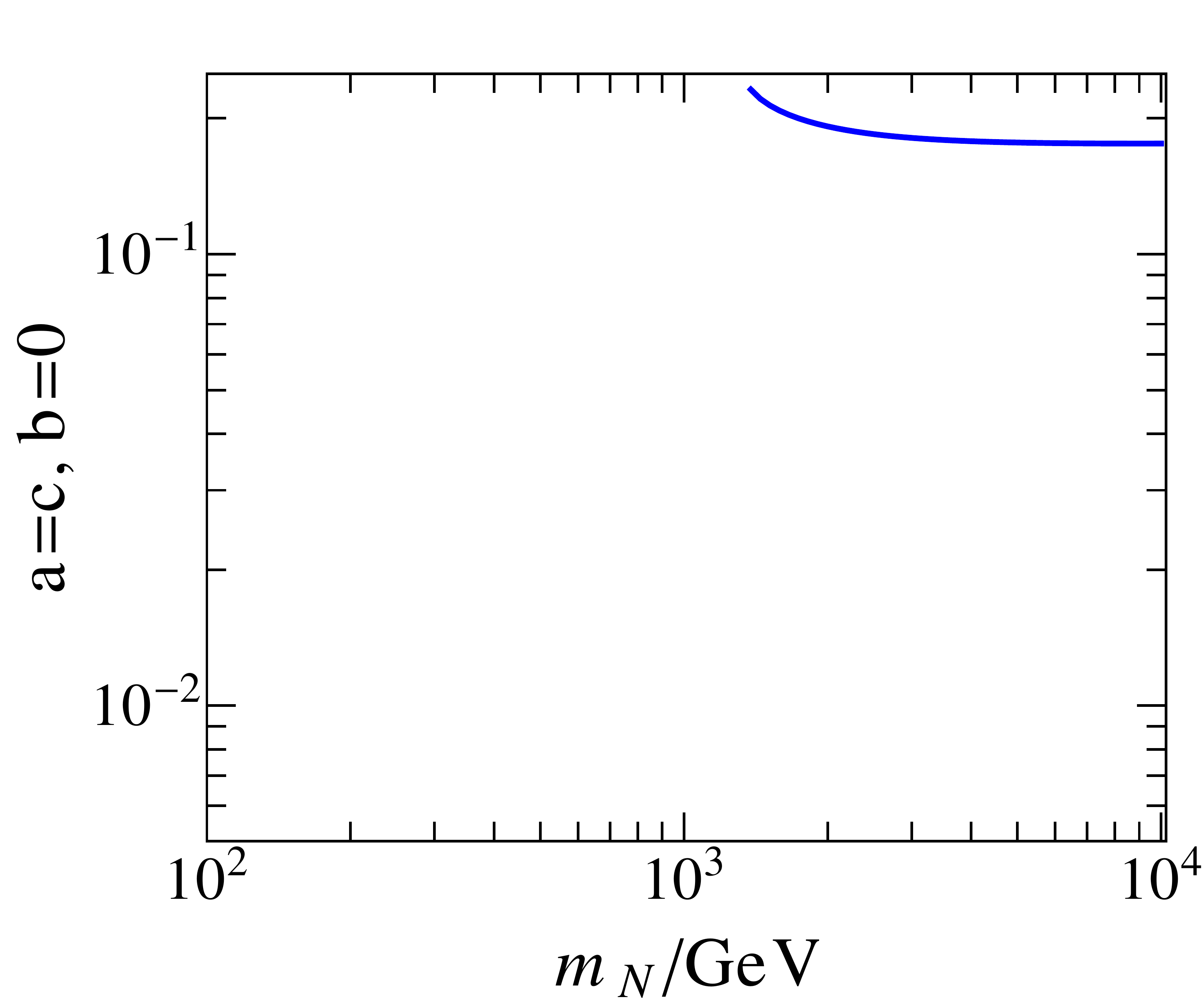}\hspace{1cm}
 \includegraphics[clip,width=0.4\textwidth]{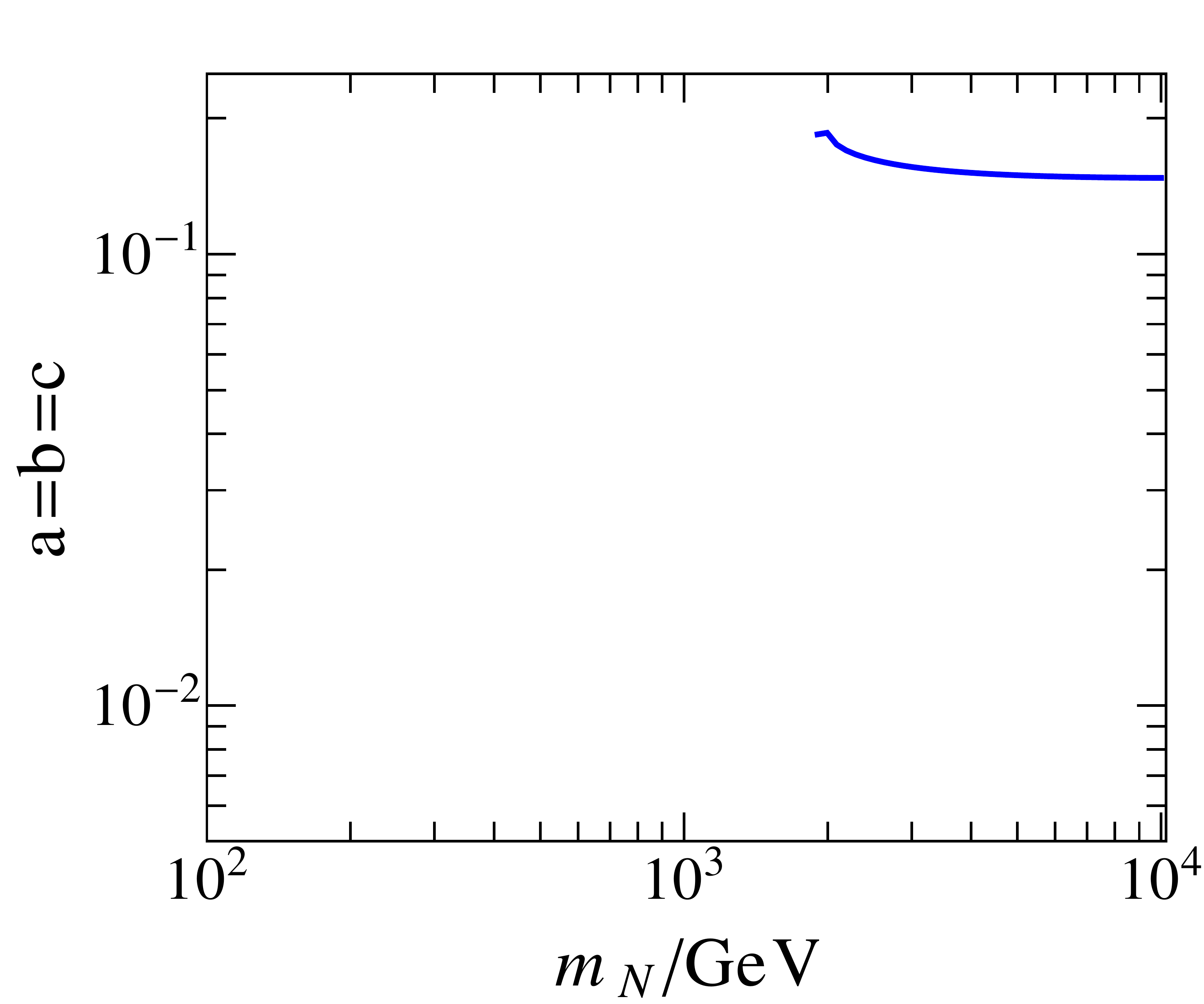}
\caption{Contours of the Yukawa parameters $(a,b,c)$ versus $m_N$, for
  $B(\tau\to e\gamma)$ [blue (solid)],
%and $B(\tau\to \mu \gamma)$ [red (dashed)],
where $\tan\beta=10$ and $a$ and $m_N$ are determined by
  the condition $B(\tau\to e\gamma)=10^{-9}$. No solutions have been
  found for $B(\tau\to eee)$ and $B(\tau\to e\mu\mu)$. }
\label{Fig9}
\end{figure}
%%%%%%%%%%%%%%%%%%%%%%%%%%%%%%%%%%%%%%%%%%%%%%%%%%%%%%%%%%%%%%%%%%%%%%

Figure \ref{Fig9} shows contours of the Yukawa parameters $(a,b,c)$, as functions
of $m_N$, for $B(\tau\to e\gamma)$ [blue (solid) line], where the parameters $a$ and
$m_N$ are determined by the condition $B(\tau\to e\gamma)=10^{-9}$. The numerical results
for $B(\tau\to \mu \gamma)$ are not given, since these  are fully complementary to the ones
given for $B(\tau\to e\gamma)$. Given the above condition on $B(\tau\to  e\gamma)$, no
solution exists for the observables $B(\tau\to eee)$ and $B(\tau\to e\mu\mu)$.

%%%%%%%%%%%%%%%%%%%%%%%%%%%%%%%%%%%%%%%%%%%%%%%%%%%%%%%%%%%%%%%%%%%%%%
\begin{figure}
 \centering
 \includegraphics[clip,width=0.4\textwidth]{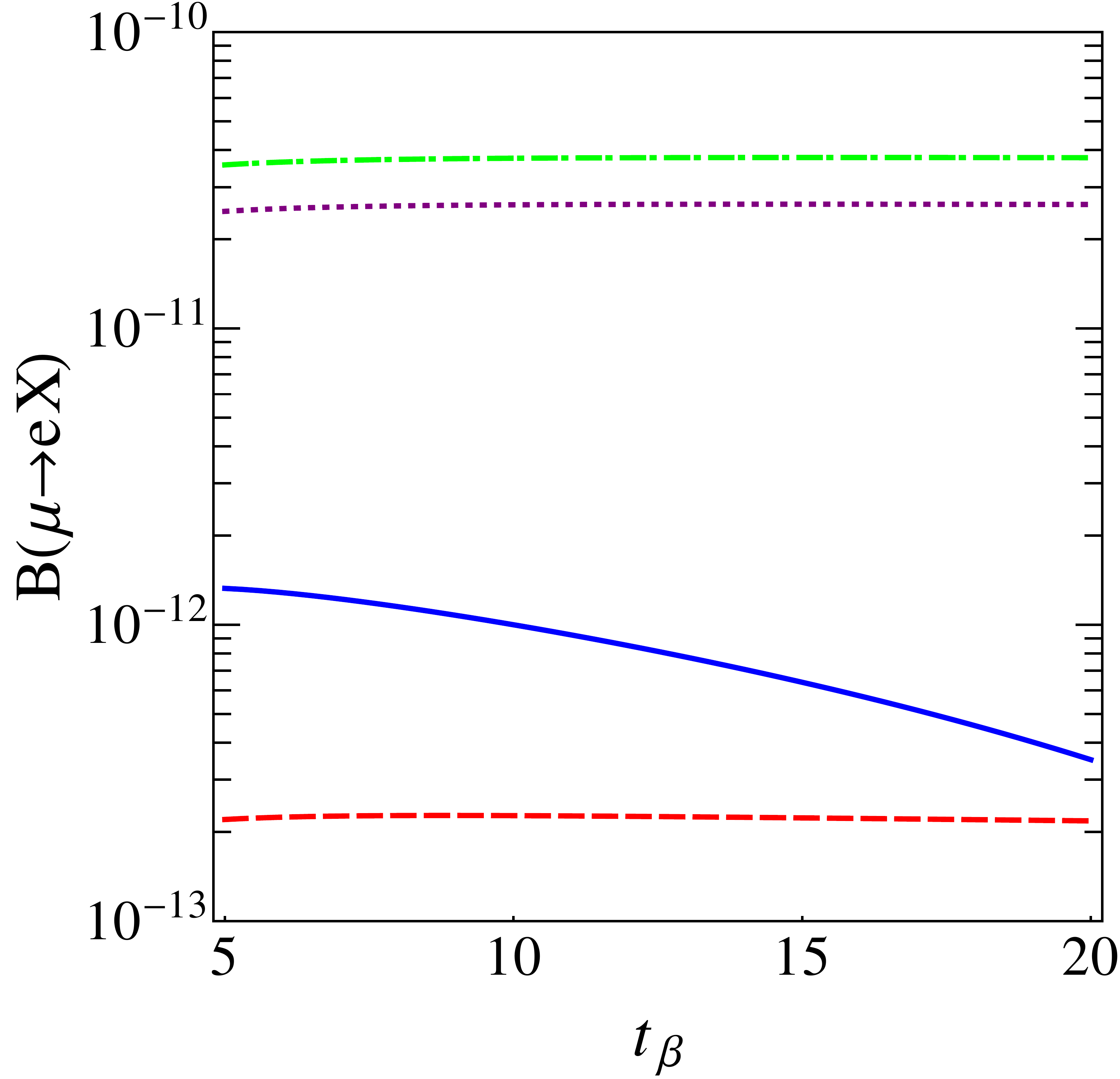}\hspace{1cm}
 \includegraphics[clip,width=0.4\textwidth]{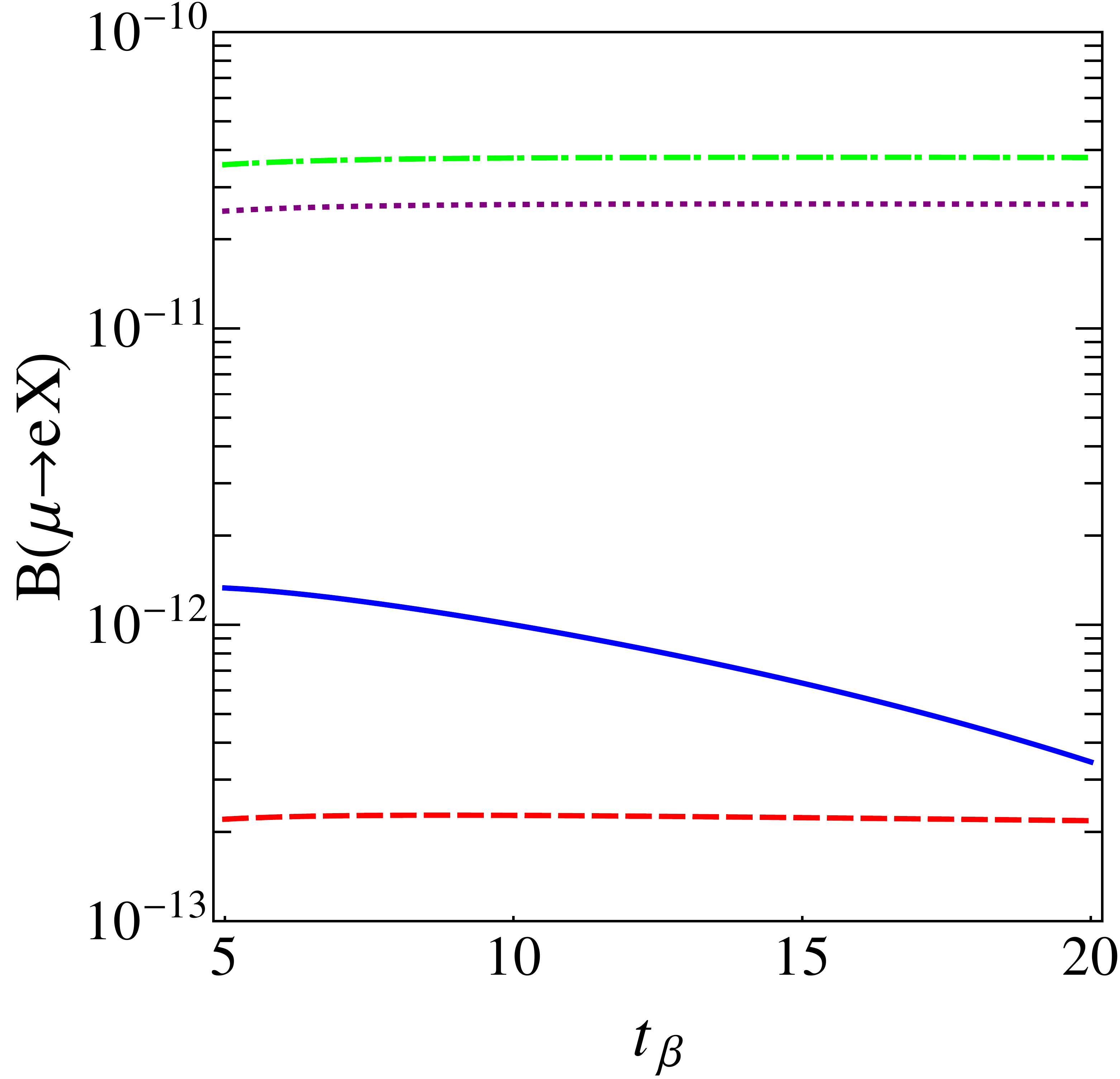}
 \\[.02\textwidth]
 \includegraphics[clip,width=0.4\textwidth]{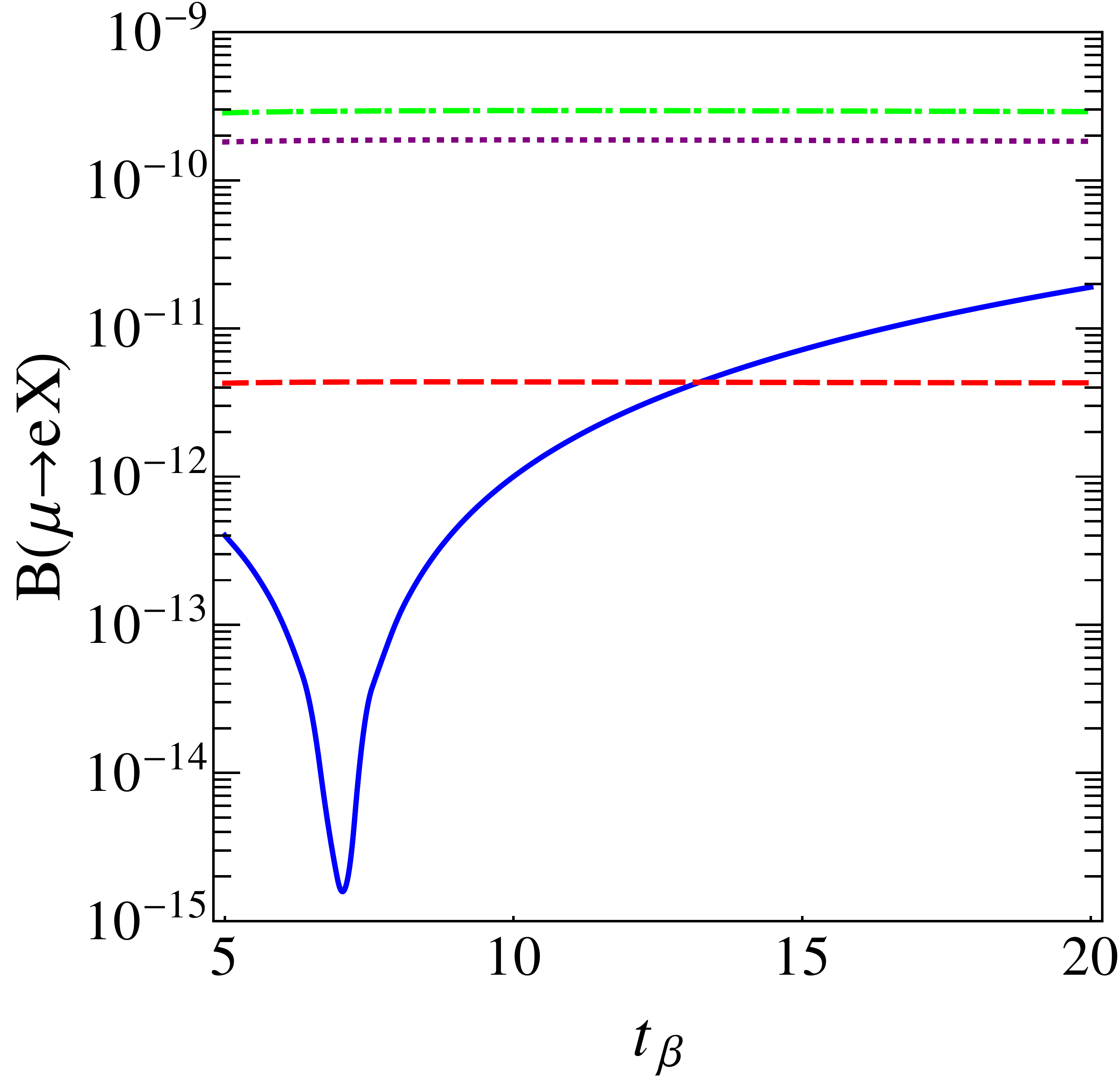}\hspace{1cm}
 \includegraphics[clip,width=0.4\textwidth]{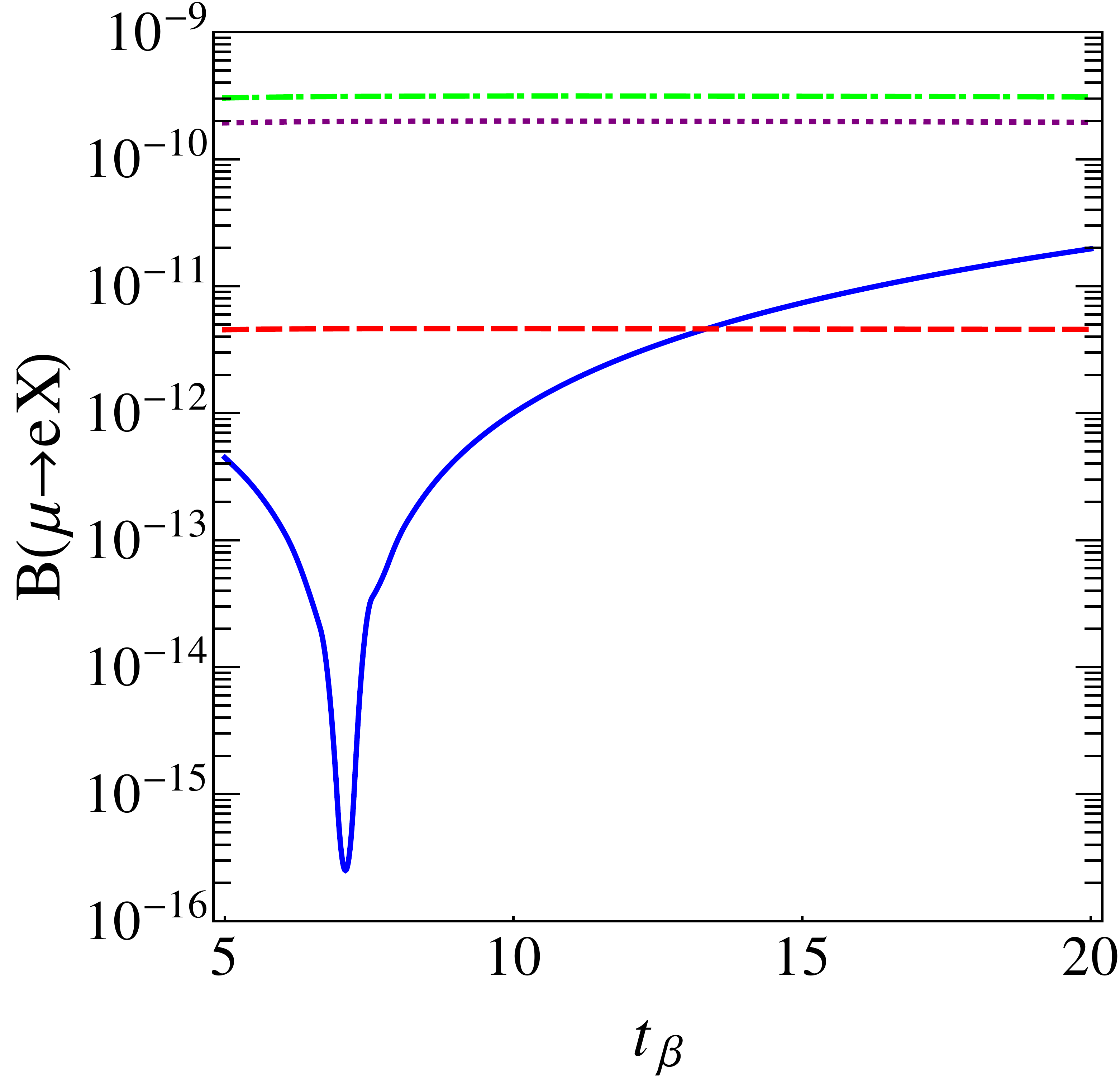}
\caption{Numerical estimates  of $B(\mu\to e  \gamma)$ [blue (solid)],
  $B(\mu\to  eee)$   [red  (dashed)],  $R^{\rm   Ti}_{\mu  e}$~[violet
    (dotted)]  and  $R^{\rm  Au}_{\mu  e}$~[green  (dash-dotted)],  as
  functions  of  $\tan\beta$.   The  upper pannels  are  obtained  for
  $m_N=400$~GeV  and  the  lower  pannels for~$m_N=1$~TeV.   The  left
  pannels use  the Yukawa texture  (\ref{YU1}), with $a=b$  and $c=0$,
  and the right pannels  the Yukawa texture (\ref{YA4}), with $a=b=c$.
  In  all pannels,  the  Yukawa  parameter $a$  is  determined by  the
  condition $B(\mu\to e \gamma)=10^{-12}$.}
\label{Fig10}
\end{figure}
%%%%%%%%%%%%%%%%%%%%%%%%%%%%%%%%%%%%%%%%%%%%%%%%%%%%%%%%%%%%%%%%%%%%%%

In the numerical analysis presented so far, the assumed value of $\tan\beta$
was fixed to its benchmark value given in \eqref{mSUGRA}, $\tan\beta =10$.
In Fig~\ref{Fig10}, this  assumption is relaxed, and $\tan\beta$ is varied in the
interval $5 \stackrel{<}{{}_\sim} \tan\beta \stackrel{<}{{}_\sim} 20$, while
maintaining agreement with a SM-like Higgs boson mass $M_H \approx  125 \textrm{ GeV}$
and taking into account that the combined experimental and theoretical errors are
currently of the order of 5--6 GeV. Specifically, in Fig~\ref{Fig10} we display the
dependence of $B(\mu\to  e  \gamma)$ [blue  (solid) line], $B(\mu\to eee)$
[red (dashed) line], $R^{\rm Ti}_{\mu e}$ [violet (dotted) line] and $R^{\rm Au}_{\mu e}$
[green (dash-dotted) line]  on $\tan\beta$. In all pannels, the Yukawa parameter $a$
is determined by the condition $B(\mu\to e \gamma)=10^{-12}$. The upper pannels in Fig~\ref{Fig10}
show numerical results for $m_N=400 \textrm{ GeV}$, while the lower pannels for
$m_N=1 \textrm{ TeV}$.  The left pannels give the predictions for the Yukawa texture
\eqref{YU1}, with $a=b$ and $c=0$, and the right pannels for the Yukawa texture \eqref{YA4},
with $a=b=c$. In the lower pannels, one can observe a suppression of $B(\mu\to e \gamma)$,
for values $\tan\beta \approx 7$, due to the cancellation between heavy neutrino, sneutrino and
soft SUSY-breaking effects.

%%%%%%%%%%%%%%%%%%%%%%%%%%%%%%%%%%%%%%%%%%%%%%%%%%%%%%%%%%%%%%%%%%%%%%
\begin{figure}
 \centering
 \includegraphics[clip,width=0.38\textwidth]{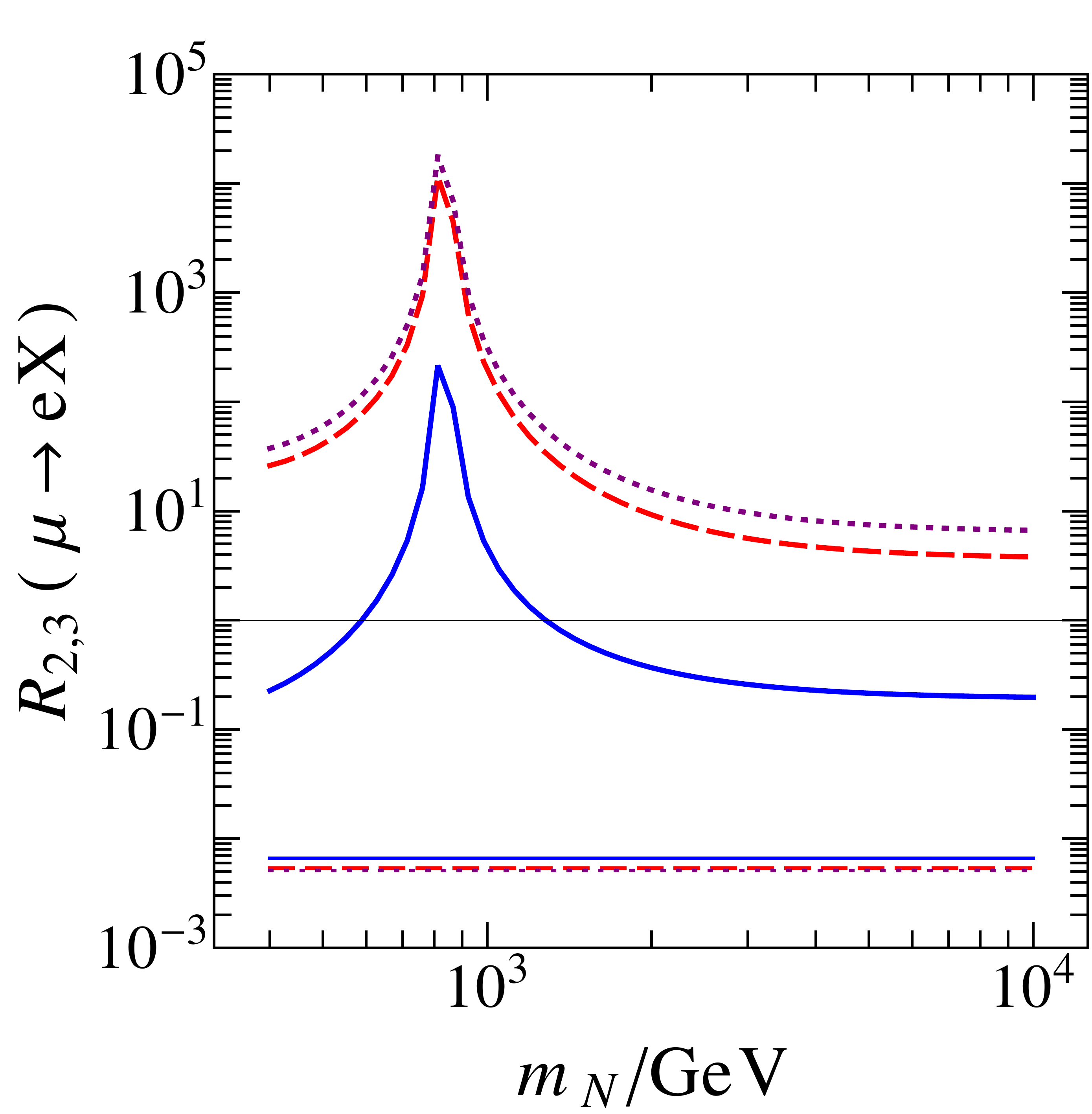} \\
 \includegraphics[clip,width=0.35\textwidth]{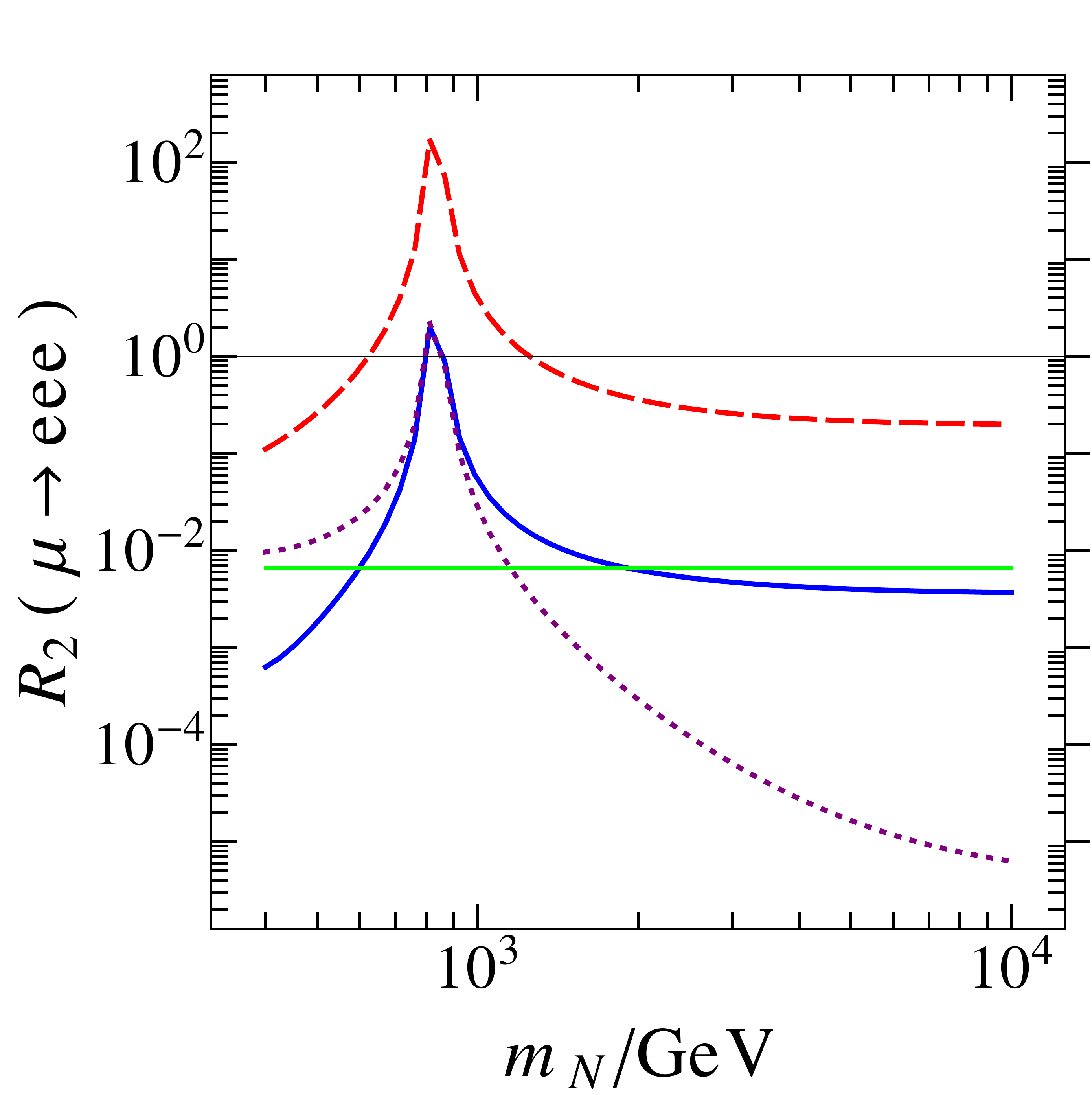}\hspace{1cm}
 \includegraphics[clip,width=0.35\textwidth]{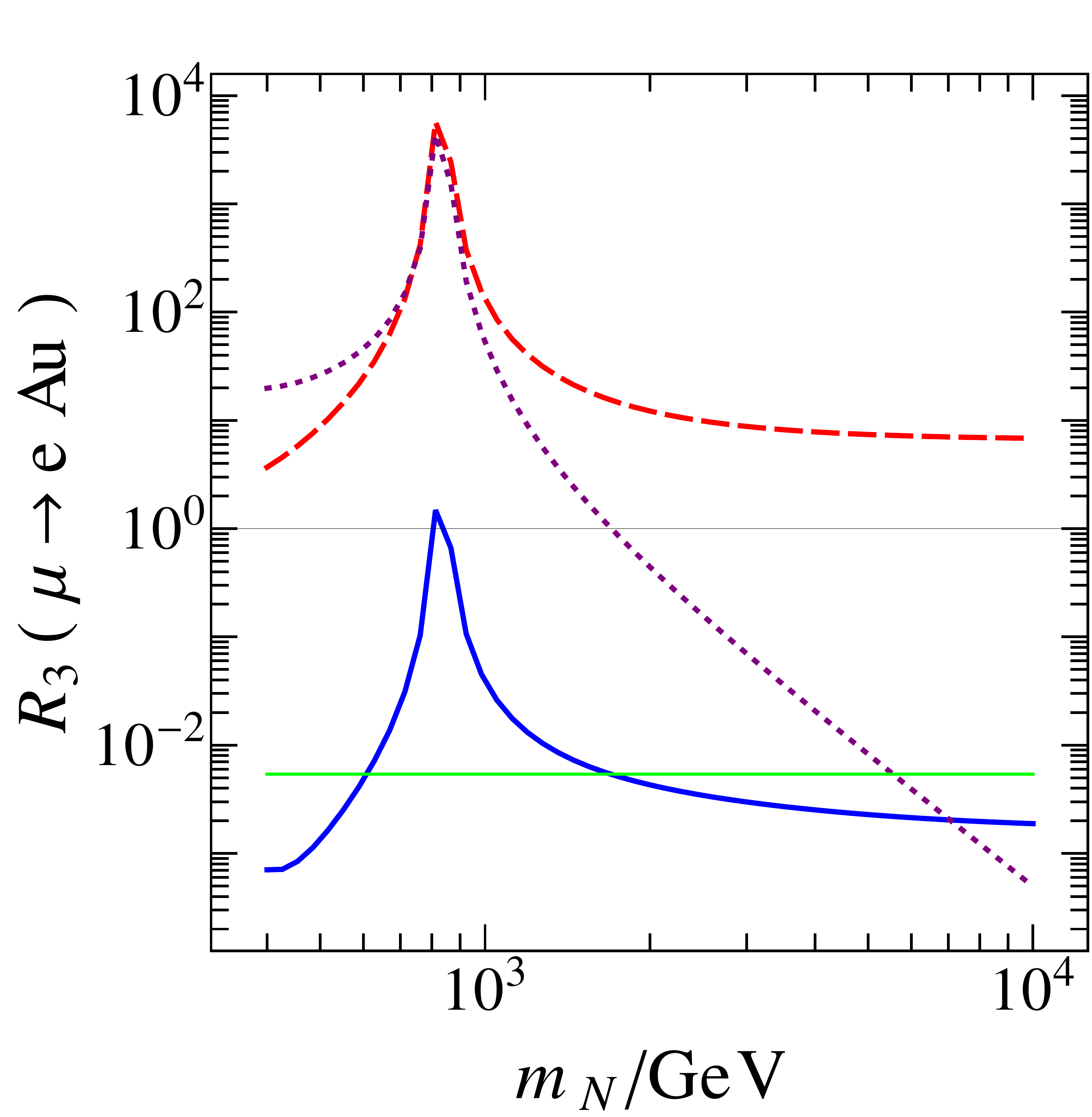} \\
 \includegraphics[clip,width=0.35\textwidth]{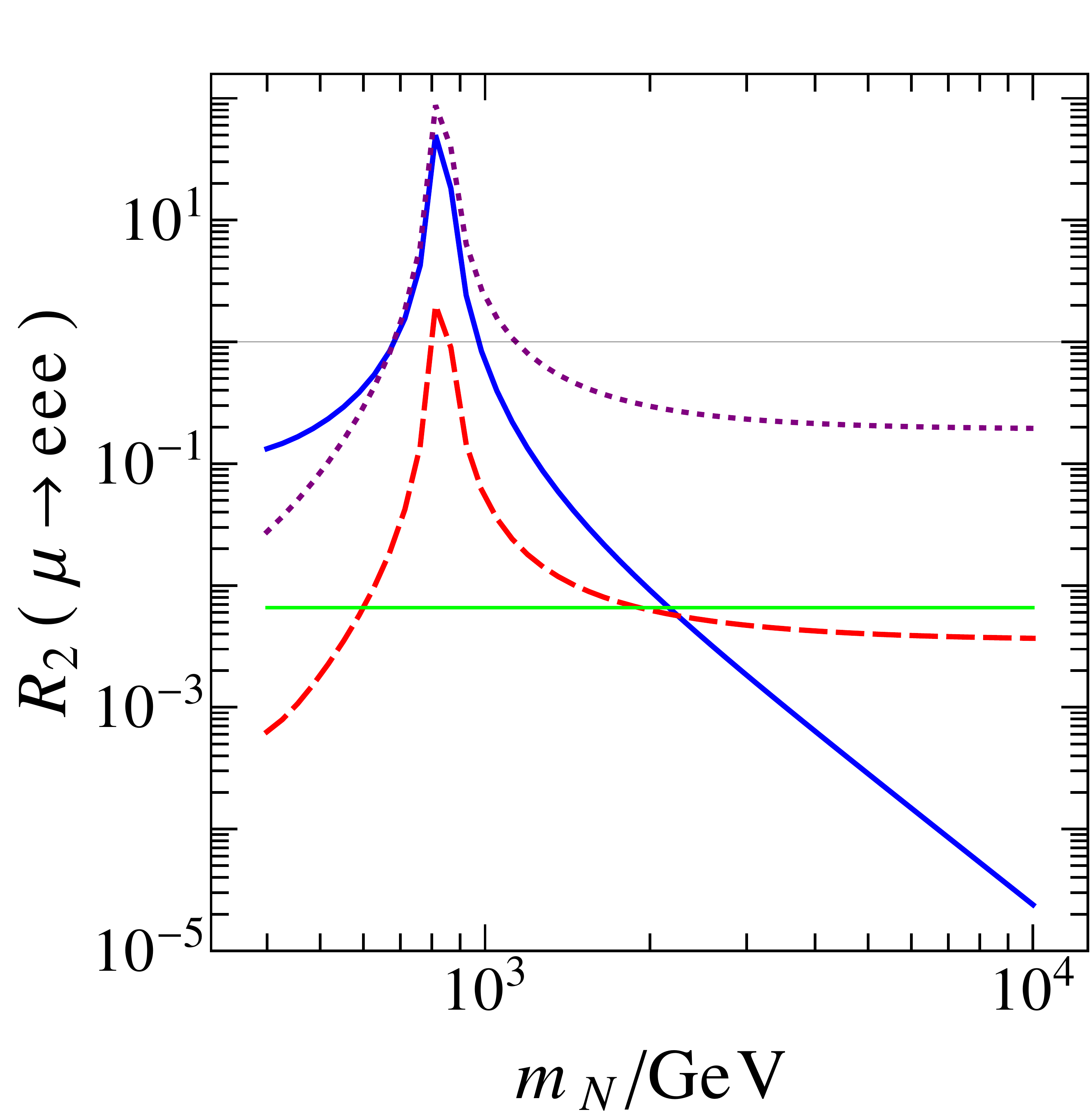}\hspace{1cm}
 \includegraphics[clip,width=0.35\textwidth]{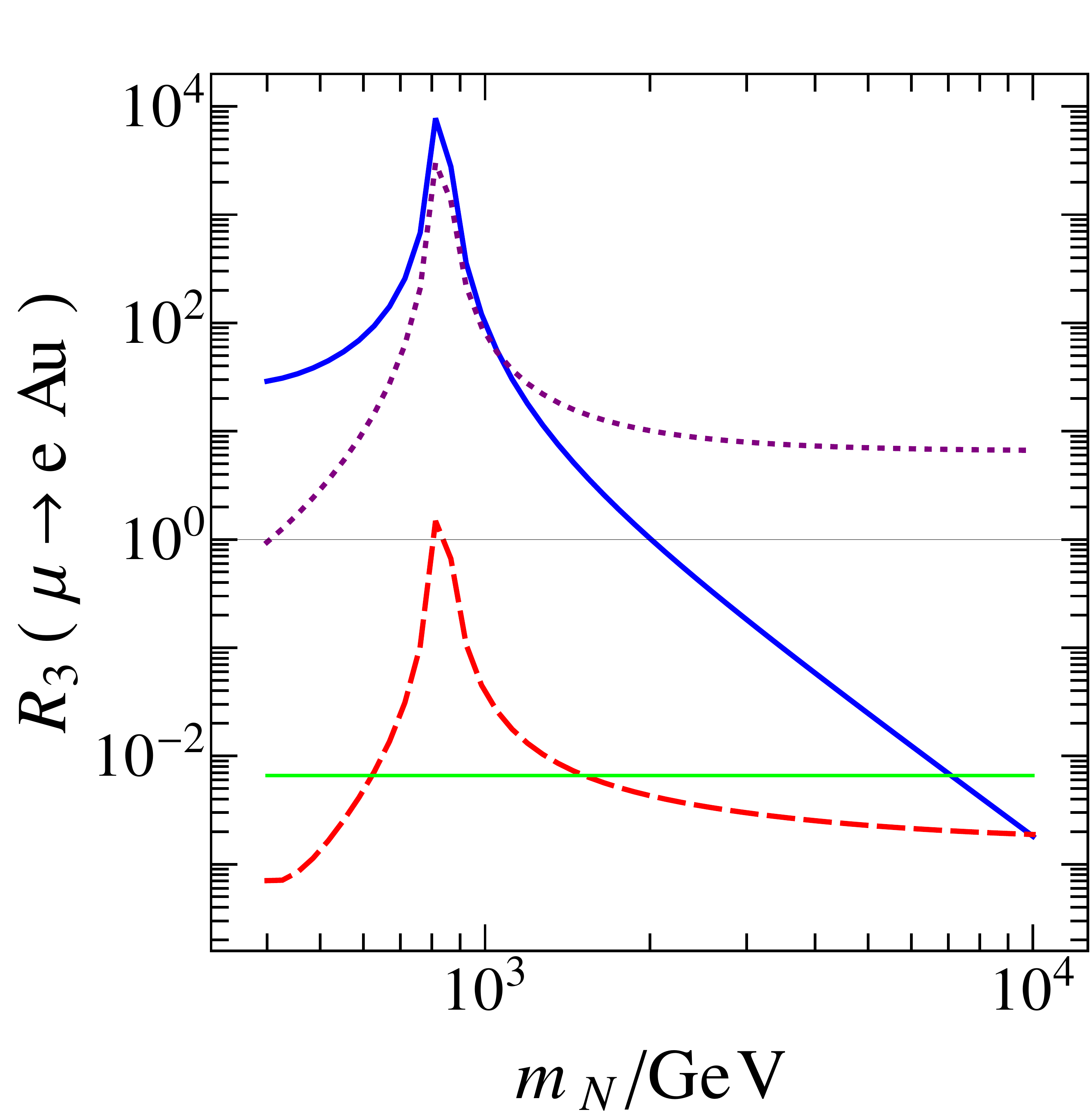}
\caption{Numerical estimates of the ratios $R_2(\mu \to eee)$, $R^{\rm
    Ti}_3$  and $R^{\rm  Au}_3$,  as functions  of  $m_N$. The  Yukawa
  parameter  $a$   is  fixed   by  the  condition   $B(\mu\to  e\gamma
  )=10^{-12}$, for  $m_N = 400$~GeV and $\tan\beta=10$.   In the upper
  pannel,  thick lines give  the complete  evaluation of  $R_2(\mu \to
  eee)$  [blue (solid)],  $R^{\rm  Ti}_3$ [red  (dashed)] and  $R^{\rm
    Au}_3$  [violet (dotted)],  while  the respective  thin lines  are
  evaluated keeping only  the magnetic dipole form factors $G_\gamma^L$
  and  $G_\gamma^R$.   The two  middle  pannels  provide a  form factor
  analysis  of $R_2(\mu  \to eee)$  and  $R^{\rm Au}_3$,  in terms  of
  contributions  due  to  $G_\gamma$  and $F_\gamma$  [blue  (solid)],
  $F_Z$~[red~(dashed)]  and box  form factors  [violet (dotted)].   The
  lower  two pannels  show  the separate  contributions  due to  heavy
  neutrinos~$N$                                         [blue~(solid)],
  sneutrinos~$\widetilde{N}$~[red~(dashed)] and soft SUSY-breaking LFV
  terms~[violet~(dotted)].  The green (horizontal) lines in the middle
  and  lower pannels give  the predicted  values obtained  by assuming
  that  only  the   $G^{L,R}_\gamma$  form factors  contribute  to  the
  amplitudes.}
  \label{Fig11}
\end{figure}
%%%%%%%%%%%%%%%%%%%%%%%%%%%%%%%%%%%%%%%%%%%%%%%%%%%%%%%%%%%%%%%%%%%%%%

It can be instructive to compare the contributions of the magnetic dipole
form factors to the CLFV observables, with those originating from the remaining
form factors. Specifically, if one assumes that only the magnetic dipole
form factors $G^{L,R}_\gamma$ contribute in \eqref{Bl3l_1}, \eqref{Bl3l_2} and
\eqref{RmueJ}, then the following analytical results are obtained for the ratios:
\begin{eqnarray}
R_1 &\equiv& \frac{B(l\to l'l_1l_1^c)}{B(l\to l'\gamma)} 
 \ =\ 
 \frac{\alpha}{3\pi}\Big(\ln\frac{m_l^2}{m_{l'}^2} - 3\Big)
\label{RBR1}
\\[1.5ex]
R_2 &\equiv&
\frac{B(l\to l'l'l'^c)}{B(l\to l'\gamma)}
 \ =\ 
 \frac{\alpha}{3\pi}\Big(\ln\frac{m_l^2}{m_{l'}^2} - \frac{11}{4}\Big)
\label{RBR2}
\\[1.5ex]
R_3 &\equiv&
 \frac{R^J_{\mu e}}{B(\mu\to e\gamma)}
 \ =\ 
 16\alpha^4 \frac{\Gamma_\mu}{\Gamma_{\rm capture}} Z Z_{eff}^4 |F(-\mu^2)|^2\  .
\label{RBR3}
\end{eqnarray}

According to the formulae \eqref{RBR1}--\eqref{RBR3}, the predicted $R_1$ values
for $\tau\to e\mu\mu$ and $\tau\to\mu ee$ are $1/90$ and $1/419$ respectively,
the predicted $R_2$  values for $\mu\to  e ee$, $\tau\to  eee$ and the
$\tau \to \mu\mu\mu$ are $1/159$, $1/91$ and $1/460$ respectively, and the predicted
$R_3$ values for Ti and Au are $1/198$  and $1/188$  respectively.

In  Fig~\ref{Fig11}, the numerical estimates are given for the ratios
$R_2(\mu \to eee)$, $R^{\rm Ti}_3$ and $R^{\rm Au}_3$, as functions of $m_N$.
The Yukawa parameter $a$ is fixed by the condition $B(\mu\to e\gamma )=10^{-12}$,
for $m_N = 400\textrm{ GeV}$ and $\tan\beta=10$. In the upper pannel, thick lines
show the predicted values obtained by a complete evaluation of $R_2(\mu \to eee)$
[blue (solid) line], $R^{\rm Ti}_3$  [red  (dashed) line] and  $R^{\rm  Au}_3$
[violet (dotted) line], while the respective thin lines are obtained by keeping only
the magnetic dipole form factors $G_\gamma^L$ and $G_\gamma^R$. Hence, we see that
going beyond the magnetic dipole moment approximation may enhance the ratios
$R_{2,3}$ by more than two orders of magnitude.

The two middle pannels of Fig~\ref{Fig11} provide a form factor analysis of $R_2(\mu  \to eee)$
and $R^{\rm  Au}_3$, by considering separately the contributions due to $G_\gamma$ and
$F_\gamma$ [blue (solid) line], $F_Z$ [red (dashed) line] and box form factors [violet (dotted)
line]. In particular, one observes that heavy neutrino contributions to the box
form factors become comparable to and even larger than the $Z$-boson mediated graphs in
$\mu \to e$ conversion in Gold, for heavy neutrino masses $m_N \stackrel{<}{{}_\sim} 1 \textrm{ TeV}$.
We have checked that for $m_N \stackrel{<}{{}_\sim} 1 \textrm{ TeV}$, box graphs due
to heavy neutrinos also dominate the process of $\mu \to e$ conversion in Titanium
(not explicitly  shown  in Fig~\ref{Fig11}). Finally, the  two lower pannels show the individual contributions due to the heavy neutrinos $N_{1,2,3}$ [blue (solid) line], sneutrinos $\widetilde{N}_{1,2,\dots,12}$ [red (dashed) line] and soft SUSY-breaking LFV terms [violet
(dotted) line]. From these two lower pannels, it is obvious that for heavy neutrino masses
$m_N \stackrel{>}{{}_\sim} 1 \textrm{ TeV}$, the  soft SUSY-breaking effects dominate the
CLFV  form factors, which are tagged with the superscripts SB in Appendix \ref{sec:olff}.
Instead, for $m_N \stackrel{<}{{}_\sim} 1 \textrm{ TeV}$, heavy-neutrino effects start becoming
the leading contribution to the CLFV observables associated with the muon. The green
(horizontal) lines in the middle and lower pannels serve as reference values obtained by
assuming that only the $G^{L,R}_\gamma$ form factors contribute to the amplitudes.

An important consistency check for this numerical analysis has been to {\em  analytically}
show that all soft SUSY-breaking effects on the form factors \eqref{gammaSB}, \eqref{ZSB} and
\eqref{ffboxSB} vanish in the limit of degenerate charged slepton masses. On the other hand,
RGE effects from $M_{\rm GUT}$ to $M_Z$ induce sizeable deviations to the charged slepton
mass matrix from the unit matrix. As a consequence, unitarity cancellations due to the
so-called GIM mechanism become less effective in this case and so render the SB  part of
the form factors, such as $F_{l'lZ}^{L,{\rm SB}}$ and $F_{l'lZ}^{R,{\rm  SB}}$, rather large.

Another essential check was to show that under the assumptions adopted in 
part of Ref~\cite{Hirsch2012},
the form factor $F_{l'lZ}^{L,\tilde{N}}$ given in Eq~\eqref{ZNt} reduces to $\frac{2c_W}{g} F_L^c$,
where $F_L^c$ is one of the form factors defined in \cite[Eq~(6)]{Hirsch2012}, which in turn
can be shown to vanish. The assumptions in \cite{Hirsch2012} are: (i) the standard seesaw mechanism
with ultra-heavy  right neutrinos, (ii) no charged wino  or higgsino mixing, and (iii) the
dominance of the wino contribution. Under these three assumptions, the interaction vertices
occurring in the form factor $F_{l'lZ}^{L,\tilde{N}}$ simplify as follows:
\begin{equation}
  \label{eq:Hirsch}
\begin{array}{l}
\tilde{B}_{lmA}^{R,1},\, \tilde{B}_{lmA}^{R,2}\ \to\ - U_{lk}\, ,
\quad
\tilde{C}_{AB}^1,\,  \tilde{C}_{AB}^2,\, \tilde{C}_{AB}^3,\, \tilde{C}_{AB}^4\
\to\ -\frac{1}{2}\, \delta_{kk'}\,, \\[1.5ex]
V_{mk}^{\tilde{\chi}^- R}\ \to\ c^2_w\, ,
\end{array}
\end{equation}
where $A,B$ now assume the restricted range of values $k,k'=1,2,3$ and $U$ is a
$3\times 3$ unitary matrix. Given the simplifications in Eq~\eqref{eq:Hirsch},
we recover the expression given in Ref \cite{Hirsch2012}, resulting in the replacement:
$F_{l'lZ}^{L,\tilde{N}}\to \frac{2c_W}{g} F_L^c$. The above non-trivial checks provide firm
support for the correctness of analytical and numerical results hereby presented. The 
full-fledged calculation in Ref~\cite{Hirsch2012} was performed without the above 
mentioned assumptions.

%% file: moments.tex
\chapter{Lepton Dipole Moments} \label{moments}

In this chapter we perform the study of anomalous magnetic and electric
dipole moments of charged leptons in $\nu_R$MSSM, under the assumption
that CP violation originates from complex soft SUSY-breaking bilinear and 
trilinear couplings associated with the  right-handed sneutrino sector.

In the first section, the conventions and notation for the lepton
dipole moments will be presented. This will be accompanied by the
description of the new sources of CP violation which are considered 
within the $\nu_R$MSSM.

Second section contains the numerical estimates for the anomalous magnetic
moment of the muon ($a_\mu$) and the electric dipole moment of the electron
($d_e$).

Technical details pertinent to the lepton-dipole moment form factors are
to be found at the end of this chapter.
%in Appendix \ref{f3.4and3.5}.

These results are presented in Ref \cite{Popov2013a}.

\newpage

\section{Magnetic and electric dipole moments}

The anomalous MDM and EDM of a charged lepton $l$ can be read off from
the Lagrangian \cite{Branco1999}:
\begin{equation}
  \label{l_effL}
{\cal L}\ =\ \bar{l}\,\Big[\gamma_\mu(i\partial^\mu + e A^\mu) 
  - m_l - \frac{e}{2m_l} \sigma^{\mu\nu} (F_l+iG_l\gamma_5) 
                                       \partial_\nu A_\mu\Big]\, l\,.
\end{equation}
In the  on-shell limit  of the photon  field $A^\mu$, the  form factor
$F_l$ defines  the anomalous magnetic dipole moment (MDM)  of the lepton 
$l$,  i.e.\ $a_l \equiv F_l$, whilst the  form factor $G_l$ defines its  
electric dipole momenr (EDM), i.e.\ $d_l \equiv
eG_l/m_l$. Using Eq~\eqref{Tll'g}, one can write down the general form-factor 
decomposition of the photonic transition amplitude,
\begin{equation}
i{\cal T}^{\gamma ll}\ =\
 i\frac{e \alpha_w}{8 \pi M_W^2}
 \Big[ (G_\gamma^L)_{ll} i\sigma_{\mu\nu}q^\nu P_L
 + (G_\gamma^R)_{ll} i\sigma_{\mu\nu}q^\nu P_R \Big]\,.
\end{equation}
The anomalous MDM ($a_l$) and the EDM ($d_l$) of a lepton $l$ are then
respectively determined by:
\begin{eqnarray}
\label{aldl_nRS}
 a_l &=& \frac{\alpha_w m_l}{8\pi M_W^2} \Big[ (G_\gamma^L)_{ll} + 
        (G_\gamma^R)_{ll} \Big]\,,\\
 d_l &=& \frac{e \alpha_w}{8\pi M_W^2} i \Big[ (G_\gamma^L)_{ll} - 
         (G_\gamma^R)_{ll} \Big]\,.
\end{eqnarray}
Here and in the following, the notation for the couplings and
the form-factors will correspond to the one used in Chapter~\ref{CLFV}.

As shown in Ref~\cite{Farzan2004}, the EMD $d_l$ of the lepton vanishes
in the MSSM with universal soft SUSY-breaking boundary conditions,
if no CP phases are introduced. This result also holds
true in the extensions of the MSSM with heavy neutrinos, as long as
the sneutrino sector is universal and CP-conserving.

As a minimal departure from the above universal scenario, let it be assumed
that {\em only}  the sneutrino sector is CP-violating due to soft CP
phases in the bilinear and trilinear soft-SUSY breaking parameters:
\begin{eqnarray}
  \label{CPBnu}
{\bf b}_\nu &\equiv& {\bf B}_\nu {\bf m}_M \ =\ B_0 e^{i\theta} m_N {\bf 1}_3\,,\\ 
  \label{CPAnu}
{\bf A}_\nu &=&  {\bf h}_\nu\, A_0 e^{i\phi}\,,
\end{eqnarray}
where $B_0$ and $A_0$ are real parameters determined at the GUT scale,
$m_N$ is a real parameter inputed at the scale $m_N$, and $\theta$ and
$\phi$ are  physical, flavor-blind CP-odd phases, and ${\bf h}_\nu$
is the $3\times 3$ neutrino Yukawa matrix given by Eq~\eqref{YA4}.
The soft SUSY breaking terms  corresponding to the ${\bf b}_\nu$ and ${\bf A}_\nu$ 
are obtained from the Lagrangian terms
\begin{equation}
 - ({\bf A}_\nu)^{ij} \tilde{\nu}^c_{iR} (  h^+_{uL}
  \tilde{e}_{jL} - h^0_{uL} \tilde{\nu}_{jL})\\  
\end{equation}
and 
\begin{equation}
({\bf b}_\nu {\bf m}_M)_{ii} \tilde{\nu}_{Ri}\tilde{\nu}_{Ri}\,,
\end{equation}
respectively. Correspondingly, $\tilde{\nu}^c_{iR}$, $\tilde{e}_{jL}$,
$h^+_{uL}$  and  $h^0_{uL}$  denote  the heavy  sneutrino,  selectron,
charged Higgs  and neutral Higgs fields.  The  $O(3)$ flavor symmetry
of the model  for the heavy neutrinos assures  that the heavy neutrino
mass matrix ${\bf m}_N$ is proportional to the unit matrix ${\bf 1}_3$
with eigenvalues $m_N$, up to small renormalization-group effects.  To
keep things simple, we also  assume that the $3\times 3$ soft bilinear
mass  matrix ${\bf  b}_\nu$ is  proportional to  ${\bf 1}_3$.   In the
standard  SUSY seesaw  scenarios  with ultra-heavy  neutrinos of  mass
$m_N$, the CP-violating sneutrino  contributions to electron EDM $d_l$
scale  as   $B_0/m_N$  and  $A_0/m_N$  at  the   one-loop  level,  and
practically decouple for heavy-neutrino  masses $m_N$ close to the GUT
scale.  Hence,  sizeable effects on  $d_e$ should only be  expected in
low-scale  seesaw  scenarios, in  which  $m_N$  can become  comparable
to~$B_0$ and~$A_0$.

Note that the bilinear soft $3\times 3$~matrix~${\bf b}_\nu$ was neglected
in the previous chapter, where it was tacitly assumed that it was
small compared to the other soft SUSY-breaking parameters in
sneutrino mass matrix given by Eq~\eqref{M2snu1}. Here, this term will be taken 
into the account, but with the restricted size of the universal bilinear mass
parameter $B_0$, such that the sneutrino masses remain always positive and 
hence physical.

The generation of a non-zero lepton EDM $d_l$ results from the soft sneutrino
CP-odd phases  $\theta$ and $\phi$,  as well as from  complex neutrino
Yukawa couplings ${\bf h}_\nu$. All these CP-odd phases are present in
the  photon  dipole form  factors  $G_{ll\gamma}^{L,\tilde{N}}$  and
$G_{ll\gamma}^{R,\tilde{N}}$,  whose  analytical  forms may  be  found
in Appendix~\ref{sec:olff}. In fact, it can be noticed that $d_l$ may be generated 
by products of vertices that are not relatively complex conjugate to each
other, since they contain the factors
\begin{equation}
  \label{CPNt}
\Delta^{LR}_{\rm CP} = \tilde{B}^{L,1}_{l k A} \tilde{B}^{R,1*}_{l k A} +
 \tilde{B}^{L,2}_{l k A} \tilde{B}^{R,2*}_{l k A}\,, \quad
\Delta^{RL}_{\rm CP} = \tilde{B}^{R,1}_{l k A} B^{L,1*}_{l k A} +
 \tilde{B}^{R,2}_{l k A} \tilde{B}^{L,2*}_{l k A}\, ,
\end{equation}
as can be seen from the Eq~\eqref{GllgRNt},

In the exact supersymmetric limit of softly-broken SUSY theories, the
anomalous MDM (as well as EDM) operator is forbidden. This comes as a 
consequence of the  Ferrara and Remiddi no-go theorem~\cite{Ferrara1974},
which is verified for every particle and its SUSY-counterpart
contribution to the anomalous MDM $a_\mu$. Besides the SM contribution,  
there are three additional contributions in the $\nu_R$MSSM, which originate
from: (i)~heavy neutrinos, (ii)~sneutrinos and~(iii) soft SUSY-breaking parameters.
In the supersymmetric limit, the latter contribution~(iii) vanishes. In the same limit, 
the heavy neutrino and sneutrino contributions read:
\begin{eqnarray}
(G_{\gamma}^{ll})^N &\to& \frac{7}{6} B_{lN_a} B_{lN_a}^*\,,
\nonumber\\
(G_{\gamma}^{ll})^{\tilde{N}} &\to& -\ \frac{7}{6} B_{lN_a} B_{lN_a}^*\,, 
\end{eqnarray}
where $B_{lN_a}$  are the lepton-to-heavy neutrino  mixings defined in
Refs \cite{Pilaftsis1992,Dev2012,Ilakovac1995}. Obviously, the sum 
$(G_{\gamma}^{ll})^N + (G_{\gamma}^{ll})^{\tilde{N}}$ vanishes, 
thereby confirming the above mentioned theorem proposed by 
Ferrara and Remiddi.

In the MSSM, the leading contribution to $a_l$ behaves as
\cite{Moroi1996,Carena1997,Stockinger2007}
\begin{eqnarray}
  \label{al_approx}
a_l^{\rm MSSM} &\propto& \frac{m_l^2}{M_{\rm SUSY}^2}\, \tan\beta\;
\mbox{sign}(\mu M_{1,2})\; , 
\end{eqnarray}
where  $M_{\rm SUSY}$  is  a typical  soft  SUSY-breaking mass  scale,
$\tan\beta=v_2/v_1$  is   the  ratio  of  the   neutral  Higgs  vacuum
expectation  values,  and  $M_{1,2}$   are  the  soft  gaugino  masses
associated  with the  ${\rm  U}(1)_{\rm Y}$  and  ${\rm SU(2)}$  gauge
groups, respectively. As will be seen in the next  section, the MSSM
contribution~(\ref{al_approx})  to  $a_\mu$  remains dominant  in  the
$\nu_R$MSSM as well.

From Eqs~(\ref{aldl_nRS}) and (\ref{al_approx}), one naively expects $d_l$
at the one-loop level to behave 
%as
\begin{eqnarray}
  \label{di_approx}
d_l^{\rm MSSM} &\propto& \sin(\phi_{\rm CP})\, \frac{m_l}{M_{\rm
    SUSY}^2}\, \tan\beta\ , 
\end{eqnarray}
where $\phi_{\rm  CP}$ is a  generic soft SUSY-breaking  CP-odd phase.
Although there are different dependencies of $d_l$ on $\tan\beta$
possible in the MSSM beyond the one-loop
approximation~\cite{Farzan2004,Pilaftsis2002} it will be shown that
within the $\nu_R$MSSM, the $\tan\beta$ dependence is linear at the
one loop level.

\section{Numerical results}

In this numerical analysis, we will adopt the procedure established
in Chapter~\ref{CLFV}. As a benchmark model, we choose a minimally
extended scenario of minimal supergravity (mSUGRA), in which we allow
for the bilinear and trilinear soft SUSY-breaking terms, ${\bf B}_\nu$
and ${\bf  A}_\nu$, to acquire  at the GUT scale overall CP-violating
phases denoted as $\theta$  and $\phi$, respectively. Like before, we
choose the sign of the $\mu$-parameter to be positive. As for the
neutrino  Yukawa  coupling  matrix  ${\bf  h}_\nu$,  we  consider  the
$A_4$-symmetric models introduced in previous chapter [see Eq~\eqref{YA4}].

For definiteness, our  numerical analysis in this section  is based on
the following baseline scenario:
\begin{equation}
\label{baseline}
\begin{array}{llll}
m_0 = 1~{\rm TeV}\,, &  M_{1/2} = 1~{\rm TeV}\,, &  
  A_0 = -4~{\rm TeV}\,, & \tan\beta = 20\,, \\
m_N = 1~{\rm TeV}\,, & B_0 = 0.1~{\rm TeV}\,, &
  a = b = c = 0.05\,, 
\end{array}
\end{equation}
where $m_0$, $M_{1/2}$  and $A_0$  are the standard universal soft  
SUSY-breaking parameters [cf Eq~\eqref{mSUGRAparam}]. All mass parameters 
except $m_N$ are defined  at the GUT scale and $m_N$ is intaken at $m_N$ scale. 
It is  understood that parameters which are not explicitly quoted  
in the text assume their default values stated in~\eqref{baseline}.   

We will analyze the deviation of $a_\mu$ from the SM value due to the
$\nu_R$MSSM, denoted by $\delta a_\mu$, as well as $d_e$ on several key
theoretical parameters, by varying them around their  baseline value
given  in \eqref{baseline},  while  keeping the  remaining  parameters
fixed. In  doing so, it will be made sure that the displayed parameters
can accommodate the LHC data for a SM-like Higgs boson with mass
$m_H=125.5\pm 2$~GeV \cite{Aad2012,Chatrchyan2012a,Chatrchyan2013}
and satisfy the current lower limits on  gluino  and   squark  masses  
\cite{Aad2012a,Chatrchyan2012a},  i.e.~$m_{\tilde{g}}>
1500$~GeV and  $m_{\tilde{t}}>500$~GeV.  

%In the  following, we present
%numerical results first for $a_\mu$ and then for $d_e$.

%\allowdisplaybreaks
\subsection{Results for $a_\mu$}
 
The  numerical  estimates  for  $\delta a_\mu$  exhibit  a  direct  quadratic
dependence on  the muon mass  $m_\mu$. In fact, one finds that for the
same set of soft  SUSY-breaking parameters $m_0$, $M_{1/2}$ and $A_0$,
the  ratio  $\delta a_\mu/ \delta a_e$  remains  constant to  a  good  approximation,
i.e.~$\delta a_\mu/ \delta a_e \approx m^2_\mu /m^2_e  \approx 42752.0$.  In order to
understand this parameter dependence, one has to carefully analyze the
soft SUSY-breaking contributions to the form-factors:
\begin{eqnarray}
  \label{GllgLSB}
G_{ll\gamma}^{L,{\rm SB}}
 &=&
 \tilde{V}^{0 \ell R}_{l m a} \tilde{V}^{0 \ell R*}_{l m a}
 \bigg[
   m_{l} \lambda_{\tilde{e}_a}
   J^1_{41}(\lambda_{\tilde{e}_a},\lambda_{\tilde{\chi}^0_m})
 \bigg]
 +
 \tilde{V}^{0 \ell L}_{l m a} \tilde{V}^{0 \ell L*}_{l m a}
 \bigg[
   m_{l} \lambda_{\tilde{e}_a} J^1_{41}(\lambda_{\tilde{e}_a},\lambda_{\tilde{\chi}^0_m})
 \bigg] 
\nonumber\\
 &&+\
 \tilde{V}^{0 \ell L}_{l m a} \tilde{V}^{0 \ell R*}_{l m a}
 \bigg[
   2 m_{\tilde{\chi}^0_m} \lambda_{\tilde{e}_a}
   J^0_{31}(\lambda_{\tilde{e}_a},\lambda_{\tilde{\chi}^0_m})
 \bigg]\; ,\\
    \label{GllgRSB}
G_{ll\gamma}^{R,{\rm SB}}
 &=&
 \tilde{V}^{0 \ell L}_{l m a} \tilde{V}^{0 \ell L*}_{l m a}
 \bigg[
   m_{l} \lambda_{\tilde{e}_a}
   J^1_{41}(\lambda_{\tilde{e}_a},\lambda_{\tilde{\chi}^0_m})
 \bigg]
 +
 \tilde{V}^{0 \ell R}_{l m a} \tilde{V}^{0 \ell R*}_{l m a}
 \bigg[
   m_{l} \lambda_{\tilde{e}_a} J^1_{41}(\lambda_{\tilde{e}_a},\lambda_{\tilde{\chi}^0_m})
 \bigg]\nonumber\\
 &&+\ 
 \tilde{V}^{0 \ell R}_{l m a} \tilde{V}^{0 \ell L*}_{l m a}
 \bigg[
   2 m_{\tilde{\chi}^0_m} \lambda_{\tilde{e}_a}
   J^0_{31}(\lambda_{\tilde{e}_a},\lambda_{\tilde{\chi}^0_m})
 \bigg]\; ,
\end{eqnarray}
where   the  different   terms  that   occur   in Eq~\eqref{GllgLSB}  and
\eqref{GllgRSB} are defined in Chapter~\ref{CLFV} as well as at the end
of this chapter. It is important to note that the neutralino
vertices induce  a term  which is not  manifestly proportional  to the
charged lepton  mass, but to  the neutralino mass. However, we have
numerically confirmed that $\delta a_\mu$ is proportional to $m_l^2$,
which means that the products of the mixing  matrices $\tilde{V}^{0 \ell
  R}_{l m  a} \tilde{V}^{0  \ell R*}_{l m  a}$ and  $\tilde{V}^{0 \ell
  R}_{l m a} \tilde{V}^{0  \ell L*}_{l m a}$, as well as 
$G_{ll\gamma}^{L,{\rm SB}}$ and $G_{ll\gamma}^{R,{\rm SB}}$, 
are themselves proportional to the charged
lepton mass  $m_l$ (cf~Ref~\cite{Stockinger2007}).  The latter provides a  
non-trivial powerful check for the correctness of the results presented in
this thesis.

\begin{figure}[!ht]
 \centering
 \includegraphics[clip,width=0.40\textwidth]{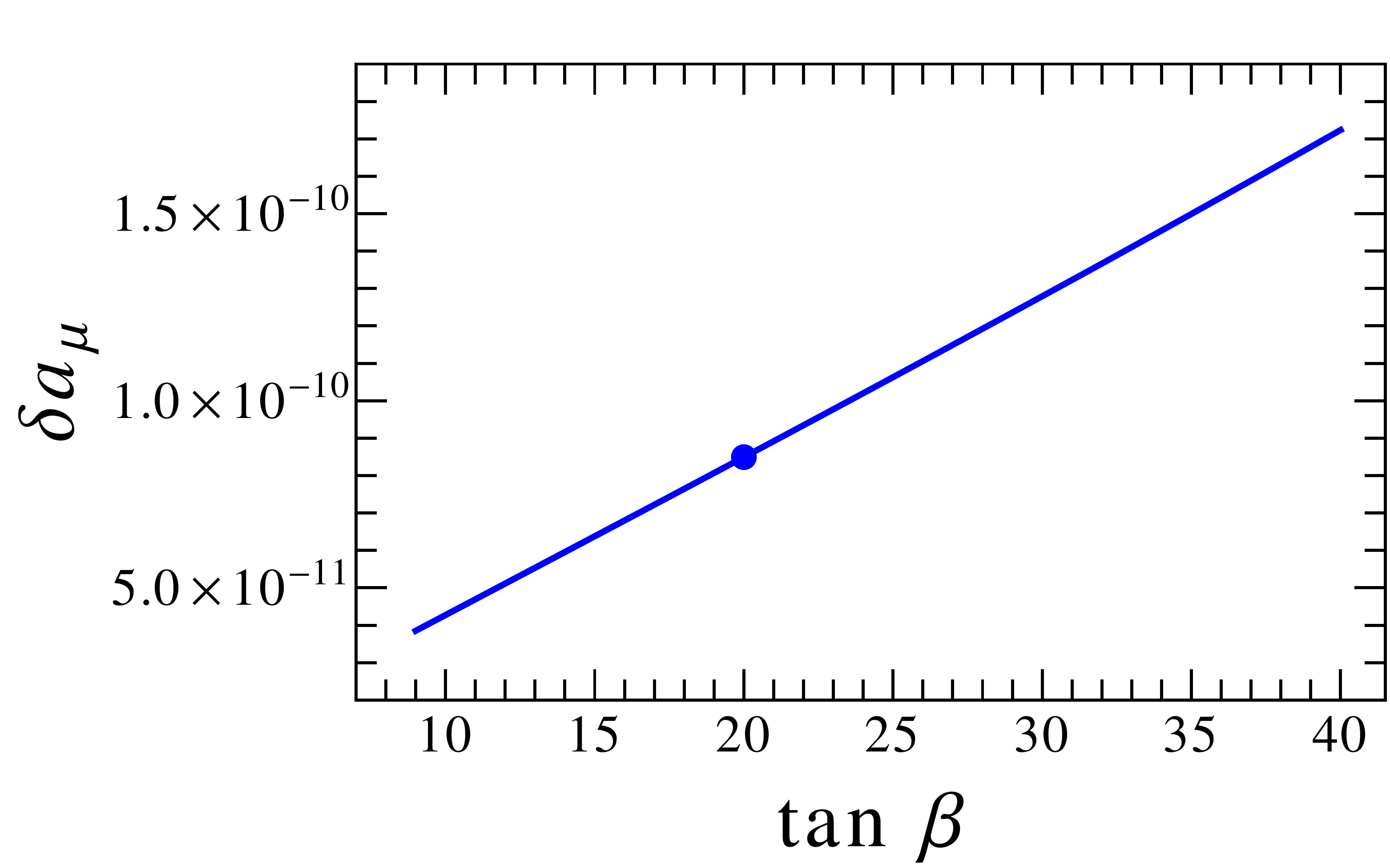} \hspace{1cm}
 \includegraphics[clip,width=0.40\textwidth]{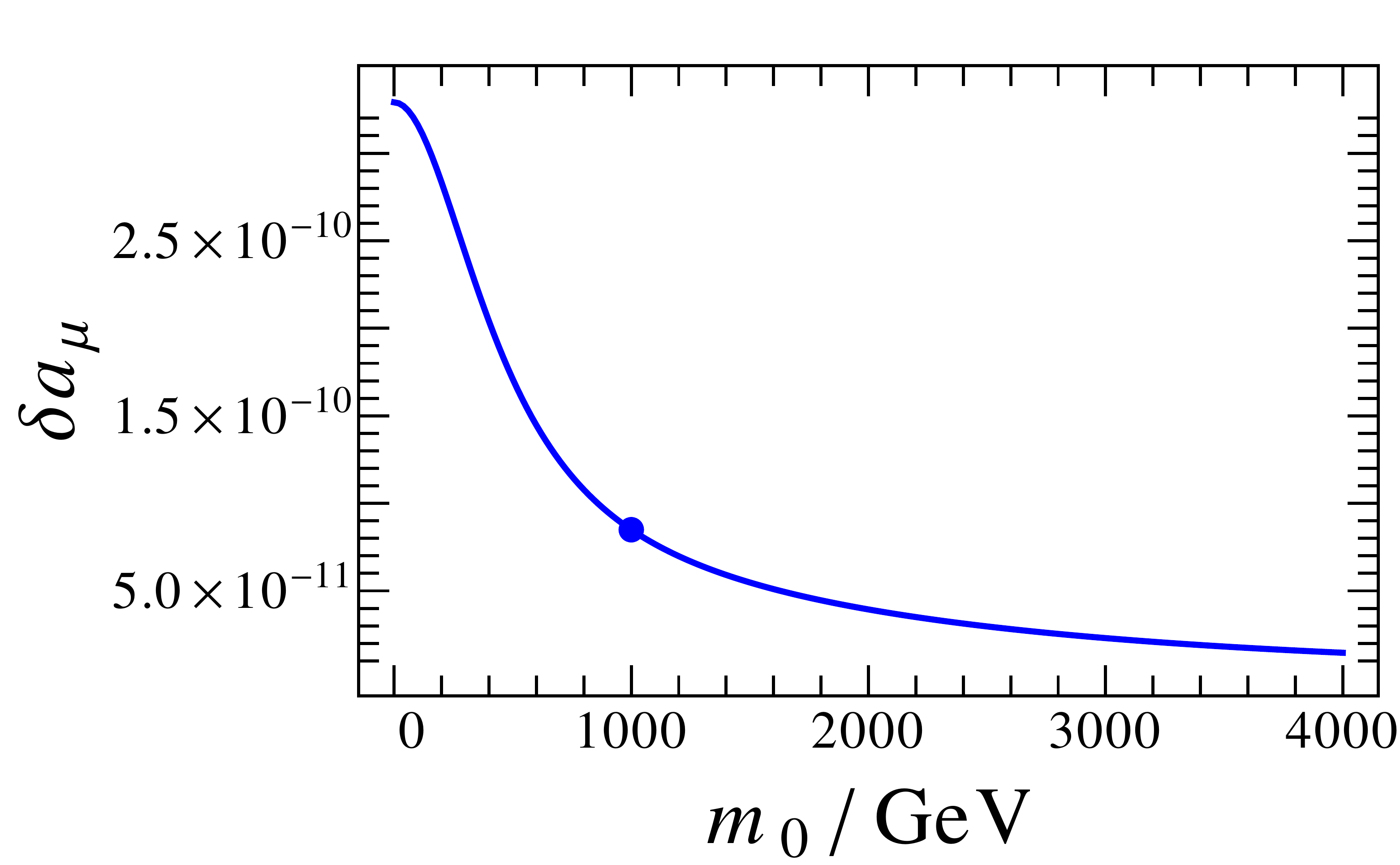}
 \vspace{-.3cm}
 \begin{flushleft}
 {\bf \footnotesize \hspace{3.7cm} (a) \hspace*{5.9cm} (b)}
 \end{flushleft}
 \vspace{2ex}
 \includegraphics[clip,width=0.40\textwidth]{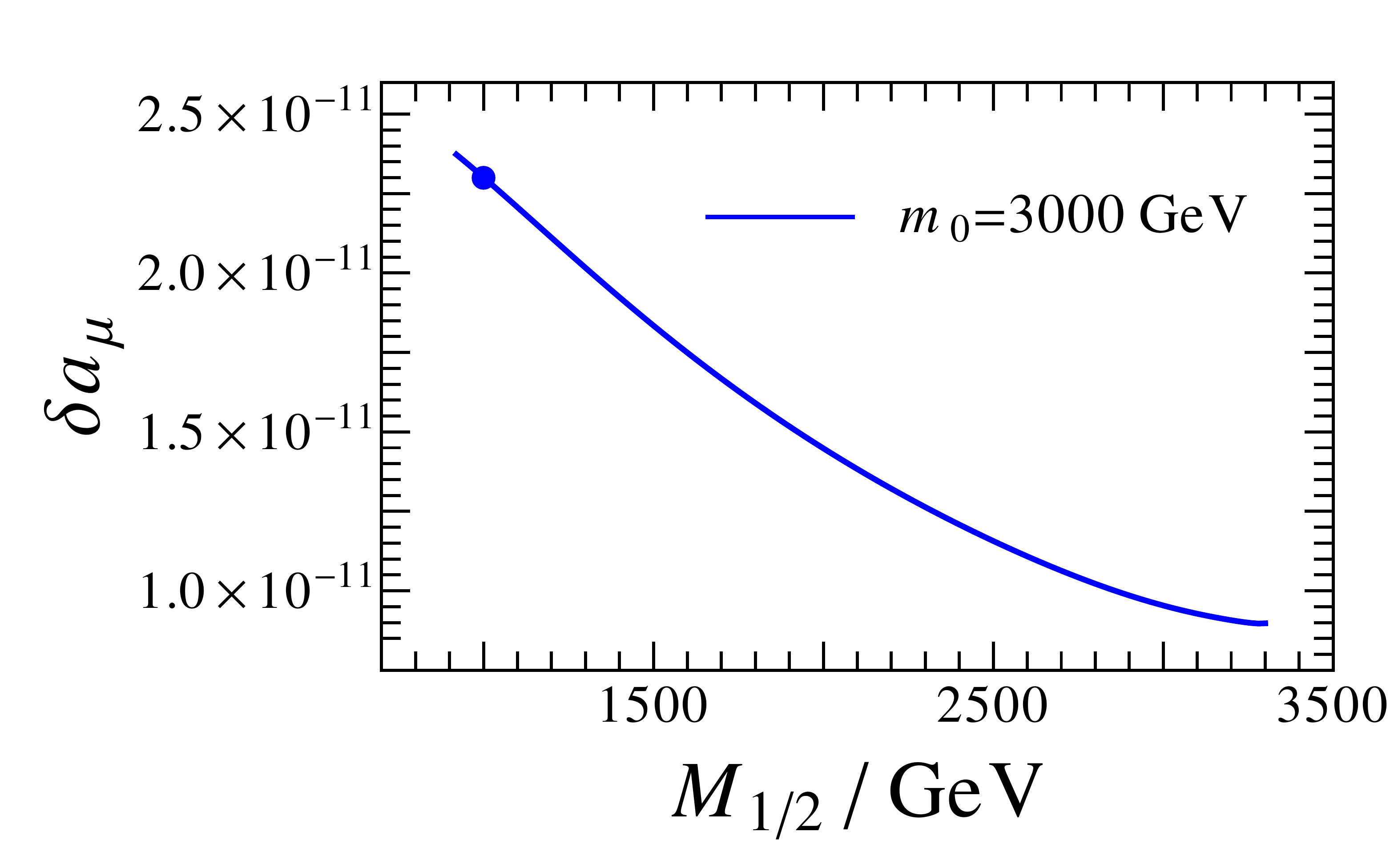}\hspace{1cm}
 \includegraphics[clip,width=0.40\textwidth]{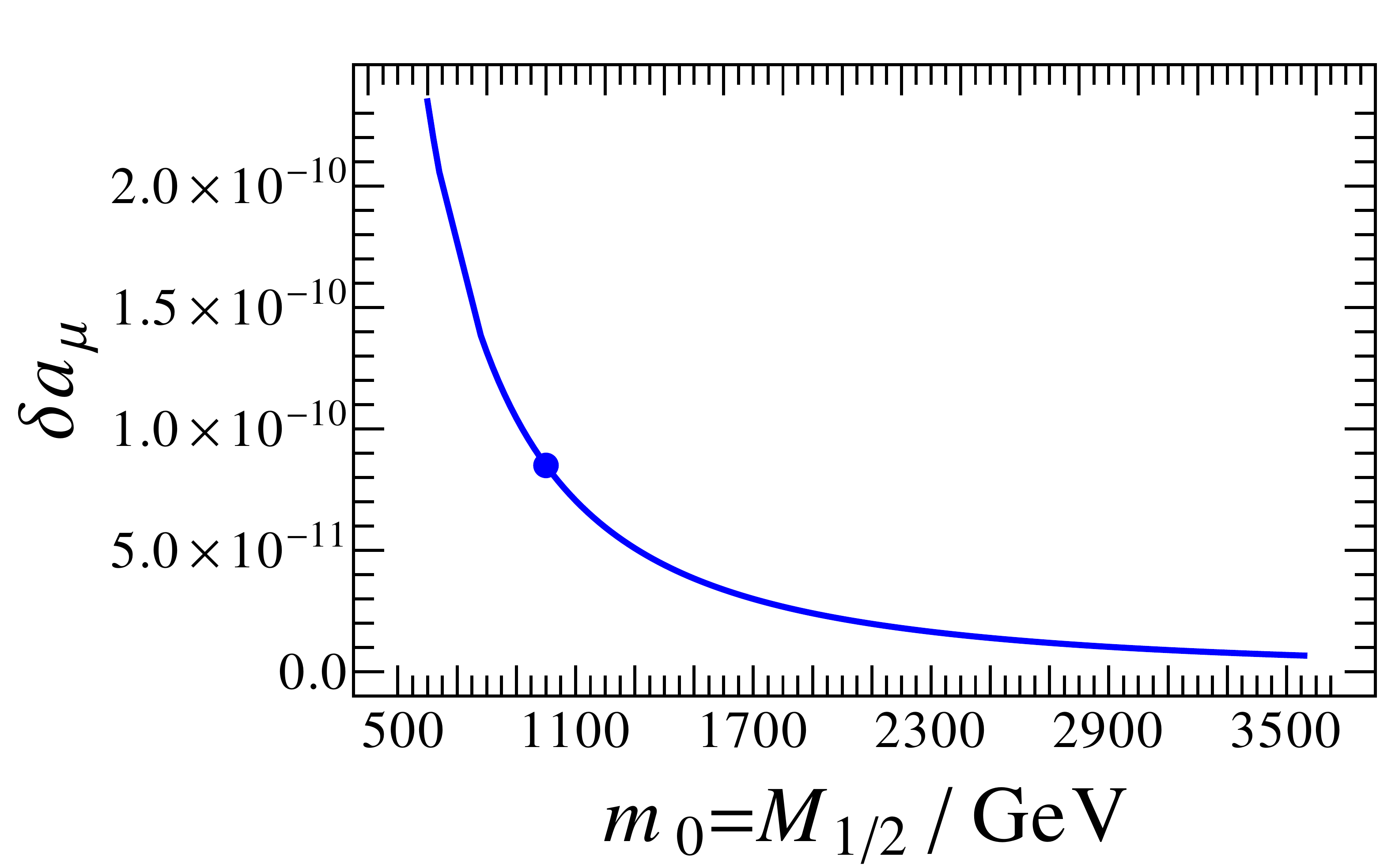}
 \vspace{-.3cm}
 \begin{flushleft}
 {\bf \footnotesize \hspace{3.7cm} (c) \hspace*{5.9cm} (d)}
 \end{flushleft}
 \vspace{2ex}
 \includegraphics[clip,width=0.40\textwidth]{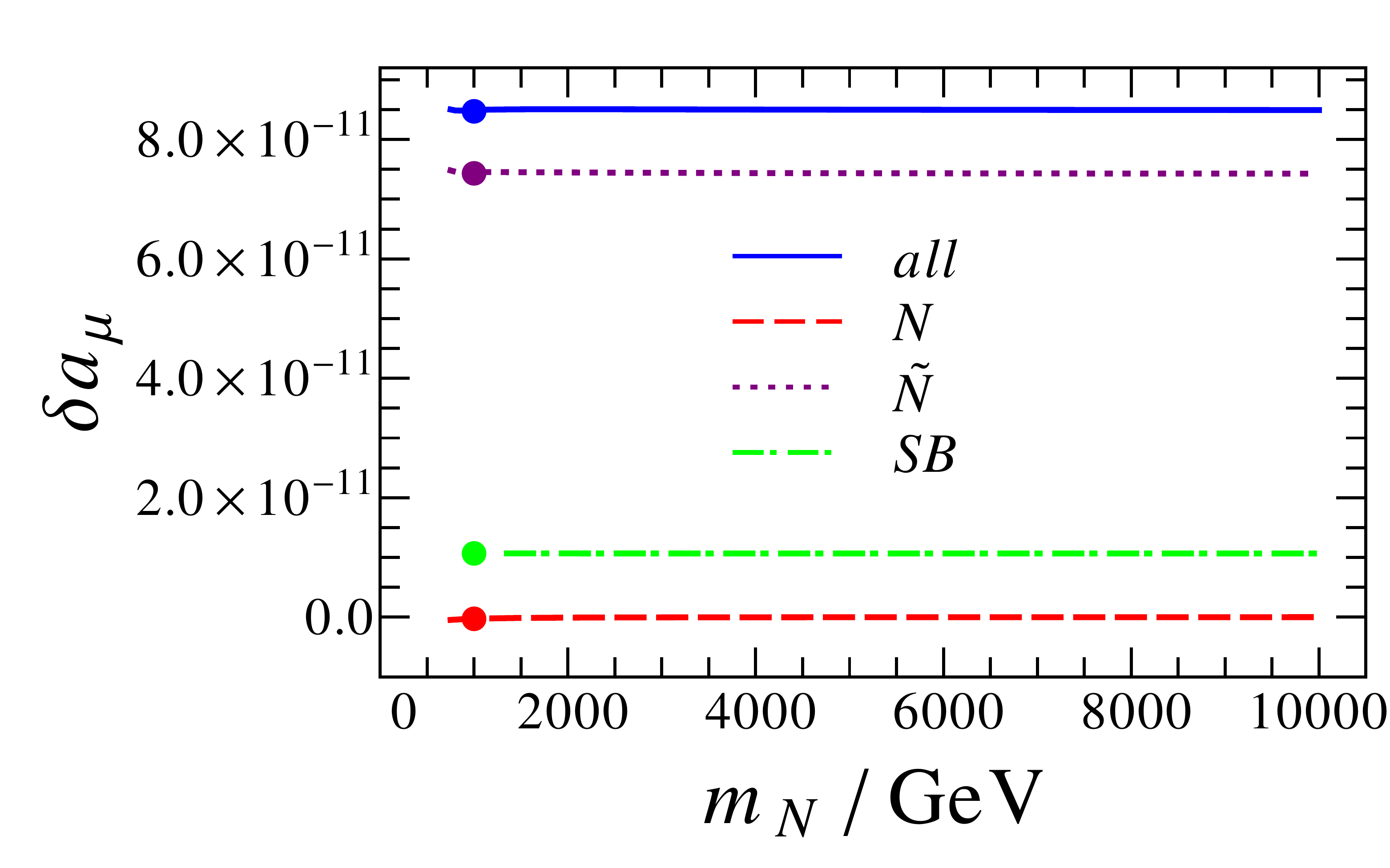}\hspace{1.75cm}
 \includegraphics[clip,width=0.36\textwidth]{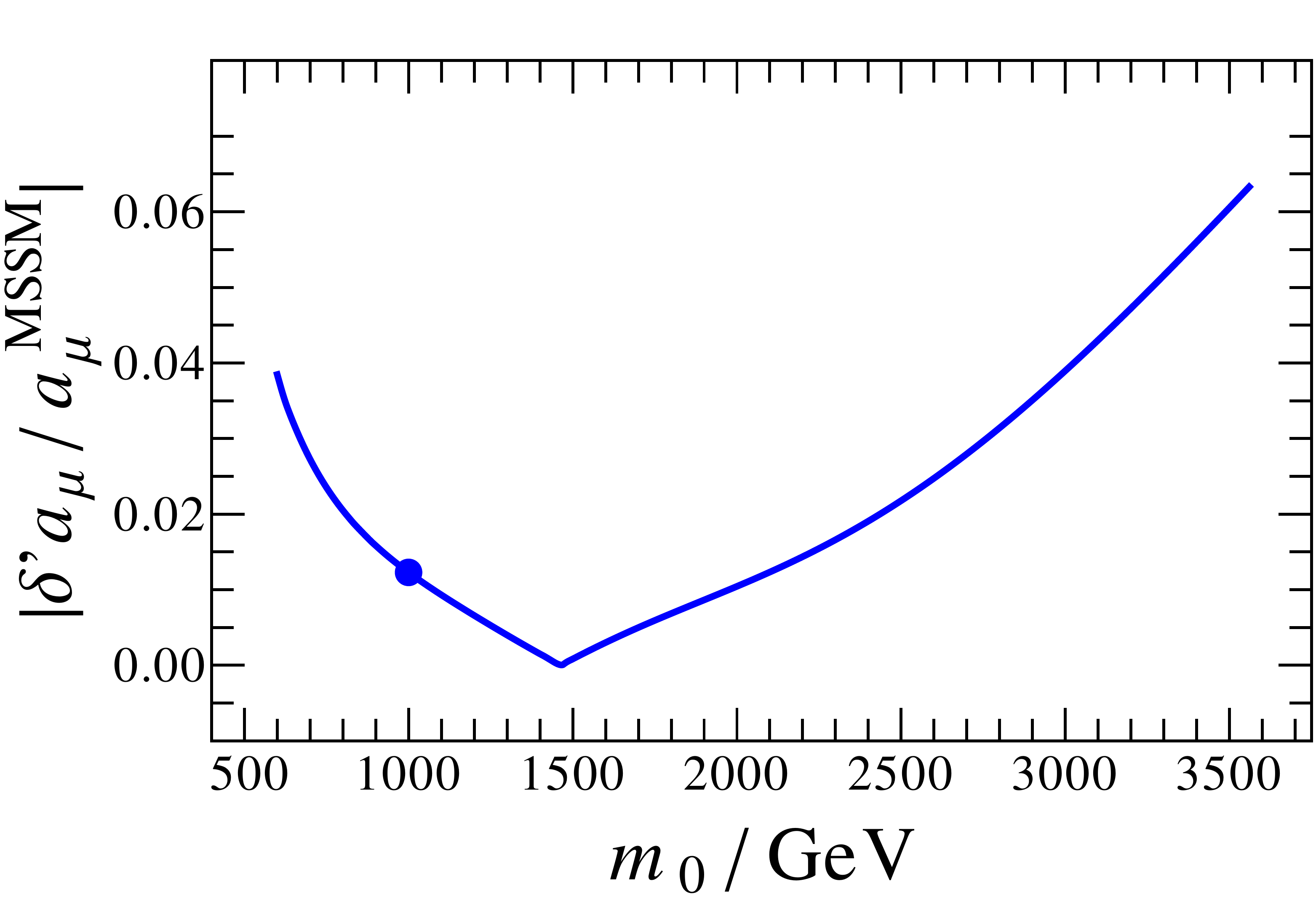}
 \vspace{-.3cm}
 \begin{flushleft}
 {\bf \footnotesize \hspace{3.7cm} (e) \hspace*{5.9cm} (f)}
 \end{flushleft}
\caption{Numerical  estimates for the contribution to the muon anomalous MDM, as
  functions  of  $\tan\beta$,  $M_{1/2}$,  $m_N$,  $m_0$  and  $m_0  =
  M_{1/2}$,  in the  $\nu_R$MSSM.  The  default parameter  set  of the
  baseline model  is given in~\eqref{baseline}.  The  pannel (e) shows
  the    heavy     neutrino~($N$),    sneutrino~($\widetilde{N}$),    soft
  SUSY-breaking~(SB)  and {\it  all}  contributions to  $\delta a_\mu$, as  a
  function of  $m_N$. The pannel  (f) displays the absolute value of the  
  relative deviation
  $\delta'  a_\mu/a_\mu$ of  the $\nu_R$MSSM  and MSSM  predictions for
  $a_\mu$ [cf.~\eqref{damu}],  as a function  of $m_0$.  The  range of
  input parameters in all plots satisfy the current LHC constraints on
  Higgs, gluino and squark masses.   The heavy dots on the curves give
  the  predicted   values  evaluated   for  the  default
  parameters \eqref{baseline}.}
\label{Fig4.1}
\end{figure}

In addition, this numerical analysis  shows that the contribution to the 
muon anomalous MDM is  
almost independent of the  neutrino-Yukawa parameters $a$,
$b$  and $c$, the  heavy neutrino  mass $m_N$  and the  soft trilinear
parameter  $A_0$.  Hence,  our  results are  almost  insensitive to  a
particular  choice  for  a  neutrino  Yukawa  texture,  e.g.~as  given
in~(\ref{YU1})  and~(\ref{YA4}), and  also independent  of  the CP-odd
phases $\theta$ and $\phi$.
 
In  Fig~\ref{Fig4.1}, the numerical estimates  for  $\delta a_\mu$ are given, as
a function of the key theoretical  parameters: $\tan\beta$, $M_{1/2}$,
$m_0$ and~$m_N$.  In the frame (a) of this figure, we see that $\delta a_\mu$
depends linearly  on $\tan\beta$, as expected from~\eqref{al_approx}. 
Likewise, in Fig~\ref{Fig4.1} we have investigated  the  dependence of
$\delta a_\mu$ on the soft  SUSY-breaking parameters $m_0$ and $M_{1/2}$, for
different kinematic  situations, and obtained  results consistent with
the scaling behaviour of~$1/M^2_{\rm SUSY}$ in~(\ref{al_approx}).

In the  pannel (e) of Fig~\ref{Fig4.1},  we observe that  the effect of
the heavy right-handed neutrinos~($N$) and sneutrinos~($\widetilde{N}$) on
$\delta a_\mu$  is negative,  but  small, in  agreement  with our  discussion
above.  The size  of their contributions alone to  $a_\mu$ ranges from
$-10^{-12}$ to  $-4.8\times 10^{-15}$, for  $m_N = 0.5 - 10~\textrm{ TeV}$.  
On the other hand, the left-handed sneutrino contributions to $a_\mu$ are
approximately independent of the heavy Majorana  mass $m_N$, reaching
values   $\approx  8.5\times   10^{-11}$.    The  soft   SUSY-breaking
contributions are also approximately independent of the heavy Majorana
mass~$m_N$ and  have values  $\approx 1.1\times 10^{-12}$.   Note that
the light  sneutrino contribution to the anomalous  magnetic moment is
the  largest in  magnitude,  and it  is  already present  in the  MSSM
contributions to  $a_\mu$.

Finally, we have checked  the dominance of the MSSM contributions by 
looking at the dependence of the parameter:
\begin{equation}
  \label{damu}
\delta' a_\mu \ =\ a^{\rm \nu_RMSSM}_\mu\ -\ a^{\rm MSSM}_\mu\,.
\end{equation}
The difference  $\delta' a_\mu$ of  the predictions for  $a_\mu$ within
the  $\nu_R$MSSM  and the  MSSM  divided by $a_\mu$ is  evaluated,  
and the absolute values of the results  are
displayed  in the  pannel (f)  of  Fig~\ref{Fig4.1}, as  a function  of
$m_0=M_{1/2}$.  The largest deviation from the MSSM is found for 
largest allowed parameter value, $m_0=3600~\textrm{ GeV}$, in which case
$\delta' a_\mu/a^{\rm MSSM}_\mu$ is as large as $6.2\times 10^{-2}$.

\subsection{Results for $d_e$}

We will now study the dependence of the electron EDM  $d_e$ on several key
model parameters, such as $m_0$, $M_{1/2}$, $B_0$, $A_0$, $\tan\beta$,
$\theta$ and  $\phi$. The predictions  for $d_\mu$ may be  obtained by
using  the naive  scaling relation:  $d_\mu \approx  (m_\mu/m_e)\, d_e
\approx 205\, d_e$.  It is found this scaling behaviour is 
numerically satisfied very well. The maximal numerical values for 
$d_e$ obtained are of the order $\sim 10^{-27}$~e$\;$cm. The
predicted values for $d_\mu$ are therefore always found to be  less than~$\sim
10^{-25}\ e\;  {\rm cm}$, which  is several orders of  magnitude below
the  present  experimental   upper  bound:  $d_\mu=  0.1\pm  0.9\times
10^{-19}\ e\; {\rm cm}$ \cite{Beringer2012}.

It is noted that heavy singlet neutrinos $N$ do  not contribute to $d_e$,
even if the  soft SUSY-breaking CP-odd phases $\phi$  and $\theta$ are
taken to be non-zero. On the  other hand, soft SUSY-breaking and 
right handed neutrino effects induce non-vanishing  $d_e$,  if  either  
$\theta$ or  $\phi$  are  non-zero. If both $\phi=0$ and $\theta=0$, 
lepton EDMs $d_l$ numerically vanish. Therefore, the complex products of 
vertices~\eqref{CPNt} emerging in the $\nu_R$MSSM do not induce the CP 
violation at one loop level, in accord with the result of 
Ref~\cite{Farzan2004} obtained in the MSSM with a high-scale seesaw mechanism.
\begin{figure}[!ht]
 \centering
 \includegraphics[clip,width=0.40\textwidth]{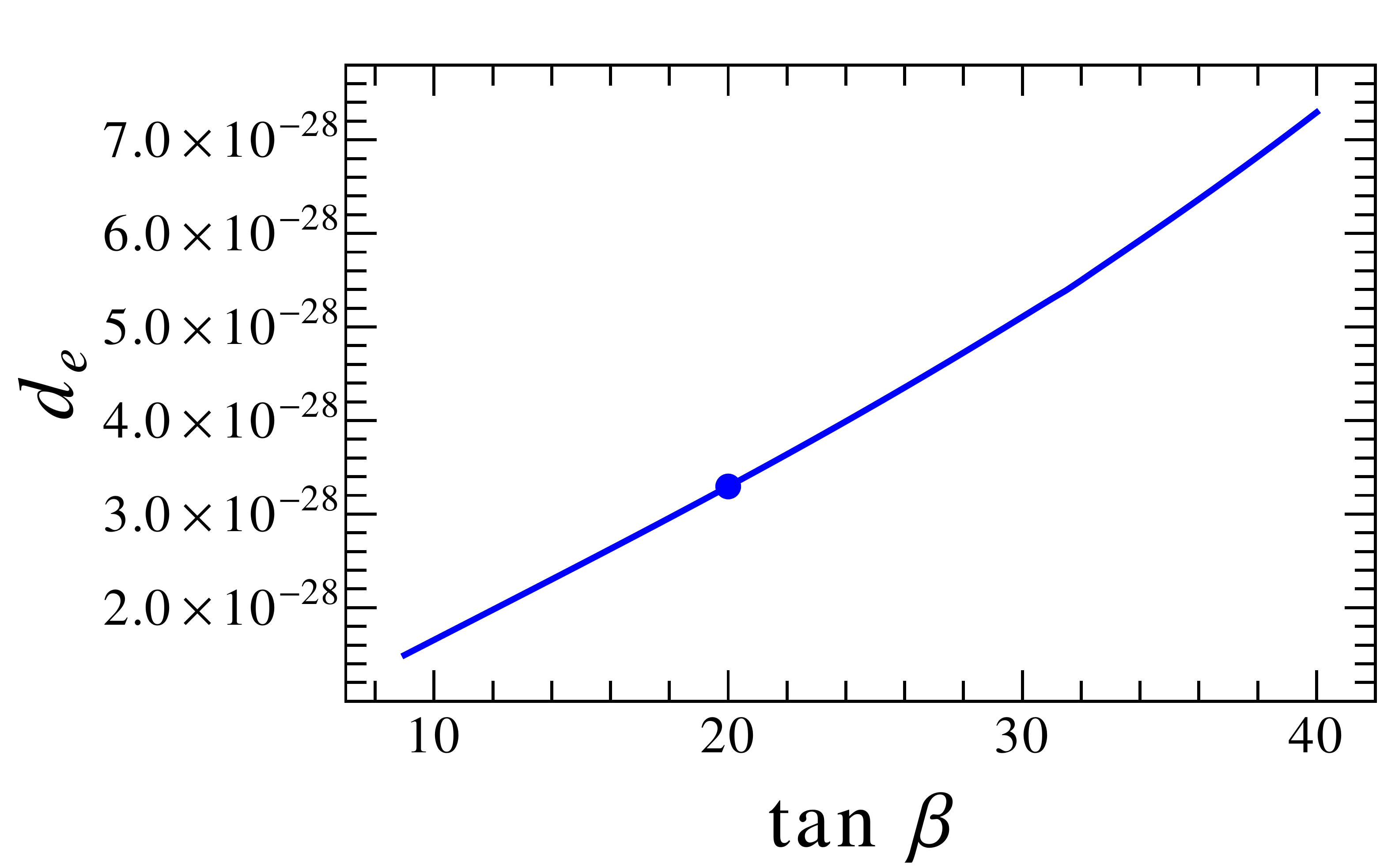} \hspace{1cm}
 \includegraphics[clip,width=0.40\textwidth]{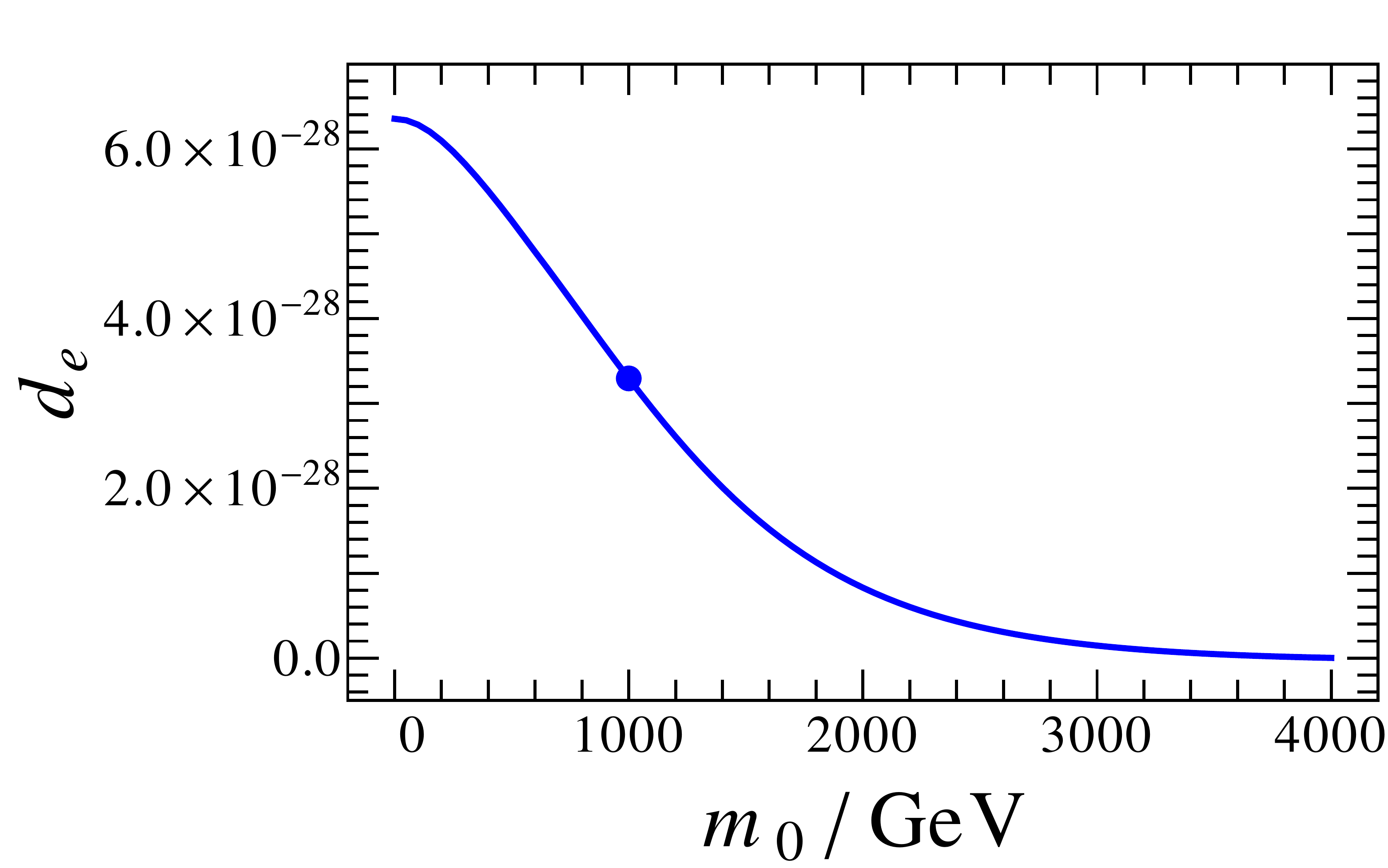}
 \vspace{-.3cm}
 \begin{flushleft}
 {\bf \footnotesize \hspace{3.7cm} (a) \hspace*{5.9cm} (b)}
 \end{flushleft}
 \vspace{2ex}
 \includegraphics[clip,width=0.40\textwidth]{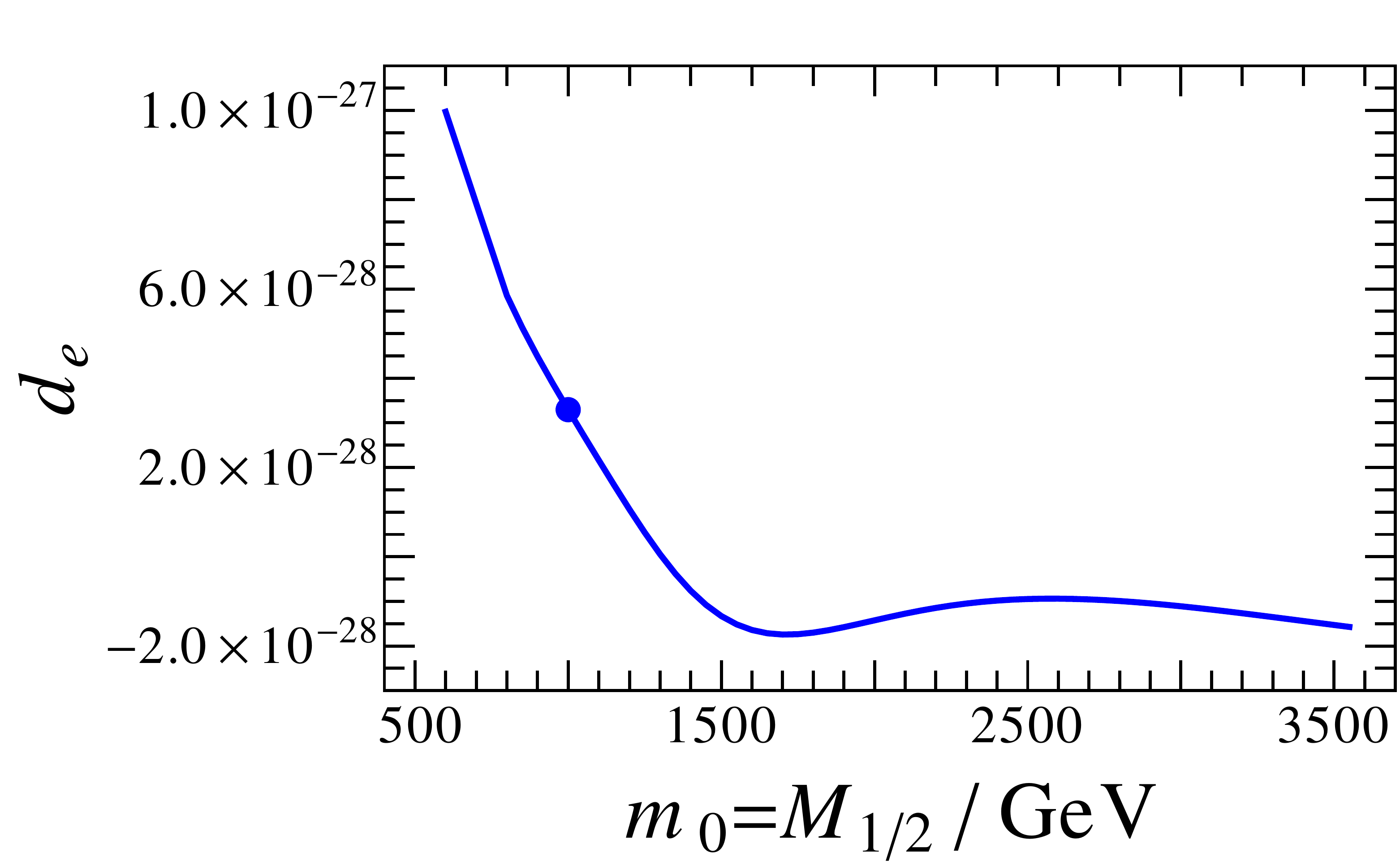} \hspace{1cm}
 \includegraphics[clip,width=0.40\textwidth]{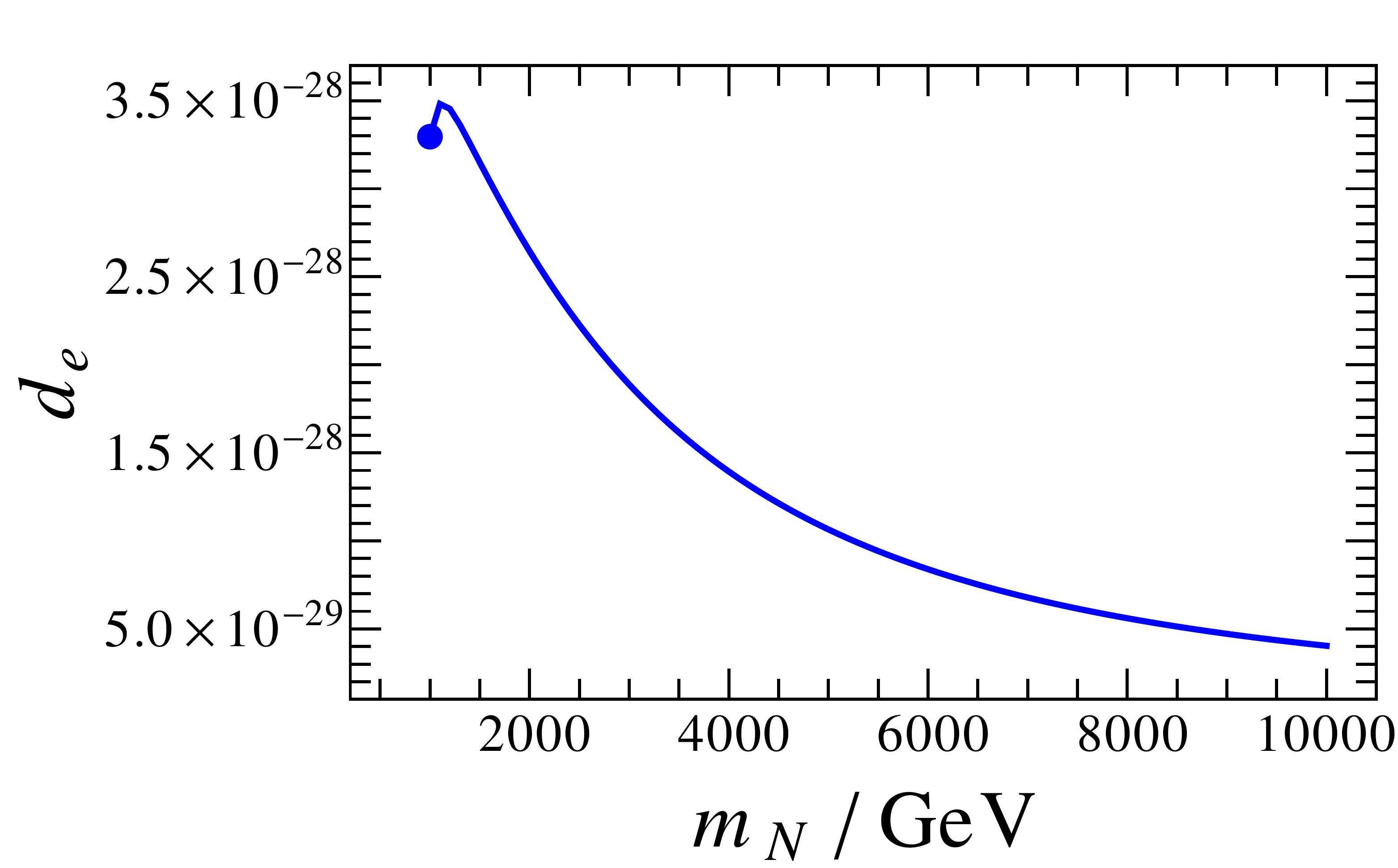}
 \vspace{-.3cm}
 \begin{flushleft}
 {\bf \footnotesize \hspace{3.7cm} (c) \hspace*{5.9cm} (d)}
 \end{flushleft}
\caption{Numerical  estimates   of  the  electron   EDM~$d_e$  in  the
  $\nu_R$MSSM, as  functions of $\tan\beta$,  $m_0$, $m_0=M_{1/2}$ and
  $m_N$, for $\phi=\pi/2$. The remaining parameters not 
  shown assume the baseline values in~(\ref{baseline}). All
  input parameters are chosen so  as to satisfy the LHC constraints on
  Higgs,  gluino and  squark masses.   The  heavy dots  on the  curves
  indicate the  predicted values for  $d_e$ evaluated for  the default
  parameters (\ref{baseline}).}
\label{Fig4.2}
\end{figure}

In  Fig~\ref{Fig4.2}, the present numerical estimates  of $d_e$  on the
$\nu_R$MSSM  parameters $\tan\beta$, $m_0$,  $M_{1/2}$ and  $m_N$, for
maximal $A_0$ phase, $\phi = \pi/2$ are presented. The value of $\theta$
is set to zero, since the dependence of $d_e$  on $B_0$ is weaker than  
the dependence on $A_0$. As shown in pannel (a) of Fig~\ref{Fig4.2} $d_e$ exhibits a linear 
dependence on $\tan\beta$ confirming the $\tan\beta$ naive scaling 
behaviour in Eq~\eqref{di_approx}.

Further, $d_e$ is a decreasing function of $m_0$. As a function of 
$m_0=M_{1/2}$, $d_e$ assumes both positive and negative values, and is roughly
proportional to $-1 -2.4\,\mathrm{TeV}/m_0 + 6.3\mathrm{TeV}^2/m_0^2$.
There is also  a small region of parameter space for  
$m_0=M_{1/2} \stackrel{<}{{}_\sim} 800~\textrm{ GeV}$,  for   which  the
prediction for $d_e$ is of the order of the  experimental upper limit on 
$d_e$ \eqref{deUB}. In addition, $d_e$ decreases with increasing $m_N$: 
for the $m_N$ values from the pannel (d) of Fig~\ref{Fig4.2} this behavior can
roughly approximated by a function $-0.13+\mathrm{TeV}^{\frac{2}{3}}m_N^{-\frac{2}{3}}$, 
in the $m_N$-range $10\,\mathrm{TeV}<m_N<100\,\mathrm{TeV}$ $d_e$ roughly scales as 
$1/m_N$, and above $m_N=100\,\mathrm{TeV}$ it becomes very slowly decreasing function 
in $m_N$. 

\begin{figure}[!t]
 \centering
 \includegraphics[clip,width=0.40\textwidth]{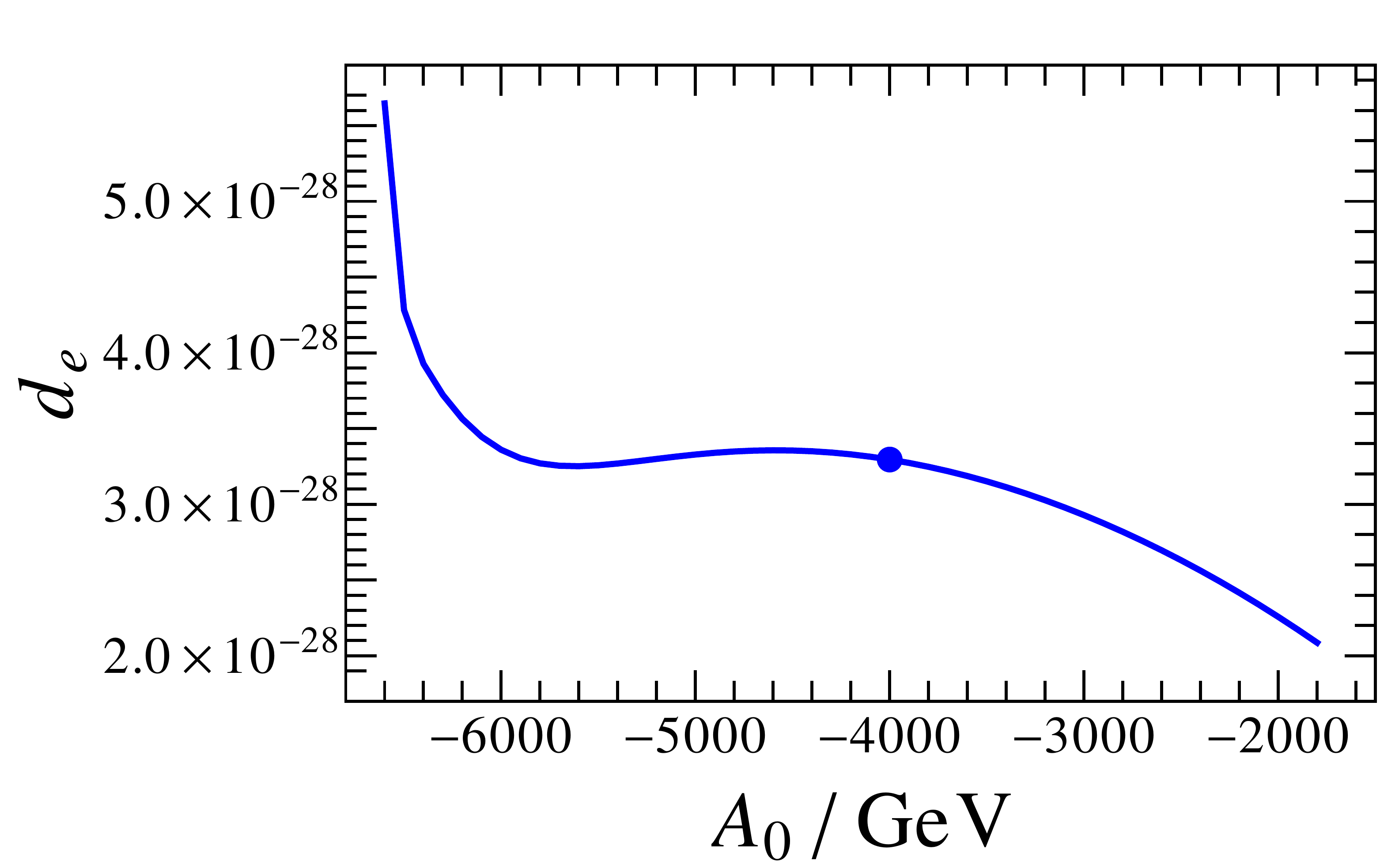}\hspace{1cm}
 \includegraphics[clip,width=0.40\textwidth]{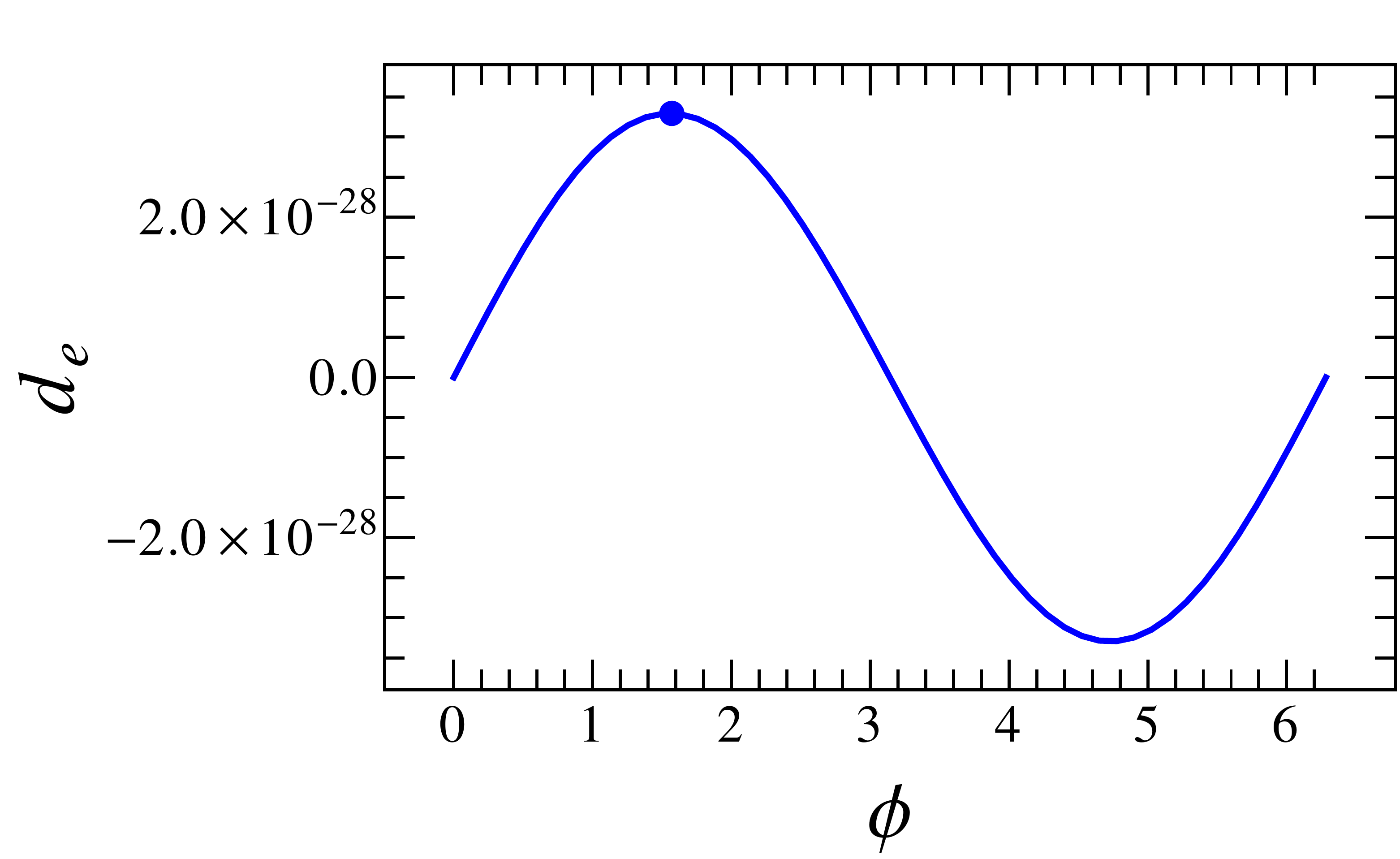} 
 \vspace{-.3cm}
 \begin{flushleft}
 {\bf \footnotesize \hspace{3.7cm} (a) \hspace*{5.9cm} (b)}
 \end{flushleft}
 \vspace{2ex}
 \includegraphics[clip,width=0.42\textwidth]{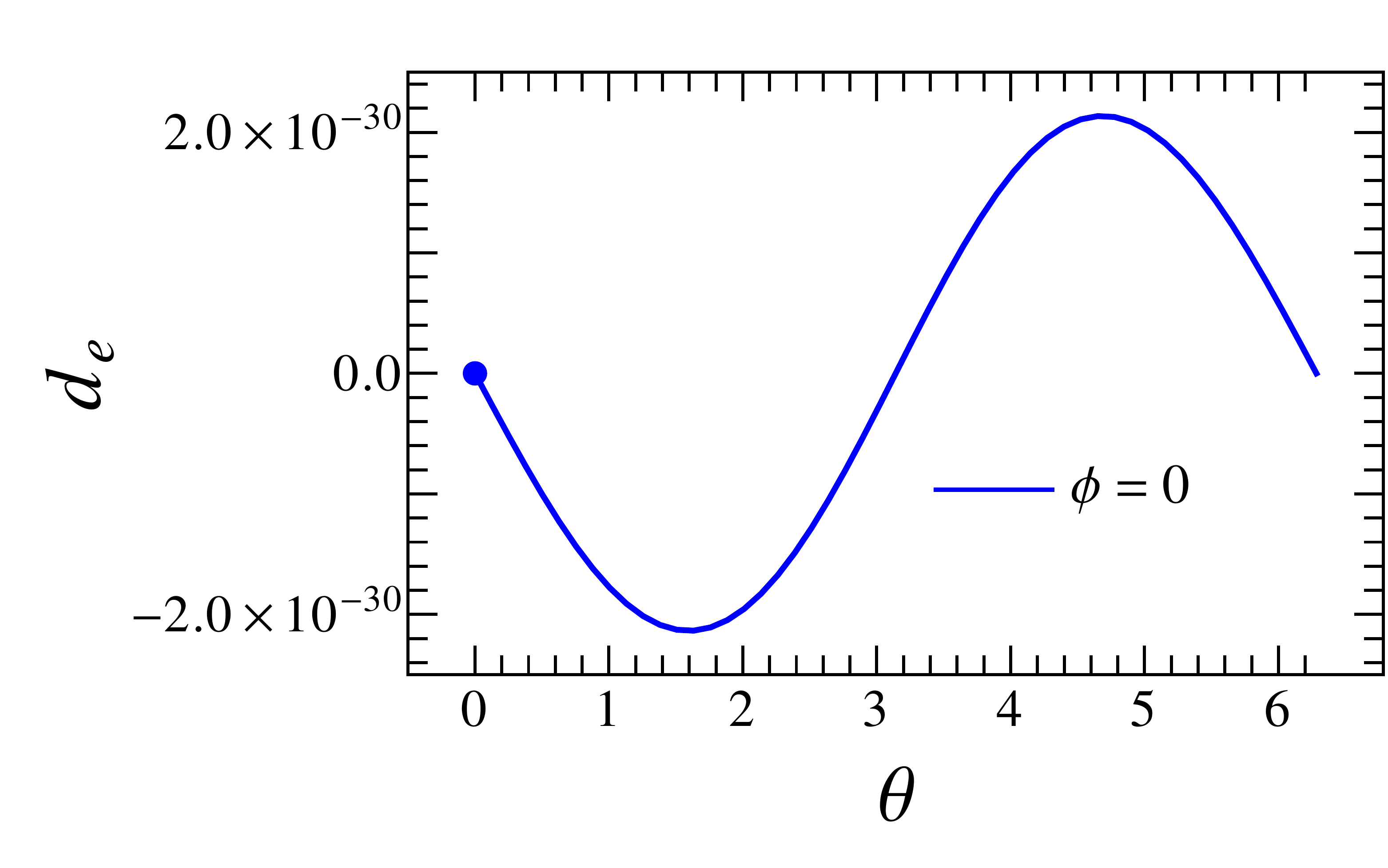}\hspace{1cm}
 \includegraphics[clip,width=0.40\textwidth]{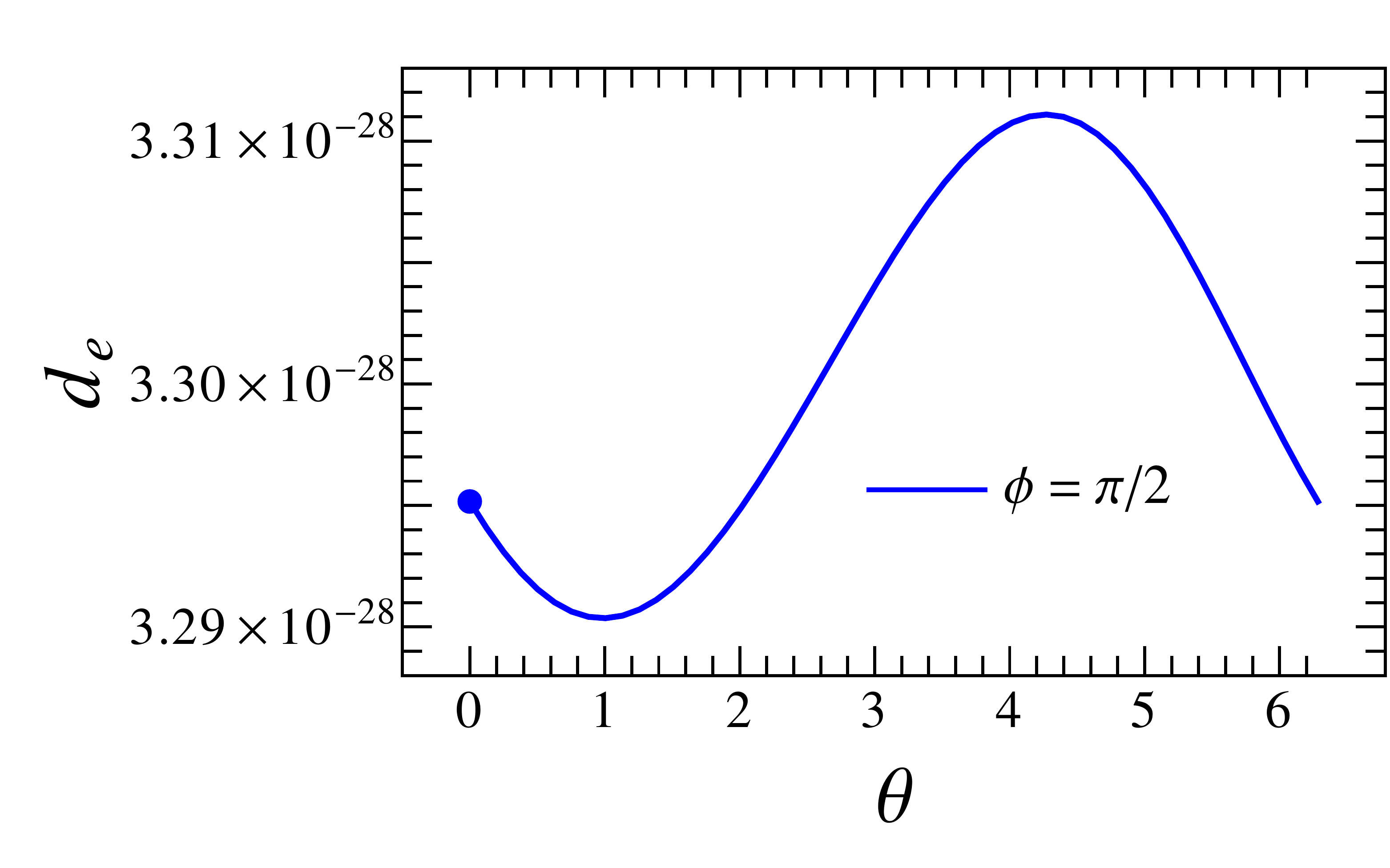}\hspace*{.5cm}
 \vspace{-.3cm}
 \begin{flushleft}
 {\bf \footnotesize \hspace{3.7cm} (c) \hspace*{5.9cm} (d)}
 \end{flushleft}
 \vspace{2ex}
 \includegraphics[clip,width=0.40\textwidth]{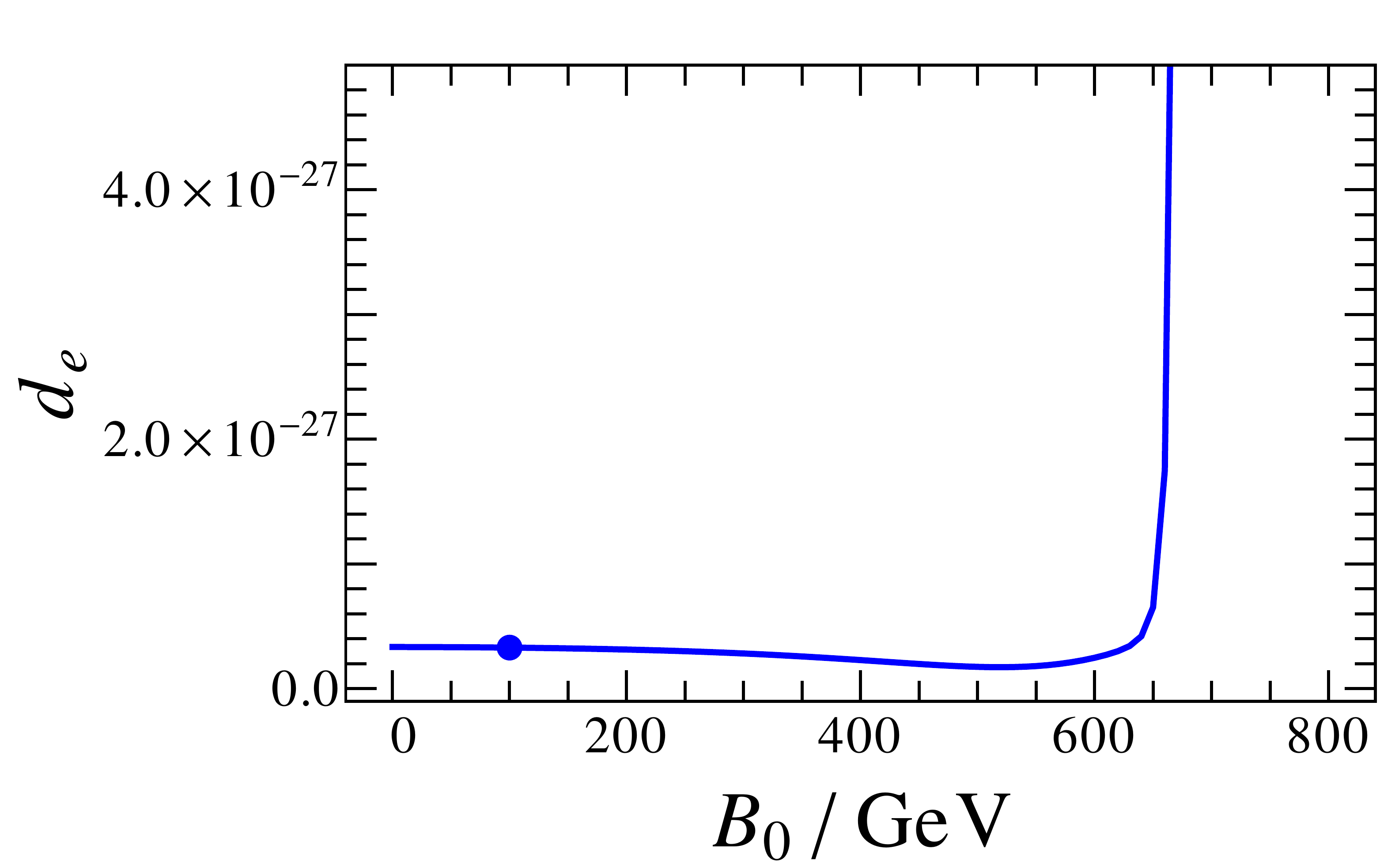}%\hspace*{7cm}
 \vspace{-.3cm}
 \begin{flushleft}
 {\bf \footnotesize \hspace{7cm} (e) }
 \end{flushleft}
\caption{Predicted numerical values  for the electron EDM~$d_e$ versus
  the  soft  SUSY  breaking  parameters  $A_0$  and  $B_0$  and  their
  corresponding  soft  CP-odd  phases   $\phi$  and  $\theta$  in  the
  $\nu_R$MSSM,  for the  baseline  scenario in~\eqref{baseline}. 
  If not shown $\phi$ assumes value $\pi/2$.   The
  range of input parameters shown  in the plots is compatible with the
  LHC constraints on  Higgs, gluino and squark masses.  The heavy dots
  show the  predicted values for  $d_e$, using the  default parameters
  (\ref{baseline}).}
\label{Fig4.3}
\end{figure}

In Fig~\ref{Fig4.3}, we show  the predicted numerical values for $d_e$,
as functions of the soft SUSY-breaking parameters $A_0$ and $B_0$, and
their corresponding CP phases $\phi$  and $\theta$. In all pannels
except the pannel (c), where $\phi=0$ and $\theta$ is a variable, 
$\phi$ assumes value $\pi/2$ or it is a variable
and  $\theta$ is taken to be equal zero. In the pannel (a)
of Fig~\ref{Fig4.3}, the soft  trilinear parameter $A_0$ is constrained
by the  LHC data  pertinent to Higgs,  gluino and squark  masses.  
The electron EDM $d_e$ is a complicated function of $|A_0|$ that 
slowly rises for $|A_0|$ between $1.8$~TeV and $4.5$~TeV, slowly 
decreases for $|A_0|$ between $4.5$~TeV and $6$~TeV, and steeply 
rises for $|A_0|>6$~TeV. This function cannot be precisely described 
by a simple Laurent series in $|A_0|$, but in the largest part of the allowed 
$|A_0|$ interval it can roughly be approximated by a constant. The $\phi$
dependence  of $d_e$  is almost  sinusoidal with  an amplitude few times
smaller than the experimental upper bound~\eqref{deUB}. Moreover,
$d_e$  is approximately constant function  of $B_0$, up  to 
$B_0\approx  600$~GeV.  For  larger values, 
i.e.~$B_0\stackrel{>}{{}_\sim} 600$~GeV,  $d_e$ steeply rises, suggesting
the numerical  instability in the diagonalization of  the
sneutrino mass  matrix, which probably makes the results in this regime invalid.
For $\phi=\pi/2$, the  electron EDM $d_e$ attains values  of order the
experimental upper limit~\eqref{deUB}, but  for $\phi=\theta = 0$, the
predictions are numerically consistent with zero.  The dependence of
$d_e$ on $\theta$ is sinusoidal with an amplitude of order few$\times
10^{-30}$,  while its  average value  strongly depends  on  the chosen
value  $\phi$.  From  Figs~\ref{Fig4.2} and \ref{Fig4.3},  the following
dependence of $d_l$ on $m_l$, $m_0=M_{1/2}$, $m_N$ and $\tan\beta$ may be 
deduced:
\begin{equation}
  \label{dlapp1}
d_l\ \propto\ \tan\beta\cdot m_l \cdot \frac{f(m_0)}{m_N^x},\qquad m_N<10\,\mathrm{TeV}, 
\end{equation}
where $x$ assumes values between $2/3$ and $1$, and $f(m_0)$ is roughly 
proportional to the function $-1 -2.4\,\mathrm{TeV}/m_0 + 6.3\mathrm{TeV}^2/m_0^2$. 
The last factor in Eq~\eqref{dlapp1} corresponds to the scaling factor $1/M_{SUSY}^2$ 
in the naive approximation \eqref{di_approx}, and in the approximate expressions 
for lepton EDM derived in Ref~\cite{Farzan2004}.

\section{Technical remarks}
\label{f3.4and3.5}

Let's end this chapter with several technical remarks, including the
detailed analytical expressions for all the quantities
that   appear    in   the   form factors    $G^{L,SB}_{ll\gamma}$   and
$G^{R,SB}_{ll\gamma}$,  given in \eqref{GllgLSB}  and \eqref{GllgRSB},
respectively.  To start with, the variables $\lambda_X$ are defined as
$\lambda_X=m_X^2/M_W^2$, for instance $\lambda_{\tilde{e}}=m^2_{\tilde{e}}/M_W^2$.  
The integrals $J^a_{bc}$ derived from loop integrations
(see Appendix \ref{sec:lf}) are UV finite. These are given by
\begin{eqnarray}
  \label{Jabc}
J^a_{bc} &=&
 (-1)^{a-n_b-n_c} 
 \int_0^\infty \frac{dx x^{1+a}}{(x+\lambda_b)^{n_b}(x+\lambda_c)^{n_c}}\,.
\end{eqnarray}
The couplings $\tilde{V}^{0 \ell L}_{l m a}$ and $\tilde{V}^{0 \ell R}_{l m a}$ 
read: 
\begin{align}
\tilde{V}^{0 \ell L}_{l m a}
 &=
 -\sqrt{2} t_w Z^*_{m1} (R_R^{\tilde{e}})^*_{al}
 - \frac{(m_e)_l}{\sqrt{2} c_\beta M_W} Z^*_{m3} (R_L^{\tilde{e}})^*_{al} \,,
\\
\tilde{V}^{0 \ell R}_{l m a}
 &=
\frac{1}{\sqrt{2} c_w}(c_w Z_{m2} +s_w Z_{m1}) (R^{\tilde{e}}_L)^*_{al}
- \frac{(m_e)_l}{\sqrt{2}c_\beta M_W} Z_{m3} (R_R^{\tilde{e}})^*_{al} \,,
\end{align}
where   $t_w=\tan\theta_w$,   $c_w=\cos\theta_w$,  $s_w=\sin\theta_w$,
$c_\beta=\cos\beta$.  The unitary matrices  ${\cal U}$ and ${\cal V}$,
which diagonalize the chargino mass matrix, and the unitary matrix $Z$
diagonalizing    the     neutralino    mass    matrix     are    taken
from Ref \cite{Drees2004}.

Finally,   the    following   lepton-slepton
disalignment matrices may be defined:
\begin{eqnarray}
R^{\tilde{e}L}_{ak} &=& U^{\tilde{e}}_{ia} U^{e_L*}_{ik}\,,
\nonumber\\
R^{\tilde{e}R}_{ak} &=& U^{\tilde{e}}_{i+3a} U^{e_R*}_{ik}\,,
\end{eqnarray}
where  $U^{e_L}$, $U^{e_R}$ and  $U^{\tilde{e}}$ are  unitary matrices
diagonalizing the lepton and slepton mass matrices, with $a=1,\dots,6$
and $i,k=1,2,3$.

%% file: conclusions.tex
\chapter{Conclusions} \label{conclusions}

In Chapter \ref{CLFV} Charged Lepton Flavor Violation was analysed in 
the MSSM extended by low-scale singlet heavy  neutrinos, paying special 
attention to the individual loop contributions due to the heavy 
neutrinos~$N_{1,2,3}$, sneutrinos~$\widetilde{N}_{1,2,\dots,12}$   
and   soft   SUSY-breaking terms. In this analysis, we 
have,  for the  first  time, included the complete  set of  box  diagrams, 
in  addition  to the  photon and  the $Z$-boson mediated  interactions.  
We have also derived the complete set of chiral amplitudes and  their 
associate form factors  related to the neutrinoless three-body  CLFV decays 
of the muon  and tau, such as $\mu \to eee$, $\tau \to  \mu\mu\mu$, 
$\tau \to e\mu\mu$ and $\tau \to ee\mu$,  and to the  coherent 
$\mu  \to e$  conversion in  nuclei. Our analytical results are  
general and can be applied to  most of the New Physics  models with  CLFV.  
In  this context,  we emphasize  that this systematic  analysis  has  
revealed  the  existence  of  two  new  box formfactors,  which have not  
been considered  before in  the existing literature of New Physics theories 
with CLFV.

This detailed  study has shown  that the soft SUSY-breaking  effects in
the  $Z$-boson-mediated  graphs  dominate  the CLFV  observables,  for
appreciable  regions of  the  $\nu_R$MSSM parameter  space in  mSUGRA.
Nevertheless, there  is a significant  portion of parameter  space for
heavy  neutrino  masses  $m_N\stackrel{<}{{}_\sim} 1$~TeV,  where  box
diagrams involving heavy  neutrinos in the loop can  be comparable to,
or even  larger than the corresponding  $Z$-boson-exchange diagrams in
$\mu   \to  eee$   and   in   $\mu  \to   e$   conversion  in   nuclei
(cf.~Fig~\ref{Fig11}). In   the  same  kinematic   regime,  due  to
accidental numerical cancellations,  we have also observed a suppression of the
branching ratios for  the photonic CLFV decays $\mu  \to e \gamma$, as
well as for  the decays $\tau \to e \gamma$  and $\tau \to \mu\gamma$.
As  was already mentioned,  such a  suppression  in low-scale seesaw models 
is a consequence  of a cancellation between particle and sparticle contributions 
due to the approximate realization of the SUSY
no-go  theorem due  to  Ferrara and  Remiddi~\cite{Ferrara1974}. Instead,  
in high-scale  seesaw models  such  cancellations can  only  occur for  a
particular   choice   of   the   neutrino-Yukawa   and   Majorana-mass
textures~\cite{Ellis2002,Arganda2006}.  Hence,  the   results   obtained  within
supersymmetric   low-scale    seesaw   type-I   models,    with   $m_N
\stackrel{<}{{}_\sim}  10~\textrm{ TeV}$,   suport  the  original  findings
in Ref~\cite{Ilakovac2009},  where  the  usual   paradigm  with   the   photon
dipole-moment operators dominating  the CLFV observables in high-scale
seesaw models~\cite{Hisano1996,Hisano1995,Hisano1999,Carvalho2001,Hisano2009}  
gets radically modified,  such that $\mu \to eee$ and $\mu  \to e$ conversion 
may also represent sensitive probes of CLFV.
 
We have found that, unlike heavy neutrinos,  CLFV effects  induced by
sneutrinos  remain subdominant  for the  entire region  of  the mSUGRA
parameter  space. In  addition, the  perturbativity constraint  on the
neutrino Yukawa  couplings~${\bf h}_\nu$ up  to the GUT  scale renders
the quartic  coupling contributions of order  $({\bf h}_\nu)^4$ small.
This study has focused on providing  numerical predictions for
relatively   small    and   intermediate   values    of   $\tan\beta$,
i.e.~$\tan\beta     \stackrel{<}{{}_\sim}    20$,     where    neutral
Higgs-mediated interactions constrained by the recent LHCb observation
of  the decay  $B_s  \to \mu\mu$  are  not expected  to give  sizeable
contributions.   A~global  analysis  that includes  large  $\tan\beta$
effects on  CLFV observables and LHC constraints is one of the goals of
the future researches.

Chapter \ref{moments} presented the systematical study of the one-loop 
contributions to the muon anomalous MDM  $a_\mu$ and the electron 
EDM~$d_e$  in the $\nu_R$MSSM. In particular, special attention was paid 
to the effect  of the sneutrino  soft  SUSY-breaking  parameters,  
${\bf B_\nu}$  and  ${\bf  A_\nu}$,  and their  universal CP  phases, 
$\theta$  and  $\phi$, on $a_\mu$ and $d_e$. As far as one can tell,
lepton dipole moments have not been  analyzed in detail before, 
within  SUSY models with low scale singlet (s)neutrinos.

For the deviation of $a_\mu$ from the SM value due to the
$\nu_R$MSSM ($\delta a_\mu$) it is found  that the
heavy  singlet neutrino  and  sneutrino contributions  to 
$\delta a_\mu$  are
small, typically one to two orders of magnitude below the muon anomaly
$\Delta a_\mu$. Instead, left-handed  sneutrinos and sleptons give the
largest effect on $\Delta a_\mu$, exactly  as is the case in the MSSM.
The dependence of  $\delta a_\mu$ on the muon mass~$m_\mu$, $\tan\beta$ and
the soft  SUSY-breaking mass scale $M_{\rm SUSY}$  have been carefully
analyzed  and their  scaling behaviour  according to Eq~\eqref{al_approx}
has  been  confirmed.  Finally,  the  dependence  of  $\delta a_\mu$ on  the
universal   soft  trilinear   parameter~$A_0$,  the   neutrino  Yukawa
couplings  ${\bf  h}_\nu$  and  the  heavy  neutrino  mass  $m_N$  are
negligible.

Furthermore, the  electron  EDM   $d_e$  in  the $\nu_R$MSSM is analysed. 
The  heavy singlet neutrinos do not  contribute to $d_e$,
and soft SUSY-breaking and sneutrino terms contribute only if the 
phases $\phi$ and/or $\theta$  have a nonzero  value. The contribution from 
the possible CP violating terms arising from the relatively complex products 
of the vertices exposed in \eqref{CPNt} is numerically shown to be equal zero. 
On  the  other  hand,  the contribution due to a non-zero value  of $\phi$ is 
the largest and may give  rise to  values for  the electron  EDM $d_e$  
comparable  to its present  experimental upper  limit.  The  effect of  
the  CP-odd phase $\theta$  on $d_e$  is approximately  one to  two orders  
of magnitude smaller than that of $\phi$. The size of $d_e$ increases with
$\tan\beta$ and mass of the lepton $m_l$, it is approximatively 
independent of $A_0$ and~$B_0$, but it generically decreases, as 
functions  of the  soft SUSY-breaking parameters $m_0$,  $M_{1/2}$. 

Based  on this  numerical  results, the  approximate
semi-analytical expressions are derived, which differ  from those presented in the
existing  literature  for SUSY  models  realizing  a high-scale  seesaw
mechanism.  Specifically, the flavor-blind CP-odd phases 
lead to a scaling of the lepton EDM $d_l  \propto m_l \tan\beta /m_N^{y}$, 
where $2/3<y<1$. While it is true that  $d_l$ generally decreases with $M_{SUSY}$, 
this dependence cannot be described with a simple scaling law. The 
dependences on SUSY breaking parameters $A_0$ and $B_0$ are weak in the 
largest part of the parameter space. The linear dependence on $\tan\beta$ and  
the dependence on heavy neutrino mass are new results arising from this study.

In comparison, the $\tan\beta$ dependence in Ref~\cite{Farzan2004} is, 
depending on its magnitude, either cubic or constant. Given  the  current 
experimental limits  on $d_e$, a significant portion of
the $\nu_R$MSSM parameter space is identified with  maximal CP phase $\phi = \pi/2$,
where the electron EDM $d_e$ can have values comparable to the present
and future experimental sensitivities.  The effect of sneutrino-sector
CP  violation  on the  neutron  and Mercury  EDMs  is  expected to  be
suppressed,  which is  a  distinctive  feature for  the  class of  the
$\nu_R$MSSM scenarios studied in this thesis.

\vspace{2.5ex}
In his brief review regarding the future of Particle Physics~\cite{Glashow2013}, 
Nobel Prize winner Sheldon Lee Glashow emphasized six points which he 
personally finds most important for the future of High Energy Physics research. 
Among others, those include charged lepton flavor violation processes, 
anomalous  magnetic dipole moment of the muon $a_\mu = g_\mu - 2$ and electric dipole 
moment of the electron $d_e$. The author of this thesis couldn't agree more.

%% file: appendices.tex
\titleformat{\section}[hang]{\Large\bfseries\filcenter}{\S\,{\small \thesection}}{10pt}{}[]

\begin{appendices}

\chapter{Interaction vertices} \label{vertices}

In this  appendix, the Lagrangians  describing the interaction
vertices required to calculate  the transition amplitudes for the CLFV
processes under study are listed. The corresponding interaction vertices for the
SM and the MSSM are obtained by adopting the conventions of the public
code   FeynArts-3.3 \cite{Hahn2001},
 {\tt FVMSSM.mod} \cite{Hahn2002,Fritzsche2013}, adapted to the notation used by
Petcov et al.~\cite{Petcov2004}.
The Lagrangians of interest to this study include:

1.  Vertices from 2HDM sector of the MSSM involving SM particles only,
\begin{equation}
{\cal L}_{\overline{d}uH^-} + \textrm{h.c.}
 =
 \frac{g_w}{\sqrt{2} M_W} V^*_{ij}\:
 \overline{d}_j
   \Big(t_\beta m_{d_j} P_L + t_\beta^{-1} m_{u_i} P_R\Big)
 u_i H^- + \textrm{h.c.}
\end{equation}
Here  $H^-$  is  the  negatively  charged Higgs  scalar,  $V$  is  the
Cabibbo--Kobayashi--Maskawa matrix, $m_{d_i}$  and $m_{u_i}$ are the quark
masses and $c_w=\cos\theta_w$.

2. Vertices of singlet neutrinos in the $\nu_R$SM sector of the MSSM,
\begin{align}
{\cal L}_{\overline{e}nG^-} + \textrm{h.c.}
 &=
 \frac{g_w}{\sqrt{2} M_W} B_{ia}\:
 \overline{e}_i
   \Big( - m_{e_i} P_L + m_{n_a} P_R \Big)
 n_a G^- + \textrm{h.c.}\,,
\\
{\cal L}_{\overline{e}nW^-} + {\rm h.c.}
 &=
 -\frac{g_w}{\sqrt{2}} B_{ia}\:
 \overline{e}_i
  \gamma^\mu P_L
 n_a W^-_\mu + {\rm h.c.}\,,
\\
{\cal L}_{\overline{n}nZ}
 &=
 -\frac{g_w}{2c_W} C_{ab}\:
 \overline{n}_a
   \gamma^\mu P_L
 n_b\, Z_\mu\,.
\end{align}
Here  $n_a$ and  $m_{n_a}$  denote the  neutrino mass-eigenstates  and
their  respective masses  and $B$  and $C$  are lepton  flavor mixing
matrices  defined in Refs \cite{Pilaftsis1992,Dev2012,Ilakovac1995}.  
The  matrices $B$ and $C$ satisfy the following set of relations:
\begin{align}
B_{la}B_{l'a}^* &= \delta_{ll'}, & C_{ac}C_{bc} &= C_{ab},
& B_{lb}C_{ba} &= B_{la}, & B_{la}^* B_{lb} = C_{ab},
\nonumber\\
m_a C_{ac} C_{bc} &= 0, & m_a B_{lb} C^*_{ba} &= 0,
& m_a B_{la}B_{l'a} &= 0 \,.
\end{align}

3.   Vertices from  the 2HDM  sector  of the  MSSM involving  Majorana
neutrinos,
\begin{equation}
{\cal L}_{\overline{e} n H^-} + {\rm h.c.}
 =
 \frac{g_w}{\sqrt{2} M_W} B_{ia} \:
 \overline{e}_i
  \Big(t_\beta m_{e_i} P_L + t_\beta^{-1} m_{n_a} P_R\Big)
 n_a H^- + {\rm h.c.}
\end{equation}

4. MSSM vertices with sparticles,
\begin{eqnarray}
{\cal L}_{\overline{d} \tilde{\chi}^- \tilde{u}} + {\rm h.c.}
 &=&
 g_w\, \overline{d}_j \Big(\tilde{V}_{jm a}^{-dL} P_L + \tilde{V}_{jm
   a}^{-dR} P_R\Big) \tilde{\chi}_m ^- \tilde{u}_a + {\rm h.c.}\,,
\\
{\cal L}_{\overline{u} \tilde{\chi}^+ \tilde{d}} + {\rm h.c.}
 &=&
 g_w\, \overline{u}_j \Big(\tilde{V}_{jm a}^{+uL} P_L + \tilde{V}_{jm
   a}^{+uL} P_R\Big) \tilde{\chi}_m ^+ \tilde{d}_a + {\rm h.c.}\,,
\\
{\cal L}_{\overline{\tilde{\chi}^-}\tilde{\chi}^-A} + {\cal
  L}_{\overline{\tilde{\chi}^+}\tilde{\chi}^+A}
 &=&
   e\:\overline{\tilde{\chi}^-_m} \gamma^\mu \tilde{\chi}^-_m A_\mu
 - e\: \overline{\tilde{\chi}^+_m} \gamma^\mu \tilde{\chi}^+_m A_\mu\,,
\\
{\cal L}_{\overline{\tilde{\chi}^-}\tilde{\chi}^-Z} + {\cal
  L}_{\overline{\tilde{\chi}^+}\tilde{\chi}^+Z}
 &=&
\frac{g_w}{c_W}\,
\overline{\tilde{\chi}_m^-}
 \gamma^\mu \Big( V_{mk}^{\tilde{\chi}^-L} P_L +
 V_{mk}^{\tilde{\chi}^-R} P_R \Big)\tilde{\chi}_k^- Z_\mu
\nonumber\\
 &-&
\frac{g_w}{c_W}\,
\overline{\tilde{\chi}_m^+}
 \gamma^\mu \Big( V_{mk}^{\tilde{\chi}^-L*} P_R +
 V_{mk}^{\tilde{\chi}^-R*} P_L \Big) \tilde{\chi}_k^+ Z_\mu\,,
\\
% change 24112012
{\cal L}_{\bar{\tilde{\chi}}^0 \tilde{\chi}^0 Z}
 &=&
 \frac{g}{c_w} \bar{\tilde{\chi}}^0_m
 ( \gamma^\mu P_L V_{mk}^{\tilde{\chi}^0 L}
 + \gamma^\mu P_R V_{mk}^{\tilde{\chi}^0 R})
 \tilde{\chi}^0_k Z_\mu \,,
\\
{\cal L}_{\tilde{e}^*\tilde{e}Z}
 &=&
 g_w \tilde{V}^{\tilde{e}}_{ab}\; \tilde{e}_a^*\,
 i \hspace{-3pt}
 \stackrel{\leftrightarrow}{\partial}\hspace{-.25em}\vspace{.6em}^\mu
\hspace{-2pt}\,
 \tilde{e}_b\: Z_\mu\,,
\\
{\cal L}_{\bar{e}\tilde{\chi}^0\tilde{e}} + {\cal
  L}_{\overline{\tilde{\chi}^0} e \tilde{e}^*}
 &=&
g_w \,
  \overline{e}_j (P_L \tilde{V}_{jma}^{0 e L} + P_R \tilde{V}_{jma}^{0
    e R})\tilde{\chi}^0 \tilde{e}_a + {\rm h.c.}\,,
\\
{\cal L}_{\bar{u}\tilde{\chi}^0\tilde{u}} + {\cal
  L}_{\overline{\tilde{\chi}^0} u \tilde{u}^*}
 &=&
g_w \,
  \overline{u}_j (P_L \tilde{V}_{jma}^{0 u L} + P_R \tilde{V}_{jma}^{0
    u R})\tilde{\chi}^0 \tilde{u}_a + {\rm h.c.}\,,
\\
{\cal L}_{\bar{d}\tilde{\chi}^0\tilde{d}} + {\cal
  L}_{\overline{\tilde{\chi}^0} d \tilde{d}^*}
 &=&
g_w \,
  \overline{d}_j (P_L \tilde{V}_{jma}^{0 d L} + P_R \tilde{V}_{jma}^{0
    d R})\tilde{\chi}^0 \tilde{d}_a + {\rm h.c.}\,,
\end{eqnarray}
\allowdisplaybreaks
where
\begin{align}
\tilde{V}_{jm a}^{-dL}
 &=
 \frac{m_{d_j}}{\sqrt{2} c_\beta M_W} {\cal U}_{m 2}^*
  V^*_{ij} (R_L^{\tilde{u}})^*_{ai}\,,
\nonumber\\
\tilde{V}_{jm a}^{-dR}
 &=
 - {\cal V}_{m 1} V^*_{ij} (R_L^{\tilde{u}})^*_{ai}
 + \frac{m_{u_i}}{\sqrt{2} s_\beta M_W} {\cal V}_{m 2}
   V^*_{ij} (R_R^{\tilde{u}})^*_{ai}\,,
\\
\tilde{V}_{jm a}^{+uL}
 &=
 \frac{m_{u_j}}{\sqrt{2} s_\beta M_W} {\cal V}_{m 2}^* V_{ji}
 (R_L^{\tilde{d}})^*_{ai}\,,
\nonumber\\
\tilde{V}_{jm a}^{+uR}
 &=
 - {\cal U}_{m 1} V_{ji} (R_L^{\tilde{d}})^*_{ai}
 + \frac{m_{d_i}}{\sqrt{2} c_\beta M_W} {\cal U}_{m 2} V_{ji}
 (R_R^{\tilde{d}})^*_{ai}\,,
\\
V_{mk}^{\tilde{\chi}^-L}
 &=
   {\cal U}_{m 1} {\cal U}^*_{k 1}
 + \frac{1}{2} {\cal U}_{m 2} {\cal U}^*_{k 2}
 - \delta_{m k} s_w^2,
\nonumber\\
V_{mk}^{\tilde{\chi}^-R}
 &=
   {\cal V}^*_{m 1} {\cal V}_{k 1}
 + \frac{1}{2} {\cal V}^*_{m 2} {\cal V}_{k 2}
 - \delta_{m k} s_w^2\,,
\\
% change 24112012
V_{mk}^{\tilde{\chi}^0 L}
 &=
 -\frac{1}{4} (Z_{m3} Z^*_{k3}-Z_{m4} Z^*_{k4})\,,
\\
V_{mk}^{\tilde{\chi}^0 R}
 &=
 \frac{1}{4} (Z^*_{m3} Z_{k3}-Z^*_{m4} Z_{k4})\,,
\\
\tilde{V}^{\tilde{e}}_{ab}
 &=
 \frac{c_{2w}}{c_w}
 ( R^{\tilde{e}}_L)_{ai} (R^{\tilde{e}}_L)_{bi}^*
 - \frac{s_w^2}{c_w} (R^{\tilde{e}}_R)_{ai} (R^{\tilde{e}}_R)_{bi}^* \,,
\\
\tilde{V}^{0 \ell L}_{j m a}
 &=
 -\sqrt{2} t_w Z^*_{m1} (R_R^{\tilde{e}})^*_{aj}
% change 24112012 Z_{m3} -> Z^*_{m3}
 - \frac{(m_e)_j}{\sqrt{2} c_\beta M_W} Z^*_{m3}
 (R_L^{\tilde{e}})^*_{aj}\,,
\\
\tilde{V}^{0 \ell R}_{j m a}
 &=
\frac{1}{\sqrt{2} c_W}(c_W Z_{m2} +s_W Z_{m1}) (R^{\tilde{e}}_L)^*_{aj}
- \frac{(m_e)_j}{\sqrt{2}c_\beta M_W} Z_{m3}
(R_R^{\tilde{e}})^*_{aj}\,,
\\
\tilde{V}^{0 u L}_{j m a}
 &=
 \frac{2\sqrt{2}}{3} t_w Z^*_{m1} (R_R^{\tilde{u}})^*_{aj}
 - \frac{(m_u)_j}{\sqrt{2} s_\beta M_W} Z^*_{m4}
 (R_L^{\tilde{u}})^*_{aj}\,,
\\
\tilde{V}^{0 u R}_{j m a}
 &=
- \frac{1}{2 c_W}(c_W Z_{m2} +\frac{1}{3} s_W Z_{m1}) (R^{\tilde{u}}_L)^*_{aj}
- \frac{(m_u)_j}{\sqrt{2}s_\beta M_W} Z_{m4}
(R_R^{\tilde{u}})^*_{aj}\,,
\\
\tilde{V}^{0 d L}_{j m a}
 &=
 -\frac{\sqrt{2}}{3} t_w Z^*_{m1} (R_R^{\tilde{d}})^*_{aj}
 - \frac{(m_d)_j}{\sqrt{2} c_\beta M_W} Z_{m3}
 (R_L^{\tilde{d}})^*_{aj}\,,
\\
\tilde{V}^{0 d R}_{j m a}
 &=
\frac{1}{2 c_W}(c_W Z_{m2} -\frac{1}{3} s_W Z_{m1}) (R^{\tilde{d}}_L)^*_{aj}
- \frac{(m_d)_j}{\sqrt{2}c_\beta M_W} Z_{m3}
(R_R^{\tilde{d}})^*_{aj}\,,
\end{align}
and  $c_{2w} = \cos{2\theta_w}$.   The unitary  matrices diagonalizing
the chargino  mass matrix  ${\cal U}$ and  ${\cal V}$ and  the unitary
matrix  diagonalizing  the  neutralino   mass  matrix  $Z$  are  taken
from \cite{Drees2004}. The matrices
\begin{equation}
R^{\tilde{f}L}_{ak}\ \equiv\ U^{\tilde{f}}_{ia} U^{f_L*}_{ik}\;,\qquad
R^{\tilde{f}R}_{ak}\ \equiv\ U^{\tilde{f}}_{i+3\,a} U^{f_R*}_{ik}\;,
\end{equation}
with $f = d,\, u,\,  e$, $a=1,2,\dots,6$ and $i,k=1,2,3$, quantify the
disalignment   between  fermions   and  sfermions.    Here  $U^{f_L}$,
$U^{f_R}$  and $U^{\tilde{f}}$ are  unitary matrices  that diagonalize
the fermion and sfermion mass matrices, respectively.

5. Sneutrino vertices in the $\nu_R$MSSM,
\begin{align}
{\cal L}_{\overline{e}\tilde{\chi}^- \tilde{N}} + {\rm h.c.}
 &=
 g_w \tilde{N}_A \overline{\ell}_l \Big(P_L  \frac{m_l}{\sqrt{2}
   c_\beta M_W} \tilde{B}_{lm A}^{L,1}
  + P_R \tilde{B}_{lm A}^{R,1}\Big) \tilde{\chi}_m^- + {\rm h.c.}
\nonumber\\
 &=
 g_w \tilde{N}_A^* \overline{\ell}_l \Big(P_L  \frac{m_l}{\sqrt{2}
   c_\beta M_W} \tilde{B}_{lm A}^{L,2}
  + P_R \tilde{B}_{lm A}^{R,2}\Big) \tilde{\chi}_m^- + {\rm h.c.}\,,
\\
{\cal L}_{\tilde{N}\tilde{N}Z}
 &=
 \frac{g_w}{c_W} \tilde{C}^1_{AB}\:
 \tilde{N}_A^* i
 \hspace{-2pt}
 \stackrel{\leftrightarrow}{\partial}\hspace{-.25em}\vspace{.6em}^\mu
\hspace{-2pt}\,
 \tilde{N}_B Z_\mu
\nonumber\\
 &=
 \frac{g_w}{c_W} \tilde{C}^2_{AB}\:
 \tilde{N}_A^* i
 \hspace{-2pt}
 \stackrel{\leftrightarrow}{\partial}\hspace{-.25em}\vspace{.6em}^\mu
\hspace{-2pt}\,
 \tilde{N}_B^* Z_\mu
\nonumber\\
 &=
 \frac{g_w}{c_W} \tilde{C}^3_{AB}\:
 \tilde{N}_A i
 \hspace{-2pt}
 \stackrel{\leftrightarrow}{\partial}\hspace{-.25em}\vspace{.6em}^\mu
\hspace{-2pt}\,
 \tilde{N}_B Z_\mu
\nonumber\\
 &=
 \frac{g_w}{c_W} \tilde{C}^4_{AB}\:
 \tilde{N}_A i
 \hspace{-2pt}
 \stackrel{\leftrightarrow}{\partial}\hspace{-.25em}\vspace{.6em}^\mu
\hspace{-2pt}\,
 \tilde{N}_B^* Z_\mu,
\end{align}
where
\begin{eqnarray}
%BL1
\tilde{B}_{lm A}^{L,1}
 &=&
 {\cal U}_{m 2} U^{\ell_R *}_{il}
  {\cal U}_{i A}^{\tilde{\nu}}\,,
\nonumber\\
%BL2
\tilde{B}_{lm A}^{L,2}
 &=&
 {\cal U}_{m 2} U^{\ell_R *}_{il}
  {\cal U}_{i+6 A}^{\tilde{\nu}*}\,,
\nonumber\\
%BR1
\tilde{B}_{lm A}^{R,1}
 &=&
 - {\cal U}_{i A}^{\tilde{\nu}} U^{\ell_L*}_{il}
   {\cal V}_{m 1}
 + \frac{m_{n_a}}{\sqrt{2} s_\beta M_W}
   {\cal V}_{m 2}
   {\cal U}_{i+9 A}^{\tilde{\nu}} U^{\nu *}_{i+3 a} B_{la}\,,
\nonumber\\
%BR2
\tilde{B}_{lm A}^{R,2}
 &=&
 - {\cal U}_{i+6 A}^{\tilde{\nu}*} U^{\ell_L*}_{il}
   {\cal V}_{m 1}
 + \frac{m_{n_a}}{\sqrt{2} s_\beta M_W}
   {\cal V}_{m 2}
   {\cal U}_{i+3 A}^{\tilde{\nu} *} U^{\nu *}_{i+3 a} B_{la}\,,
\nonumber\\
%C1
\tilde{C}^1_{AB}
 &=&
 - \frac{1}{2}
 {\cal U}_{i A}^{\tilde{\nu}*}
 {\cal U}_{i B}^{\tilde{\nu}}\,,
\nonumber\\
%C2
\tilde{C}^2_{AB}
 &=&
 - \frac{1}{2}
 {\cal U}_{i A}^{\tilde{\nu}*}
 {\cal U}_{i+6 B}^{\tilde{\nu}*}\,,
\nonumber\\
%C3
\tilde{C}^3_{AB}
 &=&
 - \frac{1}{2}
 {\cal U}_{i+6 A}^{\tilde{\nu}}
 {\cal U}_{i B}^{\tilde{\nu}}\,,
\nonumber\\
%C4
\tilde{C}^6_{AB}
 &=&
 - \frac{1}{2}
 {\cal U}_{i+6 A}^{\tilde{\nu}}
 {\cal U}_{i+6 B}^{\tilde{\nu}*}\,.
\end{eqnarray}
In  the   above,  ${\cal  U}^{\tilde{\nu}}$  is   the  unitary  matrix
diagonalizing the sneutrino mass matrix.

Notice that the weak coupling constant $g_w$ are factored out from
all   interaction   vertices  defined   above.   To better   identify
chirality-flip  mass effects  in  the CLFV  amplitudes, factor  
$m_l/(\sqrt{2} c_\beta M_W)$ is also pulled out from the interaction
vertex $\tilde{B}_{lm A}^{L}$.

\chapter{Loop functions}\label{sec:lf}

The CLFV  amplitudes are expressed in terms  of leading-order one-loop
functions.  We expand  the loop functions with respect  to the momenta
and masses  of the  external charged leptons,  while keeping  only the
leading non-zero  terms. The leading  terms may then be  expressed, in
terms of the dimensionless loop integrals
\begin{align}
\bar{J}^m_{n_1n_2\dots n_k} (\lambda_1,\lambda_2,\dots,\lambda_k)
&=
\frac{(\mu^2)^{2-D/2}}{(M^2_W)^{-D/2-m+\sum_i n_i}}
 \int\frac{d^D\ell}{(2\pi)^D} \frac{(\ell^2)^m}{\prod_{i=1}^k (\ell^2
   - m_i^2)^{n_i}}
\nonumber\\
&= \frac{i(-1)^{m-\sum_i n_i}}{(4\pi)^{D/2}\Gamma(\frac{D}{2})}
 \Big(\frac{\mu^2}{M_W^2}\Big)^{2-D/2}
 \int_0^\infty \frac{dx x^{D/2-1+m}}{\prod_{i=1}^k (x+\lambda_i)^{n_i}},
\end{align}
where $m_i$ are  loop particle masses, $n_i$ are  the exponents of the
propagator  denominators,  $\lambda_i=m_i^2/M_W^2$  are  dimensionless
mass parameters  and $\mu$ is  't~Hooft's renormalization  mass scale.
Parameter $\mu$ is chosen to be $M_W$, even though any other scale can be chosen
equally well as  a reference scale for any of  the integrals.  For the
amplitudes dealt with in this thesis,  the  integrals  are  either
divergent and satisfy $m+2-\sum_in_i = 0$, or they are convergent with
$m+2-\sum_in_i<0$.  For convergent integrals, one may set $D=4$, whilst
for  divergent integrals one takes $D=4-2\epsilon$. Factor  $i/(4\pi)^2$ 
is pulled out from  all  integrals. Thus, for  finite integrals one obtains:
\begin{equation}
\bar{J}^m_{n_1n_2\dots n_k} (\lambda_1,\lambda_2,\dots,\lambda_k)
\ \equiv\
\frac{i}{(4\pi)^2}\ J^m_{n_1n_2\dots n_k} (\lambda_1,\lambda_2,\dots,\lambda_k)\; .
\end{equation}
Instead,  the divergent  integrals  are written  down  as a  sum of  a
divergent and constant term and a finite mass-dependent term:
\begin{equation}
\bar{J}^m_{n_1n_2\dots n_k}
(\lambda_1,\lambda_2,\dots\lambda_k)\ \equiv\
 \frac{i}{(4\pi)^2}\ \big[ \frac{1}{\varepsilon} + \mbox{const} +
 J^m_{n_1n_2\dots n_k} (\lambda_1,\lambda_2,\dots,\lambda_k) \big]\,.
\end{equation}
In  the CLFV amplitudes,  the ``divergent+constant''  terms vanish  in the
total sum,  or as  a result of  a GIM-like mechanism.   Therefore, all
CLFV amplitudes  can be  expressed in terms  of finite  mass dependent
functions  $J^m_{n_1n_2\dots}  (\lambda_1,\lambda_2,\dots)$, which  we
call  the  {\it basic  integrals}. Those integrals are 
analytically calculated using \emph{Wolfram Mathematica}
package. We will now describe the procedure
used in the calculation and present the exact results thus obtained.

There are three types of $J$-functions which are used in this study:
\begin{align}
J^a_{\,bc} (x,y) & =  K \cdot (-1)^{a+b+c} \cdot
  \int \frac{dt\; t^{D/2-1+a}}{(t+x)^b \, (t+y)^c} \,, 
  \label{J3} \\
J^a_{\,bcd} (x,y,z)& =  K \cdot (-1)^{a+b+c+d} \cdot
  \int \frac{dt\; t^{D/2-1+a}}{(t+x)^b \, (t+y)^c\, (t+z)^d} \,, 
  \label{J4} \\
J^a_{\,bcde} (x,y,z,w) & =  K \cdot (-1)^{a+b+c+d+e} \cdot
  \int \frac{dt\; t^{D/2-1+a}}{(t+x)^b \, (t+y)^c\, (t+z)^d\, (t+w)^e} \,,
\label{J5}
\end{align}
where 
\begin{equation}\label{factorK}
K = \frac{ i\, 2^{-D} \pi^{-D/2} \mu^{4-D}}{\Gamma(D/2)} \,.
\end{equation}

The calculation of these integrals is performed in three steps. In the first step,
the integral is exactly calculated using \verb+Integrate+ function. In the second
step, the constant term is isolated from the expression. Because of the specific 
identities obeyed by the flavor-mixing matrices 
(see Eqs~(2.9) and (2.10) in Ref~\cite{Ilakovac1995}) the constant terms can effectively
be ignored. Finally, the third step gets rid of these constant terms, and gives 
simplified result in the zeroth order over $\epsilon$, 
with factor $i/16\pi^2$ dropped for better readability of the result.

$\blacklozenge$
For calculation of the $J$-functions type \eqref{J3}, 
these steps are performed by the following \emph{Mathematica} functions:

{\bf Step one.} 
\vspace{-3ex}
{\footnotesize
\begin{verbatim}
INT3[a_, b_, c_] := (-1)^(a + b + c)*
  Assuming[{x > 0, y > 0, D <= 2 (b + c - a - 1)}, 
   Integrate[
    t^(D/2 - 1 + a)/(((t + x)^b) ((t + y)^c) ), {t, 0, \[Infinity]}]]
\end{verbatim}}

{\bf Step two.} 
\vspace{-3ex}
{\footnotesize
\begin{verbatim}
CteExtract3[expr_] := Module[{},
  cstep1 = Collect[Expand[expr /. Log[x] -> 0], x];
  cstep2 = cstep1 /. x^n__ -> 0 /. x -> 0;
  cstep3 = cstep2 /. Log[y] -> 0 /. y^n__ -> 0]
\end{verbatim}}

{\bf Step three.} 
\vspace{-3ex}
{\footnotesize
\begin{verbatim}
ProcInt3[a_, b_, c_] := Module[{},
  step1 = 
   Series[INT3[a, b, c]*OFa[D]*(16 \[Pi]^2)/I /. 
       D -> 4 - 2 \[Epsilon], {\[Epsilon], 0, 0}] // FullSimplify // 
    Normal;
  cte = CteExtract3[step1];
  step2 = step1 - cte // FullSimplify]
\end{verbatim}}

In order to evaluate function $J^a_{\,bc} (x,y)$, one only needs to call the function defined 
in Step three, e.g.\ \verb+J241[x_, y_] = ProcInt3[2, 4, 1]+. The analogous procedure 
is applied for other types of $J$ loop-functions as well.

$\blacklozenge$ For calculation of the $J$-functions type \eqref{J4}:

{\bf Step one.} 
\vspace{-3ex}
{\footnotesize
\begin{verbatim}
INT4[a_, b_, c_, d_] := (-1)^(a + b + c + d)*
  Assuming[{x > 0, y > 0, z > 0, D <= 2 (b + c + d - a - 1)}, 
   Integrate[
    t^(D/2 - 1 + a)/((t + x)^b (t + y)^c (t + z)^d), {t, 0, \[Infinity]}]]
\end{verbatim}}
\newpage

{\bf Step two.} 
\vspace{-3ex}
{\footnotesize
\begin{verbatim}
CteExtract4[expr_] := Module[{},
  cstep1 = Collect[Expand[expr /. Log[x] -> 0], x];
  cstep2 = cstep1 /. x^n__ -> 0 /. x -> 0;
  cstep3 = Collect[Expand[cstep2 /. Log[y] -> 0], y];
  cstep4 = cstep3 /. y^n__ -> 0 /. y -> 0;
  cstep5 = cstep4 /. Log[z] -> 0 /. z^n__ -> 0]
\end{verbatim}}

{\bf Step three.} 
\vspace{-3ex}
{\footnotesize
\begin{verbatim}
ProcInt4[a_, b_, c_, d_] := Module[{},
  step1 = 
   Series[INT4[a, b, c, d]*OFa[D]*(16 \[Pi]^2)/I /. 
       D -> 4 - 2 \[Epsilon], {\[Epsilon], 0, 0}] // FullSimplify // 
    Normal;
  cte = CteExtract4[step1];
  step2 = step1 - cte // FullSimplify]
\end{verbatim}}

$\blacklozenge$ For calculation of the $J$-functions type \eqref{J5}:

{\bf Step one.} 
\vspace{-3ex}
{\footnotesize
\begin{verbatim}
INT5[a_, b_, c_, d_, e_] := (-1)^(a + b + c + d + e)*
  Assuming[{x > 0, y > 0, z > 0, w > 0, D < 1}, 
   Integrate[t^(
    D/2 - 1 + a)/((t + x)^b (t + y)^c (t + z)^d (t + w)^e), 
     {t, 0, \[Infinity]}]]
\end{verbatim}}

{\bf Step two.} 
\vspace{-3ex}
{\footnotesize
\begin{verbatim}
CteExtract5[expr_] := Module[{},
  cstep1 = Collect[Expand[expr /. Log[x] -> 0], x];
  cstep2 = cstep1 /. x^n__ -> 0 /. x -> 0;
  cstep3 = Collect[Expand[cstep2 /. Log[y] -> 0], y];
  cstep4 = cstep3 /. y^n__ -> 0 /. y -> 0;
  cstep5 = Collect[Expand[cstep4 /. Log[z] -> 0], z];
  cstep6 = cstep5 /. z^n__ -> 0 /. z -> 0;
  cstep7 = cstep6 /. Log[w] -> 0 /. w^n__ -> 0]
\end{verbatim}}
\newpage

{\bf Step three.} 
\vspace{-3ex}
{\footnotesize
\begin{verbatim}
ProcInt5[a_, b_, c_, d_, e_] := Module[{},
  step1 = 
   Series[INT5[a, b, c, d, e]*OFa[D]*(16 \[Pi]^2)/I /. 
       D -> 4 - 2 \[Epsilon], {\[Epsilon], 0, 0}] // FullSimplify // 
    Normal;
  cte = CteExtract5[step1];
  step2 = step1 - cte // FullSimplify]
\end{verbatim}}

In all these expressions, \verb+OFa[D]+ is defined as factor $K$ in the 
Eq~\eqref{factorK}.

Using this procedure, one comes out with the following results:

\begin{align}
 J^0_{\,11} (x,y) &= \frac{y \log(y)-x\log(x)}
  {x-y} \,, \nonumber \\
 J^0_{\,21} (x,y) &= \frac{y \log(x)-x+y-y\log(y)}
  {(x-y)^2} \,, \nonumber \\
 J^0_{\,31} (x,y) &= \frac{x^2 - 2xy (\log (x) - \log(y)) - y^2}
  {2 x (x-y)^3} \,, \nonumber \\
 J^1_{\,21} (x,y) &= - \frac{x(x-y)+x(x-2y)\log(x) +y^2 \log(y)}
  {(x-y)^2} \,, \nonumber \\
 J^1_{\,31} (x,y) &= \frac{2 y^2 (\log(y) - \log(x)) - (x-3y)(x-y)}
  {2 (x-y)^3} \nonumber \\
 J^1_{\,41} (x,y) &= \frac{(x-y)(x^2-5xy-2y^2)+6xy^2 (\log(x)-\log(y))}
  {6 x (x-y)^4} \,, \nonumber \\
 J^2_{\,41} (x,y) &= \frac{6 y^3 \left( \log(x)-\log(y) \right) - (x-y)
  \left( 2 x^2 - 7 x y + 11 y^2 \right)}{6 (x-y)^4} \,, 
\end{align}

\begin{align}
 J^0_{\,111} (x,y,z) &= \frac{x\log(x)(z-y) +y(x-z) \log(y) + z(y-x)\log(z)}
  {(x-y)(x-z)(y-z)} \,, \nonumber \\[1.5ex]
 J^0_{\,211} (x,y,z) &= \frac{1}{(x-y)^2 (x-z)^2 (y-z)} \; \Big[
  \log(x)(y-z)(x^2-yz)  \nonumber \\
  &{}+ z(x-y)^2\log(z)+ (x-z)(y(z-x)\log(y)-(x-y)(y-z)) \Big]
  \,, \nonumber \\[1.5ex]
 J^1_{\,111} (x,y,z) &= \frac{x^2 \log(x) (y-z) +y^2(z-x) \log(y) + 
    z^2 (x-y) \log(z)}
  {(x-y)(x-z)(z-y)} \,, \nonumber \\[1.5ex]
 J^1_{\,211} (x,y,z) &= \frac{1}{(x-y)^2 (x-z)^2 (y-z)} \nonumber\\
  &\cdot \Big[
  (x-z) ( y^2 (z-x) \log(y) + x(x-y) (z-y) ) \nonumber \\
  &\quad{}+ z^2 (x-y)^2 \log(z) + x \log(x) (y-z) (x(y+z)-2yz) \Big] \,,
\end{align}

\begin{align}
 J^0_{\,1111} (x,y,z,w) &= -\frac{w \log(w)}{(w-x)(w-y)(w-z)} +
  \frac{x\log(x)}{(w-x)(x-y)(x-z)} \nonumber\\
  & \quad {}+ \frac{y\log(y)}{(w-y)(y-x)(y-z)} +
  \frac{z\log(z)}{(w-z)(x-z)(y-z)} \,, \nonumber\\[1.5ex]
 J^1_{\,1111} (x,y,z,w) &= -\frac{w^2 \log(w)}{(w-x)(w-y)(w-z)} +
  \frac{x^2\log(x)}{(w-x)(x-y)(x-z)} \nonumber\\
  & \quad {}+ \frac{y^2\log(y)}{(w-y)(y-x)(y-z)} +
  \frac{z^2\log(z)}{(w-z)(x-z)(y-z)} \,. 
\end{align}

These functions are often evaluated in the limit in which two or more variables
are equal to each other or equal to zero. This is easily performed with the
\verb+Limit+ function, although it can be rather time consuming. To avoid time
consumption, we have evaluated all possible limits only once and then used those
results to define the one-loop form factors expressed in Appendix \ref{sec:olff}.

\chapter{One--loop form factors}\label{sec:olff}

Here we present the complete analytical form of  the CLFV form factors 
$F_\gamma$, $F_Z$ and $F_{\rm  box}$ defined in Chapter~\ref{CLFV}, in
the Feynman--'t~Hooft gauge.  In  the following, the  usual summation
convention over repeated indices is implied.  The interaction vertices
and loop  functions used  here are given  in Appendices~\ref{vertices}
and \ref{sec:lf}, respectively.

\section{Photon Form factors}

The form factors $F_\gamma^L$, $F_\gamma^R$,   $G_\gamma^L$   and
$G_\gamma^R$ may be explicitly written as follows:
\begin{eqnarray}
(F_\gamma^L)_{l'l}
 &=&
 F^N_{l'l\gamma} + F^{L,\tilde{N}}_{l'l\gamma}
 + F^{L,{\rm SB}}_{l'l\gamma}\ ,
\nonumber\\
(F_\gamma^R)_{l'l}
 &=&
 F^N_{l'l\gamma} + F^{R,\tilde{N}}_{l'l\gamma}
 + F^{R,{\rm SB}}_{l'l\gamma}\ ,
\nonumber\\
(G_\gamma^L)_{l'l}
 &=&
 m_{l'} (G^N_{l'l\gamma} + G^{L,\tilde{N}}_{l'l\gamma})
 + G^{L,{\rm SB}}_{l'l\gamma}\ ,
\nonumber\\
(G_\gamma^R)_{l'l}
 &=&
 m_{l} (G^N_{l'l\gamma} + G^{R,\tilde{N}}_{l'l\gamma})
 + G^{R,{\rm SB}}_{l'l\gamma}\ ,
\end{eqnarray}
where
\newpage
\begin{align}
F_{l'l\gamma}^{N}
 &=
 B_{l' a} B^*_{l a}
 \bigg[ 2 \bigg(J^1_{31}(1,\lambda_{n_a}) -
   \frac{1}{6}J^2_{41}(1,\lambda_{n_a})\bigg) 
 \nonumber\\
    &\qquad - \frac{1}{6} \lambda_{n_a} J^2_{41}(1,\lambda_{n_a})
    - \frac{1}{6 t^2_\beta} \lambda_{n_a} J^2_{41}(\lambda_{H^+},\lambda_{n_a})
 \bigg],
\nonumber\\
G_{l'l\gamma}^{N}
 &=
 B_{l' a} B^*_{l a}
 \bigg[ J^1_{31}(1,\lambda_{n_a}) + J^2_{41}(1,\lambda_{n_a})
    + \lambda_{n_a} \bigg(\frac{1}{2} J^1_{41}(1,\lambda_{n_a}) -
    J^0_{31}(1,\lambda_{n_a})\bigg)
\nonumber\\
   &\qquad + \lambda_{n_a} \lambda_{H^+}
      \bigg( \frac{1}{2t^2_\beta}
      J^1_{41}(\lambda_{H^+},\lambda_{n_a}) +
      J^0_{31}(\lambda_{H^+},\lambda_{n_a}) \bigg)
 \bigg],
\\
F_{l'l\gamma}^{L,\tilde{N}}
 &=
 \frac{1}{2}(\tilde{B}^{R,1}_{l' k A} \tilde{B}^{R*}_{l k A} +
 \tilde{B}^{R,2}_{l' k A} \tilde{B}^{R,2*}_{l k A})
 \bigg[
   - \frac{2}{3} J^2_{41}(\lambda_{\tilde{\chi}_k},\lambda_{\tilde{N}_A})
   + \lambda_{\tilde{\chi}_k} J^1_{41}(\lambda_{\tilde{\chi}_k},\lambda_{\tilde{N}_A})
 \bigg],
\nonumber\\
F_{l'l\gamma}^{R,\tilde{N}}
 &=
 \frac{m_l m_{l'}}{4 c_\beta^2 M_W^2}
 (\tilde{B}^{L,1}_{l' k A} \tilde{B}^{L,1*}_{l k A} +
 \tilde{B}^{L,2}_{l' k A} \tilde{B}^{L,2*}_{l k A}) \nonumber\\
 &\qquad \cdot
 \bigg[
   - \frac{2}{3} J^2_{41}(\lambda_{\tilde{\chi}_k},\lambda_{\tilde{N}_A})
   + \lambda_{\tilde{\chi}_k} J^1_{41}(\lambda_{\tilde{\chi}_k},\lambda_{\tilde{N}_A})
 \bigg],
\nonumber\\
G_{l'l\gamma}^{L,\tilde{N}}
 &=
 \frac{1}{2} (\tilde{B}^{L,1}_{l' k A} \tilde{B}^{L,1*}_{l k A} +
 \tilde{B}^{L,2}_{l' k A} \tilde{B}^{L,2*}_{l k A})
 \bigg[
   - \frac{m_l^2}{2c^2_\beta M_W^2}
   \ \lambda_{\tilde{\chi_k}} J^1_{41}(\lambda_{\tilde{\chi}_k},\lambda_{\tilde{N}_A})
 \bigg]
\nonumber\\
 &+
 \frac{1}{2}(\tilde{B}^{R,1}_{l' k A} \tilde{B}^{R,1*}_{l k A} +
 \tilde{B}^{R,2}_{l' k A} \tilde{B}^{R,2*}_{l k A})
 \bigg[
   - \lambda_{\tilde{\chi}_k}
   J^1_{41}(\lambda_{\tilde{\chi}_k},\lambda_{\tilde{N}_A})
 \bigg]
\nonumber\\
 &+
 \frac{1}{2}(\tilde{B}^{L,1}_{l' k A} \tilde{V}^{R,1*}_{l k A} +
 \tilde{B}^{L,2}_{l' k A} \tilde{V}^{R,2*}_{l k A})
 \bigg[
   \frac{\sqrt{2}}{c_\beta} \sqrt{\lambda_{\tilde{\chi_k}}}
   J^1_{31}(\lambda_{\tilde{\chi}_k},\lambda_{\tilde{N}_A})
 \bigg],
\nonumber\\
G_{l'l\gamma}^{R,\tilde{N}}
 &=
 \frac{1}{2}(\tilde{B}^{L,1}_{l' k A} \tilde{B}^{L,1*}_{l k A} +
 \tilde{B}^{L,2}_{l' k A} \tilde{B}^{L,2*}_{l k A})
 \bigg[
   - \frac{m_{l'}^2}{2c^2_\beta M_W^2}
   \ \lambda_{\tilde{\chi_k}} J^1_{41}(\lambda_{\tilde{\chi}_k},\lambda_{\tilde{N}_A})
 \bigg]
\label{GllgRNt}
\nonumber\\
 &+
 \frac{1}{2}(\tilde{B}^{R,1}_{l' k A} \tilde{B}^{R,1*}_{l k A} +
 \tilde{B}^{R,2}_{l' k A} \tilde{B}^{R,2*}_{l k A})
 \bigg[
   - \lambda_{\tilde{\chi}_k}
   J^1_{41}(\lambda_{\tilde{\chi}_k},\lambda_{\tilde{N}_A})
 \bigg]
\nonumber\\
 &+
 \frac{1}{2}(\tilde{B}^{R,1}_{l' k A} B^{L,1*}_{l k A} +
 \tilde{B}^{R,2}_{l' k A} \tilde{B}^{L,2*}_{l k A})
 \bigg[
   \frac{\sqrt{2}}{c_\beta} \sqrt{\lambda_{\tilde{\chi_k}}}
   J^1_{31}(\lambda_{\tilde{\chi}_k},\lambda_{\tilde{N}_A})
 \bigg],
\\
F_{l'l\gamma}^{L,{\rm SB}}
 &=
 \tilde{V}^{0 \ell R}_{l' m a} \tilde{V}^{0 \ell R*}_{l m a}
 \bigg[
   -\frac{1}{3} J^2_{41}(\lambda_{\tilde{e}_a},\lambda_{\tilde{\chi}^0_m})
 \bigg],
\nonumber\\
F_{l'l\gamma}^{R,{\rm SB}}
 &=
 \tilde{V}^{0 \ell L}_{l' m a} \tilde{V}^{0 \ell L*}_{l m a}
 \bigg[
   -\frac{1}{3} J^2_{41}(\lambda_{\tilde{e}_a},\lambda_{\tilde{\chi}^0_m})
 \bigg],
\nonumber\\
G_{l'l\gamma}^{L,{\rm SB}}
 &=
 \tilde{V}^{0 \ell R}_{l' m a} \tilde{V}^{0 \ell R*}_{l m a}
 \bigg[
   m_{l'} \lambda_{\tilde{e}_a}
   J^1_{41}(\lambda_{\tilde{e}_a},\lambda_{\tilde{\chi}^0_m})
 \bigg]
 +
 \tilde{V}^{0 \ell L}_{l' m a} \tilde{V}^{0 \ell L*}_{l m a}
 \bigg[
   m_{l} \lambda_{\tilde{e}_a} J^1_{41}(\lambda_{\tilde{e}_a},\lambda_{\tilde{\chi}^0_m})
 \bigg]
\nonumber\\
 &+
 \tilde{V}^{0 \ell L}_{l' m a} \tilde{V}^{0 \ell R*}_{l m a}
 \bigg[
   + 2 m_{\tilde{\chi}^0_m} \lambda_{\tilde{e}_a}
   J^0_{31}(\lambda_{\tilde{e}_a},\lambda_{\tilde{\chi}^0_m})
 \bigg],
\nonumber\\
G_{l'l\gamma}^{R,{\rm SB}}
 &=
 \tilde{V}^{0 \ell L}_{l' m a} \tilde{V}^{0 \ell L*}_{l m a}
 \bigg[
   m_{l'} \lambda_{\tilde{e}_a}
   J^1_{41}(\lambda_{\tilde{e}_a},\lambda_{\tilde{\chi}^0_m})
 \bigg]
 +
 \tilde{V}^{0 \ell R}_{l' m a} \tilde{V}^{0 \ell R*}_{l m a}
 \bigg[
   m_{l} \lambda_{\tilde{e}_a} J^1_{41}(\lambda_{\tilde{e}_a},\lambda_{\tilde{\chi}^0_m})
 \bigg]
\nonumber\\
 &+
 \tilde{V}^{0 \ell R}_{l' m a} \tilde{V}^{0 \ell L*}_{l m a}
 \bigg[
   + 2 m_{\tilde{\chi}^0_m} \lambda_{\tilde{e}_a}
   J^0_{31}(\lambda_{\tilde{e}_a},\lambda_{\tilde{\chi}^0_m})
 \bigg] .
% change 24112012
\label{gammaSB}
\end{align}

\section{$Z$-Boson Form factors}

The form factors $F_Z^L$ and $F_Z^R$ may be decomposed as follows:
\begin{eqnarray}
(F_Z^L)_{l'l}
 &=&
 F^N_{l'lZ} + F^{L,\tilde{N}}_{l'lZ}
 + F^{L,{\rm SB}}_{l'lZ}\ ,
\nonumber\\
(F_\gamma^R)_{l'l}
 &=&
 F^N_{l'lZ} + F^{R,\tilde{N}}_{l'lZ}
 + F^{R,{\rm SB}}_{l'lZ}\ ,
\end{eqnarray}
where
\begin{align}
F_{l'l Z}^{L,N}
 &=
 B_{l'a} B^*_{l a}
 \bigg[
   \frac{5}{2} \lambda_{n_a} J^0_{21}(1,\lambda_{n_a})
 \bigg]
\nonumber\\
 &{}+
 B_{l'b} C_{ba} B^*_{l a}
 \bigg[
   - \frac{1}{2} J^1_{111}(1,\lambda_{n_b},\lambda_{n_a})
   + \frac{1}{2} \lambda_{n_a}\lambda_{n_b} J^0_{111}(1,\lambda_{n_b},\lambda_{n_a})
\nonumber\\
 &\qquad \qquad \qquad \qquad + \frac{1}{2 t^2_\beta} \lambda_{n_a}\lambda_{n_b}
   J^0_{111}(\lambda_{H^+},\lambda_{n_b},\lambda_{n_a})
 \bigg],
\nonumber\\
F_{l'l Z}^{R,N}
 &=
 - \frac{m_l m_{l'} t^2_\beta}{4 M_W^2}\
 B_{l'b} C_{ba} B^*_{l a}
   J^1_{111}(\lambda_{H^+},\lambda_{n_b},\lambda_{n_a}),
\\
F_{l'l Z}^{L,\tilde{N}}
 &=
 \frac{1}{2}
 ( \tilde{B}^{R,1}_{l' m A} \tilde{V}^{\tilde{\chi}^- R}_{mk} \tilde{B}^{R,1*}_{l k A}
 + \tilde{B}^{R,2}_{l' m A} \tilde{V}^{\tilde{\chi}^- R}_{mk} \tilde{B}^{R,2*}_{l k A} )
  J^1_{111}(\lambda_{\tilde{\chi}_m},\lambda_{\tilde{\chi}_k},\lambda_{\tilde{N}_A})
\nonumber\\
 &{}-
  ( \tilde{B}^{R,1}_{l' m A} \tilde{V}^{\tilde{\chi}^- L}_{mk} \tilde{B}^{R,1*}_{l m A}
  + \tilde{B}^{R,2}_{l' m A} \tilde{V}^{\tilde{\chi}^- L}_{mk} \tilde{B}^{R,2*}_{l m A})
  \sqrt{\lambda_{\tilde{\chi}_m}\lambda_{\tilde{\chi}_k}}
  J^0_{111}(\lambda_{\tilde{\chi}_m},\lambda_{\tilde{\chi}_k},\lambda_{\tilde{N}_A})
\nonumber\\
 &{}+
 \frac{1}{2}
 ( \tilde{B}^{R,1}_{l' m A}  \tilde{B}^{R,1*}_{l k A}
 + \tilde{B}^{R,2}_{l' m A}  \tilde{B}^{R,2*}_{l k A} ) \nonumber\\
 & \qquad \cdot
 \Big(\frac{1}{2} - s_w^2\Big)
 ( J^1_{21}(\lambda_{\tilde{\chi}_k},\lambda_{\tilde{N}_A})
 - 2 J^0_{11}(\lambda_{\tilde{\chi}_k},\lambda_{\tilde{N}_A}) )
\nonumber\\
 &{}+
 \frac{1}{4}
 ( \tilde{B}^{R,1}_{l' k A} \tilde{C}^{1}_{BA} \tilde{B}^{R,1*}_{l k A}
 + \tilde{B}^{R,1}_{l' k A} \tilde{C}^{2}_{BA} \tilde{B}^{R,2*}_{l k A}
 + \tilde{B}^{R,2}_{l' m A} \tilde{C}^{3}_{BA} \tilde{B}^{R,1*}_{l k A}
\nonumber\\
 & \qquad
 {}+ \tilde{B}^{R,2}_{l' m A} \tilde{C}^{4}_{BA} \tilde{B}^{R,2*}_{l k A} )
  J^1_{111}(\lambda_{\tilde{\chi}_k},\lambda_{\tilde{N}_B},\lambda_{\tilde{N}_A}),
\nonumber\\
F_{l'l Z}^{R,\tilde{N}}
 &=
 \frac{m_l m_{l'}}{2 s_\beta^2 M_W^2}
 \bigg[
 \frac{1}{2}
 ( \tilde{B}^{L,1}_{l' m A} \tilde{V}^{\tilde{\chi}^- L}_{mk} \tilde{B}^{L,1*}_{l k A}
 + \tilde{B}^{L,2}_{l' m A} \tilde{V}^{\tilde{\chi}^- L}_{mk} \tilde{B}^{L,2*}_{l k A} )
  J^1_{111}(\lambda_{\tilde{\chi}_m},\lambda_{\tilde{\chi}_k},\lambda_{\tilde{N}_A})
\nonumber\\
 &{}-
  ( \tilde{B}^{L,1}_{l' m A} \tilde{V}^{\tilde{\chi}^- R}_{mk} \tilde{B}^{L,1*}_{l m A}
  + \tilde{B}^{L,2}_{l' m A} \tilde{V}^{\tilde{\chi}^- R}_{mk} \tilde{B}^{L,2*}_{l m A})
  \sqrt{\lambda_{\tilde{\chi}_m}\lambda_{\tilde{\chi}_k}}
  J^0_{111}(\lambda_{\tilde{\chi}_m},\lambda_{\tilde{\chi}_k},\lambda_{\tilde{N}_A})
\nonumber\\
 &{}+
 \frac{1}{2}
 ( \tilde{B}^{L,1}_{l' m A}  \tilde{B}^{L,1*}_{l k A}
 + \tilde{B}^{L,2}_{l' m A}  \tilde{B}^{L,2*}_{l k A} ) ( - s_w^2)
 ( J^1_{21}(\lambda_{\tilde{\chi}_k},\lambda_{\tilde{N}_A})
 - 2 J^0_{11}(\lambda_{\tilde{\chi}_k},\lambda_{\tilde{N}_A}) )
\nonumber\\
 &{}+
 \frac{1}{4}
 ( \tilde{B}^{L,1}_{l' k A} \tilde{C}^{1}_{BA} \tilde{B}^{L,1*}_{l k A}
 + \tilde{B}^{L,1}_{l' k A} \tilde{C}^{2}_{BA} \tilde{B}^{L,2*}_{l k A}
 + \tilde{B}^{L,2}_{l' m A} \tilde{C}^{3}_{BA} \tilde{B}^{L,1*}_{l k A}
 \nonumber\\
 &{}+
 \tilde{B}^{L,2}_{l' m A} \tilde{C}^{4}_{BA} \tilde{B}^{L,2*}_{l k A} )
  J^1_{111}(\lambda_{\tilde{\chi}_k},\lambda_{\tilde{N}_B},\lambda_{\tilde{N}_A})
 \bigg],
% change 24112012
\label{ZNt} \\
F_{l'l Z}^{L,{\rm SB}}
 &=
 -
  \tilde{V}^{0\ell R}_{l' m a} \tilde{V}^{\tilde{\chi}^0 R}_{m k}
  \tilde{V}^{0\ell R*}_{l k a}
   J^1_{111}(\lambda_{\tilde{\chi}^0_m},\lambda_{\tilde{\chi}^0_k},\lambda_{\tilde{e}_a})
\nonumber\\
 &{}+
  \tilde{V}^{0\ell R}_{l' m a} \tilde{V}^{\tilde{\chi}^0 L}_{m k}
  \tilde{V}^{0\ell R*}_{l k a}
  \ 2\sqrt{\lambda_{\tilde{\chi}_m}\lambda_{\tilde{\chi}_k}}
   J^0_{111}(\lambda_{\tilde{\chi}^0_m},\lambda_{\tilde{\chi}^0_k},\lambda_{\tilde{e}_a})
\nonumber\\
 &{}+
  \tilde{V}^{0\ell R}_{l' k a} \tilde{V}^{0\ell R*}_{l k a}
 \Big(\frac{1}{2} - s_w^2\Big)
 \Big(
   -\frac{1}{2} J^1_{21}(\lambda_{\tilde{\chi}^0_k},\lambda_{\tilde{e}_a})
   + J^0_{11}(\lambda_{\tilde{\chi}^0_k},\lambda_{\tilde{e}_a})
 \Big)
\nonumber\\
 &{}- \frac{1}{2}
 \tilde{V}^{0\ell L}_{l' k b} \tilde{V}^{\tilde{e}}_{ba} \tilde{V}^{0\ell L*}_{l k a}
   J^1_{111}(\lambda_{\tilde{\chi}^0_k},\lambda_{\tilde{e}_b},\lambda_{\tilde{e}_a})\ ,
\nonumber\\
F_{l'l Z}^{R,{\rm SB}}
 &=
 -
  \tilde{V}^{0\ell L}_{l' m a} \tilde{V}^{\tilde{\chi}^0 L}_{m k}
  \tilde{V}^{0\ell L*}_{l k a}
   J^1_{111}(\lambda_{\tilde{\chi}^0_m},\lambda_{\tilde{\chi}^0_k},\lambda_{\tilde{e}_a})
\nonumber\\
 &{}+
  \tilde{V}^{0\ell L}_{l' m a} \tilde{V}^{\tilde{\chi}^0 R}_{m k}
  \tilde{V}^{0\ell L*}_{l k a}
  \ 2\sqrt{\lambda_{\tilde{\chi}_m}\lambda_{\tilde{\chi}_k}}
   J^0_{111}(\lambda_{\tilde{\chi}^0_m},\lambda_{\tilde{\chi}^0_k},\lambda_{\tilde{e}_a})
\nonumber\\
 &+
  \tilde{V}^{0\ell L}_{l' k a} \tilde{V}^{0\ell L*}_{l k a}
 (- s_w^2)
 \Big(
   -\frac{1}{2} J^1_{21}(\lambda_{\tilde{\chi}^0_k},\lambda_{\tilde{e}_a})
   + J^0_{11}(\lambda_{\tilde{\chi}^0_k},\lambda_{\tilde{e}_a})
 \Big)
\nonumber\\
 &{}- \frac{1}{2}
 \tilde{V}^{0\ell R}_{l' k b} \tilde{V}^{\tilde{e}}_{ba} \tilde{V}^{0\ell R*}_{l k a}
   J^1_{111}(\lambda_{\tilde{\chi}^0_k},\lambda_{\tilde{e}_b},\lambda_{\tilde{e}_a})\ .
% change 24112012
\label{ZSB}
\end{align}

\section{Leptonic Box Form factors}

The  leptonic box  amplitudes are  expressed  in terms  of the  chiral
structures:   $\bar{l}'\Gamma_A^X    l\   \bar{l}_1\Gamma_A^Y   l^C_2$
[cf~Eq~(\ref{Tbl_2})].   There  are  two  distinct contributions  to  the
chiral amplitudes.   The first one  has {\it direct} relevance  to the
original  structure given  above and that one is denoted with a  subscript
$D$. The second contribution comes from a chiral amplitude of the form
$\bar{l}_1\Gamma_A^X  l\ \bar{l}'\Gamma_A^Y l^C_2$,  which contributes
to the  original amplitude $\bar{l}'\Gamma_A^X  l\ \bar{l}_1\Gamma_A^Y
l^C_2$, after performing a Fierz transformation. This Fierz-transformed 
contribution is indicated with a subscript $F$. More explicitly, the leptonic 
box form factors are given by
\begin{align}
 B_{\ell V}^{LL} &= B_{\ell V,D}^{LL} + B_{\ell V,F}^{LL} \,,
 &
 B_{\ell V}^{RR} &= B_{\ell V,D}^{RR} + B_{\ell V,F}^{RR} \,,
\nonumber\\
 B_{\ell V}^{LR} &= B_{\ell V,D}^{LR} - \frac{1}{2} B_{\ell S,F}^{LR} \,,
 &
 B_{\ell V}^{RL} &= B_{\ell V,D}^{RL} - \frac{1}{2} B_{\ell S,F}^{RL} \,,
\nonumber\\
 B_{\ell S}^{LL} &= B_{\ell S,D}^{LL} + \frac{1}{2} B_{\ell S,F}^{LL}
 + \frac{3}{2} B_{\ell T,F}^{LL},
 &
 B_{\ell S}^{RR} &= B_{\ell S,D}^{RR} + \frac{1}{2} B_{\ell
   S,F}^{RR} + \frac{3}{2} B_{\ell T,F}^{RR} \,,
\nonumber\\
 B_{\ell S}^{LR} &= B_{\ell S,D}^{LR} - 2 B_{\ell V,F}^{LR} \,,
 &
 B_{\ell S}^{RL} &= B_{\ell S,D}^{RL} - 2 B_{\ell V,F}^{RL} \,,
\nonumber\\
 B_{\ell T}^{LL} &= B_{\ell T,D}^{LL} - \frac{1}{2} B_{\ell T,F}^{LL}
 + \frac{1}{2} B_{\ell S,F}^{LL} \,,
 &
 B_{\ell T}^{RR} &= B_{\ell T,D}^{RR} - \frac{1}{2} B_{\ell
   T,F}^{RR} + \frac{1}{2} B_{\ell S,F}^{RR}\,.
\end{align}
The   {\it  direct}   and  Fierz-transformed   contributions   to  the
form factors are related by the exchange of outgoing leptons
\begin{equation}
 B_{\ell A,F}^{XY} = B_{\ell A,D}^{XY} (l'\leftrightarrow l_1)\ .
\end{equation}
The {\it direct}  contributions have {\it direct} $N$,  ${\rm SB}$ and
Fierz-transformed $\tilde{N}$ contributions:
\begin{align}
 B_{\ell V,D}^{LL} &= B_{\ell V,D}^{LL,N} + B_{\ell
   V,F}^{LL,\tilde{N}} + B_{\ell V,D}^{LL,{\rm SB}}\,,
 &
 B_{\ell V,D}^{RR} &= B_{\ell V,D}^{RR,{\rm SB}}\,,
\nonumber\\
 B_{\ell V,D}^{LR} &= B_{\ell V,D}^{LR,{\rm SB}}\,,
 &
 B_{\ell V,D}^{RL} &= -\frac{1}{2} B_{\ell S,F}^{RL,\tilde{N}} +
 B_{\ell V,D}^{RL,{\rm SB}}\,,
\nonumber\\
 B_{\ell S,D}^{LL} &= B_{\ell S,D}^{LL,{\rm SB}},
 &
 B_{\ell S,D}^{RR} &= \frac{1}{2} B_{\ell S,F}^{RR,\tilde{N}} +
 B_{\ell S,D}^{RR,{\rm SB}} \,,
\nonumber\\
 B_{\ell S,D}^{LR} &= B_{\ell S,D}^{LR,{\rm SB}}\,,
 &
 B_{\ell S,D}^{RL} &= B_{\ell S,D}^{RL,N} - 2 B_{\ell
   V,F}^{RL,\tilde{N}} + B_{\ell S,D}^{RL,{\rm SB}}\,,
\nonumber\\
 B_{\ell T,D}^{LL} &= B_{\ell T,D}^{LL,{\rm SB}}\,,
 &
 B_{\ell T,D}^{RR} &= \frac{1}{2} B_{\ell S,F}^{RR,\tilde{N}} +
 B_{\ell T,D}^{RR,{\rm SB}}\,.
\label{FF_NtNSB}
\end{align}
The form factor contributions from Eq~\eqref{FF_NtNSB} read:
\begin{align}
%A1BN BlVDLLN
B_{\ell V,D}^{LL,N}
 &=
 B^*_{l a} B^*_{l_2 b} B_{l' a} B_{l_1 b}
 \bigg[
  - \bigg( 1 + \frac{\lambda_{n_a} \lambda_{n_b}}{4} \bigg)
    J^1_{211}(1,\lambda_{n_a},\lambda_{n_b})
\nonumber \\
  &{}+ 2 \lambda_{n_a} \lambda_{n_b} J^0_{211}(1,\lambda_{n_a},\lambda_{n_b})
    - 2 \lambda_{n_a} \lambda_{n_b} t_\beta^{-2}
    J^0_{1111}(1,\lambda_{H^+},\lambda_{n_a},\lambda_{n_b})
\nonumber\\
  &{}- \frac{1}{2} \lambda_a \lambda_b t_\beta^{-2}
  J^1_{1111}(1,\lambda_{H^+},\lambda_{n_a},\lambda_{n_b})
  - \frac{1}{4} \lambda_a \lambda_b t_\beta^{-4}
J^1_{211}(\lambda_{H^+},\lambda_{n_a},\lambda_{n_b})
 \bigg]\,,
\nonumber\\
%A8BN BLSDRLN
B_{\ell V,D}^{RL,N}
 &=
 -  B^*_{l a} B^*_{l_2 b} B_{l' a} B_{l_1 b}\:
  \frac{m_l m_{l_1} t^2_\beta}{M_W^2}
\nonumber\\
 & \qquad \cdot
  \bigg(
     J^1_{1111}(1,\lambda_{H^+},\lambda_{n_a},\lambda_{n_b})
   + \lambda_a \lambda_b J^0_{1111}(\lambda_{H^+},\lambda_{n_a},\lambda_{n_b})
  \bigg)\,,
\\
\nonumber\\
%A1AN BlVFLLNt
B_{\ell V,F}^{LL,\tilde{N}}
 &=
  ( \tilde{B}^{R,1}_{l_1 k B} \tilde{B}^{R,1*}_{l_2 m B} +
\tilde{B}^{R,2}_{l_1 k B} \tilde{B}^{R,2*}_{l_2 m B} )
  ( \tilde{B}^{R,1}_{l' m A} \tilde{B}^{R,1*}_{l k A} +
\tilde{B}^{R,2}_{l' m A} \tilde{B}^{R,2*}_{l k A} )
\nonumber\\
 & \qquad \cdot
 J^1_{1111}(\lambda_{\tilde{\chi}_k},\lambda_{\tilde{\chi}_m},\lambda_{\tilde{N}_A},
\lambda_{\tilde{N}_B})\,,
\nonumber\\
%A4AN BlVFRLNt
B_{\ell V,F}^{RL,\tilde{N}}
 &=
  ( \tilde{B}^{L,1}_{l_1 k B} \tilde{B}^{R,1*}_{l_2 m B} +
\tilde{B}^{L,2}_{l_1 k B} \tilde{B}^{R,2*}_{l_2 m B} )
  ( \tilde{B}^{R,1}_{l' m A} \tilde{B}^{L,1*}_{l k A} +
\tilde{B}^{R,2}_{l' m A} \tilde{B}^{L,2*}_{l k A} )
\nonumber\\
 &\qquad \cdot
 \frac{m_{l'} m_l}{2 c_\beta^2 M_W^2}
J^1_{1111}(\lambda_{\tilde{\chi}_k},\lambda_{\tilde{\chi}_m},\lambda_{\tilde{N}_A},
\lambda_{\tilde{N}_B})\,,
\nonumber\\
%A6AN BlSFRRNt
B_{\ell S,F}^{RR,\tilde{N}}
 &=
  ( \tilde{B}^{R,1}_{l_1 k B} \tilde{B}^{L,1*}_{l_2 m B} +
\tilde{B}^{R,2}_{l_1 k B} \tilde{B}^{L,2*}_{l_2 m B} )
  ( \tilde{B}^{R,1}_{l' m A} \tilde{B}^{L,1*}_{l k A} +
\tilde{B}^{R,2}_{l' m A} \tilde{B}^{L,2*}_{l k A} )
\nonumber\\
 &\qquad \cdot
 \frac{2 m_{l_2} m_l}{c_\beta^2 M_W^2}
 \sqrt{\lambda_{\tilde{\chi}_k}\lambda_{\tilde{\chi}_m}}
 J^0_{1111}(\lambda_{\tilde{\chi}_k},\lambda_{\tilde{\chi}_m},\lambda_{\tilde{N}_A},
\lambda_{\tilde{N}_B})\,,
\nonumber\\
%A8AN BlSFRLNt
B_{\ell S,F}^{RL,\tilde{N}}
 &=
  ( \tilde{B}^{R,1}_{l_1 k B} \tilde{B}^{R,1*}_{l_2 m B} +
\tilde{B}^{R,2}_{l_1 k B} \tilde{B}^{R,2*}_{l_2 m B} )
  ( \tilde{B}^{L,1}_{l' m A} \tilde{B}^{L,1*}_{l k A} +
\tilde{B}^{L,2}_{l' m A} \tilde{B}^{L,2*}_{l k A} )
\nonumber\\
 &\qquad \cdot
 \frac{2 m_{l_1} m_l}{c_\beta^2 M_W^2}
 \sqrt{\lambda_{\tilde{\chi}_k}\lambda_{\tilde{\chi}_m}}
 J^0_{1111}(\lambda_{\tilde{\chi}_k},\lambda_{\tilde{\chi}_m},\lambda_{\tilde{N}_A},
\lambda_{\tilde{N}_B})\,, \\
\nonumber\\
% Bl1 BlVDLLSB
B_{\ell V,D}^{LL,{\rm SB}}
 &=
 - \tilde{V}^{0\ell R}_{l_1 m b} \tilde{V}^{0\ell R*}_{l m a}
 \tilde{V}^{0\ell R}_{l' n a} \tilde{V}^{0\ell R*}_{l_2 n b}  %1
  J^1_{1111}(\lambda_{\tilde{\chi}^0_m},\lambda_{\tilde{\chi}^0_n},\lambda_{\tilde{e}_a},
\lambda_{\tilde{e}_b})
\nonumber\\
 &{}-
   2 \tilde{V}^{0\ell R}_{l_2 m b} \tilde{V}^{0\ell R*}_{l m a}
   \tilde{V}^{0\ell R}_{l' n a} \tilde{V}^{0\ell R*}_{l_1 n b} %1/2
 \sqrt{\lambda_{\tilde{\chi}^0_m} \lambda_{\tilde{\chi}^0_k}}
 J^0_{1111}(\lambda_{\tilde{\chi}^0_m},\lambda_{\tilde{\chi}^0_n},\lambda_{\tilde{e}_a},
\lambda_{\tilde{e}_b})\,,
\nonumber\\
% Bl2 BlVDRRSB
B_{\ell V,D}^{RR,{\rm SB}}
 &=
 - \tilde{V}^{0\ell L}_{l_1 m b} \tilde{V}^{0\ell L*}_{l m a}
 \tilde{V}^{0\ell L}_{l' n a} \tilde{V}^{0\ell L*}_{l_2 n b} %1
  J^1_{1111}(\lambda_{\tilde{\chi}^0_m},\lambda_{\tilde{\chi}^0_n},\lambda_{\tilde{e}_a},\lambda_{\tilde{e}_b})
\nonumber\\
 &{}-
   2 \tilde{V}^{0\ell L}_{l_2 m b} \tilde{V}^{0\ell L*}_{l m a}
   \tilde{V}^{0\ell L}_{l' n a} \tilde{V}^{0\ell L*}_{l_1 n b} %1/2
 \sqrt{\lambda_{\tilde{\chi}^0_m} \lambda_{\tilde{\chi}^0_k}}
 J^0_{1111}(\lambda_{\tilde{\chi}^0_m},\lambda_{\tilde{\chi}^0_n},\lambda_{\tilde{e}_a},
\lambda_{\tilde{e}_b})\,,
\nonumber\\
% Bl3 BlVDLRSB
B_{\ell V,D}^{LR,{\rm SB}}
 &=
 2 \tilde{V}^{0\ell L}_{l_1 m b} \tilde{V}^{0\ell R*}_{l m a}
 \tilde{V}^{0\ell R}_{l' n a} \tilde{V}^{0\ell L*}_{l_2 n b} %-1/2
 \sqrt{\lambda_{\tilde{\chi}^0_m} \lambda_{\tilde{\chi}^0_k}}
 J^0_{1111}(\lambda_{\tilde{\chi}^0_m},\lambda_{\tilde{\chi}^0_n},\lambda_{\tilde{e}_a},
\lambda_{\tilde{e}_b})
\nonumber\\
 &{}+
 \tilde{V}^{0\ell L}_{l_2 m b} \tilde{V}^{0\ell R*}_{l m a}
 \tilde{V}^{0\ell R}_{l' n a} \tilde{V}^{0\ell L*}_{l_1 n b} %-1
 J^1_{1111}(\lambda_{\tilde{\chi}^0_m},\lambda_{\tilde{\chi}^0_n},\lambda_{\tilde{e}_a},
\lambda_{\tilde{e}_b})\,,
\nonumber\\
% Bl4 BlVDRLSB
B_{\ell V,D}^{RL,{\rm SB}}
 &=
 2 \tilde{V}^{0\ell R}_{l_1 m b} \tilde{V}^{0\ell L*}_{l m a}
 \tilde{V}^{0\ell L}_{l' n a} \tilde{V}^{0\ell R*}_{l_2 n b} %-1/2
 \sqrt{\lambda_{\tilde{\chi}^0_m} \lambda_{\tilde{\chi}^0_k}}
 J^0_{1111}(\lambda_{\tilde{\chi}^0_m},\lambda_{\tilde{\chi}^0_n},\lambda_{\tilde{e}_a},
\lambda_{\tilde{e}_b})
\nonumber\\
 &{}+
 \tilde{V}^{0\ell R}_{l_2 m b} \tilde{V}^{0\ell L*}_{l m a}
 \tilde{V}^{0\ell L}_{l' n a} \tilde{V}^{0\ell R*}_{l_1 n b} %-1
 J^1_{1111}(\lambda_{\tilde{\chi}^0_m},\lambda_{\tilde{\chi}^0_n},\lambda_{\tilde{e}_a},
\lambda_{\tilde{e}_b})\,,
\nonumber\\
%Bl5 BlSDLLSB
B_{\ell S,D}^{LL,{\rm SB}}
 &=
 -2 \tilde{V}^{0\ell L}_{l_1 m b} \tilde{V}^{0\ell R*}_{l m a}
 \tilde{V}^{0\ell L}_{l' n a} \tilde{V}^{0\ell R*}_{l_2 n b} %1/2
 \sqrt{\lambda_{\tilde{\chi}^0_m} \lambda_{\tilde{\chi}^0_k}}
 J^0_{1111}(\lambda_{\tilde{\chi}^0_m},\lambda_{\tilde{\chi}^0_n},\lambda_{\tilde{e}_a},
\lambda_{\tilde{e}_b})
\nonumber\\
 &{}-
 2 \tilde{V}^{0\ell R}_{l_2 m b} \tilde{V}^{0\ell R*}_{l m a}
 \tilde{V}^{0\ell L}_{l' n a} \tilde{V}^{0\ell L*}_{l_1 n b} %1/2
 \sqrt{\lambda_{\tilde{\chi}^0_m} \lambda_{\tilde{\chi}^0_k}}
 J^0_{1111}(\lambda_{\tilde{\chi}^0_m},\lambda_{\tilde{\chi}^0_n},\lambda_{\tilde{e}_a},
\lambda_{\tilde{e}_b})\,,
\nonumber\\
%Bl6 BlSDRRSB
B_{\ell S,D}^{RR,{\rm SB}}
 &=
 -2 \tilde{V}^{0\ell R}_{l_1 m b} \tilde{V}^{0\ell L*}_{l m a}
 \tilde{V}^{0\ell R}_{l' n a} \tilde{V}^{0\ell L*}_{l_2 n b} %1/2
 \sqrt{\lambda_{\tilde{\chi}^0_m} \lambda_{\tilde{\chi}^0_k}}
 J^0_{1111}(\lambda_{\tilde{\chi}^0_m},\lambda_{\tilde{\chi}^0_n},\lambda_{\tilde{e}_a},
\lambda_{\tilde{e}_b})
\nonumber\\
 &{}-
  2 \tilde{V}^{0\ell L}_{l_2 m b} \tilde{V}^{0\ell L*}_{l m a}
  \tilde{V}^{0\ell R}_{l' n a} \tilde{V}^{0\ell R*}_{l_1 n b} %1/2
 \sqrt{\lambda_{\tilde{\chi}^0_m} \lambda_{\tilde{\chi}^0_k}}
 J^0_{1111}(\lambda_{\tilde{\chi}^0_m},\lambda_{\tilde{\chi}^0_n},\lambda_{\tilde{e}_a},
\lambda_{\tilde{e}_b})\,,
\nonumber\\
%Bl7 BlSDLRSB
B_{\ell S,D}^{LR,{\rm SB}}
 &=
 2 \tilde{V}^{0\ell R}_{l_1 m b} \tilde{V}^{0\ell R*}_{l m a}
 \tilde{V}^{0\ell L}_{l' n a} \tilde{V}^{0\ell L*}_{l_2 n b} %-2
  J^1_{1111}(\lambda_{\tilde{\chi}^0_m},\lambda_{\tilde{\chi}^0_n},\lambda_{\tilde{e}_a},
\lambda_{\tilde{e}_b})
\nonumber\\
 &{}+
  2 \tilde{V}^{0\ell L}_{l_2 m b} \tilde{V}^{0\ell R*}_{l m a}
  \tilde{V}^{0\ell L}_{l' n a} \tilde{V}^{0\ell R*}_{l_1 n b} %-2
 J^1_{1111}(\lambda_{\tilde{\chi}^0_m},\lambda_{\tilde{\chi}^0_n},\lambda_{\tilde{e}_a},
\lambda_{\tilde{e}_b})\,,
\nonumber\\
%Bl8 BlSDRLSB
B_{\ell S,D}^{RL,{\rm SB}}
 &=
 2 \tilde{V}^{0\ell L}_{l_1 m b} \tilde{V}^{0\ell L*}_{l m a}
 \tilde{V}^{0\ell R}_{l' n a} \tilde{V}^{0\ell R*}_{l_2 n b} %-2
  J^1_{1111}(\lambda_{\tilde{\chi}^0_m},\lambda_{\tilde{\chi}^0_n},\lambda_{\tilde{e}_a},
\lambda_{\tilde{e}_b})
\nonumber\\
 &{}+
 2 \tilde{V}^{0\ell R}_{l_2 m b} \tilde{V}^{0\ell L*}_{l m a}
 \tilde{V}^{0\ell R}_{l' n a} \tilde{V}^{0\ell L*}_{l_1 n b} %-2
 J^1_{1111}(\lambda_{\tilde{\chi}^0_m},\lambda_{\tilde{\chi}^0_n},\lambda_{\tilde{e}_a},
\lambda_{\tilde{e}_b})\,,
\nonumber\\
%Bl9 BlTDLLSB
B_{\ell T,D}^{LL,{\rm SB}}
 &=
 -2 \tilde{V}^{0\ell L}_{l_1 m b} \tilde{V}^{0\ell R*}_{l m a}
 \tilde{V}^{0\ell L}_{l' n a} \tilde{V}^{0\ell R*}_{l_2 n b} %1/2
 \sqrt{\lambda_{\tilde{\chi}^0_m} \lambda_{\tilde{\chi}^0_k}}
 J^0_{1111}(\lambda_{\tilde{\chi}^0_m},\lambda_{\tilde{\chi}^0_n},\lambda_{\tilde{e}_a},
\lambda_{\tilde{e}_b})
\nonumber\\
 &{}+
  2 \tilde{V}^{0\ell R}_{l_2 m b} \tilde{V}^{0\ell R*}_{l m a}
  \tilde{V}^{0\ell L}_{l' n a} \tilde{V}^{0\ell L*}_{l_1 n b} %1/2
 \sqrt{\lambda_{\tilde{\chi}^0_m} \lambda_{\tilde{\chi}^0_k}}
 J^0_{1111}(\lambda_{\tilde{\chi}^0_m},\lambda_{\tilde{\chi}^0_n},\lambda_{\tilde{e}_a},
\lambda_{\tilde{e}_b})\,,
\nonumber\\
%Bl10 BlTDRRSB
B_{\ell T,D}^{RR,{\rm SB}}
 &=
 - 2 \tilde{V}^{0\ell R}_{l_1 m b} \tilde{V}^{0\ell L*}_{l m a}
 \tilde{V}^{0\ell R}_{l' n a} \tilde{V}^{0\ell L*}_{l_2 n b} %1/2
 \sqrt{\lambda_{\tilde{\chi}^0_m} \lambda_{\tilde{\chi}^0_k}}
 J^0_{1111}(\lambda_{\tilde{\chi}^0_m},\lambda_{\tilde{\chi}^0_n},\lambda_{\tilde{e}_a},
\lambda_{\tilde{e}_b})
\nonumber\\
 &{}+
  2 \tilde{V}^{0\ell L}_{l_2 m b} \tilde{V}^{0\ell L*}_{l m a}
  \tilde{V}^{0\ell R}_{l' n a} \tilde{V}^{0\ell R*}_{l_1 n b} %1/2
 \sqrt{\lambda_{\tilde{\chi}^0_m} \lambda_{\tilde{\chi}^0_k}}
 J^0_{1111}(\lambda_{\tilde{\chi}^0_m},\lambda_{\tilde{\chi}^0_n},\lambda_{\tilde{e}_a},
\lambda_{\tilde{e}_b})\,.
\label{ffboxSB}
\end{align}

\section{Semileptonic Box Form factors}

Semileptonic form  factors have only {\it  direct} contributions, with
the following $N$, $\tilde{N}$ and ${\rm SB}$ content:
\begin{eqnarray}
B_{dV}^{LL} &=& B_{dV}^{LL,N} + B_{dV}^{LL,\tilde{N}} +B_{dV}^{LL,{\rm SB}}\ ,
\nonumber\\
B_{uV}^{LL} &=& B_{uV}^{LL,N} + B_{uV}^{LL,\tilde{N}} +  B_{uV}^{LL,{\rm SB}}\ ,
\end{eqnarray}
and
\begin{equation}
\quad\, B_{dA}^{XY}\ =\ B_{dA}^{XY,{\rm SB}}\;,\qquad
B_{uA}^{XY}\ =\ B_{uA}^{XY,{\rm SB}}\ ,
\end{equation}
for $(X,Y,A)\neq (L,L,V)$.  The  $N$ and $\tilde{N}$ contributions are
given by
\begin{align}
% A1dN BdVLLN
B_{dV}^{LL,N}  &=  B_{l' a} B^*_{l a} (V^*)_{bd_1}(V)_{bd_2}
 \bigg[
    - \bigg( 1 + \frac{\lambda_{n_a} \lambda_{u_b}}{4} \bigg)
    J^1_{211}(1,\lambda_{n_a},\lambda_{u_b})
\nonumber\\
 &{} + 2 \lambda_{n_a} \lambda_{u_b}
 J^0_{211}(1,\lambda_{n_a},\lambda_{u_b})
 + \frac{1}{2 t_\beta^2} \lambda_{n_a}\lambda_{u_b}
    J^0_{1111}(1,\lambda_{H^+},\lambda_{n_a},\lambda_{u_b})
\nonumber\\
 &{} - \frac{1}{2 t_\beta^2} \lambda_{n_a}\lambda_{u_b}
  J^1_{1111}(1,\lambda_{H^+},\lambda_{n_a},\lambda_{u_b})
 - \frac{1}{4 t_\beta^4} \lambda_{n_a}\lambda_{u_b}
J^1_{211}(\lambda_{H^+},\lambda_{n_a},\lambda_{u_b})
 \bigg],
\nonumber\\
% A1uN BuVLLN
B_{uV}^{LL,N}
 &=
 B_{l' a} B^*_{l a} (V^*)_{d_2b}(V)_{d_1b}
 \bigg[
    \bigg( 4 + \frac{\lambda_{n_a} \lambda_{d_b}}{4} \bigg)
    J^1_{211}(1,\lambda_{n_a},\lambda_{d_b})
\nonumber\\
 &{}- 2 \lambda_{n_a} \lambda_{d_b} J^0_{211}(1,\lambda_{n_a},\lambda_{d_b})
 + \frac{1}{2}\lambda_{n_a}\lambda_{d_b}
    J^0_{1111}(1,\lambda_{H^+},\lambda_{n_a},\lambda_{d_b})
\nonumber\\
 &{} - \frac{1}{2} \lambda_{n_a}\lambda_{d_b}
  J^1_{1111}(1,\lambda_{H^+},\lambda_{n_a},\lambda_{d_b})
 + \frac{1}{4} \lambda_{n_a}\lambda_{d_b}
J^1_{211}(\lambda_{H^+},\lambda_{n_a},
\lambda_{d_b})
 \bigg],
\\
% A1dNt BdVLLNt
B_{dV}^{LL,\tilde{N}}
 &=
 - \frac{1}{2} \tilde{V}_{d_1ka}^{-dR}\tilde{V}_{d_2ma}^{-dR*}
   ( \tilde{V}_{lkA}^{-\ell R,1*}\tilde{V}_{l'mA}^{-\ell R,1}
   + \tilde{V}_{lkA}^{-\ell R,2*}\tilde{V}_{l'mA}^{-\ell R,2})
\nonumber\\
 & \qquad \cdot J^1_{1111}(\lambda_{\lambda_{\tilde{\chi}_k}},\lambda_{\tilde{\chi}_m},
\lambda_{\tilde{N}_A},\lambda_{\tilde{u}_a}),
\nonumber\\
% A1uNt BuVLLNt
B_{uV}^{LL,\tilde{N}}
 &=
 - \tilde{V}_{u_1ka}^{-uR}\tilde{V}_{u_2ma}^{-uR*}
   ( \tilde{V}_{lkA}^{-\ell R,1*}\tilde{V}_{l'mA}^{-\ell R,1}
   + \tilde{V}_{lkA}^{-\ell R,2*}\tilde{V}_{l'mA}^{-\ell R,2})\:
   \sqrt{\lambda_{\tilde{\chi}_k} \lambda_{\tilde{\chi}_m}}
\nonumber\\
 & \qquad \cdot
   J^0_{1111}(\lambda_{\lambda_{\tilde{\chi}_k}},\lambda_{\tilde{\chi}_m},
\lambda_{\tilde{N}_A},\lambda_{\tilde{d}_a})\ .
\end{align}
The   ${\rm    SB}$   form   factors    $B_{dA}^{XY,{\rm   SB}}$   and
$B_{uA}^{XY,{\rm  SB}}$,  with  $X=L,R$,  $Y=L,R$ and  $A=V,S,T$,  are
obtained   from  the  {\it   direct}  leptonic   form factors  $B_{\ell
  A}^{XY,{\rm SB}}$, by making  the replacements: $\ell\to d$, $l_1\to
d$, $l_2\to  d$, $\tilde{e}\to\tilde{d}$ and $\ell\to  u$, $l_1\to u$,
$l_2\to u$, $\tilde{e}\to\tilde{u}$,  in both the interaction vertices
and the arguments  of the $J$-loop functions that  carry the index $b$
in Eq~\eqref{ffboxSB}.

\input{FFanalysis}

\end{appendices}

%% file: FFanalysis.tex
% this should be part of the appendices.tex

\chapter{Form factor analysis} \label{app:ff}

In order to calculate three-body decays given by Eqs~\eqref{Bl3l_1} and
\eqref{Bl3l_2}, we have developed the \emph{model independent} procedure
for calculation of CLFV three-body decay rates. The analytical
calculus is for the most part performed using \emph{Wolfram Mathematica},
with the aid of \verb+FeynCalc+ package \cite{Mertig1991} which was found
to be very useful in dealing with the Dirac algebra (traces of $\gamma$
matrices, kinematics, etc.).

The starting point of the calculation are the most general photon mediated, 
$Z$-boson mediated and box-diagram effective operators inducing 
$l\to l'l_1 l_2^c$ LFV transitions,

\allowdisplaybreaks
\begin{align}
  \mathcal{T}_\gamma^{l \to l' l_1 l_2} &=  \frac{ \alpha_w^2 }{ M_W^2 } \cdot \Big\{
 \bar{l}' \gamma_\alpha P_L l \;\; \bar{l_1} \gamma^\alpha P_L l_2 \cdot P_1 + 
 \bar{l}' \gamma_\alpha P_R l \;\; \bar{l_1} \gamma^\alpha P_R l_2 \cdot P_2 \nonumber \\
 & \quad {}+
 \bar{l}' \gamma_\alpha P_L l \;\; \bar{l_1} \gamma^\alpha P_R l_2 \cdot P_3 + 
 \bar{l}' \gamma_\alpha P_R l \;\; \bar{l_1} \gamma^\alpha P_L l_2 \cdot P_4 \nonumber \\
 & \quad {}+
 \bar{l}' i \sigma_{\alpha\beta} q^\beta P_L  l \;\; \bar{l_1} \gamma^\alpha P_L l_2 
	\cdot \frac{P_{11}}{q^2} + 
 \bar{l}' i \sigma_{\alpha\beta} q^\beta P_R  l \;\; \bar{l_1} \gamma^\alpha P_R l_2 
	\cdot \frac{P_{12}}{q^2} \nonumber \\
 & \quad {}+ 
 \bar{l}' i \sigma_{\alpha\beta} q^\beta P_L  l \;\; \bar{l_1} \gamma^\alpha P_R l_2 
	\cdot \frac{P_{13}}{q^2} + 
 \bar{l}' i \sigma_{\alpha\beta} q^\beta P_R  l \;\; \bar{l_1} \gamma^\alpha P_L l_2 
	\cdot \frac{P_{14}}{q^2} \; \Big\} \,, \\[2ex]
 \mathcal{T}_Z^{l \to l' l_1 l_2} & = \frac{ \alpha_w^2 }{ M_W^2 } \cdot \Big\{
 \bar{l}' \gamma_\alpha P_L l \;\; \bar{l_1} \gamma^\alpha P_L l_2 \cdot Z_1 + 
 \bar{l}' \gamma_\alpha P_R l \;\; \bar{l_1} \gamma^\alpha P_R l_2 \cdot Z_2 \nonumber \\
 & \quad{}+
 \bar{l}' \gamma_\alpha P_L l \;\; \bar{l_1} \gamma^\alpha P_R l_2 \cdot Z_3 + 
 \bar{l}' \gamma_\alpha P_R l \;\; \bar{l_1} \gamma^\alpha P_L l_2 \cdot Z_4 \; \Big\} \,,
\end{align}
\begin{align}
  \mathcal{T}_{box}^{l \to l' l_1 l_2} & = \frac{ \alpha_w^2 }{ M_W^2 } \cdot \Big\{
 \bar{l}' \gamma_\alpha P_L l \;\; \bar{l_1} \gamma^\alpha P_L l_2 \cdot B_1 + 
 \bar{l}' \gamma_\alpha P_R l \;\; \bar{l_1} \gamma^\alpha P_R l_2 \cdot B_2 \nonumber \\
 & \quad{}+
 \bar{l}' \gamma_\alpha P_L l \;\; \bar{l_1} \gamma^\alpha P_R l_2 \cdot B_3 + 
 \bar{l}' \gamma_\alpha P_R l \;\; \bar{l_1} \gamma^\alpha P_L l_2 \cdot B_4 \nonumber \\
 & \quad{}+
 \bar{l}' P_L l \;\; \bar{l_1} P_L l_2 \cdot B_5 + 
 \bar{l}' P_R l \;\; \bar{l_1} P_R l_2 \cdot B_6 \nonumber \\
 & \quad{}+
 \bar{l}' P_L l \;\; \bar{l_1} P_R l_2 \cdot B_7 + 
 \bar{l}' P_R l \;\; \bar{l_1} P_L l_2 \cdot B_8 \nonumber \\
 & \quad{}+
 \bar{l}' \sigma_{\alpha\beta} P_L  l \;\; \bar{l_1} \sigma^{\alpha\beta} P_L l_2 \cdot B_9 + 
 \bar{l}' \sigma_{\alpha\beta} P_R  l \;\; \bar{l_1} \sigma^{\alpha\beta} P_R l_2 \cdot B_{10} 
  \; \Big\} \,,
\end{align}
expressed in terms of the form factors $P_i$, $Z_i$ and $B_i$ multiplied by the 
correspoding four-lepton operators. The Higgs mediated contributions were not 
included since we assured their smallness assuming small $\tan\beta$ 
($\tan\beta \lesssim 20$).

Note that the four-lepton operators are all written in the form 
$\bar{l}'\cdots l \;\; \bar{l_1} \cdots l_2$. This is achieved by
applying the Fiertz transformations
\begin{eqnarray}
(\gamma^\mu P_L \times \gamma_\mu P_L)_{1234} & = &
        (\gamma^\mu P_L \times \gamma_\mu P_L)_{1432} \,, \nonumber \\
(\gamma^\mu P_R \times \gamma_\mu P_R)_{1234} & = &
        (\gamma^\mu P_R \times \gamma_\mu P_R)_{1432} \,, \nonumber \\
(\gamma^\mu P_L \times \gamma^\mu P_R)_{1234} & = &
        -2 (P_R \times P_L)_{1432}  \,, \nonumber \\
(\gamma^\mu P_R \times \gamma_\mu P_L)_{1234} & = &
        -2 (P_L \times P_R)_{1432}  \,, \nonumber \\
(P_R \times P_L)_{1234} & = &
        - \frac{1}{2} (\gamma_\mu P_L \times \gamma^\mu P_R)_{1432} \,, \nonumber \\
(P_L \times P_R)_{1234} & = &
        - \frac{1}{2} (\gamma_\mu P_R \times \gamma^\mu P_L)_{1432} \,, \nonumber \\
(P_R \times P_R)_{1234} & = &
        \left[ -\frac{1}{2} (P_R \times P_R) + \frac{1}{8} \sigma_{\mu\nu}
        P_R \times \sigma^{\mu\nu} P_R \right]_{1432}  \,, \nonumber \\
(P_L \times P_L)_{1234} & = &
        \left[ -\frac{1}{2} (P_L \times P_L) + \frac{1}{8} \sigma_{\mu\nu}
        P_L \times \sigma^{\mu\nu} P_L \right]_{1432} \,.
\end{eqnarray}
to the terms with different ordering of the lepton fields.

The total effective operator is a sum of the photon-mediated, 
$Z$-boson mediated and box contributions,
\begin{align}
 \mathcal{T}_{TOT}^{l \to l' l_1 l_2} & =  \frac{ \alpha_w^2 }{ M_W^2 } \cdot \Big\{ 
 \bar{l}' \gamma_\alpha P_L l \;\; \bar{l_1} \gamma^\alpha P_L l_2 \cdot [ P_1 + Z_1 + B_1 ]
  \nonumber \\ 
 & \quad {}+
 \bar{l}' \gamma_\alpha P_R l \;\; \bar{l_1} \gamma^\alpha P_R l_2 \cdot [ P_2 + B_2 + Z_2 ] 
  \nonumber \\ 
 & \quad {}+
 \bar{l}' \gamma_\alpha P_L l \;\; \bar{l_1} \gamma^\alpha P_R l_2 \cdot [ P_3 + Z_3 + B_3 ]
  \nonumber \\ 
 & \quad {}+
 \bar{l}' \gamma_\alpha P_R l \;\; \bar{l_1} \gamma^\alpha P_L l_2 \cdot [ P_4 + Z_4 + B_4 ]
  \nonumber \\ 
 & \quad {}+
 \bar{l}' P_L l \;\; \bar{l_1} P_L l_2 \cdot B_5 + 
 \bar{l}' P_R l \;\; \bar{l_1} P_R l_2 \cdot B_6 
  \nonumber \\ 
 & \quad {}+
 \bar{l}' P_L l \;\; \bar{l_1} P_R l_2 \cdot B_7 + 
 \bar{l}' P_R l \;\; \bar{l_1} P_L l_2 \cdot B_8  
  \nonumber \\ 
 & \quad {}+
 \bar{l}' \sigma_{\alpha\beta} P_L  l \;\; \bar{l_1} \sigma^{\alpha\beta} P_L l_2 \cdot B_9 + 
 \bar{l}' \sigma_{\alpha\beta} P_R  l \;\; \bar{l_1} \sigma^{\alpha\beta} P_R l_2 \cdot B_{10}
  \nonumber \\ 
 & \quad {}+
 \bar{l}' i \sigma_{\alpha\beta} q^\beta P_L  l \;\; \bar{l_1} \gamma^\alpha P_L l_2
        \cdot \frac{P_{11}}{q^2} + 
 \bar{l}' i \sigma_{\alpha\beta} q^\beta P_R  l \;\; \bar{l_1} \gamma^\alpha P_R l_2
        \cdot \frac{P_{12}}{q^2}
  \nonumber \\ 
 & \quad {}+
 \bar{l}' i \sigma_{\alpha\beta} q^\beta P_L  l \;\; \bar{l_1} \gamma^\alpha P_R l_2
        \cdot \frac{P_{13}}{q^2} + 
 \bar{l}' i \sigma_{\alpha\beta} q^\beta P_R  l \;\; \bar{l_1} \gamma^\alpha P_L l_2
        \cdot \frac{P_{14}}{q^2} \; \Big\} \,.
\end{align}
This is the most general form factor CLFV $l\to l'l_1l_2^c$ 
structure valid in any model. The four-lepton operators with the 
Lorentz structure $P_L \times P_R$ and $P_R \times P_L$ are novelty, 
since they have not been considered in the previous publications
\cite{Hisano1996,Arganda2006}.

The total amplitude can be written down in the more compact form,
\begin{eqnarray} \label{TTOTcompact}
\mathcal{T}_{TOT}^{l \to l' l_1 l_2} & = & \frac{ \alpha_w^2 }{ M_W^2 } \cdot
	\sum_{i=1}^{14} \, \bar{l}' \, \Gamma_{1\,i} \, l \;\; \bar{l_1} \, 
  \Gamma_{2\,i} \, l_2 \cdot \mathcal{F}_i \,,
\end{eqnarray}
where $\Gamma_1$, $\Gamma_2$ and $\mathcal{F}_i$ are given by
\begin{align}
\Gamma_{1\,i} &=  \Big\{ \gamma_\alpha P_L , \gamma_\alpha P_R , \gamma_\alpha P_L , \gamma_\alpha P_R ,
	P_L , P_R , P_L , P_R , \sigma_{\alpha\beta} P_L , \sigma_{\alpha\beta} P_R , \\
 & \qquad i \sigma_{\alpha\beta} q^\beta P_L , i \sigma_{\alpha\beta} q^\beta P_R ,
	i \sigma_{\alpha\beta} q^\beta P_L , i \sigma_{\alpha\beta} q^\beta P_R \Big\} 
  \,,\\[1.5ex]
\Gamma_{2\,i} &= \Big\{ \gamma_\alpha P_L , \gamma_\alpha P_R , \gamma_\alpha P_R , \gamma_\alpha P_L ,
        P_L , P_R , P_R , P_L , \sigma_{\alpha\beta} P_L , \sigma_{\alpha\beta} P_R , \\
 & \qquad \gamma^\alpha P_L, \gamma_\alpha P_R , \gamma_\alpha P_R , \gamma_\alpha P_L \Big\} 
  \,, \\[1.5ex]
\mathcal{F}_i &= \Big\{ F_1 , F_2 , F_3 , F_4 , F_5 , F_6 , F_7 , F_8 , F_9 , F_{10} , 
 	\frac{F_{11}}{q^2} , \frac{F_{12}}{q^2} , \frac{F_{13}}{q^2} , \frac{F_{14}}{q^2} 
  \Big\} \,.
\end{align}

Evaluating the operator~\eqref{TTOTcompact} between initial and final 
lepton states  of the $l\to l'l_1l_2^c$ transition, one arrives at the 
amplitude written in the terms of spinors,
\begin{eqnarray} \label{TTOT}
\mathcal{T}_{TOT}^{e_1 \to e_2 e_3 e_4^c } & = & \frac{ \alpha_w^2 }{ M_W^2 } \cdot \Bigg\{
 \sum_{i=1}^{14} \, \bar{u}_2 \, \Gamma_{1\,i}^A \, u_1 \;\; \bar{u}_3 \, 
  \Gamma_{2\,i}^A \, v_4 \cdot \mathcal{F}_i^A \nonumber \\
 & & {}-\ \sum_{i=1}^{14} \, \bar{u}_3 \, \Gamma_{1\,i}^B \, u_1 \;\; \bar{u}_2 \, 
  \Gamma_{2\,i}^B \, v_4 \cdot \mathcal{F}_i^B 
 \Bigg\}\,.
\end{eqnarray}

The contents of the arrays denoted by $\mathcal{F}_i^A$ and $\mathcal{F}_i^B$ depends 
on the leptons in the final state. There are three possible cases:

{\bf 1)} ~{\boldmath $l'=l_1=l_2$}
  ~{\small 
  ($\tau^- \to \mu^- \mu^- \mu^+ \;; \tau^- \to e^- e^- \mu^+ \;; \mu^- \to e^- e^- e^+$)} :
\begin{align*}
\mathcal{F}_i^A &= \Big\{ F_1^A , F_2^A , F_3^A , F_4^A , F_5^A , F_6^A , F_7^A , F_8^A , 
  F_9^A , F_{10}^A , \frac{F_{11}^A}{s_{12}} , \frac{F_{12}^A}{s_{12}} , \frac{F_{13}^A}{s_{12}} , 
  \frac{F_{14}^A}{s_{12}} \Big\} \,, \\[1.5ex]
\mathcal{F}_i^B &= \Big\{ F_1^B , F_2^B , F_3^B , F_4^B , F_5^B , F_6^B , F_7^B , F_8^B , 
  F_9^B , F_{10}^B , \frac{F_{11}^B}{s_{13}} , \frac{F_{12}^B}{s_{13}} , 
  \frac{F_{13}^B}{s_{13}} , \frac{F_{14}^B}{s_{13}} \Big\} \,.
\end{align*}

{\bf 2)} ~{\boldmath $l'\neq l_2 \,, l_1=l_2$}
  ~{\small 
  ($\tau^- \to e^- \mu^- \mu^+ \;; \tau^- \to e^- \mu^- e^+$)} :
\begin{align*}
\mathcal{F}_i^A & =  \Big\{ F_1^A , F_2^A , F_3^A , F_4^A , F_5^A , F_6^A , F_7^A , F_8^A , 
  F_9^A , F_{10}^A , \frac{F_{11}^A}{s_{12}} , \frac{F_{12}^A}{s_{12}} , 
  \frac{F_{13}^A}{s_{12}} , \frac{F_{14}^A}{s_{12}} \Big\} \,, \\[1.5ex]
\mathcal{F}_i^B & =  \Big\{ F_1^B , F_2^B , F_3^B , F_4^B , F_5^B , F_6^B , F_7^B , F_8^B , 
  F_9^B , F_{10}^B ,  0, 0, 0, 0 \Big\} \,.
\end{align*}

{\bf 3)} ~{\boldmath $l'\neq l_2 \,, l_1 \neq l_2$}
  ~{\small 
  ($\tau^- \to \mu^- \mu^- e^+ \;; \tau^- \to e^- e^- \mu^+$)} :
\begin{align*}
\mathcal{F}_i^A & = \Big\{ F_1^A , F_2^A , F_3^A , F_4^A , F_5^A , F_6^A , F_7^A , F_8^A , 
  F_9^A , F_{10}^A ,  0, 0, 0, 0 \Big\} \,, \\[1.5ex]
\mathcal{F}_i^B & = \Big\{ F_1^B , F_2^B , F_3^B , F_4^B , F_5^B , F_6^B , F_7^B , F_8^B , 
  F_9^B , F_{10}^B , 0, 0, 0, 0 \Big\} \,.
\end{align*}

Here, $s_{12}$ and $s_{13}$ are Mandelstem variables defined in~\eqref{mandelstam}.

Absolute square of the amplitude~\eqref{TTOT} reads
{\small
\begin{align} \label{TOTtr}
\left| \mathcal{T}_{TOT}^{e_1 \to e_2 e_3 e_4^c } \right| ^2 & =  \frac{ \alpha_w^4 }{ M_W^4 } 
  \cdot \Bigg\{ \nonumber \\[1.3ex]
 & \hskip-4em \sum_{i,j}  \Tr \left[ (\Slash{p}_2 + m_2) \Gamma_{1\,i}^A (\Slash{p}_1 + m) 
  \bar{\Gamma}_{1\,j}^A \right]
	\cdot \Tr \left[ (\Slash{p}_3 + m_3) \Gamma_{2\,i}^A (\Slash{p}_4 - m_4) 
   \bar{\Gamma}_{2\,j}^A \right]
	\cdot \mathcal{F}_i^A \mathcal{F}_j^{A\,*}  \nonumber \\
 &  \hskip-5em + \sum_{i,j} \Tr \left[ (\Slash{p}_3 + m_3) \Gamma_{1\,i}^B 
  (\Slash{p} + m) \bar{\Gamma}_{1\,j}^B \right]
        \cdot \Tr \left[ (\Slash{p}_2 + m_2) \Gamma_{2\,i}^B 
  (\Slash{p}_4 - m_4) \, \bar{\Gamma}_{2\,j}^B \right]
        \cdot \mathcal{F}_i^B \mathcal{F}_j^{B\,*}  \nonumber \\
 &  \hskip-5em - \sum_{i,j} \Tr \left[ (\Slash{p}_2 + m_2) \, 
  \Gamma_{1\,i}^A \, (\Slash{p}_1 + m) \, \bar{\Gamma}_{1\,j}^B
        (\Slash{p}_3 + m_3) \, \Gamma_{2\,i}^A \, (\Slash{p}_4 - m_4) \, \bar{\Gamma}_{2\,j}^B \right]
        \cdot \mathcal{F}_i^A \mathcal{F}_j^{B\,*}  \nonumber \\
 &  \hskip-5em - \sum_{i,j} \Tr \left[ (\Slash{p}_3 + m_3) \, 
  \Gamma_{1\,i}^B \, (\Slash{p}_1 + m) \, \bar{\Gamma}_{1\,j}^A
        (\Slash{p}_2 + m_2) \, \Gamma_{2\,i}^B \, (\Slash{p}_4 - m_4) \, \bar{\Gamma}_{2\,j}^A \right]
        \cdot \mathcal{F}_i^{A\,*} \mathcal{F}_j^B \Bigg\} \,.
\end{align}}

After evaluating the traces in Eq~\eqref{TOTtr} , one imposes the kinematics 
of the three-body process:
\begin{eqnarray}
p_1 \cdot p_2 & = & \frac{1}{2} \left( m^2 + m_2^2 - s_{12} \right) \,, \nonumber \\
p_1 \cdot p_3 & = & \frac{1}{2} \left( m^2 + m_3^2 - s_{13} \right) \,, \nonumber \\
p_3 \cdot p_4 & = & \frac{1}{2} \left( s_{12} - m_3^2 - m_4^2 \right) \,, \nonumber \\
p_2 \cdot p_4 & = & \frac{1}{2} \left( s_{13} - m_2^2 - m_4^2 \right) \,, \nonumber \\
p_1 \cdot p_4 & = & \frac{1}{2} \left( m^2 + m_4^2 - s_{14} \right) = 
	\frac{1}{2} \left( s_{12} + s_{13} - m_2^2 - m_3^2 \right) \,, \nonumber \\
p_2 \cdot p_3 & = & \frac{1}{2} \left( s_{14} - m_2^2 - m_3^2 \right) =
        \frac{1}{2} \left( m^2 + m_4^2 - s_{12} - s_{13}  \right) \,,
\end{eqnarray}
where $s_{12}$, $s_{13}$ and $s_{14}$ are well-known Mandelstam variables,
\begin{eqnarray} \label{mandelstam}
s_{12} & \equiv & (p_1-p_2)^2 = s_{34} \,, \nonumber \\
s_{13} & \equiv & (p_1-p_3)^2 = s_{24} \,, \nonumber \\
s_{14} & \equiv & (p_1-p_4)^2 = s_{23} \,,
\end{eqnarray}
satisfying $s_{12} + s_{13} + s_{14} = m^2 + m_2^2 + m_3^2 + m_4^2$.

The last step in the evaluation of the branching ratios is an evaluation 
of the three-body phase-space integral \cite{Beringer2012}
\begin{equation} \label{PhSpaceInt}
\Gamma ( e_1 \to e_2 e_3 e_4^c ) \;=\; 
	\frac{1}{(2\pi)^3} \frac{1}{32 m^3}  
	\int_{4 \varepsilon^2}^{m^2} d s_{12} \;
	\int_{s_{13}^-}^{s_{13}^+} d s_{13} \; 
	\left| \mathcal{T}_{TOT}^{e_1 \to e_2 e_3 e_4^c } \right| ^2 \,,
\end{equation}
where
\begin{eqnarray}
s_{13}^{\pm} & = & \varepsilon^2 + \frac{m^2-s_{12}}{2} \left[ 1 \pm
 \left( 1 - \frac{4\varepsilon^2}{s_{12}} \right) ^{1/2} \right] \,,
\end{eqnarray}
and $\varepsilon$ is equal to the external masses ($\varepsilon=m_2=m_3=m_4$ for $l'=l_1=l_2$
and $\varepsilon=m_3=m_4$ for $l'\neq l_1 \,, l_1=l_2$). Since the decaying particle is
much more massive than the resulting particles, we take $\varepsilon \to 0$ whenever
possible.

In Eq~\eqref{PhSpaceInt} there are seven types of integrals which are divergent in the 
$\varepsilon \to 0$ limit. These integrals were evaluated partly by hand and partly
using \emph{Mathematica}, keeping the leading terms in $\varepsilon$ expansion,
\begin{equation*}
 \begin{array}{rclcl}
\mathcal{I}_1 & = & \displaystyle \int_{4 \varepsilon^2}^{m^2} \frac{d s_{12}}{s_{12}}
	\int_{s_{13}^-}^{s_{13}^+} d s_{13} 
	& \simeq & \displaystyle m^2 \left( \ln\frac{m^2}{\varepsilon^2} - 3 \right)\,, \\[2.5ex]
\mathcal{I}_2 & = & \displaystyle \int_{4 \varepsilon^2}^{m^2} \frac{d s_{12}}{s_{12}}
        \int_{s_{13}^-}^{s_{13}^+} s_{13} \, d s_{13} 
	& \simeq & \displaystyle m^4 \left( \frac{1}{2} \ln\frac{m^2}{\varepsilon^2} 
  - \frac{7}{4} \right) \,, \\[2.5ex]
\mathcal{I}_3 & = & \displaystyle \int_{4 \varepsilon^2}^{m^2} \frac{d s_{12}}{s_{12}}
        \int_{s_{13}^-}^{s_{13}^+} s_{13}^2 \, d s_{13}
        & \simeq & \displaystyle m^6 \left( \frac{1}{3} \ln\frac{m^2}{\varepsilon^2} - 
  \frac{4}{3} \right) \,,\\[2.5ex]
\mathcal{I}_4 & = & \displaystyle \int_{4 \varepsilon^2}^{m^2} \frac{d s_{12}}{s_{12}}
        \int_{s_{13}^-}^{s_{13}^+} \frac{d s_{13}}{s_{13}}
	& \simeq & \displaystyle - \frac{\pi^2}{6} + \frac{1}{2} 
  \ln^2\frac{m^2}{\varepsilon^2} \,, \\[2.5ex]
\mathcal{I}_5 & = & \displaystyle \int_{4 \varepsilon^2}^{m^2} \frac{d s_{12}}{s_{12}^2}
        \int_{s_{13}^-}^{s_{13}^+} d s_{13}
        & \simeq & \displaystyle \frac{m^6}{6\varepsilon^2} 
  - \ln\frac{m^2}{\varepsilon^2} + 1 \,, \\[2.5ex]
\mathcal{I}_6 & = & \displaystyle \int_{4 \varepsilon^2}^{m^2} \frac{d s_{12}}{s_{12}^2}
        \int_{s_{13}^-}^{s_{13}^+} s_{13}^2 \, d s_{13}
        & \simeq & \displaystyle m^4 \left( \frac{m^2}{20\varepsilon^2} + \frac{17}{6} - 
  \ln\frac{m^2}{\varepsilon^2} \right) \,, \\[2.5ex] 
\mathcal{I}_7 & = & \displaystyle \int_{4 \varepsilon^2}^{m^2} \frac{d s_{12}}{s_{12}^2}
        \int_{s_{13}^-}^{s_{13}^+} s_{13} \, d s_{13}
        & \simeq & \displaystyle m^2 \left( \frac{m^2}{12\varepsilon^2} + \frac{13}{3} - 
  \ln\frac{m^2}{\varepsilon^2} \right)\,.
\end{array}
\end{equation*}
The leading order divergences in the integrals $\mathcal{I}_4$, $\mathcal{I}_5$, 
$\mathcal{I}_6$ and $\mathcal{I}_7$ cancel out in the final expression, 
leaving only $\varepsilon$-divergent therms which comprise 
$\ln\frac{m^2}{\varepsilon^2}$. 

The final result again depends on the lepton content of the final state:

{\bf 1)} ~{\boldmath $l'=l_1=l_2$}
\begin{align} \label{totres1}
\left| \mathcal{T}_{TOT}^{e_1 \to e_2 e_2 e_2^c } \right| ^2 & = 
 \frac{m_l^6}{12} \cdot \Big\{ \nonumber \\
  & \hskip-3em (|B_5| + |B_6|^2)\, m_l^2 
  + 2 \,(|B_7|^2 + |B_8|^2) \, m_l^2 
  + 144 \,(|B_9|^2 +|B_{10}|^2) \, m_l^2 \nonumber\\
  & \hskip-3em - 4 \,[\, (B_3 + P_3 + Z_3)^* B_7 + (B_4 + P_4 + Z_4)^* B_8 + 
    \textrm{c.c} \,]\, m_l^2 \nonumber \\ 
  & \hskip-3em + 8 \,(\, |B_3 + P_3 + Z_3| + |B_4 + P_4 + Z_4| \,)\, m_l^2 \nonumber\\
  & \hskip-3em - 12 \,( B_9^* B_5 + B_{10}^* B_6 + \textrm{c.c} \,)\, m_l^2 \nonumber\\
  & \hskip-3em + 16 \,(\, |B_1 + P_1 + Z_1| + |B_2 + P_2 + Z_2| \,)\, m_l^2 \nonumber \\
  & \hskip-3em + 8 \,(\, P_{12}^* B_7 + P_{11}^* B_8 +\textrm{c.c.} \,)\, m_l \nonumber \\
  & \hskip-3em - 16 \,[\, (B_4 + P_4 + Z_4)^* P_{11} + (B_3 + P_3 + Z_3)^* P_{12}
    + \textrm{c.c} \,]\, m_l \nonumber \\
  & \hskip-3em - 32 \,[\, (B_1 + P_1 + Z_1)^* P_{12} + (B_2 + P_2 + Z_2)^* P_{11}
    + \textrm{c.c} \,]\, m_l \nonumber \\
  & \hskip-3em + 64 \,( |P_{11}|^2 + |P_{12}|^2 )\, 
  \left( \ln\frac{m^2}{\varepsilon^2} - \frac{11}{4} \right) \Big\} \,.
\end{align}

{\bf 2)} ~{\boldmath $l'\neq l_2 \,, l_1=l_2$}
\begin{align} \label{totres2}
\left| \mathcal{T}_{TOT}^{e_1 \to e_2 e_3 e_3^c } \right| ^2 & =
 \frac{m_l^6}{12} \cdot \Big\{ \nonumber \\
  & \hskip-3em (|B_5| + |B_6|^2)\, m_l^2 
  + 2 \,(|B_7|^2 + |B_8|^2) \, m_l^2 
  + 4 \,(|B_3|^2 + |B_4|^2) \, m_l^2 \nonumber\\
  & \hskip-3em + 8 \,(|B_1|^2 + |B_2|^2) \, m_l^2
  + 144 \,(|B_9|^2 +|B_{10}|^2) \, m_l^2 \nonumber\\
  & \hskip-3em + 8 \, ( B_1 + P_1 + Z_1)^* B_1 m_l^2 
  + 8 \,( B_2 + P_2 + Z_2)^* B_2 m_l^2 
  - 4 B_7^* B_3 m_l^2 \nonumber\\
  & \hskip-3em + 4 \, (B_3 + P_3 + Z_3)^* B_3 m_l^2 
  - 4 B_8^* B_4 m_l^2 
  + 4 \, ( B_4 + P_4 + Z_4)^* B_4 m_l^2 \nonumber\\
  & \hskip-3em - 12 B_9^* B_5 m_l^2
  - 12 B_{10}^* B_5 m_l^2 
  - 2 B_3^* B_7 m_l^2 \nonumber\\
  & \hskip-3em - 2 \, (B_3 + P_3 + Z_3)^* B_7 m_l^2 
  - 2 B_4^* B_8 m_l^2 
  + 2 (B_4 + P_4 + Z_4)^* B_8 m_l^2 \nonumber\\
  & \hskip-3em - 12 B_5^* B_9 m_l^2 
  - 12 B_6^* B_{10} m_l^2 
  + 4 B_1^* P_1 m_l^2 \nonumber\\
  & \hskip-3em + 4 \, ( B_1 + P_1 + Z_1 )^* P_1 m_l^2 
  + 4 B_2^* P_2 m_l^2 
  + 4 \, (B_2 + P_2 + Z_2)^* P_2 m_l^2 \nonumber\\
  & \hskip-3em - 2 B_7^* P_3 m_l^2 
  + 4 (B_3 + P_3 + Z_3)^* P_3 m_l^2 
  - 2 B_8^* P_4 m_l^2 \nonumber\\
  & \hskip-3em + 4 \,( B_4 + P_4 + Z_4)^* P_4 m_l^2 
  + 4 B_1^* Z_1 m_l^2 
  + 4 \, (B_1 + P_1 + Z_1)^* Z_1 m_l^2 \nonumber\\
  & \hskip-3em + 4 B_2^* Z_2 m_l^2 
  + 4 \, (B_2 + P_2 + Z_2)^* Z_2 m_l^2
  - 2 B_7^* Z_3 m_l^2 \nonumber\\
  & \hskip-3em + 4 \, (B_3 + P_3 + Z_3)^* Z_3 m_l^2
  - 2 B_8^* Z_4 m_l^2
  + 4\, (B_4 + P_4 + Z_4)^* Z_4 m_l^2 \nonumber \\
  & \hskip-3em -16 P_{12}^* B_1 m_l 
  - 16 P_{11}^* B_2 m_l 
  - 8 P_{12}^* B_3 m_l \nonumber\\
  & \hskip-3em - 8 P_{11}^* B_4 m_l 
  + 4 P_{12}^* B_7 m_l
  + 4 P_{11}^* B_8 m_l \nonumber\\
  & \hskip-3em - 8 P_{12}^* P_1 m_l
  - 8 P_{11}^* P_2 m_l
  - 8 P_{12}^* P_3 m_l \nonumber \\
  & \hskip-3em - 8 P_{11}^* P_4 m_l
  - 8 B_2^* P_{11} m_l
  + 4 B_8^* P_{11} m_l \nonumber\\
  & \hskip-3em - 8 \, ( B_2 + P_2 + Z_2 )^* P_{11} m_l
  - 8 \, (B_4 + P_4 + Z_4)^* P_{11} m_l 
  - 8 B_1^* P_{12} m_l \nonumber \\
  & \hskip-3em + 4 B_7^* P_{12} m_l 
  - 8\, (B_1 + P_1 + Z_1)^* P_{12} m_l
  - 8\, (B_3 + P_3 + Z_3)^* P_{12} m_l \nonumber \\
  & \hskip-3em - 8 P_{12}^* Z_1 m_l
  - 8 P_{11}^* Z_2 m_l
  - 8 P_{12}^* Z_3 m_l 
  -8 P_{11}^* Z_4 m_l\nonumber \\
  & \hskip-3em + 32 \,(\, |P_{11}|^2  |P_{12}|^2 \,)\, 
  \left( \ln\frac{m^2}{\varepsilon^2} - 3 \right) \Big\} \,.
\end{align}

{\bf 3)} ~{\boldmath $l'\neq l_2 \,, l_1 \neq l_2$}
\begin{align} \label{totres3}
\left| \mathcal{T}_{TOT}^{e_1 \to e_2 e_3 e_3^c } \right| ^2 & =
 \frac{m_l^8}{12} \cdot \Big\{ \nonumber \\
  & \hskip-3em 16 \,( |B_1|^2 + |B_2|^2 )
  + 8 \,( |B_3|^2 + |B_4|^2 ) 
  (|B_5|^2 + |B_6|^2) \nonumber \\
  & \hskip-3em + 2 \, ( |B_7|^2 + |B_8|^2 ) 
  + 144\, (|B_9|^2 + |B_{10}|^2) \nonumber \\
  & \hskip-3em + 4 \,[\, B_7^* B_3 + B_8^* B_4 + 3 B_9^* B_5 + 3 B_{10}^* B_6 
    + \textrm{c.c.} \,] \; \Big \} \,.
\end{align}

The results \eqref{totres1}, \eqref{totres2} and \eqref{totres3} were tested in 
several different manners. One of the main tests was to reproduce the result from 
Ref~\cite{Ilakovac1995}, which was performed with success.

%% file: sazetak.tex
\titleformat{\section}[hang]{\Large\bfseries\filcenter}{\S\,\oldstylenums{\thesection}}{10pt}{}[]
\fancyhead[RO]{{\small \nouppercase{\rightmark}}}

\chapter*{Pro\s ireni sa\z etak\markboth{Pro\s ireni sa\z etak}{}}
\addcontentsline{toc}{chapter}{Pro\s ireni sa\z etak}
\allowdisplaybreaks

\renewcommand{\theequation}{\arabic{equation}}
\setcounter{equation}{0}

\allowdisplaybreaks
\begin{sloppypar}

Ova disertacija izla\z e minimalni supersimetri\cc ni standardni model
sa modelom njihalice na niskoj skali. U okviru tog modela napravljena
je detaljna studija naru\s enja leptonskog okusa u nabijenom leptonskom
sektoru. Izveden je cjelovit skup kiralnih amplituda i pridru\z enih im form-faktora
povezanih sa tro\cc esti\cc nim CLFV raspadima miona i tau-leptona bez neutrina,
kao \s to su $\mu\to eee$, $\tau\to \mu\mu\mu$, $\tau\to \mu ee$ i 
$\tau \to ee\mu$ te $\mu\to e$ prijelazi na atomskim jezgrama. Dobiveni
analiti\cc ki rezultati su op\'{c}eniti i mogu se primijeniti na ve\'{c}inu
modela nove fizike koji uklju\cc uju naru\s enje nabijenog leptonskog broja.

Osim toga, u istom su modelu sustavno izu\cc eni doprinosi na razini jedne petlje
anomalnom magnetskom dipolnom momentu miona $a_\mu$ i elektri\cc nom
dipolnom momentu elektrona $d_e$.

\newpage

\section*{Pregled teku\'{c}ih i budu\'{c}ih eksperimenata
  \markright{Pregled teku\'{c}ih i budu\'{c}ih eksperimenata}}
\addcontentsline{toc}{section}{Pregled teku\'{c}ih i budu\'{c}ih eksperimenata}

Kada neutrino nastane u nekom slabo-interakcijskom procesu i propagira se
putem neke kona\cc ne udaljenosti, postoji kona\cc na vjerojatnost da \'{c}e
promijeniti okus. Ova opa\z ena i dobro utvr\dj ena \cc injenica poznata je
pod nazivom \emph{neutrinske oscilacije} 
\cite{Pontecorvo1957,Pontecorvo1958,Maki1962}, poradi oscilatorne ovisnosti
vjerojatnosti promjene okusa u odnosu na energiju neutrina i udaljenost 
propagacije.

Nebrojeni eksperimenti s neutrinima izvje\s tavaju o naru\s enju leptonskog
okusa u neutrinskom sektoru, bilo da je rije\cc\ o nestanku ili nastanku
pojedinog okusa \cite{Cleveland1998,Fukuda1996,Abdurashitov2009,
Anselmann1992,Hampel1999,Altmann2005,Fukuda2002,Ahmad2001,Ahmad2002,
Fukuda1998,Ashie2004,Eguchi2003,Araki2005,Abe2012,Michael2006,Adamson2008,
Ahn2006,An2012,An2013,Ahn2012,Abe2011,Adamson2011}.

Ti su eksperimenti pru\z ili nedvojbeni dokaz postojanja neutrinskih oscilacija
uzrokovane kona\cc nim neutrinskim masama i, posljedi\cc no, parametrima
mije\s anja neutrina. Budu\'{c}i su neutrini masivni, prijelaz izme\dj u 
neutrinskih polja napisanih u bazi okusa ($\nu_e$, $\nu_\mu$, $\nu_\tau$)
u neutrinska polja napisanih u masenoj bazi ($\nu_1$, $\nu_2$, $\nu_3$)
postaje netrivijalan,
\begin{equation}
\nu_l (x) = \sum_{i=1}^3 U_{l\,i} \nu_i (x) \;, \quad l=e,\mu,\tau \,.
\end{equation}
Unitarna matrica $U$ poznata je kao Pontecorvo-Maki-Nakagawa-Sakata matrica
\cite{Pontecorvo1957,Pontecorvo1958,Maki1962} i obi\cc no se parametrizira
na sljede\'{c}i na\cc in:
\begin{equation}
U_{PMNS} = \begin{pmatrix}
  c_{12} c_{13} & s_{12} c_{13} & s_{13} e^{-i \delta} \\
  -s_{12} c_{23}-c_{12} s_{23} s_{13}e^{i \delta}
    & c_{12} c_{23}-s_{12} s_{23} s_{13} e^{i \delta} & s_{23} c_{13} \\
  s_{12} s_{23}-c_{12} c_{23} s_{13}e^{i \delta}
    & -c_{12} s_{23}-s_{12} c_{23} s_{13} e^{i \delta} & c_{23} c_{13} \\
\end{pmatrix} \cdot P \,,
\end{equation}
gdje su $P=\textrm{diag}(1,e^{i\alpha},e^{i\beta})$, $c_{ij}\equiv\cos\theta_{ij}$ i
$s_{ij}\equiv\sin\theta_{ij}$. $\theta_{12}$ ozna\cc ava solarni kut mije\s anja,
$\theta_{23}$ atmosferski kut mije\s anja, a $\theta_{13}$ reaktorski kut mije\s anja.
Faze $\delta$, $\alpha$ i $\beta$ ozna\cc avaju Diracovu, odnosno dvije Majoranine 
faze koje naru\s avaju CP simetriju.

Nedavno izvije\s \'{c}e reaktorskih neutrinskih eksperimenata 
\cite{Abe2012,An2012,Ahn2012} o kona\cc noj vrijednosti parametra $\theta_{13}$,
sna\z no upu\'{c}uje na netrivijalnu okusnu strukturu neutrinskog sektora,
kao i mogu\'{c}nost postojanja CP naru\s enja.

Naru\s enje leptonskog okusa (LFV) u neutrinskom sektoru mo\z e implicirati 
mogu\'{c}nost postojanja LFV-a i nabijenom sektoru. Me\dj utim, unato\cc\
intenzivnoj eksperimentalnoj potrazi
\cite{Adam2011,Bellgardt1988,Dohmen1993,Bertl2006, Miyazaki2011,Miyazaki2013,Aubert2009,
Hayasaka2010,Lees2010,Miyazaki2010,Aubert2010,Beringer2012}
jo\s\ uvijek nije prona\dj en dokaz naru\s enja leptonskog okusa u nabijenom
leptonskom sektoru standardnog modela (SM). Teku\'{c}i i budu\'{c}i eksperimenti
usmjereni na detekciju naru\s enja leptonskog okusa u nabijenom sektoru (CLFV)
uspjeli su odrediti neke gornje granice na pripadaju\'{c}e opservable. U donjoj
tablici navedene su neke trenutne gornje granice, kao i one koje se o\cc ekuju
u budu\'{c}nosti, tokom idu\'{c}e dvije dekade.

\begin{center}
\footnotesize{
\begin{tabular}{rlll}
\hline\hline
 Br.~ & Opservabla & Gornja granica & O\cc ekivana budu\'{c}a osjetljivost \\
\hline
 1.~  & $B (\mu \to e \gamma)$ & $2.4 \times 10^{-12}$ \cite{Adam2011}
      &$1$--$2\times 10^{-13}$ \cite{Golden2012,Adam2012}, $10^{-14}$ \cite{Hewett2012}\\
 2.~  & $B (\mu \to eee)$ & $10^{-12}$ \cite{Bellgardt1988}
      &$10^{-16}$ \cite{Berger2013}, $10^{-17}$  \cite{Hewett2012} \\
 3.~  & $R_{\mu e}^{\rm Ti}$ & $4.3\times 10^{-12}$ \cite{Dohmen1993}
      &$3$--$7\times 10^{-17}$ \cite{Kurup2011,Abrams2012,Kutschke2011,Dukes2011},
         $10^{-18}$ \cite{Hewett2012,Kuno2005,Barlow2011}\\
 4.~  & $R_{\mu e}^{\rm Au}$ & $7 \times 10^{-13}$ \cite{Bertl2006}
      &$3$--$7\times 10^{-17}$ \cite{Kurup2011,Abrams2012,Kutschke2011,Dukes2011},
         $10^{-18}$ \cite{Hewett2012,Kuno2005,Barlow2011}\\
 5.~  & $B(\tau \to e \gamma)$ & $3.3 \times 10^{-8}$
    \cite{Miyazaki2011,Miyazaki2013,Aubert2009,
    Hayasaka2010,Lees2010,Miyazaki2010,Aubert2010,Beringer2012}
      & $1$--$2\times 10^{-9}$ \cite{Bona2007,Hayasaka2009}\\
 6.~  & $B(\tau \to \mu \gamma)$ & $4.4 \times 10^{-8}$
    \cite{Miyazaki2011,Miyazaki2013,Aubert2009,
    Hayasaka2010,Lees2010,Miyazaki2010,Aubert2010,Beringer2012}
      & $2\times 10^{-9}$ \cite{Bona2007,Hayasaka2009} \\
 7.~  & $B(\tau \to eee)$ & $2.7 \times 10^{-8}$
    \cite{Miyazaki2011,Miyazaki2013,Aubert2009,
    Hayasaka2010,Lees2010,Miyazaki2010,Aubert2010,Beringer2012}
      & $2 \times 10^{-10}$ \cite{Bona2007,Hayasaka2009} \\
 8.~  & $B(\tau \to e\mu\mu)$ & $2.7 \times 10^{-8}$
    \cite{Miyazaki2011,Miyazaki2013,Aubert2009,
    Hayasaka2010,Lees2010,Miyazaki2010,Aubert2010,Beringer2012}
      & $10^{-10}$ \cite{Hayasaka2009} \\
 9.~  & $B(\tau \to \mu\mu\mu)$ & $2.1 \times 10^{-8}$
    \cite{Miyazaki2011,Miyazaki2013,Aubert2009,
    Hayasaka2010,Lees2010,Miyazaki2010,Aubert2010,Beringer2012}
      & $2 \times 10^{-10}$ \cite{Bona2007,Hayasaka2009} \\
10.~  & $B(\tau \to \mu ee)$ & $1.8 \times 10^{-8}$
    \cite{Miyazaki2011,Miyazaki2013,Aubert2009,
    Hayasaka2010,Lees2010,Miyazaki2010,Aubert2010,Beringer2012}
      & $10^{-10}$ \cite{Hayasaka2009}\\
\hline\hline
\end{tabular}}
\end{center}

Budu\'{c}i da je CLFV zabranjen u okviru standardnog modela, opa\z anje
takve pojave bio bi jasan signal postojanja nove fizike, \s to ovo
podru\cc je istra\z ivanja \cc ini posebno zanimljivim.

Uz opservable koje uklju\cc uju CLFV, korisno je i osvrnuti se na
leptonske dipolne momente, posebno anomalni magnetski dipolni moment
(MDM) miona i elektri\cc ni dipolni moment (EDM) elektrona.

Trenutna eksperimentalna vrijednost MDM-a miona $a_\mu$, prema podacima 
navedenim u \emph{Particle Data Group} \cite{Beringer2012} iznosi
\begin{equation}
a_\mu^\textrm{exp} = (116592089 \pm 63) \times 10^{-11} \,.
\end{equation}

S druge strane, teorijska vrijednost koja proizlazi iz standardnog 
modela glasi
\begin{equation}
a_\mu^\textrm{SM} = (116591820 \pm 49) \times 10^{-11} \,.
\end{equation}

Razlike izme\dj u izmjerene i predvi\dj ene vrijednosti,
\begin{equation}
\Delta a_\mu \equiv a_\mu^\textrm{exp} - a_\mu^\textrm{SM}
  = (287 \pm 80) \times 10^{-11}
\end{equation}
nalazi se na nivou pouzdanosti od $3.6\sigma$ pa se stoga naziva
\emph{mionska anomalija}. Ova vrijednost ograni\cc ava dozvoljene
doprinose nove fizike, pa se kao takva \cc esto upotrebljava kao
ograni\cc avaju\'{c}i faktor u gra\dj enju novih modela, a ponekad
\cc ak i kao argument za eliminaciju nekih od predlo\z enih modela 
nove fizike.

U skoroj budu\'{c}nosti o\cc ekujemo znatno preciznija mjerenja
ove opservable. Tako Fermilab eksperiment E989 najavljuje pove\'{c}anje
preciznosti mjerenja za faktor 4 
\cite{Roberts2011,Roberts2006,Venanzoni2012,Venanzoni2012a,E989}.

Sli\cc no tomu, EDM elektrona $d_e$ mo\z e slu\z iti kao iznimno 
precizan test postojanja naru\s enja CP simetrije induciranog
novim CP fazama koje mogu biti prisutne u fizici izvan standardnog
modela. Trenutna gornja granica na $d_e$ \cite{Beringer2012,Hudson2011,Jung2013} 
iznosi
\begin{equation}
d_e < 10.5 \times 10^{-28} \;e\,\textrm{cm} \,.
\end{equation}

Neki budu\'{c}i ekperimenti mogli bi znatno pove\'{c}ati ovu osjetljivost,
\cc ak do reda veli\cc ine $10^{-29} - 10^{-31}\; e\,\textrm{cm}$
\cite{Jung2013,Amini2008,Kittle2004,Weiss2003,Sakemi2011,Wundt2012,
Kara2012,Raidal2008}. S druge strane, predvi\dj anja standardnog
modela za $d_e$ kre\'{c}u se izme\dj u  $10^{-38}\; e\,\textrm{cm}$ i
$10^{-33}\; e\,\textrm{cm}$, ovisno o tome jesu li Diracove CP faze u
matricama koje opisuju mije\s anje lakih neutrina razli\cc ite od nule
ili ne (za detalje vidi ref.~\cite{Pospelov2005}). Prema tome, svako
opa\z anje EDM-a razli\cc itog od nule, tj.\ opa\z anje vrijednosti 
ve\'{c}e od $10^{-33}\; e\,\textrm{cm}$, zna\cc ilo bi postojanje 
fizike izvan standardnog modela koja u sebi sadr\z i naru\s enje
CP simetrije.

Iz svega navedenog, vidimo da su ove opservable od velikog interesa
za istra\z ivanje mogu\'{c}ih scenarija u okvirima nove fizike. Za vi\s e
detalja, preporu\cc amo konzultirati neke od izvrsnih preglednih radova 
navedenih u referencama \cite{Fukuyama2012,Jung2013a,Raidal2008}.

\section*{Teorijski okvir
  \markright{Teorijski okvir}}
\addcontentsline{toc}{section}{Teorijski okvir}
 
Osnovna ideja iza svih supersimetri\cc nih modela jest postojanje
simetrije (prikladno nazvane \emph{supersimetrija}) koja transformira
fermion u bozon i obratno. \emph{Minimalni supersimetri\cc ni standardni
model} (MSSM) supersimetrizira standardni model (SM) uz minimalno
pro\s irenje standardnomodelskog \cc esti\cc nog spektra: svakoj \cc estici
iz SM-a pridru\z ena je jedna \emph{super\cc estica} ili \emph{superpartner}.
Superpartneri fermiona materije su \cc estice spina nula, nazvani
\emph{sfermioni}. Oni se dalje mogu klasificirati u skalarne leptone ili
\emph{sleptone} i skalarne kvarkove ili \emph{skvarkove}. Fermioni materije i 
njihovi superpartneri opisani su \emph{kiralnim superpoljima}. Superparneri
ba\z darnih bozona SM-a \cc estice su spina 1/2 i zovemo ih 
\emph{gejd\z ini} (engl.\ \emph{gauginos}). Oni se dalje mogu klasificirati
u jako interagiraju\'{c}i \emph{gluino} i elektroslabo interagiraju\'{c}i \emph{zino}
i \emph{vino} (engl.\ \emph{wino}) (superpartneri $Z$ odnosno $W$ bozona). Zajedno sa ba\z darnim
bozonima SM-a, oni su opisani \emph{vektorskim superpoljima}. Superpatneri
Higgsovih bozona \cc estice su spina 1/2 nazvani \emph{higgsini} i skupa s njima
opisani su kiralnim superpoljima. Lom elektroslabe simetrije mije\s a
elektroslabe gejd\z ine sa higgsinima, \s to rezultira fizikalnim \cc esticama
koje nazivamo \emph{\cc ard\z ini} (engl.\ \emph{charginos}) i \emph{neutralini}. 
Donja tablica prikazuje sadr\z aj \cc estica i polja MSSM-a, zajedno sa
pripadaju\'{c}im kvantnim brojevima.

\begin{center}
\tiny{
\begin{tabular}{cccccccc}
\hline
\multicolumn{8}{c}{\bf Sadr\z aj \cc estica MSSM-a} \\
\hline
Superpolje & \multicolumn{2}{c}{Bozoni} & \multicolumn{2}{c}{Fermioni} & 
	$SU_c(3)$ & $SU_L(2)$ & $U_Y(1)$ \\
\hline\hline
baľdarno \\
$\mathbf{G^a}$ & gluon & $g^a$ & gluino & $\tilde{g}^a$ & 8 & 0 & 0 \\
$\mathbf{V^k}$ & slabi & $W^k\,\,(W^\pm,Z)$ & vino, zino & $\tilde{w}^k\,\,
	(\tilde{w}^\pm,\tilde{z})$ & 1 & 3 & 0 \\
$\mathbf{V'}$ & hipernabojni & $B\,\,(\gamma)$ & bino & $\tilde{b}\,\,
	(\tilde{\gamma})$ & 1 & 1 & 0 \\
\hline
materije \\
$\mathbf{L_i}$ & {} & \multicolumn{1}{l}{
	$\tilde{L}_i = (\tilde{\nu},\tilde{e})_L$} & {} &
	\multicolumn{1}{l}{$L_i = (\nu,e)_L$} & 1 & 2 & -1 \\
$\mathbf{E_i}$ & \raisebox{1.5ex}[0pt]{sleptoni} & \multicolumn{1}{l}{
	$\tilde{E}_i = \tilde{e}_R$} & \raisebox{1.5ex}[0pt]{leptoni}
	& \multicolumn{1}{l}{$E_i = e_R$} & 1 & 1 & 2 \\
$\mathbf{Q_i}$ & {} & \multicolumn{1}{l}{
	$\tilde{Q}_i = (\tilde{u},\tilde{d})_L$} & {} &
	\multicolumn{1}{l}{$Q_i = (u,d)_L$}  & 3 & 2 & 1/3 \\
$\mathbf{U_i}$ & skvarkovi & \multicolumn{1}{l}{
	$\tilde{U}_i = \tilde{u}_R$}
	& kvarkovi & 
	\multicolumn{1}{l}{$U_i = u_R^c$} & $3^*$ & 1 & -4/3 \\
$\mathbf{D_i}$ & {} & \multicolumn{1}{l}{
	$\tilde{D}_i = \tilde{d}_R$} & {} & 
	\multicolumn{1}{l}{$D_i = d_R^c$} & $3^*$ & 1 & 2/3 \\
\hline 
Higgs \\
$\mathbf{H_1}$ & {} & \multicolumn{1}{l}{$H_1$} & {} & $\tilde{H}_1$ &
	1 & 2 & -1 \\
$\mathbf{H_2}$ & \raisebox{1.5ex}[0pt]{Higgsovi} 
	& \multicolumn{1}{l}{$H_2$} & \raisebox{1.5ex}[0pt]{higgsini} & 
	$\bar{H}_2$ & 1 & 2 & 1 \\
\hline
\end{tabular}}
\end{center}

Kao \s to se vidi iz tablice, u MSSM-u postoje dva Higgsova superpolja,
koja se mogu napisati kao
\begin{equation}
 H_1 = \begin{pmatrix} H_1^1 \\ H_1^2 \end{pmatrix} \;, \quad
 H_2 = \begin{pmatrix} H_2^1 \\ H_2^2 \end{pmatrix} \,.
\end{equation}
Polje $H_1$ ponekad se naziva donje Higgsovo superpolje ($Y=-1$),
a sastoji se od polja $h_1$ and $\tilde{h}_{1L}$. Polje $H_2$ tako\dj er
se naziva gornje Higgsovo superpolje, a sastoji se od polja $h_2$ and 
$\tilde{h}_{2L}$. Komponentna polja ozna\cc ena malim tiskanim slovima
mogu se dalje napisati kao
\begin{eqnarray}
h_1 \equiv \begin{pmatrix} h_1^1 \\ h_1^2 \end{pmatrix}
  = \begin{pmatrix} h_1^0 \\ h_1^- \end{pmatrix} & \; ; \quad &
h_2 \equiv \begin{pmatrix} h_2^1 \\ h_2^2 \end{pmatrix}
  = \begin{pmatrix} h_2^+ \\ h_2^0 \end{pmatrix} \,, \\[1.5ex]
\tilde{h}_{1L} \equiv \begin{pmatrix} \tilde{h}_1^1 \\ \tilde{h}_1^2 \end{pmatrix}
  = \begin{pmatrix} \tilde{h}_1^0 \\ \tilde{h}_1^- \end{pmatrix}_L & \; ; \quad &
\tilde{h}_{2L} \equiv \begin{pmatrix} \tilde{h}_2^1 \\ \tilde{h}_2^2 \end{pmatrix}
  = \begin{pmatrix} \tilde{h}_2^+ \\ \tilde{h}_2^0 \end{pmatrix}_L \,.
\end{eqnarray}

Nakon spontanog loma elektroslabe simetrije, vakuumske o\cc ekivane vrijednosti
dane su realnim i pozitivnim vrijednostima $v_1$ i $v_2$,
\begin{equation}
\left< h_1 \right> = \frac{1}{\sqrt{2}}
  \begin{pmatrix} v_1 \\ 0 \end{pmatrix} \,; \quad
\left< h_2 \right> = \frac{1}{\sqrt{2}}
  \begin{pmatrix} 0 \\ v_2 \end{pmatrix} \,,
\end{equation}
koji dolaze od minimalizacije Higgsovog potencijala. Omjer ovih vrijednosti,
\begin{equation} 
 \frac{v_2}{v_1} \equiv \tan\beta \,,
\end{equation}
smatra se slobodnim parametrom teorije, barem \s to se ti\cc e fermionskih masa.

Lagran\z ijan MSSM-a mo\z e se napisati kao zbroj dvaju dijelova: prvi koji
dolazi od egzaktne supersimetrizacije standardnog modela, i drugi koji
eksplicite lomi supersimetriju,
\begin{equation}
 \mathcal{L}_{\textrm{MSSM}} = \mathcal{L}_{\textrm{SUSY}}
 + \mathcal{L}_{\textrm{SSB}} \,.
\end{equation}

Prvi \cc lan mo\z emo dalje pisati po komponentama,
\begin{equation}
\mathcal{L}_{\textrm{SUSY}} = \mathcal{L}_{g} +
  \mathcal{L}_{M} + \mathcal{L}_{H} \,,
\end{equation}
gdje su $\mathcal{L}_{g}$, $\mathcal{L}_{M}$ i $\mathcal{L}_{H}$ 
lagran\z ijani koji sadr\z e ba\z darna polja, polja materije
te Higgsova polja. Detaljan prikaz ovih komponenti mo\z e se
na\'{c}i u literaturi \cite[str.~171-172]{Drees2004}. U kontekstu
ove disertacije najvi\s e \'{c}e nas zanimati tzv.~\emph{superpotencijal}, 
koji \cc ini va\z an dio lagran\z ijana $\mathcal{L}_{H}$ i glasi
\begin{equation}
\mathcal{W}_{\textrm{MSSM}} = \mu \, H_1 \cdot H_2 
  + \bar{E}_i \, {\mathbf h}_{ij}^e\, H_1 \cdot L_j 
  + \bar{D}_i \, {\mathbf h}_{ij}^d\, H_1 \cdot Q_j  
  + \bar{U}_i \, {\mathbf h}_{ij}^u\, H_2 \cdot Q_j \,.
\end{equation}
Matrice ${\mathbf h}$ dane su sa
\begin{eqnarray}
{\mathbf h}_{ij}^{e\,\dagger} & = & \frac{g_2}{\sqrt{2} M_W \cos\beta}
  \left( \mathbf{m}_e \right)_{ij} \,,\\
{\mathbf h}_{ij}^{d\,\dagger} & = & \frac{g_2}{\sqrt{2} M_W \cos\beta}
  \left( \mathbf{m}_d \right)_{ij} \,,\\
{\mathbf h}_{ij}^{u\,\dagger} & = & \frac{g_2}{\sqrt{2} M_W \cos\beta}
  \left( \mathbf{m}_u \right)_{ij} \,.
\end{eqnarray}
Matrice $\mathbf{m}_e$, $\mathbf{m}_d$ i $\mathbf{m}_u$ su dimenzije
$3 \times 3$, a predstavljaju leptonsku masenu matricu, te masene
matrice donjih odnosno gornjih kvarkova. Skalarni produkti definirani
su u dvokomponentnoj notaciji \cite{Derendinger,Bajc} kao
$A \cdot B \equiv \epsilon_{\alpha\beta} A^\alpha B^\beta$ 
($\epsilon_{12} \equiv +1$). Drugi, tre\'{c}i i \cc etvrti \cc lan
na desnoj strani izraza za superpotencijal samo su supersimetri\cc no
popo\'{c}enje Yukawinih vezanja u lagran\z ijanu standardnog modela
\cite{Herrero1998}. Prvi \cc lan me\dj utim predstavlja novost te
o njemu mo\z emo razmi\s ljati kao o supersimetri\cc nom poop\'{c}enju
masenih \cc lanova Higgsovog polja. Mo\z e se pokazati da konzistentno
provo\-\dj enje loma elektroslabe simetrije zahtjeva da parametar $\mu$
bude reda veli\cc ine mase $W$ bozona.

Ovdje je jo\s\ potrebno re\'{c}i da je u cijelom ovom razmatranju implicitno
pretpostavljeno sa\cc uvanje $R$-pariteta, definiranog kvantnim brojem
$R_p$,
\begin{equation}
R_p = (-1)^{3(B-L) + 2S} \,,
\end{equation}
pri \cc emu su $B$, $L$ i $S$ barionski i leptonski broj, odnosno spin
dane \cc estice. Sa\cc uvanje ovog kvantnog broja u MSSM-u mo\z e
se smatrati prirodnom pretpostavkom u minimalnom supersimetri\cc nom
pro\s irenju standardnog modela, s obzirom na \cc injenicu da su
barionski i leptonski broj sa\cc uvani u lagran\z ijanu SM-a.

Pogledajmo sada sadr\z aj drugog dijela lagran\z ijana MSSM-a, onog
koji eksplicitno lomi supersimetriju ($\mathcal{L}_{\textrm{SSB}}$).
Koriste\'{c}i Symanzikovo pravilo \cite[str.~107-8]{Coleman1988},
mo\z e se pokazati da, ukoliko \z elimo zadr\z ati po\z eljno 
konvergentno pona\s anje supersimetri\cc ne teorije, \cc lanovi
u $\mathcal{L}_{\textrm{SSB}}$ moraju biti \emph{mekani}
\cite{Witten1981,Dimopoulos1981,Sakai1981,Kaul1982}, \s to
zna\cc i da operatori polja u tom lagran\z ijanu moraju biti dimenzije
strogo manje od \cc etiri. Osim toga, o\cc ekujemo da ti \cc lanovi
budu mali u odnosu na \cc lanove u $\mathcal{L}_{\textrm{SUSY}}$.

Kada to uzmemo u obzir, mo\z emo pisati \cite[str.~185]{Drees2004}
\begin{eqnarray}
- {\cal L}_{SOFT} & = &
 \tilde{q}_{iL}^* ( \mathcal{M}_{\tilde{q}}^2 )_{ij} \tilde{q}_{jL}
    + \tilde{u}_{iR}^* ( \mathcal{M}_{\tilde{u}}^2 )_{ij} \tilde{u}_{jR}
    + \tilde{d}_{iR}^* ( \mathcal{M}_{\tilde{d}}^2 )_{ij} \tilde{d}_{jR} \nonumber \\
 && {} + \tilde{l}_{iL}^* ( \mathcal{M}_{\tilde{l}}^2 )_{ij} \tilde{l}_{jL}
    + \tilde{e}_{iR}^* ( \mathcal{M}_{\tilde{e}}^2 )_{ij} \tilde{e}_{jR} \nonumber \\
 && {} + \Big[ h_1 \cdot \tilde{l}_{iL} (A^e)^T_{ij} \tilde{e}_{jR}^* +
    h_1 \cdot \tilde{q}_{iL} (A^d)^T_{ij} \tilde{d}_{jR}^* \nonumber \\
 && {} + \tilde{q}_{iL} \cdot h_2 (A^u)^T_{ij} \tilde{u}_{jR}^* + \textrm{h.c.} \Big]
  \nonumber \\
 && {} + m_1^2 |h_1|^2 + m_2^2 |h_2|^2 + (B \mu h_1 \cdot h_2 + \textrm{h.c.}) \nonumber \\
 && {} + \frac{1}{2} ( M_1 \bar{\tilde{\lambda}}_0 P_L \tilde{\lambda}_0
    + M_1^* \bar{\tilde{\lambda}}_0 P_R \tilde{\lambda}_0 ) \nonumber \\
 && {} + \frac{1}{2} ( M_2 \bar{\vec{\tilde{\lambda}}} P_L \vec{\tilde{\lambda}}
    + M_2^* \bar{\vec{\tilde{\lambda}}} P_R \vec{\tilde{\lambda}} ) \nonumber \\
 && {} + \frac{1}{2} ( M_3 \bar{\tilde{g}}^a P_L \tilde{g}^a
    + M_3^* \bar{\tilde{g}}^a P_R \tilde{g}^a )
\end{eqnarray}

Prakti\cc ni izra\cc uni unutar MSSM-a obi\cc no uklju\cc uju nekoliko
pojednostavljuju\'{c}ih pretpostavki, kako bi se smanji veliki broj parametara
koje smo morali dodati u teoriju. Takve pretpostavke rezultiraju razli\cc itim
ina\cc icama ograni\cc enog minimalnog supersimetri\cc nog standardnog modela
ili CMSSM-a.

U ovoj disertaciji usvojit \'{c}emo jednu takovu ina\cc icu, tzv.\ minimalni
supergravitacijski model, skra\'{c}eno mSUGRA. Budu\'{c}i da polja MSSM-a ne
mogu sama spontano slomiti supersimetriju na skalama karakteriziranim masom
$W$ bozona, spontani lom supersimetrije mora se odviti u sektoru polja koji
su singleti u odnosu na ba\z darnu grupu standardnog modela. Jedan od 
najekonomi\cc nijih mehanizama ove vrste koristi gravitacijsku interakciju
koja se temelji na lokalnoj supersimetriji poznatoj kao \emph{supergravitacija}
\cite{Nilles1984,Arnowitt1984}.

Velika korist od ovog modela sastoji se u tome da dodatnih 105 parametara
uspjeva svesti na samo pet parametara,
\begin{equation} 
\{p\} = \{\textrm{sign}(\mu), m_0, M_{1/2}, A_0, \tan\beta \} \,.
\end{equation}
Ovdje $\textrm{sign}(\mu)$ ozna\cc ava predznak parametra $\mu$ koji se nalazi
u superpotencijalu, $m_0$ ozna\cc ava mase skalara ($m_{ij} = m_0 \delta_{ij}$),
$M_{1/2}$ zajedni\cc ku masu svih MSSM gejd\z ina, $A_0$ zajedni\cc ku konstantute
trilinearnog vezanja (higgs-sfermion-sfermion), a $\tan\beta$ omjer vakuumskih
o\cc ekivanih vrijednosti kojeg smo definirali ranije. Ove parametre nazivamo
\emph{parametrima loma supersimetrije}. Njihove vrijednosti obi\cc no se 
postavljaju na skali velikog ujedinjenja (GUT), te se putem renormalizacijskih
grupnih jednad\z bi \cite{Petcov2004} prenose do skale karakterizirane masom
$W$ bozona.

Recimo i to da postoji vi\s e teorijskih motivacija za rad unutar MSSM-a.
MSSM nudi stabilno kvantnomehani\cc ko rje\s enje problema hijerarhije
u ba\z darnom sektoru i daje prili\cc no preciznu predikciju ujedinjenja
ba\z darnih vezanja SM-a na skali bliskoj GUT skali. Najlak\s a supersimetri\cc na
\cc estica je stabilna i, ako bi bila neutralna poput neutralina, mo\z e 
predstavljati dobrog kandidata za konstituenta tamne tvari u svemiru. Osim toga,
MSSM tipi\cc no predvi\dj a da je masa standardnomodelskog Higgsa manja od
135~GeV, \s to je u skladu s nedavnim opa\z anjima od strane kolaboracija
ATLAS \cite{Aad2012} i CMS \cite{Chatrchyan2012,Chatrchyan2013}.

U minimalnom supersimetri\cc nom standardnom modelu sa sadr\z ajem polja
kakav je dan u gornjoj tablici ne dolazi do naru\s enja leptonskog okusa
u nabijenom leptonskom sektoru (CLFV). To je posljedica odsustva desnih neutrina,
\s to rezultira trivijalnom okusnom strukturom. Jedan od na\cc ina da unutar
MSSM-a omogu\'{c}imo procese koji uklju\cc uju CLFV jest da navedeni \cc esti\cc ni
spektar pro\s irimo s desnim neutrinima \cc ija je masa reda veli\cc ine 
1~TeV. Takav model ozna\cc it \'{c}emo sa $\nu_R$MSSM.

Pro\s irenje \cc esti\cc nog spektra vr\s i se putem mehanizma njihalice
(engl.\ \emph{seesaw mechanism}) na niskoj skali. Za razliku od uobi\cc ajenog
mehanizma njihalice u kojem te\s ki neutrinski singleti poprimaju mase
$m_N \sim 10^{12-14}$~GeV, mehanizam njihalice na niskoj skali omogu\'{c}ava
da desni neutrini imaju znatno ni\z e mase, ve\'{c}  od 100~GeV. 
Dok je u obi\cc ajenom mehanizmu
njihalice vezanje izme\dj u te\s kih i lakih neutrina reda veli\cc ine
$\xi_{\nu N} \sim \sqrt{m_\nu / m_N} \sim 10^{-12}$ (za $m_\nu \sim 10^{-1}$~GeV),
u mehanizmu njihalice na niskoj skali $\xi_{\nu N}$ postaje slobodan parametar. Sve
te pogodnosti omogu\'{c}avaju nam da masu desnih neutrina postavimo na skalu
eksperimentalne dohvatljivosti, kao i da pove\'{c}amo u\cc inak CLFV-a uslijed
nezanemarive jakosti vezanja lakih i te\s kih neutrina.

Realizacija mehanizma njihalice na niskoj skali posti\z e se uvo\dj enjem
dodatnih leptonskih simetrija 
\cite{Wyler1983,Mohapatra1986,Mohapatra1986a,Nandi1986,Branco1989,
Pilaftsis1992,Dev2012} u teoriju koje razultiraju time da laki neutrini
postaju bezmaseni, dok te\s ki neutrini mogu biti na skali $\sim 1$~TeV. Ukoliko
bismo \z eljeli reproducirati niskoenergetski maseni spektar lakih neutrina,
te simetrije se mogu blago naru\s iti pa u tom slu\cc aju govorimo o 
\emph{aproksimativnim leptonskim simetrijama}.

Ovakav model mo\z e biti zanimljiv iz vi\s e razloga. Novo uvedeni singletni
neutrini mogu biti kandidati za hladnu tamnu tvar 
\cite{Arina2008,Deppisch2008,Josse-Michaux2011,An2012a,Dumont2012}. Uz to,
mehanizam rezonantne leptogeneze na niskoj skali
\cite{Pilaftsis2005,Pilaftsis2005a,Deppisch2011,Pilaftsis1997,Pilaftsis2004}
mo\z e ponuditi obja\-\s njenje za opa\z enu barionsku asimetriju u svemiru,
\s to je posebno aktualno u svjetlu \cc injenice da se parametarski prostor
za elektroslabu bariogenezu svakoga dana sve vi\s e su\z ava novim
podacima koji pristi\z u sa LHC-a \cite{Cohen2012,Carena2013}.

Leptonski dio superpotencijala u $\nu_R$MSSM-u glasi
\begin{equation}
  W_{\rm lepton} =  \widehat{E}^C {\bf h}_e \widehat{H}_d
\widehat{L} + \widehat{N}^C {\bf h}_\nu \widehat{L} \widehat{H}_u
+ \frac{1}{2}\,\widehat{N}^C {\bf m}_M \widehat{N}^C \,,
\end{equation}
gdje $\widehat{H}_{u,d}$, $\widehat{L}$, $\widehat{E}$ i $\widehat{N}^C$
ozna\cc avaju dva superpolja Higgsovih dubleta, tri nabijena leptonska
superpolja lijeve i desne kiralnosti te tri neutrinska superpolja 
desne kiralnosti, redom. Yukawina vezanja ${\bf  h}_{e,\nu}$ i
Majoranini maseni parametri ${\bf m}_M$ \cc ine kompleksnu $3\times 3$
matricu. Za matricu ${\bf m}_M$ uzeli smo da na skali $m_N$ bude simetri\cc na 
s obzirom na okusnu grupu SO(3), tj.~$\mathbf{m}_M = m_N\, {\bf 1}_3$.

U ovoj disertaciji posebno \'{c}emo razmatrati dva scenarija neutrinskih
Yukawinih vezanja. Prvi realizira U(1) leptonsku simetriju 
\cite{Pilaftsis2005,Pilaftsis2005a,Deppisch2011} i dan je sa
\begin{eqnarray} \label{saz:YU1}
{\bf h}_\nu &=&
 \left(\begin{array}{lll}
 0 & 0 & 0 \\
 a\, e^{-\frac{i\pi}{4}} & b\, e^{-\frac{i\pi}{4}} & c\, e^{-\frac{i\pi}{4}} \\
 a\, e^{ \frac{i\pi}{4}} & b\, e^{ \frac{i\pi}{4}} & c\, e^{ \frac{i\pi}{4}}
 \end{array}\right)\; .
\end{eqnarray}

U drugom scenariju, struktura neutrinske Yukawine matrice ${\bf h}_\nu$
motivirana je diskretnom grupom simetrija $A_4$ i poprima sljede\'{c}i
oblik \cite{Kersten2007}:
\begin{eqnarray} \label{saz:YA4}
{\bf h}_\nu &=&
 \left(\begin{array}{lll}
 a & b & c \\
 a e^{-\frac{2\pi i}{3}} & b\, e^{-\frac{2\pi i}{3}} & c\, e^{-\frac{2\pi i}{3}} \\
 a e^{ \frac{2\pi i}{3}} & b\, e^{ \frac{2\pi i}{3}} & c\, e^{ \frac{2\pi i}{3}}
\end{array}\right)\;.
\end{eqnarray}

U ovim izrazima pretpostavljamo da su Yukawini parametri $a$, $b$ i $c$ realni.
Opservable koje uklju\cc uju CLFV ne ovise o niskoenergetskom masenom spektru
lakih neutrina, pa \'{c}emo iz prakti\cc nih razloga uzeti da su mase lakih neutrina
jednake nuli, tj.\ da su leptonske simetrije egzaktno realizirane.

Prema tome, u $\nu_R$MSSM-u postoje tri relevantna doprinosa naru\s enju
leptonskog okusa u nabijenom sektoru. Jedan dolazi od te\s kih neutrina ($N$),
drugi od sneutrina ($\widetilde{N}$), a tre\'{c}i od sektora koji mekano lomi
supersimetriju (SB). Svaki od tih doprinosa bit \'{c}e zasebno analiziran.

\section*{CLFV opservable
  \markright{CLFV opservable}}
\addcontentsline{toc}{section}{CLFV opservable}

Na razini jedne petlje, efektivna  $\gamma l'l$ i $Zl'l$ vezanja 
generirana su Feynmanovim dijagramima prikazanima na slici 
\ref{f1} (str.~\pageref{f1}). Op\'{c}i oblici amplituda prijelaza
povezan s ovim efektivnim vezanjima dani su sa
\begin{eqnarray}
{\cal T}^{\gamma l'l}_\mu
 \!&=&\!
 \frac{e\, \alpha_w}{8\pi M^2_W}\: \bar{l}'
 \Big[ (F^L_\gamma)_{l'l}\, (q^2\gamma_\mu-\Slash{q}q_\mu) P_L
     + (F^R_\gamma)_{l'l}\, (q^2\gamma_\mu-\Slash{q}q_\mu) P_R
\nonumber\\&&
     +  (G^L_\gamma)_{l'l}\, i \sigma_{\mu\nu}q^\nu P_L
     +  (G^R_\gamma)_{l'l}\, i \sigma_{\mu\nu}q^\nu P_R \Big]\: l,
\\
{\cal T}^{Z l'l}_\mu
 \!&=&\!
 \frac{g_w\, \alpha_w}{8\pi \cos\theta_w}\: \bar{l}'
 \Big[ (F_Z^L)_{l'l}\, \gamma_\mu P_L
     + (F_Z^R)_{l'l}\, \gamma_\mu P_R \Big]\: l,
\end{eqnarray}

gdje je $P_{L(R)}   =  \frac{1}{2}\,[1-\!(+)\,\gamma_5]$,
$\alpha_w  = g^2_w/(4\pi)$, $e$ je elektromagnetska konstanta vezanja,
$M_W = g_w \sqrt{v^2_u  +v^2_d}/2$ masa $W$ bozona, $\theta_w$
slabi kut vezanja, a $q =  p_{l'} - p_l$ impuls fotona. Form-faktori
$(F^L_\gamma)_{l'l}$,      $(F^R_\gamma)_{l'l}$
$(G^L_\gamma)_{l'l}$,    $(G^R_\gamma)_{l'l}$,   $(F_Z^L)_{l'l}$  i
$(F_Z^R)_{l'l}$ dobijaju doprinose od te\s kih neutrina $N_{1,2,3}$,
te\s kih sneutrina $\widetilde{N}_{1,2,3}$ i \cc lanova koji mekano
lome supersimetriju induciranih putem renormalizacijskih 
grupnih jednad\z bi (RGE). Analiti\cc ki izrazi za svaki od ovih
doprinosa mogu se na\'{c}i u Dodatku~\ref{sec:olff}.

Iz amplitude mo\z emo izra\cc unati i pripadaju\'{c}e omjere
grananja,
\begin{eqnarray}
B(l\to l'\gamma)
 \!&=&\!
 \frac{\alpha_w^3 s_w^2}{256\pi^2} \frac{m_l^3}{M_W^4\Gamma_l}
 \Big( |(G^L_\gamma)_{l'l}|^2 + |(G^R_\gamma)_{l'l}|^2 \Big) \,,\\
B(Z\to l'l^C+l'^Cl)
 \!&=&\!
 \frac{\alpha_w^3 M_W}{768 \pi^2 c_w^3\Gamma_Z} \Big(|(F_Z^L)_{l'l}|^2 +
 |(F_Z^R)_{l'l}|^2\Big) \,.
\end{eqnarray}

Svi ovi izrazi napisani su u aproksimaciji razvoja do vode\'{c}eg reda
u vanjskim masama i impulsima, kao i pretpostavci da je masa $Z$ bozona $M_Z$
znatno ispod supersimetri\cc ne skale $M_{SUSY}$ i mase te\s kog neutrina
$m_N$.

U idu\'{c}em koraku prelazimo na tro\cc esti\cc ne CLFV raspade
$l\to l' l_1 l_2^C$, pri \cc emu $l$ mo\z e biti mion ili tau-lepton,
a  $l',\, l_1,\, l_2$ ozna\cc avaju drugi nabijeni lepton u kojeg se
lepton $l$ mo\z e raspasti s obzirom na kinematiku.

Prijelazna amplituda za $ l\to  l' l_1 l_2^C$ sadr\z i doprinose od
fotonskih i $Z$-bozonskih dijagrama prikazanih na slici~\ref{f1}
(str.~\pageref{f1}), ali i od pravokutnih dijagrama prikazanih
na slici~\ref{f2} (str.~\pageref{f2}). Amplitude za ova tri doprinosa
glase:
\begin{eqnarray}
{\cal T}_{\gamma}^{ll'l_1l_2}
 \!&=&\!
 \frac{\alpha_w^2 s_w^2}{2 M_W^2}
 \Big\{ \delta_{l_1l_2}\, \bar{l}'\,
  \Big[ (F_\gamma^L)_{l'l}\, \gamma_\mu P_L + (F_\gamma^R)_{l'l}\,
    \gamma_\mu P_R  +
 \frac{(\Slash{p}-\Slash{p}')}{(p-p')^2}
  \nonumber\\
 & \cdot &
         \Big( (G_\gamma^L)_{l'l}\, \gamma_\mu P_L +
         (G_\gamma^R)_{l'l}\, \gamma_\mu P_R \Big) \Big]\, l
      \, \bar{l}_1\gamma^\mu l_2^C
  - [ l'\leftrightarrow l_1 ] \Big\} \,,\\
%%%
{\cal T}_Z^{ll'l_1l_2}
 \!&=&\!
 \frac{\alpha_w^2}{2 M_W^2}
 \Big[ \delta_{l_1l_2}\, \bar{l}'\Big( (F_Z^L)_{l'l}\,\gamma_\mu P_L
+ (F_Z^R)_{l'l}\, \gamma_\mu P_R\Big) l \nonumber\\
 & \cdot & \bar{l}_1 \Big(g_L^l\, \gamma^\mu P_L + g_R^l\, \gamma^\mu P_R\Big) l_2^C
  - ( l'\leftrightarrow l_1 ) \Big] \,,\\
{\cal T}_{\rm box}^{ll'l_1l_2}
 \!&=&\!
 - \frac{\alpha_w^2}{4 M_W^2}
 \Big(
    B_{\ell V}^{LL}\, \bar{l}'\gamma_\mu P_L l\ \bar{l}_1\gamma^\mu P_L l_2^C
  + B_{\ell V}^{RR}\, \bar{l}'\gamma_\mu P_R l\ \bar{l}_1\gamma^\mu P_R l_2^C
\nonumber\\
  &+& B_{\ell V}^{LR}\, \bar{l}'\gamma_\mu P_L l\ \bar{l}_1\gamma^\mu P_R l_2^C
  + B_{\ell V}^{RL}\, \bar{l}'\gamma_\mu P_R l\ \bar{l}_1\gamma^\mu P_L l_2^C
  \nonumber\\
  &+& B_{\ell S}^{LL}\, \bar{l}' P_L l\ \bar{l}_1 P_L l_2^C
  + B_{\ell S}^{RR}\, \bar{l}' P_R l\ \bar{l}_1 P_R l_2^C
  \nonumber\\
  &+& B_{\ell S}^{LR}\, \bar{l}' P_L l\ \bar{l}_1 P_R l_2^C
  + B_{\ell S}^{RL}\, \bar{l}' P_R l\ \bar{l}_1 P_L l_2^C
\nonumber\\
  &+& B_{\ell T}^{LL}\, \bar{l}' \sigma_{\mu\nu} P_L l\ \bar{l}_1
  \sigma^{\mu\nu} P_L l_2^C
  + B_{\ell T}^{RR}\, \bar{l}' \sigma_{\mu\nu} P_R l\ \bar{l}_1
  \sigma^{\mu\nu} P_R l_2^C  \Big) \,.
% \\
% \!&\equiv&\!
% - \frac{\alpha_w^2}{4 M_W^2} \sum_{X,Y=L,R}\ \sum_{A=V,S,T}
%     B_{\ell A}^{XY}\, \bar{l}'\Gamma^X_{A} l\ \bar{l}_1\Gamma^Y_{A}
%     l_2^C \,.
\end{eqnarray}
U gornjim izrazima, $g_L^l = -1/2+s_w^2$ i $g_R^l=s_w^2$ su
$Z$-bozonska leptonska vezanja, a $s_w=\sin\theta_w$. Kompozitni 
form-faktori pravokutnih dijagrama $B_{\ell  A}^{XY}$ dani su u 
Dodatku~\ref{sec:olff}. Oznake $V$, $S$ i $T$ ozna\cc avaju 
form-faktore vektorskih, skalarnih i tenzorskih kombinacija struja,
dok $L$ i $R$ razlikuju lijeve i desne kiralnosti tih struja. Form-faktori
iz pravokutnih dijagrama sadr\z e izravne i Fiertz-transformirane doprinose,
\s to se mo\z e vidjeti u Dodatku~\ref{app:ff}.

S obzirom na tri leptonske generacije, prijelazna amplituda za raspad
$l\to l'l_1l_2^C$ mo\z e upasti u jednu od tri klase ili 
kategorije~\cite{Ilakovac1995}: (i)~$l'\neq l_1=l_2$, (ii)~$l'= l_1=l_2$, 
(iii)~$l'= l_1\neq l_2$. Za prve dvije klase, ukupni leptonski broj je
sa\cc uvan, dok je u tre\'{c}oj klasi ukupni leptonski broj na razini 
struja naru\s en za
dvije jedinice. Budu\'{c}i da se za predikcije za opservable iz klase (iii)
ispostavlja da su neopazivo malene u $\nu_R$MSSM-u, ove procese \'{c}emo
ignorirati. Omjeri grananja za klase (i) i (ii) glase:
\begin{eqnarray}
B(l\to l'l_1l_1^C)
 &=&
 \frac{m_l^5\alpha_w^4}{24576\pi^3 M_W^4\Gamma_l}
 \Bigg\{\bigg[
  \Big|2s_w^2 (F_\gamma^L + F_Z^L) - F_Z^L - B_{\ell V}^{LL}\Big|^2 \nonumber\\
 &+& \Big|2s_w^2 (F_\gamma^R + F_Z^R) - B_{\ell V}^{RR}\Big|^2
  + \Big|2s_w^2 (F_\gamma^L + F_Z^L) - B_{\ell V}^{LR} \Big|^2
  \nonumber\\
 &+&\Big|2s_w^2 (F_\gamma^R + F_Z^R) - F_Z^R - B_{\ell V}^{RL} \Big|^2
  \bigg]
\nonumber\\
 &+& \frac{1}{4} \Big( |B_{\ell S}^{LL} |^2 + |B_{\ell S}^{RR} |^2 +
  |B_{\ell S}^{LR} |^2 + |B_{\ell S}^{RL} |^2 \Big)
\nonumber\\
 &+& 12 \Big( |B_{\ell T}^{LL}|^2  +  |B_{\ell T}^{RR}|^2 \Big)
\nonumber\\
 &+& \frac{32s_w^4}{m_l} \Big[
 \mbox{Re} \Big( (F_\gamma^R + F_Z^R ) G_\gamma^{L*}\Big)
 + \mbox{Re} \Big( (F_\gamma^L + F_Z^L ) G_\gamma^{R*}\Big)
 \Big]
\nonumber\\
 &-& \frac{8s_w^2}{m_l} \Big[
 \mbox{Re}\Big( (F_Z^R + B_{\ell V}^{RR}  + B_{\ell V}^{RL} ) G_\gamma^{L*}\Big)
\nonumber\\
 &+& \mbox{Re}\Big( (F_Z^L + B_{\ell V}^{LL} + B_{\ell V}^{LR} )
  G_\gamma^{R*}\Big)\Big]
\nonumber\\
 &-& \frac{32 s_w^4}{m_l^2} \Big(|G_\gamma^L|^2 + |G_\gamma^R|^2\Big)
 \bigg(\ln\frac{m^2_l}{m^2_{l'}} - 3\bigg)
\Bigg\}\,,
\end{eqnarray}
\begin{eqnarray}
B(l\to l'l'l'^C)
 &=&
 \frac{m_l^5\alpha_w^4}{24576\pi^3 M_W^4\Gamma_l}\:
 \Bigg\{ 2 \bigg[
  \Big|2s_w^2 (F_\gamma^L + F_Z^L) - F_Z^L - \frac{1}{2}B_{\ell V}^{LL} \Big|^2
  \nonumber\\
  &+& \Big|2s_w^2 (F_\gamma^R + F_Z^R) - \frac{1}{2}B_{\ell V}^{RR} \Big|^2 \bigg]
  + \Big|2s_w^2 (F_\gamma^L + F_Z^L) - B_{\ell V}^{LR} \Big|^2
  \nonumber\\
  &+& \Big|2s_w^2 (F_\gamma^R + F_Z^R) - (F_Z^R + B_{\ell V}^{RL} )\Big|^2
  + \frac{1}{8} \Big( |B_{\ell S}^{LL}|^2 + |B_{\ell S}^{RR}|^2 \Big)
\nonumber\\
 &+& 6 \Big( |B_{\ell T}^{LL}|^2 +  |B_{\ell T}^{RR}|^2 \Big)
\nonumber\\
 &+& \frac{48s_w^4}{m_l} \Big[
 \mbox{Re} \Big( ( F_\gamma^R + F_Z^R ) G_\gamma^{L*}\Big)
 + \mbox{Re} \Big( ( F_\gamma^L + F_Z^L ) G_\gamma^{R*}\Big)
 \Big]
\nonumber\\
 &-& \frac{8s_w^2}{m_l} \Big[
 \mbox{Re} \Big( \big(F_Z^R + B_{\ell V}^{RR}
     + B_{\ell V}^{RL} \big) G_\gamma^{L*}\Big)
\nonumber\\
 &+& \mbox{Re} \Big( \big(2 F_Z^L + B_{\ell V}^{LL}
     + B_{\ell V}^{LR} \big) G_\gamma^{R*}\Big)\Big]
\nonumber\\
 &+& \frac{32 s_w^4}{m_l^2} \Big(|G_\gamma^L|^2 + |G_\gamma^R|^2\Big)
 \bigg(\ln\frac{m^2_l}{m^2_{l'}} - \frac{11}{4}\bigg)
 \Bigg\}\; ,
\end{eqnarray}
gdje smo radi jednostavnosti izostavili univerzalne indekse $l'l$ koji
se pojavljuju u fotonskim i $Z$-bozonskim form-faktorima. U gornjim
izrazima, $m_{l'}$, $m_{l_1}$ i $m_{l_2}$ predstavljaju mase ulaznih i
izlaznih nabijenih leptona, a $\Gamma_l$ ukupnu \s irinu raspada nabijenog
leptona $l$.

Prijelaz $\mu\to e$ na atomskim jezgrama odgovara procesu $J_\mu\to e^-J^+$,
pri \cc emu je $J_\mu$ atom jezgre $J$ u kojem je jedan orbitalni elektron
zamjenjen mionom, a $J^+$ odgovaraju\'{c}i ion bez miona. Prijelazna
amplituda za ovaj proces,
\begin{eqnarray}
{\cal T}^{\mu e;J}
 &=&
   \langle J^+ e^- | {\cal T}^{d\mu \to de} | J_\mu\rangle
 + \langle J^+ e^- | {\cal T}^{u\mu \to ue} | J_\mu\rangle\; ,
\end{eqnarray}
ovisi o dva efektivna operatora pravokutnih dijagrama:
\begin{eqnarray}
{\cal T}_{\rm box}^{d\mu \to de}
 &=&
 - \frac{\alpha_w^2}{2 M_W^2}\,
 (d^\dagger d)\;\, \bar{e}\, ( V_d^R\, P_R + V_d^L\, P_L )\, \mu\; ,\\
{\cal T}_{\rm box}^{u\mu \to ue}
 &=&
  - \frac{\alpha_w^2}{2 M_W^2}\,
 (u^\dagger u)\;\, \bar{e}\, (V_u^R\, P_R + V_u^L\, P_L)\, \mu\; .
\end{eqnarray}
U gornjem izrazu, $\mu$ i $e$ predstavljaju mionsku i elektronsku valnu
funkciju, a $d$ i $u$ su operatori polja koji djeluju na $J_\mu$ odnosno
$J^+$ stanja. Kompozitni form-faktori $V_d^{L}$,  $V_u^{L}$, $V_d^{R}$ 
i $V_u^{R}$ mogu se napisati kao
\begin{eqnarray}
V_d^L
 &=&
 - \frac{1}{3} s_w^2 \Big(F^L_\gamma - \frac{1}{m_\mu} G^R_\gamma\Big)
 + \Big(\frac{1}{4} - \frac{1}{3} s_w^2\Big) F_Z^L \nonumber\\
 &&{}+\, \frac{1}{4}\Big(B_{d V}^{LL}+B_{d V}^{LR} + B_{d S}^{RR} + B_{d S}^{RL}\Big)\,,
 \\[1.5ex]
V_d^R
 &=&
 -  \frac{1}{3} s_w^2 \Big(F^R_\gamma - \frac{1}{m_\mu} G^L_\gamma\Big)
 + \Big(\frac{1}{4} - \frac{1}{3} s_w^2\Big) F_Z^R \nonumber\\
 &&{}+\, \frac{1}{4}\Big(B_{d V}^{RR}+B_{d V}^{RL} + B_{d S}^{LL} + B_{d S}^{LR}\Big)\,,
 \\[1.5ex]
V_u^L
 &=&
 \frac{2}{3} s_w^2 \Big(F^L_\gamma - \frac{1}{m_\mu} G^R_\gamma\Big)
 + \Big(- \frac{1}{4} + \frac{2}{3} s_w^2\Big) F_Z^L \nonumber\\
 &&{}+ \frac{1}{4}\Big(B_{u V}^{LL}+B_{u V}^{LR} + B_{u S}^{RR} + B_{u S}^{RL}\Big)\,,
 \\[1.5ex]
V_u^R
 &=&
  \frac{2}{3} s_w^2 \Big(F^R_\gamma - \frac{1}{m_\mu} G^L_\gamma\Big)
 + \Big(- \frac{1}{4} + \frac{2}{3} s_w^2\Big) F_Z^R \nonumber\\
 &&{}+ \frac{1}{4}\Big(B_{u V}^{RR}+B_{u V}^{RL} + B_{u S}^{LL} + B_{u S}^{LR}\Big)\,.
\end{eqnarray}

Nukeonski matri\cc ni elementi operatora $u^\dagger u$ i $d^\dagger d$
glase
\begin{eqnarray}
\langle J^+e^- |u^\dagger u| J_\mu \rangle &=& (2Z+N) F(-m_\mu^2)\,,
\nonumber\\
\langle J^+e^- |d^\dagger d| J_\mu \rangle &=& (Z+2N) F(-m_\mu^2)\,,
\end{eqnarray}
pri \cc emu $F(q^2)$ ozna\cc ava odboj iona $J^+$ \cite{Chiang1993},
a faktori $2Z+N$ i $Z+2N$ broj $u$ odnosno $d$ kvarkova u jezgri $J$.
Matri\cc ni element za $J_\mu\to J^+\mu^-$ se prema tome mo\z e napisati
kao:
\begin{eqnarray}
T^{J_\mu\to J^+ e^-}
 &=&
 -\frac{\alpha_w^2}{2 M_W^2}\, F(-m_\mu^2)\: \bar{e}
 \,(Q_W^L\, P_R + Q_W^R\, P_L )\,
 \mu\,,
\end{eqnarray}
uz
\begin{eqnarray}
Q_W^L &=& (2Z+N) V_u^L +(Z+2N)V_d^L\, ,
\nonumber\\
Q_W^R &=& (2Z+N) V_u^R +(Z+2N)V_d^R\, .
\end{eqnarray}

Koriste\'{c}i gore navedene izraze, dolazimo do izraza za brzinu
prijelaza $J_\mu\to J^+e^-$
\begin{eqnarray}
R^J_{\mu    e}     &=&    \frac{\alpha^3\alpha_w^4    m_\mu^5}{16\pi^2
  M_W^4\Gamma_{{\rm  capture}}} \frac{Z^4_{\rm  eff}}{Z} |F(-m_\mu^2)|^2
\: \Big(\,|Q_W^L|^2+|Q_W^R|^2\,\Big)\, ,
\end{eqnarray}
gdje je $\Gamma_{{\rm capture}}$ brzina uhvata miona od jezgre, a
$Z_{\rm eff}$ efektivni naboj koji uzima u obzir koherentne u\cc inke
koji se javljaju u jezgri $J$ uslijed njezine kona\cc ne dimenzije.
Vrijednosti za $Z_{\rm eff}$ uzete su iz ref.~\cite{Kitano2002}.

Na osnovi gore izlo\z enih analiti\cc kih rezultata napravljena je
numeri\cc ka ana\-liza. mSUGRA parametri odabrani su tako da zadovoljavaju
ograni\cc enja koja je postavio LHC: 
$m_H=125.5\pm 2$~GeV \cite{Aad2012,Chatrchyan2012a,Chatrchyan2013},
$m_{\tilde{g}}>1500$~GeV i $m_{\tilde{t}}>500$~GeV \cite{Aad2012a,Chatrchyan2012a},
gdje je $m_H$ masa standardnomodelskog Higgsa, a $m_{\tilde{g}}$
i $m_{\tilde{t}}$ mase gluina odnosno stop kvarka:
\begin{equation}
\begin{array}{ll}
 \tan\beta = 10\,, \qquad &  m_0 = 1000~{\rm GeV}\,, \\
 A_0 = -3000~{\rm GeV}\,, \qquad & M_{1/2} = 1000~{\rm GeV}\,.
\end{array}
\end{equation}

Kori\s tenjem renormalizacijske grupne jednad\z be danih u referencama
\cite{Chankowski2002,Petcov2004}, radimo evoluciju ba\z darnih vezanja
te Yukawinih matrica za kvarkove i nabijene leptone od skale $M_Z$
do GUT skale, dok neutrinska masena matrica (${\bf m}_M$) i Yukawina matrica
(${\bf h}_\nu$) matrica evoluiraju od skale odre\dj ene masom te\s kog 
neutrina $m_N$, do GUT skale.

Analiza je provedena u dva razli\cc ita scenarija. Prvi realizira U(1)
simetriju \eqref{saz:YU1} i za njega uzimamo jednu od tri opcije:
(i)~$a=b$ i $c=0$ za $\mu\to eX$, (ii)~$a=c$ i $b=0$ za $\tau\to eX$,
(iii)~$b=c$ i $a=0$ za $\tau\to\mu X$, pri \cc emu je
$X=\gamma,\, e^+e^-,\,  \mu^+\mu^-,\, q\overline{q}$. Drugi scenarij
realizira simetriju grupe $A_4$ \eqref{saz:YA4} i tu smo uzeli $a=b=c$.

Rezultati numeri\cc ke analize prikazani su na slikama
\ref{Fig3} -- \ref{Fig11} na str.~\pageref{Fig3}--\pageref{Fig11}.

\section*{Dipolni momenti leptona
  \markright{Dipolni momenti leptona}}
\addcontentsline{toc}{section}{Dipolni momenti leptona}

S obzirom na trenutno stanje eksperimenata, mo\z e se vidjeti da
anomalni magnetski moment (MDM) miona i elektri\cc ni dipolni moment
(EDM) elektrona zavrije\dj uju posebnu pa\z nju. Zbog toga se u\cc inilo
uputnim unutar modela s kojim smo izu\cc avali procese s naru\s enjem
leptonskog okusa u nabijenom sektoru izvrijedniti i ove opservable.

Anomalni magnetski dipolni moment i elektri\cc ni dipolni moment leptona
$l$ mo\z e se i\s\cc itati iz lagran\z ijana \cite{Branco1999}:
\begin{equation}
{\cal L}\ =\ \bar{l}\,\Big[\gamma_\mu(i\partial^\mu + e A^\mu) 
  - m_l - \frac{e}{2m_l} \sigma^{\mu\nu} (F_l+iG_l\gamma_5) 
                                       \partial_\nu A_\mu\Big]\, l\,.
\end{equation}

U podru\cc ju u kojem se fotonsko polje $A^\mu$ nalazi na ljusci mase, 
form-faktor $F_l$ definira magnetski dipolni moment
leptona $l$, t.j.\ $a_l \equiv F_l$,
dok form-faktor $G_l$ definira njegov elektri\cc ni dipolni moment,
t.j.\ $d_l \equiv eG_l/m_l$. Iz prethodnog analiti\cc kog izraza za
fotonsku amplitudu mo\z emo napisati op\'{c}u form-faktorsku dekompoziciju
prijelazne amplitude,
\begin{equation}
i{\cal T}^{\gamma ll}\ =\
 i\frac{e \alpha_w}{8 \pi M_W^2}
 \Big[ (G_\gamma^L)_{ll} i\sigma_{\mu\nu}q^\nu P_L
 + (G_\gamma^R)_{ll} i\sigma_{\mu\nu}q^\nu P_R \Big]\,.
\end{equation}
Tako dolazimo do izraza za anomalni MDM ($a_l$) i EDM ($d_l$) leptona $l$,
\begin{eqnarray}
 a_l &=& \frac{\alpha_w m_l}{8\pi M_W^2} \Big[ (G_\gamma^L)_{ll} + 
        (G_\gamma^R)_{ll} \Big]\,,\\
 d_l &=& \frac{e \alpha_w}{8\pi M_W^2} i \Big[ (G_\gamma^L)_{ll} - 
         (G_\gamma^R)_{ll} \Big]\,,
\end{eqnarray}
gdje notacija za vezanja i form-faktore odgovara onoj koju smo koristili i
ranije.

Kao \s to je pokazano u ref.~\cite{Farzan2004}, EDM leptona $d_l$
i\s\cc ezava u u MSSM-u sa univerzalnim rubnim uvjetima mekanog loma
supersimetrije bez dodatno uvedenih CP faza. Ovaj rezultat vrijedi i 
u pro\s irenjima MSSM-a te\s kim neutrinima, dok god sneutrinski
sektor \cc uva CP simetriju.

Kao minimalno odstupanje od ovog scenarija, pretpostavimo da samo
sneutrinski sektor naru\s ava CP simetriju i to putem mekanih CP
faza u bilinearnim i trilinearnim parametrima u lagran\z ijanu
mekanog loma CP simetrije,
\begin{eqnarray}
{\bf b}_\nu &\equiv& {\bf B}_\nu {\bf m}_M \ =\ B_0 e^{i\theta} m_N {\bf 1}_3\,,\\ 
{\bf A}_\nu &=&  {\bf h}_\nu\, A_0 e^{i\phi}\,.
\end{eqnarray}
pri \cc emu su $B_0$ i $A_0$ realni parametri odre\dj eni na GUT skali, $m_N$
realni parametar unesen na skali $m_N$, $\theta$ i $\phi$ fizikalne CP faze,
a ${\bf h}_\nu$ neutrinska Yukawina $3\times 3$ matrica dana sa~\eqref{saz:YA4}.
Ovi parametri nalaze se u sljede\'{c}im \cc lanovima lagran\z ijana koji mekano
lomi supersimetriju:
\begin{equation}
 - ({\bf A}_\nu)^{ij} \tilde{\nu}^c_{iR} (  h^+_{uL}
  \tilde{e}_{jL} - h^0_{uL} \tilde{\nu}_{jL})\,,
\end{equation}
\begin{equation}
({\bf b}_\nu {\bf m}_M)_{ii} \tilde{\nu}_{Ri}\tilde{\nu}_{Ri}\,.
\end{equation}
Radi jednostavnosti pretpostavljamo da je matrica ${\bf b}_\nu$ 
proporcionalna sa jedini\cc nom matricom ${\bf 1}_3$. U uobi\cc ajenim
SUSY scenarijima sa mehanizmom njihalice uz ultra-te\s ke neutrine
mase $m_N$, doprinosi CP naru\s enja u sneutrinskom sektoru elektri\cc nom
dipolnom momentu $d_l$ pona\s aju se kao $B_0/m_N$ i $A_0/m_N$ na razini
jedne petlje. A kako su mase $m_N$ u tim scenarijima
velike (blizu GUT skale), doprinos
EDM-u prakti\cc ki je zanemariv. Uo\cc ljivi doprinosi EDM-u mogu se
dakle o\cc ekivati u scenarijima sa mehanizmom njihalice na niskoj skali, 
u kojem masa $m_N$ mo\z e biti usporediva sa $B_0$ i $A_0$.

Bilinearnu mekanu matricu ${\bf b}_\nu$ zanemarili smo kada smo izu\cc avali
procese s naru\s enjem leptonskog okusa u nabijenom sektoru. Tada smo
pre\s utno pretpostavili da je taj parametar malen u odnosu na druge parametre
mekanog loma supersimetrije. Ovdje \'{c}emo ga ipak uzeti u obzir, ali tako
da senutrinske mase uvijek ostanu pozitivne, dakle i fizikalne.

U MSSM-u, vode\'{c}i doprinos anomalnom magnetskom momentu leptona $a_l$ 
ima sljede\'{c}e pona\s anje~\cite{Moroi1996,Carena1997,Stockinger2007}:
\begin{eqnarray}
  \label{saz:al_approx}
a_l^{\rm MSSM} &\propto& \frac{m_l^2}{M_{\rm SUSY}^2}\, \tan\beta\;
\mbox{sign}(\mu M_{1,2})\; , 
\end{eqnarray}
pri \cc emu je $M_{\rm SUSY}$ tipi\cc na masena skala mekanog loma
supersimetrije, $\tan\beta=v_2/v_1$ omjer o\cc ekivanih vakuumskih
vrijednosti Higgsovih dubleta, a $M_{1,2}$ mase gejd\z ina
povezanih sa ${\rm  U}(1)_{\rm Y}$ odnosno ${\rm SU(2)}$ ba\z darnom
grupom. Kao \s to \'{c}emo vidjeti, doprinos MSSM-a anomalnom magnetskom
momentu miona ostaje dominantan i u $\nu_R$MSSM-u.

Uspore\dj uju\'{c}i izraze za $d_l$ i $a_l$, mo\z emo dati naivnu procjenu
pona\s anja elektri\cc nog dipolnog momenta leptona $l$ na razini jedne 
petlje,
\begin{eqnarray}
d_l^{\rm MSSM} &\propto& \sin(\phi_{\rm CP})\, \frac{m_l}{M_{\rm
    SUSY}^2}\, \tan\beta\ , 
\end{eqnarray}
gdje je $\phi_{\rm  CP}$ generi\cc ka CP faza iz sektora koji mekano
lomi supersimetriju. Iako su u MSSM-u 
mogu\'{c}e i druga\cc ije ovisnosti $d_l$-a o $\tan\beta$ 
\cite{Farzan2004,Pilaftsis2002}, pokazat \'{c}e se da je u okviru
$\nu_R$MSSM-a ova ovisnost uvijek linearna na razini jedne petlje.

Numeri\cc ka analiza napravljena je u ovisnosti o mSUGRA parametrima,
i to u okolici to\cc ke odre\dj ene parametrima
\begin{equation}
\begin{array}{llll}
m_0 = 1~{\rm TeV}\,, &  M_{1/2} = 1~{\rm TeV}\,, &  
  A_0 = -4~{\rm TeV}\,, & \tan\beta = 20\,, \\
m_N = 1~{\rm TeV}\,, & B_0 = 0.1~{\rm TeV}\,, &
  a = b = c = 0.05\,, 
\end{array}
\end{equation}
Kao i ranije, parametri su odbrani tako da zadovoljavaju ekperimentalna
ograni\cc enja koja je postavio LHC ($m_H=125.5\pm 2$~GeV, $m_{\tilde{g}}>1500$~GeV, 
$m_{\tilde{t}}>500$~GeV).

Rezultati numeri\cc ke analize prikazani su na slikama \ref{Fig4.1} (str.~\pageref{Fig4.1}), 
\ref{Fig4.2} (str.~\pageref{Fig4.2}) i \ref{Fig4.3} (str.~\pageref{Fig4.3}).

\section*{Zaklju\cc ak
  \markright{Zaklju\cc ak}}
\addcontentsline{toc}{section}{Zaklju\cc ak}

Naru\s enje leptonskog okusa (CLFV) izu\cc avano je u okviru minimalnog
supersimetri\cc nog standardnog modela (MSSM) pro\s irenog sa singletnim
te\s kim neutrinima na niskoj skali, pri \cc emu je posebna pa\z nja posve\'{c}ena
pojedinim doprinosima petlji koji dolaze od te\s kih neutrina $N_{1,2,3}$,
sneutrina $\widetilde{N}_{1,2,\dots,12}$ i \cc lanova koji mekano lome
supersimetriju. U ovoj analizi, po prvi put smo uklju\cc ili potpuni 
skup pravokutnih dijagrama, uz dijagrame sa fotonskim i $Z$-bozonskim
doprinosima. Tako\dj er smo izveli potpun skup kiralnih amplituda i 
pridru\z enih im form-faktora koji su povezani sa CLFV raspadima miona
i tau-leptona bez neutrina, kao \s to su $\mu \to eee$, $\tau \to  \mu\mu\mu$, 
$\tau \to e\mu\mu$ i $\tau \to ee\mu$, te $\mu \to e$ prijelazima na atomskim
jezgrama. Dobiveni analiti\cc ki rezultati su op\'{c}eniti i mogu se primijeniti
na ve\'{c}inu modela nove fizike koji uklju\cc uju CLFV. U tom kontekstu
valja naglasiti da je sustavna analiza ovih procesa pokazala postojanje dvaju
novih form faktora iz pravokutnih dijagrama, koji se ne navode u postoje\'{c}oj
literaturi koja se bavi teorijama sa CLFV-om.

Ova detaljna studija pokazala je da efekti mekanog loma supersimetrije u
dijagramima sa $Z$ bozonom dominira u CLFV opservablama, i to u velikom dijelu
$\nu_R$MSSM parametarskog prostora u okviru mSUGRA scenarija. Ipak, postoji
zna\cc ajno podru\cc je parametarskog prostora (za neutrinske mase
$m_N\stackrel{<}{{}_\sim} 1$~TeV) u kojima pravokutni dijagrami koji uklju\cc uju
te\s ke neutrine u petlji mogu biti usporedivi, pa \cc ak i ve\'{c}i, od odgovaraju\'{c}ih
doprinosa iz dijagrama sa $Z$ bozonom u procesima $\mu \to eee$ i $\mu  \to   e$
prijelazima na atomskim jezgrama (v.~sliku \ref{Fig11} na str.~\pageref{Fig11}).
U istom kinemati\cc kom re\z imu, uslijed slu\cc ajnih numeri\cc kih poni\s tenja,
opa\z amo i potisnu\'{c}e omjera grananja za fotonske CLFV raspade $\mu  \to e \gamma$,
kao i za raspade $\tau \to e \gamma$ i $\tau \to \mu\gamma$. Kao \s to je ve\'{c} 
re\cc eno, takva potisnu\'{c}a u mehanizmu njihalice na niskoj skali dolaze kao
posljedica poni\s tenja izme\dj u \cc esti\cc nih i s-\cc esti\cc nih doprinosa
uslijed aproksimativne realizacije supersimetri\cc nog \emph{no-go} teorema
kojeg su postavili Ferrara i Remiddi~\cite{Ferrara1974}. U mehanizmu njihalice na
visokoj skali ova se poni\s tenja mogu pojaviti samo za odre\dj eni izbor neutrinskih
Yukawinih matrica i Majoraninih masenih matrica, kao \s to je pokazano u 
ref.~\cite{Ellis2002,Arganda2006}. Stoga mo\z emo re\'{c}i da rezultati koje
smo dobili unutar supersimetri\cc nog modela njihalice na niskoj skali 
(uz $m_N \stackrel{<}{{}_\sim}  10~\textrm{ TeV}$), podr\z avaju izvorni rezultat
iznesen u ref.~\cite{Ilakovac2009}, gdje uobi\cc ajena paradigma po
kojoj fotonski operatori dipolnog momenta dominiraju CLFV opservablama u
modelima njihalice na visokoj 
skali~\cite{Hisano1996,Hisano1995,Hisano1999,Carvalho2001,Hisano2009}
do\z ivljava radikalnu promjenu. Iz toga mo\z emo zaklju\cc iti da raspadi 
$\mu \to eee$ i $\mu\to e$ prijelazi na jezgrama tako\dj er mogu slu\z iti
kao precizan test naru\s enja leptonskog okusa u nabijenom sektoru.

Otkrili smo da CLFV efekti inducirani od strane sneutrina, za razliku od
onih induciranih te\s kim neutrinima, ostaju potisnuti u cijelom prostoru
mSUGRA parametarskog prostora. Uz to, perturbacijsko ograni\cc enje na neutrinska
Yukawina vezanja~${\bf h}_\nu$ do GUT skale \cc ini kvadri\cc ne doprinose
reda~$({\bf h}_\nu)^4$ malima. Ova studija usmjerena je na davanje numeri\cc kih
predikcija za male i umjerene vrijednosti parametra $\tan\beta$
($\tan\beta  \stackrel{<}{{}_\sim} 20$), pri \cc emu se o\cc ekuje da 
neutralne interakcije s Higgsovim bozonima ograni\cc ene nedavnim opa\z anjima 
raspada $B_s \to \mu\mu$ ne daju zna\cc ajan doprinos ovakvim procesima.
Globalna analiza koja bi uklju\cc ivala velike $\tan\beta$ na CLFV
opservablama uz LHC ograni\cc enja jedan je od ciljeva budu\'{c}ih istra\z ivanja.

Uz navedeno, disertacija sadr\z i sustavnu studiju doprinosa jedne petlje
mionskom anomalnom magnetskom momentu (MDM)~$a_\mu$ te elektri\cc nom
dipolnom momentu (EDM)~$d_e$ u okviru $\nu_R$MSSM modela. Posebna pa\z nja
dana je u\cc inku sneutrniskih parametara koji mekano lome supersimetriju,
${\bf B_\nu}$ i ${\bf  A_\nu}$, kao i njihovim univerzalnim CP fazama,
$\theta$ i $\phi$. Koliko znamo, leptonski dipolni momenti u prija\s njoj
literaturi nisu detaljno analizirani u okviru superisimetri\cc nih modela
sa singletnim (s)neutrinima na niskoj skali.

Za anomalni MDM miona $a_\mu$ pokazali smo da su doprinosi te\s kih
neutrinskih i sneutrinskih singleta MDM-u maleni, tipi\cc no jedan
ili dva reda veli\cc ine ispod mionske anomalije $\Delta a_\mu$. S
druge strane, sneutrini i sleptoni lijeve kiralnosti daju najve\'{c}i
efekt na $\Delta a_\mu$, to\cc no kao i u MSSM-u.
Ovisnost MDM-a o masi miona~$m_\mu$,  $\tan\beta$ i masenoj skali
mekanog loma supersimetrije $M_{\rm SUSY}$ pa\z ljivo su analizirani
i potvr\dj eno je njihovo pona\s anje u skladu s 
jednad\z bom~\eqref{saz:al_approx}. Kona\cc no smo utvrdili i to da
ovisnost $a_\mu$ o univerzalnom mekom trilinearnom parameteru~$A_0$ te 
neutrinskim Yukawinim vezanjima ${\bf  h}_\nu$ i masi te\s kog neutrina
$m_N$ zanemariva.

Nadalje, u okviru istog $\nu_R$MSSM modela napravljena je analiza 
EDM-a elektrona~$d_e$. Te\s ki singletni neutrini ne doprinose EDM-u,
a \cc lanovi koji mekano lome supersimetriju iz sneutrinskog sektora
doprinose samo ako su faze $\phi$ i/ili $\theta$ razli\cc iti od nule.
Numeri\cc ki je pokazano da je mogu\'{c}i doprinos naru\s enju iz CP simetrije 
koji bi dolazili od relativno kompleksnih produkata vrhova 
(v.~jednad\z bu \eqref{CPNt} na str.~\pageref{CPNt}) jednak nuli. S druge 
strane, doprinos koji dolazi od kona\cc nih vrijednosti faze 
$\phi$ najve\'{c}i je i mo\z e rezultirati vrijednostima za EDM koji su usporedivi
s trenutno postavljenom eksperimentalnom gornjom granicom. Efekt CP faze
$\theta$ na $d_e$ je od prilike jedan do dva reda veli\cc ine manji nego onaj
koji dolazi od faze $\phi$. EDM elektrona $d_e$ linearno raste sa $\tan\beta$ i
masom leptona $m_l$, pribli\z no je neovisan o parametrima $A_0$ i $B_0$, 
ali op\'{c}enito pada sa mSUGRA parametrima $m_0$ i $M_{1/2}$.

Na temelju ovih numeri\cc kih rezultata, izvedeni su pribli\z ni 
poluanaliti\cc ki izrazi, koji se razlikuju od onih iz postoje\'{c}e
literature o supersimeti\cc nim modelima s mehanizmom njihalice na visokoj
skali. Specifi\cc no, dodavanje CP faza vodi na pona\s anje EDM-a 
tipa  $d_l \propto m_l \tan\beta /m_N^{y}$, gdje je $2/3<y<1$. Dok je
istina da $d_e$ op\'{c}enito pada sa $M_{SUSY}$, ova ovisnost ne mo\z e
se opisati jednostavnim potencijskim padom. Ovisnost o mSUGRA parametrima
$A_0$ i $B_0$ su slabe u najve\'{c}em dijelu parametarskog prostora. Linearna
ovisnost o $\tan\beta$ kao i ovisnost o masi te\s kog neutrina predstavljaju
nove rezultate koji proizlaze iz ove studije.

Za usporedbu, ovisnost o $\tan\beta$ navedena u ref.~\cite{Farzan2004} je,
ovisno o veli\cc ini, ili kubi\cc na ili konstantna. Uzev\s i u obzir
trenutne eksperimentalne granice na $d_e$ zna\cc ajan dio $\nu_R$MSSM
parametarskog prostora identificiran je maksimalnom vrijedno\s\'{c}u 
CP faze $\phi = \pi/2$, pri \cc emu EDM elektrona $d_e$ mo\z e poprimiti
vrijednosti usporedive sa trenutnom i budu\'{c}om eksperimentalnom
osjetljivo\s\'{c}u. U\cc inak CP naru\s enja iz 
sneutrinskog sektora na elektri\cc ne dipolne momente neutrona i
\z ive trebao bi biti potisnut, pa ovakav tip studije mo\z e
slu\z iti kao razlikovni kriterij za $\nu_R$MSSM scenarije koje smo
razmatrali u ovoj disertaciji.

%\vspace{2.5ex}
\newpage
U svom kratkom osvrtu o budu\'{c}nosti fizike elementarnih 
\cc estica~\cite{Glashow2013}, dobitnik Nobelove nagrade
Sheldon Lee Glashow iznio je \s est to\cc aka koje on osobno smatra
najva\z nijim za budu\'{c}nost teorijskog istra\z ivanja u fizici
visokih energija. Izme\dj u ostalog, ove to\cc ke uklju\cc uju
procese s naru\s enjem leptonskog okusa u nabijenom sektoru, anomalni
magnetski dipolni moment miona $a_\mu = g_\mu - 2$, kao i 
elektri\cc ni dipolni moment elektrona $d_e$. Autor ove disertacije
rado bi se s time slo\z io.

\end{sloppypar}

%% file: cv.tex
\chapter*{Curriculum Vit\ae\markboth{Curriculum Vit\ae}{Curriculum Vit\ae}}
\addcontentsline{toc}{chapter}{Curriculum Vit\ae}

\pagenumbering{roman}

\noindent
{\large \textbf{First name $|$ Surname} : \ Luka Popov}\\

\noindent
{\large \textbf{Address}}

\noindent
Physics Department\\
Faculty of Science\\
University of Zagreb\\
Bijeni\v cka cesta 32, P.O.B.~331\\
HR-10002 Zagreb, Croatia

\noindent
\textsf{e-mail: lpopov@phy.hr} $\mathbf |$ 
\textsf{phone: +385 1 460 5605} $\mathbf |$ 
\textsf{fax: +385 1 460 5606}\\

\noindent
{\large \textbf{Born}} \ February 19 1982 (Split Croatia)\\

\noindent
{\large \textbf{Education}}

\noindent
\begin{tabular}{cl}
1989-1996 & Elementary school (\emph{Spinut} Split)\\
1996-2000 & High school (\emph{IV. Gimnazija} Split)\\
2000-2007 & Faculty of Science, University of Zagreb\\[1.5ex]
Nov 2007 & \textbf{dipl. ing.}, Diploma thesis:
 \emph{Neutral mesons mixings} \\
 & \emph{in the MSSM}, advisor: Prof Amon Ilakovac\\[1.5ex]
2009-2013 & PhD student, University of Zagreb\\[1.5ex]
Dec 2013 & \textbf{PhD} thesis completed and submited\\
\end{tabular}

\newpage
\noindent
{\large \textbf{Research interests}}

\noindent
Physics beyond standard model; Supersymmetry; Lepton flavor violation\\

\noindent
{\large \textbf{Position}}

\noindent
Since 2008 research and teaching assistant at Faculty of Science, University of
Zagreb.\\

\vspace{0.5cm}

% \noindent
% {\large \textbf{Research stay}}\\
% 
% \noindent
% From October 2008 to February 2009, one semester research stay at the University
% of Vienna.\\
% 
% \vspace{0.5cm}

\noindent
{\large \textbf{Teaching assistant experience}}
\begin{itemize}
\item Classical electrodynamics \ (3th year courses)
\item Relativistic quantum physics \ (4th year course)
\item Theory of fields I \ (4th year course)
\item Theory of fields II \ (5th year course)
\item General physics \ (Faculty of Electrical Engineering, 1st year course)
\end{itemize}

\vspace{0.5cm}

\newpage

\noindent
{\large \textbf{List of publications}}

\noindent
1. Ilakovac A and Popov L 2010 {Two-step Lorentz Transformation of Force}
  {\em Fizika} {\bf A 19} 3

\noindent
2. Popov L 2013 Newtonian-Machian analysis of the neo-Tychonian model of planetary motions
   {\em Eur.~J.~Phys.} {\bf 34} 383 (\textit{Preprint} \texttt{arXiv:1301.6045})

\noindent
3. Ilakovac A, Pilaftsis A and Popov L 2013 {Charged Lepton Flavour Violation in
  Supersymmetric Low-Scale Seesaw Models} {\em Phys.~Rev.\/} {\bf D 87} 053014
  (\textit{Preprint} \texttt{arXiv:1212.5939})

\noindent
4. Ilakovac A, Pilaftsis A and Popov L 2013 {Lepton Dipole Moments in
  Supersymmetric Low-Scale Seesaw Models} (\textit{Preprint}
  \texttt{arXiv:1308.3633}) In press\\

\noindent
{\large \textbf{Conference proceedings}}

\noindent 
1. Popov L 2009 Einsten-Podolsky-Rosen (EPR) argument and consequences
  {\em Teorija relativnosti i filozofija. Povodom 100. obljetnice Einsteinove 
  Specijalne teorije relativnosti} ed Petkovi\c\ T (Zagreb: Hrvatsko 
  filozofsko dru\s tvo)\\

\noindent
{\large \textbf{Active participation in conferences}}

\noindent
1. Ilakovac A, Pilaftsis A and Popov L 2012 
  {Charged Lepton Flavour Violation in Supersymmetric Low-Scale Seesaw Models}
  {\em LHC Days in Split} Poster

\noindent